\documentclass[default, iicol]{class/sn-jnl}

\usepackage{graphicx}
\usepackage{multirow}
\usepackage{amsmath,amssymb,amsfonts}
\usepackage{amsthm}
\usepackage{mathrsfs}
\usepackage[title]{appendix}
\usepackage{xcolor}
\usepackage{textcomp}
\usepackage{manyfoot}
\usepackage{booktabs}
\usepackage{algorithm}
\usepackage{algorithmicx}
\usepackage{algpseudocode}
\usepackage{listings}
\usepackage{xspace}
\usepackage{tabularx}
\usepackage{makecell}
\usepackage{caption}
\usepackage{subcaption}
\usepackage{placeins}
\usepackage{bookmark}
\usepackage[shortlabels]{enumitem}
\usepackage[capitalise]{cleveref}
\usepackage{orcidlink}
\usepackage[absolute]{textpos}

\newboolean{wordcount}
\setboolean{wordcount}{false} 

\usepackage{ifthen} 

\Crefname{figure}{Fig.\,}{Figs.\,}
\Crefname{table}{Tab.\,}{Tabs.\,}
\Crefname{equation}{Eq.\,}{Eqs.\,}

\newcommand\thirdwidth{0.325}
\newcommand\halfwidth{0.475}
\newcommand\fullwidth{0.975}

\def \belletwo {Belle\,II\xspace}
\def \superkekb {SuperKEKB\xspace}
\def \pyg {\texttt{PyTorch Geometric}\xspace}

\newcommand{\gev}{\ensuremath{\mathrm{\,Ge\kern -0.1em V}}\xspace}
\newcommand{\mev}{\ensuremath{\mathrm{\,Me\kern -0.1em V}}\xspace}
\def \opengen {\ensuremath{\alpha^{3D}_{\mathrm{gen}}}\xspace}
\def \pgen {\ensuremath{p_{\mathrm{gen}}}\xspace}
\def \ptgen {\ensuremath{p_{\mathrm{t,gen}}}\xspace}
\def \thetagen {\ensuremath{\theta_{\mathrm{gen}}}\xspace}
\def \phigen {\ensuremath{\phi_{\mathrm{gen}}}\xspace}

\def \rhogenthreed {\ensuremath{r^{3D}_{\mathrm{gen}}}\xspace}
\def \ks {\ensuremath{K^0_S}\xspace}
\def \kshortsingle {\ensuremath{K^0_S}\xspace}
\def \kshort {\ensuremath{K^0_S\to\pi^+\pi^-}\xspace}
\def \darkhiggs {\ensuremath{h\to\mu^+\mu^-}\xspace}

\def \kkmc {\ensuremath{\mu^- \mu^+ (\gamma)}\xspace}
\def \dh {\ensuremath{h\to\mu^+\mu^-}\xspace}

\def \legendre {\ensuremath{\text{\textit{Baseline Finder}}}\xspace}

\def \cat {\ensuremath{\text{\textit{CAT Finder}}}\xspace}
\def \databackground {\ensuremath{\text{\textit{high data beam backgrounds}}}\xspace}
\def \hiteff {\ensuremath{\varepsilon_{\text{hit}}}\xspace}
\def \hitpur {\ensuremath{\mathfrak{p}_{\text{hit}}}\xspace}
\def \trackeff {\ensuremath{\varepsilon_{\text{trk}}}\xspace}
\def \trackchareff {\ensuremath{\varepsilon_{\text{trk,ch}}}\xspace}
\def \trackpur {\ensuremath{\mathfrak{p}_{\text{trk}}}\xspace}

\def \fakerate {\ensuremath{\mathfrak{r}_{\text{fake}}}\xspace}
\def \clonerate {\ensuremath{\mathfrak{r}_{\text{clone}}}\xspace}
\def \wrongchargerate {\ensuremath{\mathfrak{r}_{\text{wrong ch.}}}\xspace}

\begin{document}

\title{\centering End-to-End Multi-Track Reconstruction \\using Graph Neural Networks at \belletwo}

\author{L.~Reuter\,\orcidlink{0000-0002-5930-6237}}
\author{G.~De~Pietro\,\orcidlink{0000-0001-8442-107X}}
\author{S.~Stefkova\,\orcidlink{0000-0003-2628-530X}}
\author{T.~Ferber\,\orcidlink{0000-0002-6849-0427}}
\author{V.~Bertacchi\,\orcidlink{0000-0001-9971-1176}}
\author{G.~Casarosa\,\orcidlink{0000-0003-4137-938X}}
\author{L.~Corona\,\orcidlink{0000-0002-2577-9909}}
\author{P.~Ecker\,\orcidlink{0000-0002-6817-6868}}
\author{A.~Glazov\,\orcidlink{0000-0002-8553-7338}}
\author{Y.~Han\,\orcidlink{0000-0001-6775-5932}}
\author{M.~Laurenza\,\orcidlink{0000-0002-7400-6013}}
\author{T.~Lueck\,\orcidlink{0000-0003-3915-2506}}
\author{L.~Massaccesi\,\orcidlink{0000-0003-1762-4699}}
\author{S.~Mondal\,\orcidlink{0000-0002-3054-8400}}
\author{B.~Scavino\,\orcidlink{0000-0003-1771-9161}}
\author{S.~Spataro\,\orcidlink{0000-0001-9601-405X}}
\author{C.~Wessel\,\orcidlink{0000-0003-0959-4784}}
\author{L.~Zani\,\orcidlink{0000-0003-4957-805X}}

\abstract{We present the study of an end-to-end multi-track reconstruction algorithm for the central drift chamber of the \belletwo experiment at the \superkekb collider using Graph Neural Networks for an unknown number of particles.
The algorithm uses detector hits as inputs without pre-filtering to simultaneously predict the number of track candidates in an event and their kinematic properties.
In a second step, we cluster detector hits for each track candidate to pass to a track fitting algorithm.
Using a realistic full detector simulation including beam-induced backgrounds and detector noise taken from actual collision data, we find significant improvements in track finding efficiencies for tracks in a variety of different event topologies compared to the existing baseline algorithm used in \belletwo.
For events involving a hypothetical long-lived massive particle with a mass in the GeV-range, decaying uniformly along its flight direction into two charged particles, the GNN achieves a combined track finding and fitting charge efficiency of 85.4\% per track, with a fake rate of 2.5\%, averaged over the full detector acceptance. 
In comparison, the baseline algorithm achieves 52.2\% efficiency and a fake rate of 4.1\%.
This is the first end-to-end multi-track machine learning algorithm for a drift chamber detector that has been utilized in a realistic particle physics environment.
}


\maketitle
\markboth{}{} 

\section{Introduction}\label{sec_intro}
Experimental particle physics experiments rely on the measurement of charged particles' kinematics, namely their production point location and their momenta at the production point.
These measurements are performed by tracking detectors that provide position measurements of energy depositions (or detector hits) left by charged particles ionizing material along their trajectories, commonly named tracks.
In this paper, we describe a new track finding algorithm \cat~(\textit{CDC AI Track}) using  Graph Neural Networks~(GNNs) in the \belletwo central drift chamber~(CDC) to reconstruct charged tracks in electron-positron collisions.
The \cat simultaneously detects an unknown number of objects, and infers their momenta and their point of origin.
In a second step we associate detector hits to each of those objects that are then used as starting point for a subsequent conventional track fitting algorithm.
We find significant improvements in track finding efficiencies for displaced tracks that originate from a position separated from the interaction point by a macroscopic distance of a few centimetres up to a meter.
Such particles appear in theories beyond the Standard Model from decays of long-lived neutral mediators like dark photons~\cite{Ferber:2022ewf}, or even displaced decay vertices that involve invisible particles in addition to a pair of charged particles~\cite{Duerr:2019dmv, Duerr:2020muu}.
At the same time, the efficiency and fake rate for prompt tracks from the electron-positron interaction point~(IP) is comparable to established methods in the central detector region but significantly better in the forward and backward detector regions.

The \belletwo experiment is located at the high-intensity asymmetric $e^+e^-$ collider SuperKEKB in Tsukuba, Japan. 
SuperKEKB collides primarily 4\,\gev positron and 7\,\gev electron beams at a center-of-mass energy of the $\Upsilon(4S)$ resonance at approximately 10.58\,\gev to investigate rare B-meson decays and new physics phenomena. 
To achieve a higher sensitivity to very rare processes at the \belletwo experiment, SuperKEKB has the goal of increasing the instantaneous luminosity significantly compared to its predecessor, KEKB. 
However, this increase in luminosity also results in a significant increase in beam-induced background (called beam background in the following) that manifests in a high number of background detector hits, and a large number of charged and neutral background particles not originating from the IP~\cite{Natochii:2022vcs}.
On average an $\Upsilon(4S)\to B\bar{B}$ decay in \belletwo will produce about 11 charged particles that typically feature momenta ranging from tens of MeV to a few GeV, while most direct searches for new physics feature lower charged track multiplicities.

This paper is organized as follows: 
Section\,\ref{sec_related} gives an overview of related work on machine learning~(ML) for track finding and end-to-end reconstruction.
Section\,\ref{sec_cdc} describes the \belletwo central drift chamber. 
The event simulation and details of the beam background simulation and data are reported in Section\,\ref{sec_dataset}.
The metrics used to quantify track finding and track fitting performance are defined in Section\,\ref{sec_Metrics}.
The existing and our new GNN-based track reconstruction algorithms are described in Section\,\ref{sec_reconstruction}. 
The main performance studies are discussed in Section\,\ref{sec_results}. 
The results are summarized in Section\,\ref{sec_conclusion}.
\FloatBarrier
\section{Related work}\label{sec_related}
While machine learning is widely used in high energy physics for event selection and analysis, the computationally intensive task of track reconstruction still largely relies on traditional algorithms.
GNNs are recognized as a potential solution for handling irregular detector structures in high energy physics\,\cite{Shlomi:2020gdn}.
GNN architectures in particular have the ability to learn a latent space representation of the detector structure itself~\cite{Wang:2018nkf, Qasim:2019otl}, which is a key ingredient of the work presented in this paper and which has been applied to the \belletwo calorimeter reconstruction \cite{BelleII:2023egc}.
These architectures have been proven highly effective in handling the complex and irregular spatial structures of particle detectors, enabling more accurate and efficient analysis of high energy physics data.
Graph segmentation in our work relies on object condensation\,\cite{Kieseler:2020wcq} which has been used, e.g., in end-to-end calorimeter reconstruction studies for the CMS HGCAL\,\cite{Qasim:2022rww}.

In the context of LHC track reconstruction in high pile-up events, the TrackML challenge~\cite{Amrouche:2019wmx, Amrouche:2021nbs} has led to a significant increase of development activities in the area of ML-based track reconstruction\,\cite{ExaTrkX:2020apx, Caillou:2024dcn, ExaTrkX:2021abe, Lieret:2023aqg}.
The events from the TrackML challenge have a significantly higher number of tracks compared to \belletwo, a higher fraction of sensor hits belonging to signal tracks, and simpler track kinematics from particles produced at the interaction point with high transverse momentum.
GNN-based tracking pipelines aiming for rather detector-unspecific solutions have been developed by the Exa.TrkX project for the HL-LHC\,\cite{ExaTrkX:2021abe}, and by the ETX4VELO project for LHCb\,\cite{Correia:2024ogc}.
Previous work has usually focused on simplified and idealized detector structures and simulations, and on tracks without significant displacement.

Comparisons of GNN track finding to conventional algorithms in realistic HL-LHC scenarios have been shown for the ATLAS inner tracking pixel detector\,\cite{Caillou:2024smf}.

GNNs for gaseous detectors have been studied for edge classification for the PANDA experiment\,\cite{Akram:2022zmj} and BES~III\,\cite{Jia:2024rbx}.
Utilizing deep learning techniques of semantic segmentation inspired by so called U-Nets, hit classification in high background environments has been demonstrated for the drift chamber of the COMET experiment ~\cite{Kaneko:2024ixs}.
However, none of these works feature end-to-end ML-based solutions or conclusive studies of complex event typologies.
Additional challenges arise from the differing input features between drift chambers as used in \belletwo, and the silicon pixels and strips used in HL-LHC. 
HL-LHC detectors, such as silicon pixel detectors \cite{Lieret:2023aqg, Lieret:2023ydc} that provide 3D spatial information differ from drift chambers, which rely on indirect measurements of drift time and wire positions. 
This makes track reconstruction in drift chambers more complex due to the lack of direct spatial information.
In addition to GNNs, a wide range of ML-algorithms are currently being investigating for usage in track reconstruction, including e.g. large language models\,\cite{Huang:2024voo} or transformers\,\cite{Caron:2024cyo}.

Modern implementations of traditional track reconstruction algorithms are often enhanced with ML methods for specific tasks. 
For instance, the \belletwo experiment incorporates gradient boosted decision trees~\cite{Keck:2017gsv} into its track reconstruction pipeline to improve beam background filtering and track-candidate search, and feed-forward neural networks for real-time reconstruction of the $z$ position in the CDC~\cite{Bahr:2024dzg}.

For an up-to-date list of works in particle physics that utilize ML, we refer to the living review \cite{hepmllivingreview}.

\FloatBarrier
\section{The Belle~II central drift chamber tracking detector}\label{sec_cdc}
The \belletwo detector is a charged particle spectrometer surrounded by particle-identification detectors, an electromagnetic calorimeter, and a $K^0_L$ and muon detector, arranged around the beam pipe in a cylindrical structure~\cite{Belle-II:2010dht}.
The positive $z$ direction is pointing in the direction of the electron beam.
The $x$ axis is horizontal and points away from the accelerator center, while the $y$ axis is vertical and points upwards.
The longitudinal direction, the transverse plane with azimuthal angle $\phi$, and the polar angle $\theta$ are defined with respect to the detector’s solenoid axis.\\
The charged particle spectrometer consists of silicon-based pixel and silicon-strip detectors that are not used for this work, and a gas-filled CDC.
The CDC  is 2.3\,m long, and has in total  14,336 sense wires and about 36,000 field wires forming drift cells with a size of about \hbox{$1\times1$\,cm$^{2}$} in the inner wire layers, to about \hbox{$2\times2$\,cm$^{2}$} elsewhere. 
The CDC covers the polar angle range $17^{\circ} < \theta < 150^{\circ}$ and the full azimuthal angle $\phi$ range.
Particles with a polar angle between $17^{\circ} < \theta < 35.4^{\circ}$ leave the CDC early in the forward endcap, particles with $35.40^{\circ} < \theta < 123.04^{\circ}$ traverse the full detector, defined as barrel region, and particles $123.04^{\circ} < \theta < 150^{\circ}$ leave the CDC early in the backward endcap.
The sense wires cover a radius between about 17\,cm to 110\,cm, and are arranged in 56 layers that are grouped in nine superlayers: 
the innermost superlayer consists of 8 layers with 160 sense wires each; the outer eight superlayers consist of 6 layers with 160 to 384 sense wires each. 
All superlayers alternate between wires aligned with the solenoid magnetic field, called axial layers A, and superlayers skewed by an angle between 66.8 and 74.1\,mrad in the positive and -58 to -78.6\,mrad in the negative direction, called stereo layers U and V.
The resulting superlayer arrangement, numbered from inward to outward, is A1, U2, A3, V4, A5, U6, A7, V8, A9.
The electrical field in each drift cell is approximately radial.
The drift distance resolution of the CDC is about 120\,$\mu$m.

A superconducting solenoid generates a magnetic field of about 1.5~T, which is directed along the central axis of the CDC support cylinder. 
Near the IP, there is a system of final focusing quadrupole and compensating solenoid magnets. 
The magnetic field is relatively uniform, with variations of approximately 1\% throughout the entire tracking volume.
More details can be found in Ref.~\cite{BelleIITrackingGroup:2020hpx,Belle-II:2010dht,Kou:2018nap}.

\begin{figure}[ht!]
 \centering
        \begin{subfigure}[b]{\halfwidth\textwidth}
         \centering
         \includegraphics[width=\fullwidth\textwidth]{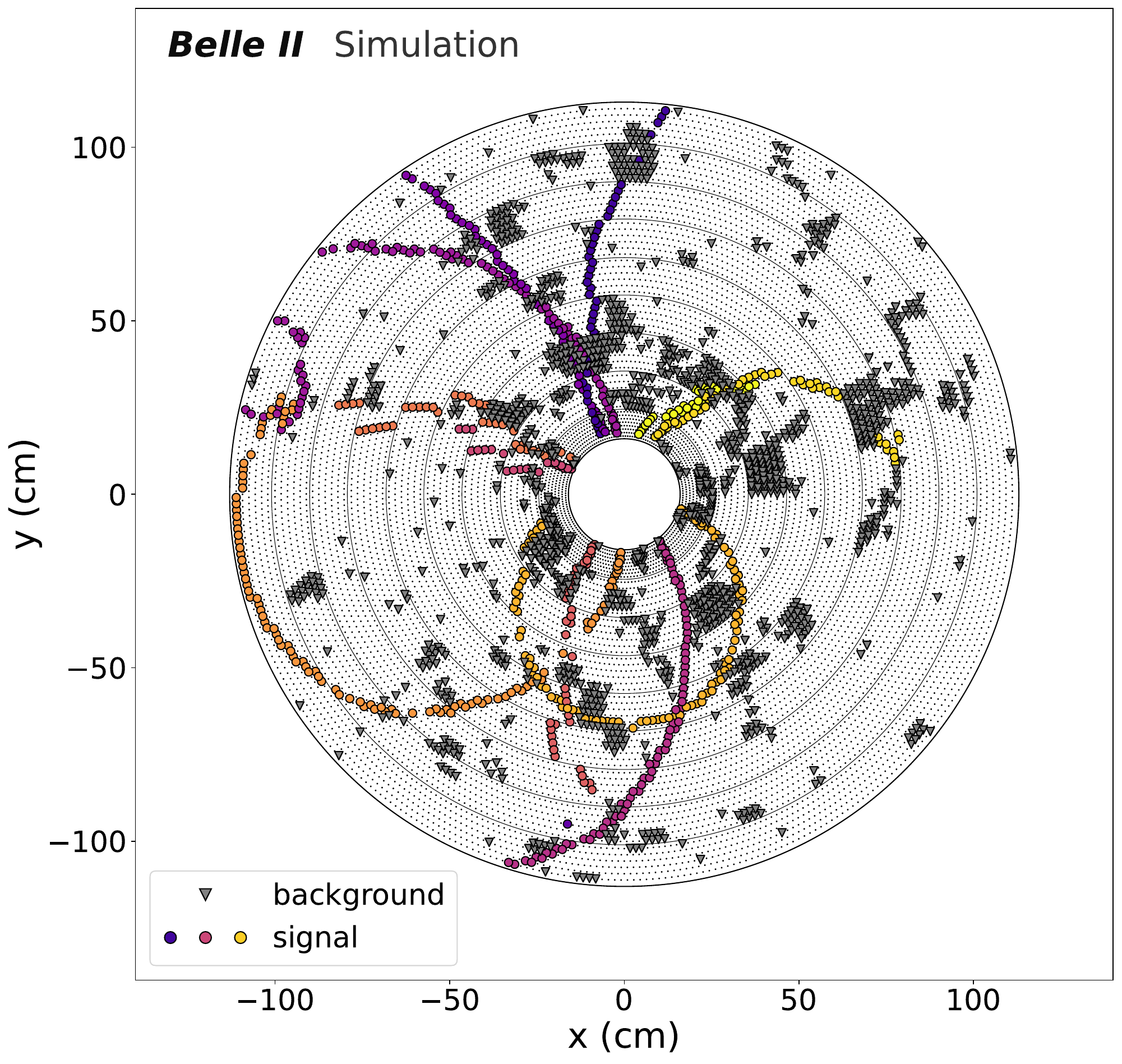}
         \caption{Event display in the $x$-$y$ plane.}
         \label{fig:example_cdc:xy}
     \end{subfigure}\\
     \begin{subfigure}[b]{\halfwidth\textwidth}
         \centering
         \includegraphics[width=\fullwidth\textwidth]{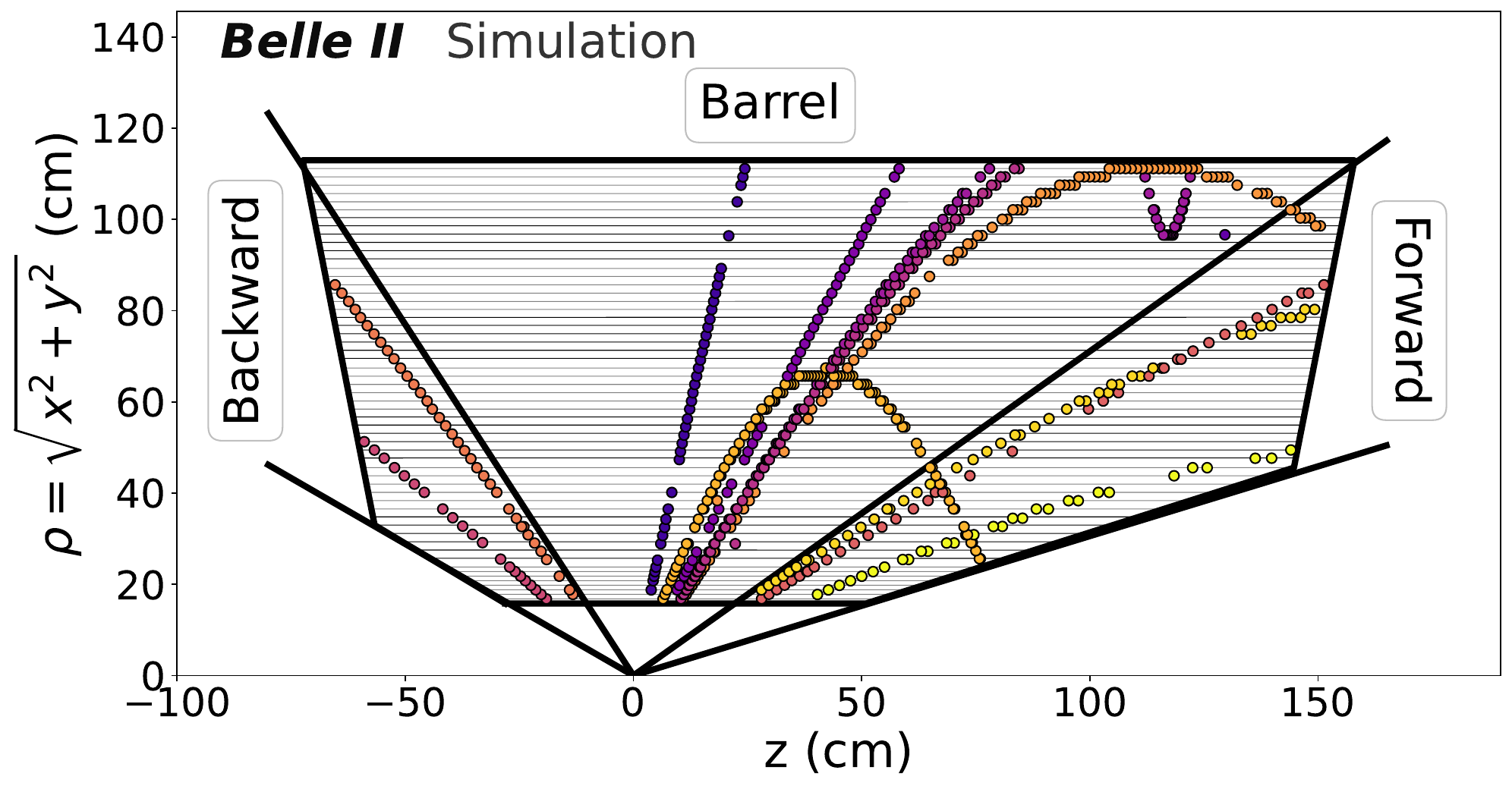}
         \caption{Event display in the $z$-$\rho$-plane.}
         \label{fig:example_cdc:zrho}
     \end{subfigure}
\caption{
Typical event display in the $x$-$y$ plane (\cref{fig:example_cdc:xy}) and the $z$-$\rho$ plane (\cref{fig:example_cdc:zrho}) for a simulated $\Upsilon(4S) \to B^+ B^-$ event with \textit{high data beam backgrounds}. In the $x$-$y$ plane, filled colored circular markers represent signal hits, while filled gray triangular markers represent background hits. These markers correspond to the locations of the sense wires at the $z$ position of the wire center, for wires with recorded ADC signals. In the $z$-$\rho$ plane, where $\rho = \sqrt{x^2 + y^2}$, only the signal hits are shown. The three detector regions, forward endcap, barrel, and backward endcap, are also indicated in the $z$-$\rho$ plane.}\label{fig1}
\label{fig:example_cdc}
\end{figure}

The  CDC provides spatial information in the plane perpendicular to the sense wire axis, which aligns with the \belletwo $x-y$ plane for the axial layers and is tilted by their respective skew angles in the stereo layers. There is no information available for the $z$-coordinate on the wires, because the readout is only done on one side.
This spatial information is encoded as the wire position and the signal's drift time (TDC), recorded when the energy deposition crosses the readout threshold, relative to an unknown global (common for all hits) time offset at the start of track finding. In addition, the digitized signal amplitude~(ADC) and the time over the readout threshold~(TOT) are also provided. 
The digitized signal amplitude is proportional to the energy deposition of a particle.
A typical event display for the CDC is shown \cref{fig:example_cdc}.

\section{Data set}\label{sec_dataset}

We use simulated \belletwo events for the training and evaluation of the reconstruction algorithms.  
The full detector geometry and interactions of final state particles with detector material are simulated using \texttt{GEANT4}~\cite{GEANT4:2002zbu}, which is combined with the simulation of a detector response to create digitized detector hits using the \belletwo Analysis Software Framework~\texttt{basf2}~\cite{basf21, basf22}.

There are three key signatures with qualitatively different behaviour relevant for tracking:
\begin{enumerate}
    \item Low transverse momentum tracks forming circles in the CDC ($p_{t} \lesssim 0.4$\,GeV) versus high momentum tracks moving straight through the CDC ($p_{t} > 0.4$\,GeV);
    \item Particles traversing all CDC layers versus those exiting through the endcaps, creating shorter tracks;
    \item Decay vertices where the decay particles have a small opening angle with potentially overlapping tracks, versus those with a larger opening angle and well isolated tracks.
\end{enumerate}
To effectively train our model on a comprehensive physics phase space, we utilize samples that do not follow conservation laws, but instead are drawn from a parameter space as defined in \cref{tab:samples}. 
To ensure a sufficient number of events with challenging signatures, we enriched our samples by generating events in the direction of the endcaps, including low momentum particles, and those with very small opening angles. 
In addition, we included a transition sample between displaced vertices and prompt tracks.
This sample increased the tracking performance for displaced vertex samples and provided a faster model convergence. 
In general, the model performance improved for continuous transitions between different topologies in comparison to strictly independent topologies, which is the reason we also included a sample where we combine all of the above signatures together in single events in the mix.

All events in categories~1-11 feature muons as primary charged particles, with their charges randomly chosen with equal probability.
The \textit{displaced} samples (categories~4-7) have a starting point $v_x$, $v_y$, $v_z$, that is displaced in 3D in the momentum direction of the respective charged particle.
For the \textit{displaced angled} sample~(category~7) we generate a new momentum direction with a rotation angle $\opengen$ with respect to the vector connecting the origin and the starting point, along a randomly selected perpendicular direction.
The \textit{prompt} and \textit{displaced} samples are generated independently in the forward (fwd), barrel (brl), and backward (bwd) detector regions as well as the full detector acceptance region based on the particle's polar angle at their production point. 
The \textit{vertex} samples~(categories~9-10) consist of two displaced angled particles with opposite charge, generated at the same starting point.
This approach covers both large~(category~9) and small~(category~10) opening angles, effectively enriching the training sample with events with small opening angles, for which the reconstruction is more difficult.

The generated quantities $\ptgen$ (prompt samples), $\pgen$ (displaced and vertex samples), $\thetagen$,  $\phigen$, $\rhogenthreed$, $\opengen$ are drawn randomly from independent uniform distributions for each charged particle. 
The displacement 
\begin{equation}
    \rhogenthreed = \sqrt{v_x^2 + v_y^2 + v_z^2}
\end{equation}
is calculated in 3D and not just in the plane transverse to the $z$-axis.
Each event in category~1-8 contains 1-6 charged particles, this number is drawn from an independent uniform distribution.
Each sample of category~1-8 contains 60,000 events with \hbox{$0.05 < p_t < 6.0$\,GeV}~(categories~1-3 and 8) and \hbox{$0.05 < p < 6.0$\,GeV}~(categories~4-7).
To enrich the events in categories~1-3 and 8 with low momentum particles, we add a random number of prompt low momentum charged particles with $0.05 < p_t < 0.4$\,GeV and all other quantities as above to each event.
For the events in categories~4-7 we enrich with displaced low momentum charged particles with $0.05 < p < 0.4$\,GeV.
The number of low momentum charged particles is drawn from a Poisson distribution with mean $\lambda=1$.
On average the events contain 4.5 particles, resulting in 276,000 particles for each sample in categories~1-8.

The events in categories~9 and 10 contain two, four or six charged particles, where the number is drawn from an independent uniform distribution. 
Example event displays of the different event categories are shown in \cref{app:displays}.
Each sample of categories~9 and 10 contains 120,000 events with 480,000 particles.
This results in 240,000 events in each major category-group of prompt (categories~1-3 and category~8), displaced (categories~4-7), and vertex events (categories~9 and 10). 

Category~11 contains a mix of tracks from category~8 and category~10.
For this we generate a number of charged particles drawn from a Poisson distribution with $\lambda=1.5$, and enrich the sample with low momentum particles (see above) drawn from a Poisson distribution with $\lambda=1.5$.
We finally add vertex events with small opening angles, where the number of decay vertices is drawn from a Poisson distribution with $\lambda=1.5$.
This sample contains 300,000 events.
With this training setup, we observe a maximum of 15 charged particles per event in our training and evaluation categories.

For the evaluation in \cref{sec_results} we use three additional samples that are not used for the GNN training.
We generate radiative muon pairs $e^+e^-\to\mu^+\mu^-(\gamma)$ using the KKMC event generator~\cite{Jadach:1999vf}.
We generate dark Higgs $e^+e^-\to A'h (\to\mu^+\mu^-)$ events using \textsc{MadGraph5@NLO}\,\cite{Alwall:2014hca} with an inelastic dark matter model~\cite{Duerr:2020muu} with on-shell two-body kinematics, with one dark Higgs $h\to\mu^+\mu^-$, and a fully invisible decay of a light dark photon $A'$ with the dark Higgs masses \hbox{$m_h=[0.5, 2.0, 4.0]$~GeV}.
The dark Higgs decay vertex position is drawn randomly from a uniform $\rhogenthreed$-distribution to populate the parameter space of very displaced vertices.
We generate neutral kaon $\kshort$ events containing one \kshortsingle each.
The \kshortsingle decay vertex is calculated from the nominal \kshortsingle lifetime~\cite{ParticleDataGroup:2022pth} with a uniformly generated transverse momentum of $p_t(\kshortsingle)=[0.05-3]$~GeV.
The average transverse decay distance is $v_{\rho}=8.24$\,cm.

\begin{table*}[ht!]
    \centering
    \begin{tabular}{llcccc}
        category & name&  $\thetagen$~[${}^{\circ}$]  & $\rhogenthreed$ [cm] & $\opengen$~[${}^{\circ}$]\\
        \toprule
        1& prompt fwd&  17.0 -- 35.4 & 0 & 0\\
        2& prompt brl & 35.4 -- 123.04 & 0 &  0\\
        3& prompt bwd& 123.04 -- 150.0 & 0 &  0\\
        4& displaced fwd&  17.0 -- 35.4 &  0 -- 100 & 0\\
        5& displaced brl&  35.4 -- 123.04 &  0 -- 100 & 0\\
        6& displaced bwd& 123.04 -- 150.0 &  0 -- 100 & 0\\
        7& displaced angled&  17.0 -- 150.0 & 0 -- 100& 0 -- 30 \\
        8& prompt full  &  17.0 -- 150.0 &  0 -- 100 & 0\\
        9& vertex large &   17.0 -- 150.0 &  0 -- 100 & 0 -- 90\\
        10& vertex small &  17.0 -- 150.0 &  0 -- 100 & 0 -- 25\\
        11& mix 8+10 & - & - & -  \\
    \end{tabular}
    \caption{Event samples used for training and validation. See text for details.}
    \label{tab:samples}
\end{table*}

As part of the simulation, we overlay randomly triggered events from data with a very low probability of containing actual collision events.
The overlay events are taken from the last data-taking period of run~I and correspond to high beam backgrounds (\textit{high data beam backgrounds}) recorded at an instantaneous luminosity of about \hbox{$\mathcal{L}_{\text{beam}}=3.53\times10^{34}$\,cm$^{-2}$s$^{-1}$}.

To speed-up the training by using pre-trained GNN models (see \cref{sec_gnn}) and to evaluate the robustness of the GNN inference against varying beam background conditions (see \cref{sec_results_beambackground}), we also use simulated beam background events approximating the collider conditions in 2021 to our signal particles~\cite{Liptak:2021tog, Natochii:2022vcs}  (\textit{low simulated beam backgrounds}).
The simulated beam backgrounds correspond to an instantaneous luminosity of \hbox{$\mathcal{L}_{\text{beam}}=1.06\times10^{34}$\,cm$^{-2}$s$^{-1}$}.
Additionally, we include \textit{cross-talk} noise simulation to model the behavior of the CDC readout chips, where neighboring channels may be triggered by a large charge deposit in an adjacent channel.
\textit{Beam backgrounds} can either leave track signatures, or single wire hits due to low-energy photon conversions, with electron-positron pairs trapped in the magnetic field.
\textit{Cross-talk} hits on the other hand  typically leave extended cluster-like hit patterns as visible in~\cref{fig:example_cdc}.
The low simulated beam backgrounds contain on average 370 hits not belonging to signal particles, while the high data beam backgrounds contain on average 1230 such hits per event.
We include inactive signal wires and signal wires with reduced hit-efficiency corresponding to the respective average detector conditions for the two beam background scenarios (see \cref{app:wireeff}).

Each CDC hit is matched to up to one simulated particle which is then used as training label in our supervised learning.
If multiple simulated particles deposit energy in the same drift cell, we match to the simulated particle that leaves the first hit in time.

The total number of events in our training and validation sample is 1,120,000 before removal of about 2\% of the events that did not contain any particle with enough matched signal hits in the CDC, or because they contained more than 15 particles.
We use 80\% of our combined sample for training, and 20\% for validation of our models.
The performance evaluation described in later sections of this paper is performed on statistically independent additional samples that were not used for training or validation.
Each of these evaluation samples consist of 90,000 to 150,000 events, resulting in a total of 1~million evaluated events. 
\section{Metrics}\label{sec_Metrics}
For the evaluation of the track finding algorithms we first determine the \textit{hit efficiency} and \textit{hit purity} for each found track.

The \textit{hit efficiency} \hiteff per track is defined as the number of CDC hits \textit{matched} to a simulated particle and included in a found track, divided by the number of all CDC hits \textit{matched} to the same simulated particle in the whole event:
\begin{equation}
    \hiteff = \frac{n_{\text{hits}}(\text{matched and} \in \text{track})}{n_{\text{hits}}(\text{matched})}
\end{equation}
A perfect \textit{hit efficiency} is 1.0, indicating that all \textit{matched} CDC hits are included in this track and no other found track contains hits \textit{matched} to this simulated particle.

The \textit{hit purity} \hitpur per track is defined as the number of CDC hits \textit{matched} to a simulated particle and included in a found track, divided by the number of all CDC hits included in the found track:
\begin{equation}
    \hitpur = \frac{n_{\text{hits}}(\text{matched and} \in \text{track})}{n_{\text{hits}}(\in\text{track)}}
\end{equation}

A perfect \textit{hit purity} is 1.0, indicating that all hits included in the found track are \textit{matched} to the same particle.

We use the \textit{hit efficiency} and \textit{hit purity}, and a minimal number of hits to define if a found track is \textit{related} to a simulated particle:
We require $\hiteff > 0.05$, $\hitpur > 0.66$, and $n_{\text{hits}}(\in\text{track)}\geq 7$.
The hit efficiency criterion is chosen so low to account for tracks that curl inside the tracking volume and leave many hits behind. 
The hit purity criterion ensures proper matching by requiring that at least 66\% of the hits be associated with a single unique particle, even when hits from one track are coming from multiple simulated particles.
If more than one found track can be related to the same simulated particle, we choose the found track with the highest hit purity as \textit{matched} to a simulated particle. 
In this case we call all other tracks \textit{clone tracks}.
If a track does not achieve the purity or efficiency requirements, it is defined as a \textit{fake track}.
If several tracks have the same  hit purity, we choose the track with the highest hit efficiency as the correct match.

We define the \textit{track efficiency} as the ratio of the number of matched tracks~(trks) to the number of all simulated particles that are matched to at least one hit:
\begin{equation}
\label{eq:trackeff}
    \trackeff = \frac{n_{\text{trks}}(\text{matched to part.})}{n_{\text{simulated}}(\geq 1 \text{matched hit})}.
\end{equation}
We define the \textit{track charge efficiency} as the ratio of the number of matched tracks reconstructed with the correct charge, to the number of all simulated particles that are matched to at least one hit:
\begin{equation}
\label{eq:trackchargeeff}
    \trackchareff = \frac{n_{\text{trks}}(\text{matched to part., corr. charge})}{n_{\text{simulated}}(\geq 1 \text{matched hit})}.
\end{equation}

We define the \textit{track finding purity} as the ratio of 
matched tracks to the number of all found tracks:
\begin{equation}
\label{eq:trackpur}
    \trackpur = \frac{n_{\text{trks}}(\text{matched to part.})}{n_{\text{trks}}}.
\end{equation}

We define the \textit{clone rate} as the ratio of number of \textit{clone tracks} to the number of all tracks that are related to a particle,
\begin{equation}
    \clonerate = \frac{n_{\text{clone trks}}}{n_{\text{tracks}}(\text{related to part.})},
\end{equation}
and the \textit{fake rate} as the ratio of \textit{fake tracks} to the number of all found tracks
\begin{equation}
    \fakerate = \frac{n_{\text{fake trks}}}{n_{\text{trks}}}.
\end{equation}

We define the \textit{wrong charge rate} as the ratio of number of matched tracks with the wrong charge, to the number of all tracks that are matched to a particle,
\begin{equation}
    \wrongchargerate = \frac{{n_{\text{trks}}(\text{matched to part., wrong ch.})}}{n_{\text{trks}}(\text{matched to part.})}.
\end{equation}

Since tracks may be found, but then fail the track fitting step, we distinguish between \textit{track finding efficiency}, \textit{track charge finding efficiency}, \textit{track finding clone rate},  \textit{track finding fake rate}, and \textit{wrong finding charge rate}, to indicate track objects after track finding, and \textit{track fitting efficiency}, \textit{track charge fitting efficiency}, \textit{track fitting clone rate}, \textit{track fitting fake rate}, and  \textit{wrong fitting charge rate}, to indicate tracks after track finding and track fitting.
In the following, we will refer to these parameters as performance metrics.
We evaluate the normalized residuals of track momentum components $p_t$ and $p_z$, by comparing the reconstructed parameters with the simulated ones 
\begin{align}\label{eq:etap}
     \eta(p_{t,z}) & = \frac{p_{t,z\,\text{rec}}-p_{t,z\,\text{simulated}}}{p_{t,z,\,\text{simulated}}}
\end{align}
for matched tracks.

These distributions are expected to peak at zero for an unbiased reconstruction.

We then define the resolution $r(p_{t,z})$ for each of these normalized residuals $\eta(p_{t,z})$ as the 68\% coverage 
\begin{align}\label{eq:resolution}
     r(p_{t,z}) = P_{68\%} \left(\left|\eta(p_{t,z}) - P_{50\%}(\eta(p_{t,z}))\right|\right),
\end{align}
where $P_q$ is the $q$--th quantile of the distribution of $p_{t,z}$, and $P_{50\%}$ is the median of $\eta(p_{t,z})$~\cite{BelleIITrackingGroup:2020hpx}. 
For a normal distribution, $r(p_{t,z})$ is identical to the standard deviation.

\section{Track reconstruction algorithms}
\label{sec_reconstruction}
Track reconstruction at \belletwo is  performed in two steps.
In the first step track finding algorithms assign tracking detector hits into subsets of hits belonging to the same charged particle.
The track finding algorithms also provide a first estimate of the track kinematics. 
In a second step a dedicated track fitting algorithm is using these starting values and the set of identified hits to perform a track fit.
Since the \cat performance is optimized using the \textit{track charge finding efficiency} (see \cref{sec_Metrics}), we first describe the track fitting procedure, then the baseline finder and finally the \cat.

\subsection{Track fitting}\label{sec_fitting}
The track finding algorithms need to provide three sets of information to the subsequent track fitting algorithm: 
\begin{itemize}
    \item an initial estimate of the particle kinematics and the particle charge;
    \item a set of ordered identified hits belonging to this track;
    \item an initial estimate of the covariance matrix of the track parameters. 
\end{itemize}

The track fitting  is performed using Kalman Filter algorithms implemented in GENFIT2~\cite{Hoppner:2009af,Rauch:2014wta,Bilka:2019ang, GENFIT_zenodo}, that uses a Deterministic Annealing Filter~(DAF) to downweight hits far away from the fitted trajectory.
It is possible that a found track fails the track fitting if too many hits are rejected by the DAF.
In the nominal \belletwo reconstruction, three mass hypotheses (pions, kaons, and protons) are used during the fitting step, while the finding process is independent of the mass assumption. The mass hypothesis is used when calculating energy loss and material effects, with the kaon and proton hypotheses improving momentum resolution for kaons and protons. Since this work focuses on pions and muons, we have chosen to use only the pion hypothesis.\\

\subsection{Baseline track finding}\label{sec_baseline}
The baseline track-finding algorithm is described in detail in Ref.~\cite{BelleIITrackingGroup:2020hpx} and implemented in \texttt{basf2}.
The baseline track-finding algorithm is based on the Legendre transformation~\cite{Alexopoulos:2008zza} with a main focus on tracks that originate from the close proximity of the IP (called \legendre in the following).
Hits with ADC and TOT values compatible with background hits are removed with minimal loss of signal hits in a pre-processing step.
The \legendre is rather insensitive to missing hits in a track and starts by using axial-layers only to find 2D tracks in the $\rho$-$\phi$-plane.
In a second step, hits in stereo layers are added that allow $z$ determination.
A cellular automaton~\cite{Glazov:1993ur} is used to find locally connected wire hits into track segments, followed by boosted decisions trees to add missing tracks-segments to tracks found by the \legendre.
The \legendre uses a fit of a two-dimensional circle to all identified axial hits to obtain an initial estimate of the track curvature and hence its transverse momentum. 
Using this previous 2D fit result, the initial estimate of the track longitudinal momentum is obtained from a linear fit to the track skew line in the $\rho$-$z$-plane.

\subsection{Graph Neural Network track finding}\label{sec_gnn}
In the following section, we first describe the GNN architecture and the model inference of the \cat.
We then describe the post-processing steps to choose the final track candidates and extract the track parameter information, the hits assigned to each tracks, and the hit ordering.

\paragraph{Graph Neural Network architecture}
Due to the sparsity of the wire hits in the CDC, with an average hit occupancy of up to 15\% per event for the high data beam backgrounds, the variable input size, and the non-uniform arrangement of the drift wires, we utilize a GNN architecture. 
The implementation of this GNN is done in \pyg~\cite{Fey/Lenssen/2019}.
Each node in the graph corresponds to a wire hit. 
The input features for our model are the $x$- and $y$-positions at the $z$ center of the corresponding sense wire, the layer, and superlayer, and layer within a superlayer information, the ADC count, and the TDC count information. 
Using cartesian coordinates as input features and prediction targets was crucial for the model's performance. 
Polar coordinates and angle-based predictions caused issues due to the discontinuity when the angle resets from 360 \(^\circ \) to 0, which the model struggled to learn effectively. 
Additionally, using the track helix radius as a training target proved ineffective, as it becomes nearly infinite for high-momentum, close to straight tracks, and very small for low-momentum tracks, making it impossible to scale the model prediction target to a reasonable range and therefore impossible for the model to learn.
All input features except the position are normalized by dividing by their maximum to a range between 0 and 1.
The ADC count is clipped to a maximum of 600 before normalization to remove the influence of rare anomalous values. 
For signal hits the majority of ADC counts is between 25 and 300.
The position coordinates range between -1.11 and 1.11. One of the targets for our model is the starting position in the same coordinate range. Therefore, we do not scale either the input or the target coordinates.

The objective of our algorithm is to provide the number of tracks, and the three-momentum, the starting point, the charge, and the hits associated with each track.
Given the inherent challenge of not knowing the precise number of tracks in advance, we use an object condensation loss~\cite{Kieseler:2020wcq} for this task.

\begin{figure*}[ht!]
\centering
{\includegraphics[width=\fullwidth\textwidth]{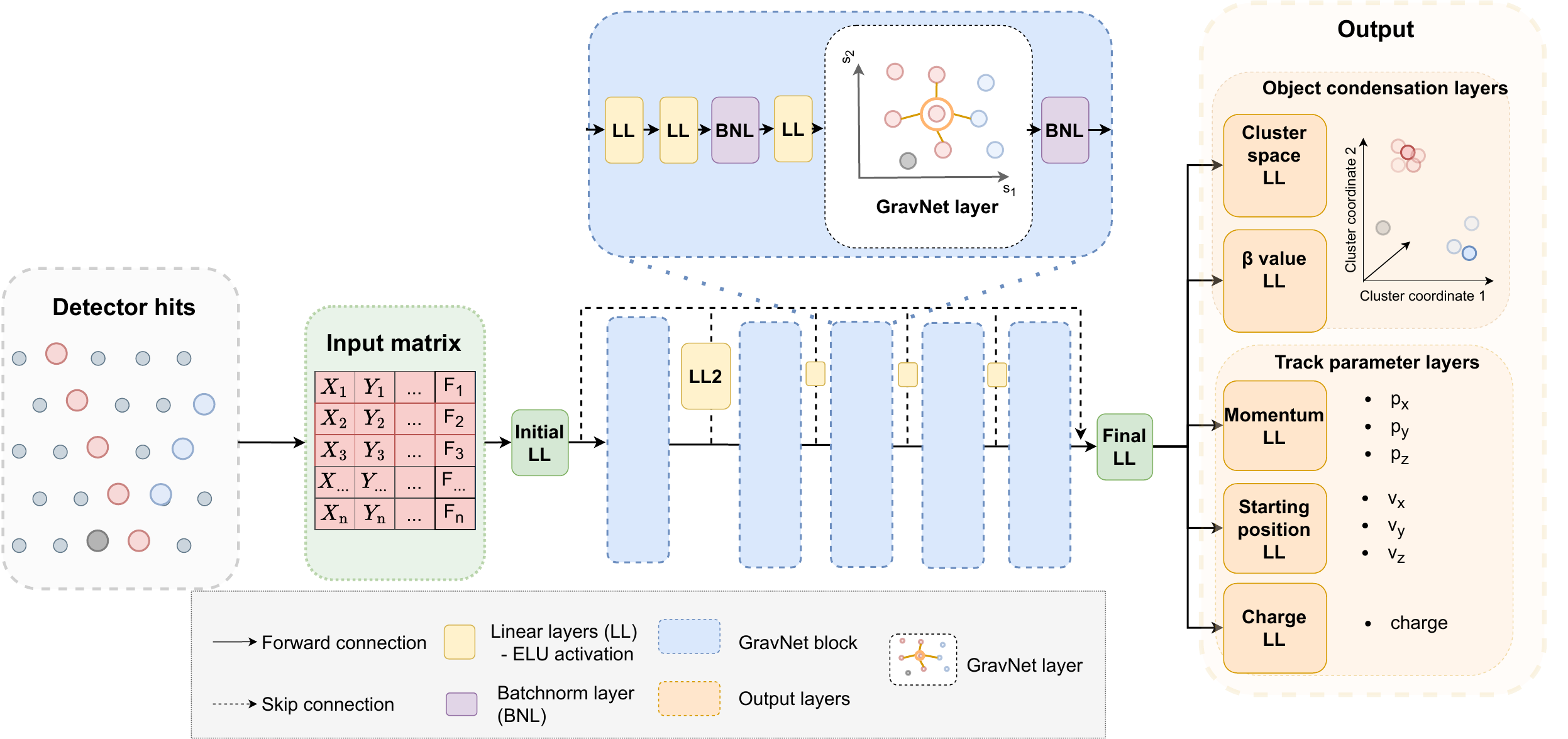}}
\caption{An illustration of the GNN architecture.}\label{fig:architecture}
\end{figure*}

The architecture of our model is illustrated in \cref{fig:architecture} using \hbox{$N=4$} GravNet blocks~\cite{Qasim:2019otl}.
The output of each GravNet block is used for both the subsequent block and directly for the final layer through concatenation.
In a final step, four parallel linear layers are responsible for generating the model's outputs as described in detail below.

Each GravNet block starts with global average pooling \cite{globalpooling}, where the mean value for each feature of the graph is calculated. 
This averaged representation is then added to the original individual node features.  
This pooling technique enables the network to incorporate a collective understanding of the graph, complementing the information from individual nodes. 
This is followed by a sequence of two linear layers (LL), a batch normalization layer \cite{batchnorm} and another linear layer.
Each of the linear layers use an exponential linear unit (ELU) activation functions \cite{elu}. 
The GravNet layer is responsible for building the graph and for message passing between nodes.
In its first step, the GravNet translates the input features into learned representation spaces encoding spatial information, called $S$, and learned features $F_{LR}$.
Undirected edges are then built between each node $j$ and its $k$ nearest neighbours in the representation space $S$ using an $n$-dimensional Euclidean distance $d=\sqrt{\sum_{i=1,..,n}(X_{i,j} - X_{i,k})^2}$, where $X_j$ is the position of the $j$-th node, and $X_k$ is the position of the $k$-th node.
The learned features of the connected nodes are weighted dependent on their Euclidean distance and then aggregated by a summation, resulting in updated features for each node. 
These features are concatenated with the initial node features.
Following the GravNet layer, the feature extraction process continues with batch normalization.
The resulting output is then forwarded to the next GravNet block and directly to the final linear layer, being passed through an extra linear layer (LL2) on this path.

The five output layers in our model serve a dual purpose, addressing both object condensation and parameter prediction tasks. 
Each node of the graph is assigned one object to identify:
if a wire hit is caused by an energy deposition of a signal particle, this node is assigned a unique integer particle ID$>0$. 
In contrast, if the hit is not created by a signal particle, an ID of 0 is assigned.
We only use particles that have at least 7 matched hits in the event as signal particles.
One output of the object condensation layers is a linear layer with a single node and a sigmoid activation function that generates a single output value $\beta$. 
This $\beta$ value is used as a measure of a node being a condensation point. 
Another linear layer with the cluster space dimension of $CS=3$ output nodes provides coordinates within a learned cluster space for each node. 
This is also where the model learns to attract the nodes from the same objects together to the node with the highest $\beta$ value of the object and repels nodes that are from different objects. 
The primary objective is to have a single node with a high $\beta$ value per signal particle.

For the track parameter prediction, we use a linear layer with three output nodes for predicting the three-momentum vector components for each node. 
A second  linear layer with again three output nodes is used to predict the track starting position for each node. 
And finally, a linear layer with one output node and a sigmoid activation function is used for the charge prediction.
We apply a threshold of 0.5 on the charge prediction, where predictions that exceed this threshold correspond to a positive charge of 1 and predictions below to a negative charge of -1.
The target truth information for these predictions is taken from the simulated particle matched to the node.
The loss for the parameter prediction is weighed with the $\beta$ value, since we only require the actual condensation point to predict the particle parameters, condensing all information about the object on this one node. 
The total loss is then given by an unweighted sum of the different loss terms for the attraction, repulsion, $\beta$ value which includes a component to enforce only one condensation point per object and a component to suppress background, and the model parameter predictions.
The details of the loss function are described in~\cite{Kieseler:2020wcq}.

\begin{figure*}[ht]
     \centering
     \begin{subfigure}[t]{0.45\textwidth}
         \centering
         \includegraphics[width=\textwidth]{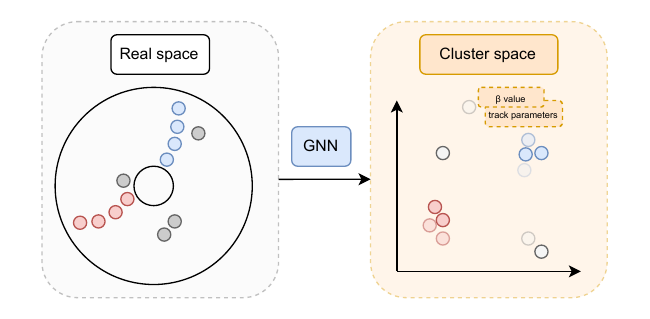}
         \caption{Translation from real space to latent cluster space.}
         \label{fig:gnn_1}
     \end{subfigure}
     \hfill
     \begin{subfigure}[t]{0.45\textwidth}
         \centering
         \includegraphics[width=\textwidth]{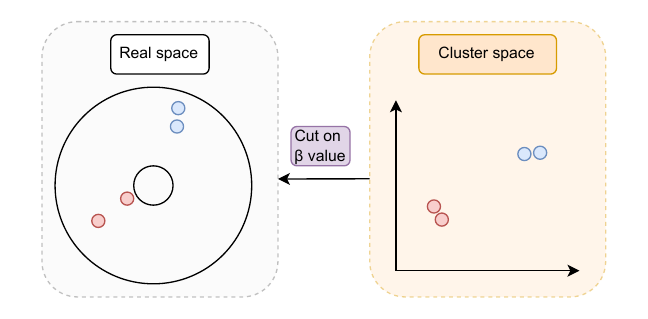}
         \caption{Condensation point candidate selection based on $\beta$ threshold.}
         \label{fig:gnn_2}
     \end{subfigure}\\

     \begin{subfigure}[t]{0.45\textwidth}
         \centering
         \includegraphics[width=\textwidth]{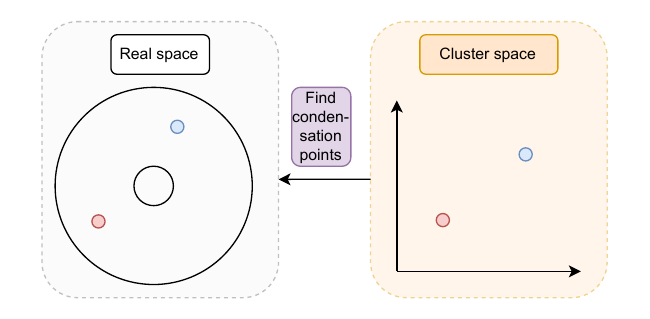}
         \caption{Condensation point selection based on isolation criteria.}
         \label{fig:gnn_3}
     \end{subfigure}
     \hfill
     \begin{subfigure}[t]{0.45\textwidth}
         \centering
         \includegraphics[width=\textwidth]{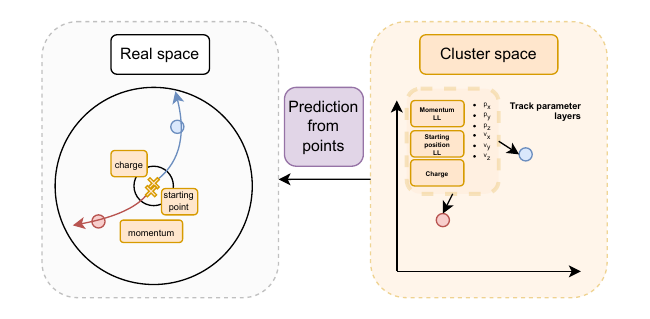}
         \caption{Parameter extraction for condensation points.}
         \label{fig:gnn_4}
     \end{subfigure}\\

      \begin{subfigure}[t]{0.45\textwidth}
         \centering
         \includegraphics[width=\textwidth]{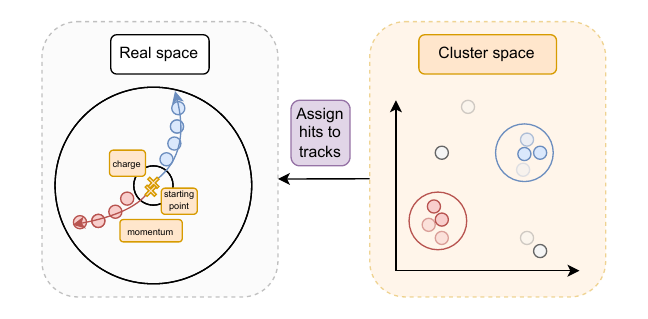}
         \caption{Clusterisation based on condensation points.}
         \label{fig:gnn_5}     
     \end{subfigure}
    \hfill
     \begin{subfigure}[t]{0.45\textwidth}
         \centering
         \includegraphics[width=0.8\textwidth]{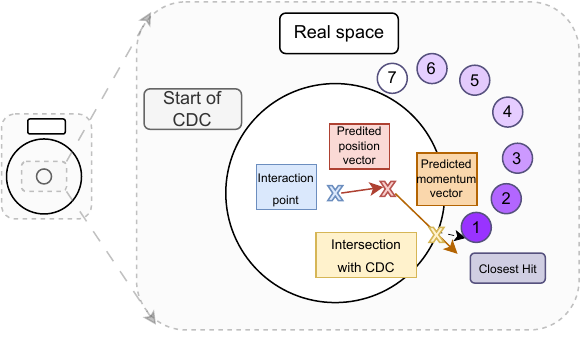}
         \caption{Hit ordering in real space, zoom to start of the CDC.}
         \label{fig:gnn_6}
         
     \end{subfigure}

    \caption{Track finding using object condensation: (\subref{fig:gnn_1}) Latent space, (\subref{fig:gnn_2}) condensation point candidate selection based on $\beta$ threshold, (\subref{fig:gnn_3}) condensation point selection based on isolation, (\subref{fig:gnn_4}) parameter extraction, (\subref{fig:gnn_5}) clustering, and (\subref{fig:gnn_6}) hit ordering in real space.}
    \label{fig:gnn_educational}
\end{figure*}

\begin{figure}[ht!]
     \centering
        \begin{subfigure}[b]{\halfwidth\textwidth}
         \centering
         \includegraphics[width=\fullwidth\textwidth]{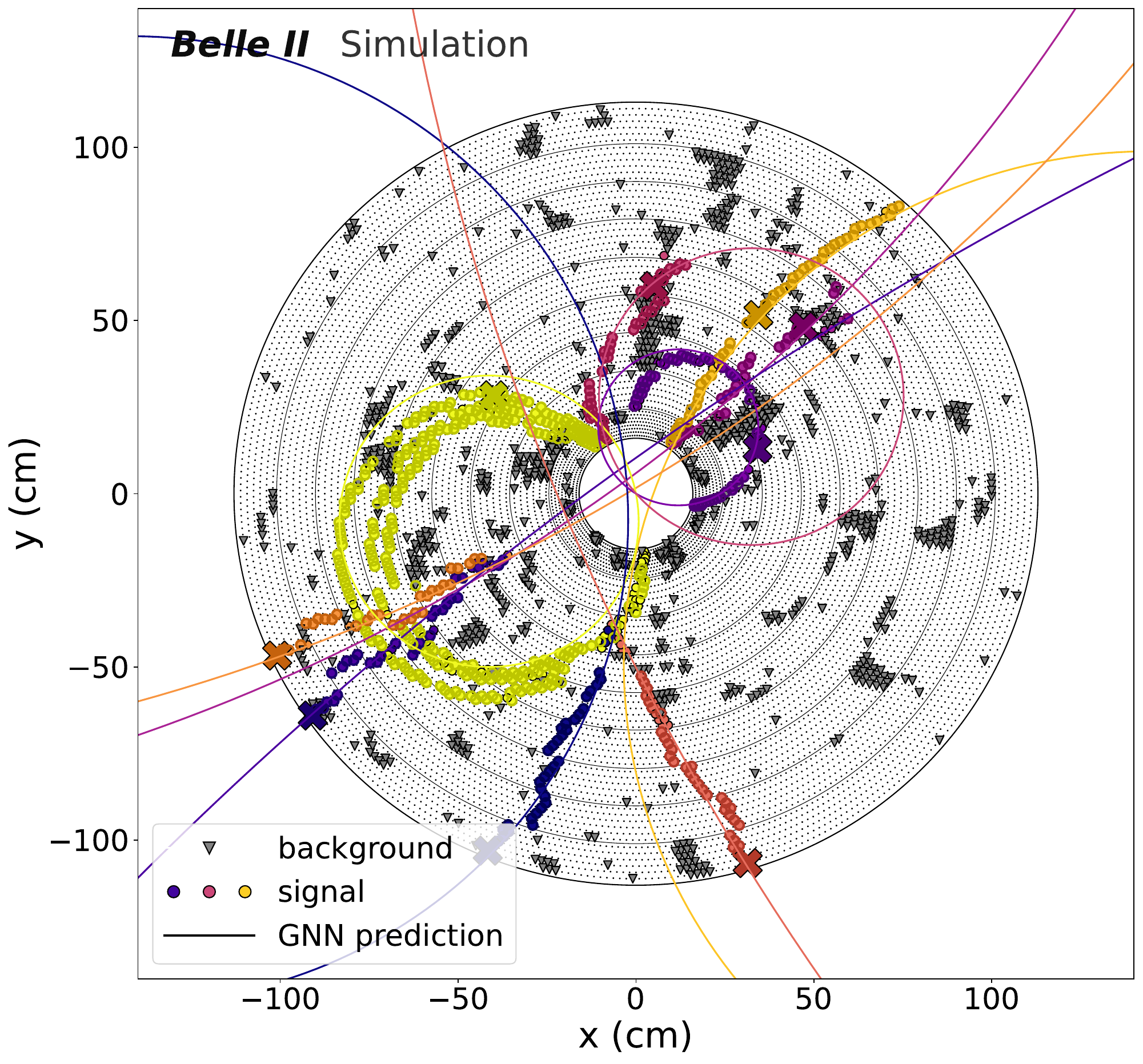}
         \caption{Example event display in the $x-y$-plane. 
         Filled colored circular markers show signal hits, filled gray triangular markers show background hits (see \cref{fig:example_cdc} for details). 
Markers with colored outlines are found by the GNN to belong to the same track object.
The GNN predictions (colored lines) are drawn using the predicted starting point and three momentum for the predicted particle charge, and the corresponding condensation point is marked by a colored cross.}
         \label{fig:latent_space:display}
     \end{subfigure}\\
     \begin{subfigure}[b]{\halfwidth\textwidth}
         \centering
         \includegraphics[height=0.25\textheight]{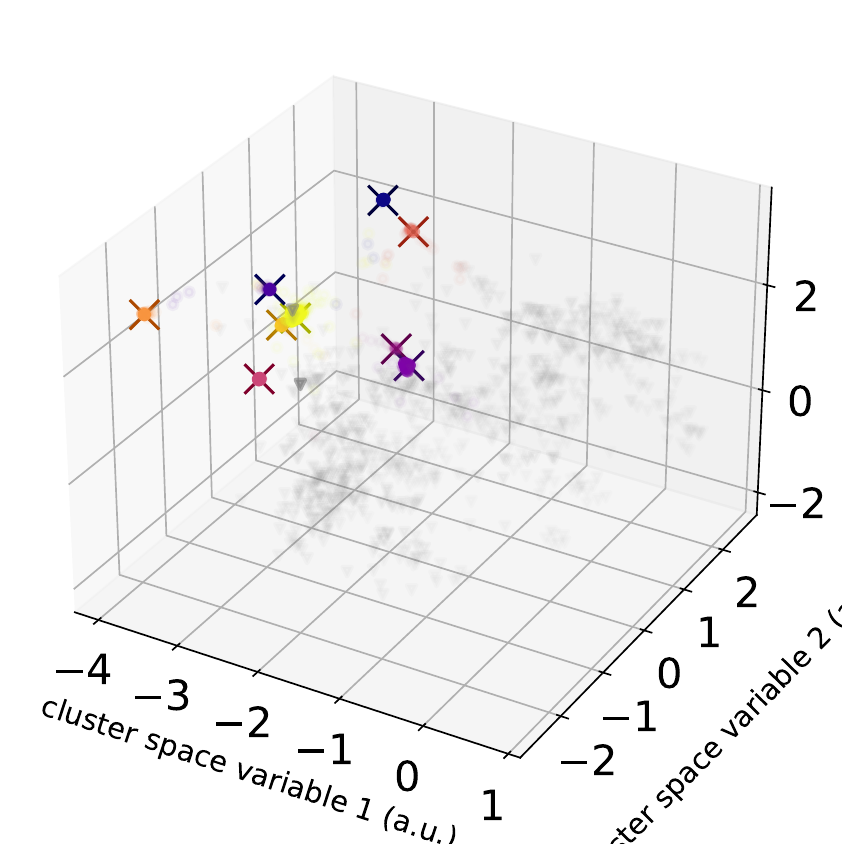}
     \end{subfigure} \\
     \begin{subfigure}[b]{\halfwidth\textwidth}
         \centering
         \includegraphics[height=0.1\textheight]{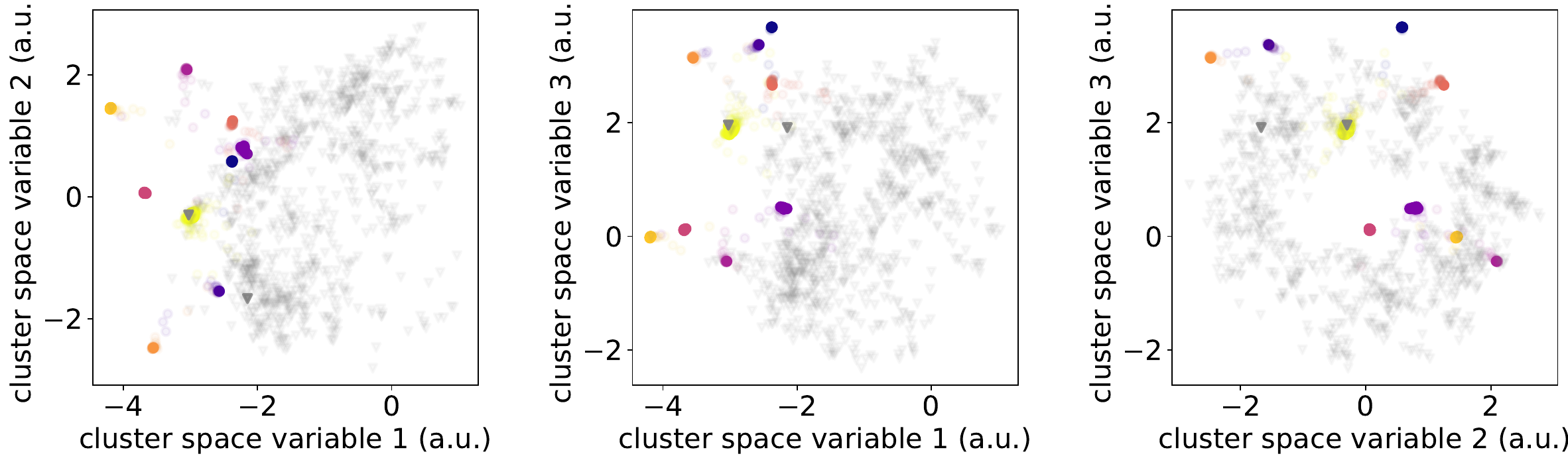}
         \caption{Cluster space representation (top) in 3D with condensation points marked by a cross, and (bottom) 2D projections. The colors are identical to those in \cref{fig:latent_space:display}.}
         \label{fig:latent_space:projections_visible}
     \end{subfigure}\\

\caption{(\cref{fig:latent_space:display}) Event display and (\cref{fig:latent_space:projections_visible}) cluster space representation of one example event from category~11 (\cref{tab:samples}) for \textit{high data beam backgrounds}.
} 
\label{fig:latent_space}
\end{figure}

\paragraph{Graph neural network post-processing}
The procedure to retrieve the track information from the inference step of the trained model is shown schematically in \cref{fig:gnn_educational}.
The following steps are performed in this order:
\begin{enumerate}[label=(\alph*)]
    \item Each event is inferred by the model so each node has a predicted position in the latent cluster space, a $\beta$ value, and parameter predictions for all seven track parameters.
    \item We initiate the track finding process by introducing a threshold $t_{\beta}$,\ to the $\beta$ values resulting in condensation point candidates, as illustrated in \cref{fig:gnn_2}.
    \item In the subsequent step we find isolated condensation points among the condensation point candidates: We compute distances $r$ between the condensation point candidate in the event with the highest $\beta$ value and all other condensation point candidates within the latent cluster space.
    We introduce distance threshold $t_d$, so that condensation point candidates located within a radius $r < t_d$ from the candidate with the highest $\beta$ value are removed. 
    This process iterates until only condensation points remain, each separated by distances exceeding the radius $t_d$.
    The result of these operations is a set of condensation points as shown in \cref{fig:gnn_3}, each corresponding to a found track.
    \item The parameters of each found track are given by the predicted momentum, position, and charge parameters of the respective condensation point (see \cref{fig:gnn_4}).
    \item To assign nodes to each condensation point we cluster nodes in the cluster space: we first calculate the distances between each condensation point and every other node within the cluster space.
    Any node with \hbox{$r < t_h$} is assigned to the found condensation point, shown in \cref{fig:gnn_5}.
    We constrain the $t_h<t_d/2$, ensuring that each hit is used exclusively for at most one track.
    The found condensation point with its set of nodes is equivalent to a found track with assigned hits.
    To ensure that the found track can be fitted, we require at least seven hits assigned to the track, otherwise we reject the track and count it as not found.
    \item Finally, we order the assigned hits. The hits are ordered based on their positions in the $x$-$y$ plane of the detector.
    The process begins by selecting the hit closest to the predicted starting point if the starting point is within the CDC, or closest to the intersection between the predicted particle direction and the inner CDC cylinder surface if the starting point is before the CDC.
    We then calculate the Euclidean distance between the starting point $x$ and $y$ to the $x$ and $y$ positions for all hits.
    Subsequently, the closest hit to the previous one is determined iteratively until all hits are ordered.
    We note that this procedure is not working for low $p_t$ tracks that re-enter the CDC several times and we discuss these cases in \cref{subsec:prompt}.
\end{enumerate}
An example event with corresponding learned latent space representation is shown in \cref{fig:latent_space}.
Unlike the \legendre, the \cat does not provide an initial covariance matrix for the track fitting.
All initial covariance matrix entries for the \cat are set to 0.1.
We observe that the track fitting time depends on the initial parameters of the covariance matrix.
However, the impact on the track finding efficiency is negligible and we leave performance optimization of this to future work.\\

The hyperparameter optimization of the GNN, and the optimization of the node parameters $\beta$, $t_d$ and $t_h$ are described in \cref{subsec:hyper}.

\subsection{Hyperparameter optimization}
\label{subsec:hyper}
The \cat requires to optimize both the model hyperparameters itself as usual, but also the track finding hyperparameters $t_{\beta}$, $t_d$, and $t_h$.
We optimize the model parameters and the track finding parameters in two subsequent steps and focus on tracks coming from the interaction point.
An optimal solution would be to co-optimize the full track finding and fitting for a wide range of physics processes in \belletwo.
We anticipate that this is an area for further exploration and future development.\\

The hyperparameter optimization of the model parameters is done using \texttt{Weights and Biases}~\cite{wandb} with respect to the model loss.
We generated a new train dataset with the same samples as for training (see \cref{tab:samples}), but with a reduced dataset size of only 6\% (about 62,000 events in total) with the low simulated beam backgrounds.
The range of tested hyperparameters and the final values are summarized in \cref{tab:wandbhyperparam}. 
The optimal model has 797,812 trainable parameters.
For the final training with the optimal hyperparameters, the learning rate is reduced by a factor of 2 once the learning stagnates for 30 epochs. 

We use a two-phase training strategy to speed up model training. 
First, the model was trained on the simulated low beam background dataset to learn track signatures, which is the most time-consuming part that takes about 500 epochs to converge. 
Then, we retrain the model on a \databackground dataset, with a factor 10 smaller learning rate focusing on the background suppression component of the loss, allowing the model to fine-tune the performance. 
Despite each epoch on the high data beam background taking over three times longer because of the much higher hit occupancy, our two-phase approach led to an overall significantly faster convergence in just 50 additional epochs.

\begin{table*}[ht!]
\centering
    \caption{Hyperparameters of the GNN Model with their examined range and the result after the optimization, ranked according to their relevance as given by \cite{wandb}. 
    They are trained on the event categories described in \cref{tab:samples}, using an independent dataset with 6\% of the full data sample size. 
    We use 80\% of the data for training and 20\% for validation. 
    The hyperparameters are chosen according to the minimal loss on the validation set.}
\begin{tabular}{lrr}
\toprule
Hyperparameter                                      & Examined range & Result \\ \midrule
Number of GravNet blocks $N$                          & 2-7            & 4     \\
Number of nearest neighbours in GravNet     $k$        & 2-100           & 54     \\
Momentum       & 0.1-0.8            & 0.77      \\
GravNet spacial information space dimension   $S$      & 3-6            & 4      \\
Width of the linear layer LL                           & 32-128         & 126     \\
Dimension of the Object Condensation cluster coordinate space $CS$  & 2-5            & 3      \\
Width of the linear layer LL2                            & 16-64          & 16     \\
\bottomrule
\end{tabular}
\label{tab:wandbhyperparam}
\end{table*}

\begin{figure}[ht!]
     \centering
             \begin{subfigure}[b]{0.4\textwidth}
         \centering
         \includegraphics[width=\fullwidth\textwidth]{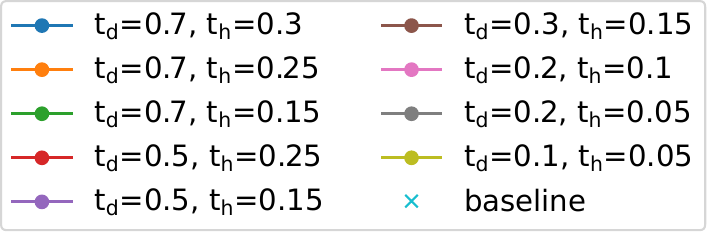}
         \label{fig:fit_p_eff:c}
     \end{subfigure}
     \\
     \begin{subfigure}[b]{\halfwidth\textwidth}
         \centering
         \includegraphics[width=\fullwidth\textwidth]{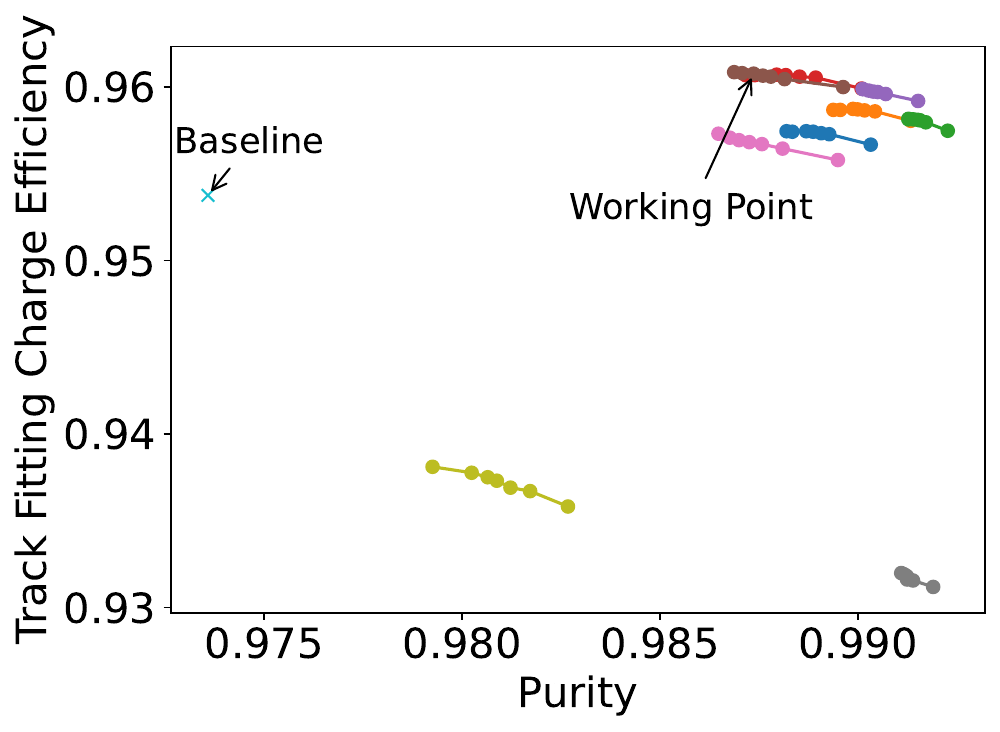}
         \caption{Category 2.}
         \label{fig:fit_p_eff:a}
     \end{subfigure}\\
        \begin{subfigure}[b]{\halfwidth\textwidth}
         \centering
         \includegraphics[width=\fullwidth\textwidth]{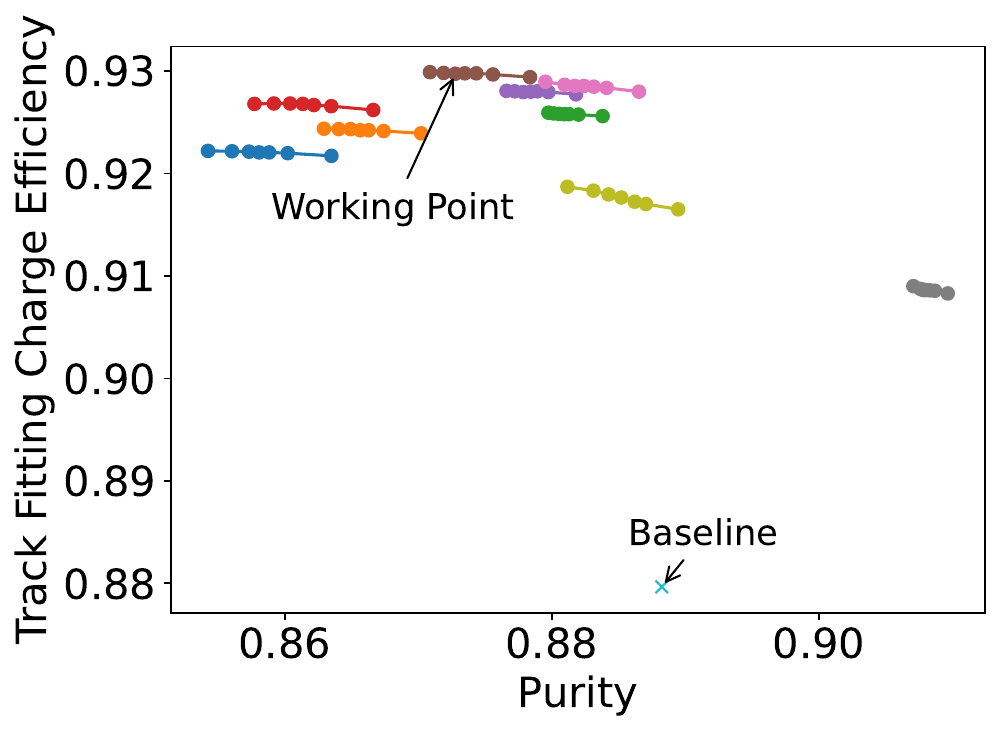}
         \caption{\kshort (barrel).}
         \label{fig:fit_p_eff:b}
     \end{subfigure}\hfill
     \\
        
\caption{Combined track finding and fitting charge efficiency as function of purity for the \cat, and the respective value for the \legendre for (\subref{fig:fit_p_eff:a}) category 2 and (\subref{fig:fit_p_eff:b}) \kshort for \textit{high data beam backgrounds}. See text for details.}
\label{fig:fitting_parameter_roc_curves}
\end{figure}

The three tracking hyperparameters $t_{\beta}$, $t_d$ and $t_h$ are optimized using samples from category~2 and \kshort events with the \kshortsingle momentum pointing in the CDC barrel region (see \cref{sec_dataset} for details). 
The track finding and fitting efficiency \trackeff, and the purity \trackpur are calculated for $t_{\beta}=[0.01,0.1,0.3,0.5,0.7,0.9, 0.95]$, to achieve a ROC curve showing the trade-off between purity and efficiency. 
This is done for several combinations of condensation point distances $t_d=[0.1,0.2,0.3,0.5,0.7]$ and hit radii $t_h=[0.05,0.1,0.15,0.25,0.3]$.
The results of this optimization are shown in \cref{fig:fitting_parameter_roc_curves} for the combined track finding and fitting charge efficiency.
The working point $\mathfrak{w}=(t_{\beta}, t_d, t_h$) is chosen so that first
\begin{equation}
    \trackeff(\mathfrak{w}_i)+ \trackpur(\mathfrak{w}_i)\geq\trackeff(baseline)+\trackpur(baseline),
\end{equation}
and then
\begin{equation}
    \max_{i} \left( \trackeff(\mathfrak{w_i})_{\text{category 2}}+\trackeff(\mathfrak{w_i})_{\kshort} \right),
\end{equation}
where $\trackeff(\mathfrak{w_i})_{x}$ is the track finding and fitting efficiency on the category 2 or the \kshort sample.
This results in the optimal values $t_{\beta}=0.3$, $t_d=0.3$ and $t_h=0.15$.
We note that a different choice of events to optimize these hyperparameters resulted in slightly different optimal values, but these optimal values were always rather close to the ones finally chosen.

\FloatBarrier
\section{Results}\label{sec_results}
In this section we show a comparison of the \cat and the \legendre, with and without the GENFIT2 track fits, for prompt and displaced tracks.
We first discuss the track finding and track fitting efficiency in \cref{sec_results_efficiency} and then compare the momentum resolution of the track finding and fitting algorithms in \cref{subsec:resolution}.
We discuss the performance of the \cat to infer the starting position of a track in \cref{subsec:position}.
We show an analysis of the \cat robustness against different beam background conditions in training and evaluation in \cref{sec_results_beambackground}.
Finally, we list various lessons we learned while training the model in \cref{sec_results_lessonslearned}.

\subsection{Track finding and track fitting efficiency}
\label{sec_results_efficiency}
The track finding efficiency (see \cref{eq:trackeff}) depends on the fraction of matched signal hits, the fraction of beam background hits that are wrongly assigned to the track, and wrongly assigned signal hits from other particles.
Particles that leave a small number of true signal hits are harder to reconstruct by the track finding algorithms.
This affects particles with very small or very large polar angles in the forward or backward regions, or particle that are produced at a large distance from the IP.
Particles that are close to other particles in real space often lead to a correlated efficiency loss, meaning that if one particle is lost, the other is lost, too. 
In our evaluation samples, this makes the light mass \dh samples the most challenging event samples for track finding.

The baseline finder begins tracking in the $x-y$ plane, requiring a sufficient number of hits in the axial layers. 
This approach reduces efficiency for short tracks in the endcap regions. 
In the overall \belletwo tracking chain, efficiency is restored through the use of tracks identified by the silicon vertex detector. 
However, for displaced vertex signatures, the CDC remains the sole tracking detector, with no additional support to recover efficiency.

\subsubsection{Prompt tracks}
\label{subsec:prompt}
We evaluate the track finding efficiency for prompt tracks using the track categories 1-3 (see \cref{tab:samples}).
The events have between 1 and 12 muons per event in the respective detector regions, with transverse momenta $0.05 < p_t < 6 \,\gev$.

\begin{figure}[ht!]
     \centering
        \begin{subfigure}[b]{\halfwidth\textwidth}
         \centering
         \includegraphics[width=\fullwidth\textwidth]{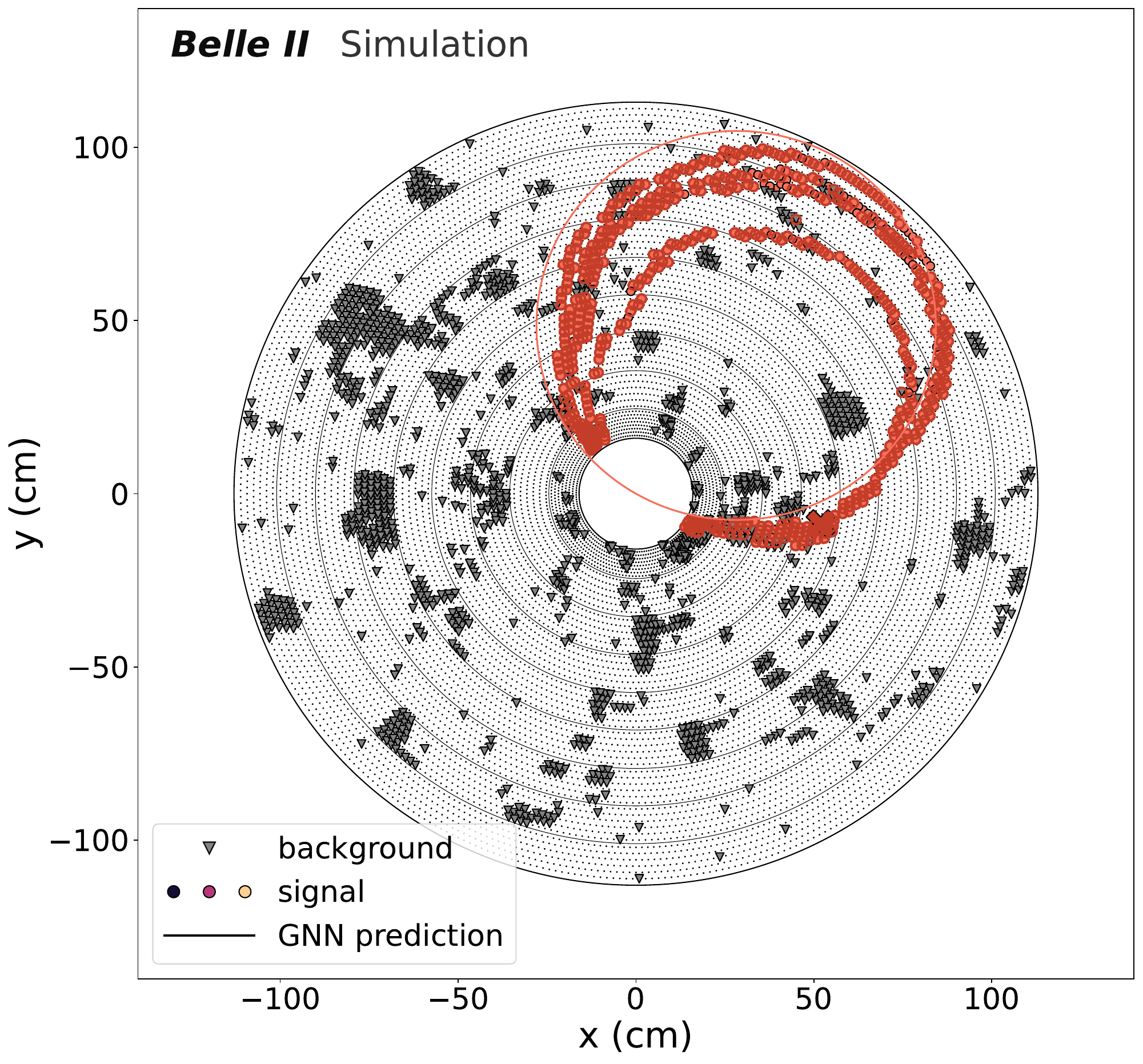}
         \caption{Inner Curler.}
         \label{fig:curler:inner}
     \end{subfigure}\\
     \begin{subfigure}[b]{\halfwidth\textwidth}
         \centering
         \includegraphics[width=\fullwidth\textwidth]{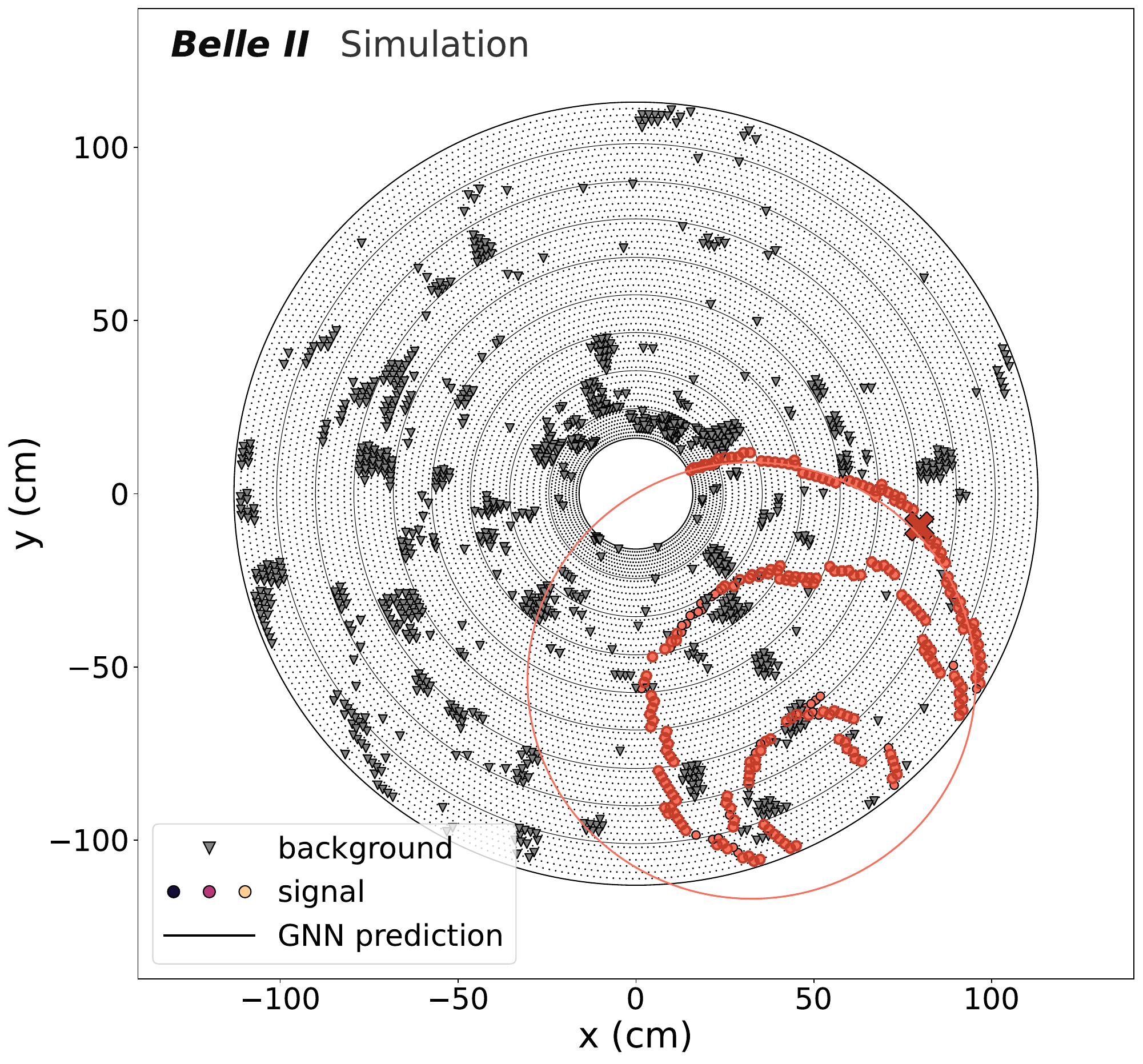}
         \caption{Outer Curler.}
         \label{fig:curler:outer}
     \end{subfigure}
\caption{Example event displays in the $x$-$y$-plane for (\cref{fig:curler:inner}) \textit{inner curler} and (\cref{fig:curler:outer}) \textit{outer curler}  for \textit{high data beam background}.
         Filled colored circular markers show signal hits, filled gray triangular markers show background hits (see \cref{fig:example_cdc} for details). 
Markers with colored outlines are found by the GNN to belong to the same track object.
The GNN predictions (colored lines) are drawn using the predicted starting point and three momentum for the predicted particle charge, and the corresponding condensation point is marked by a colored cross.}
\label{fig:curler}
\end{figure}

Two categories of prompt tracks are particularly difficult in the \belletwo tracking environment:
\begin{itemize}
    \item Particles that re-enter the CDC several times without significant energy loss if their momenta is $p_t\lesssim 0.25\,\gev$ (so called \textit{inner curler}), see \cref{fig:curler:inner};
    \item Minimum-ionizing\footnote{Electrons, protons, and heavier mesons with such low transverse momentum usually loose too much energy in the calorimeter to produce a re-entering track.} charged particles with momenta around $p_t\approx 0.3\,\gev$ with polar angles pointing in the central part of the barrel that leave the CDC, travel through passive material or outer detectors, loose a significant amount of energy by ionization, and re-enter the CDC (so called \textit{outer curler}), see \cref{fig:curler:outer}.
\end{itemize}
We  split the discussion in this subsection into tracks that are non-curling, and tracks that curl.

To remove \textit{inner curler} and \textit{outer curler}, we exclude prompt particles with a distance between two consecutive hits larger than 20\,cm, and all particles with more than 32 signal hits in the CDC's first superlayer A1.
We stress that we only remove such tracks from the evaluation, but not the events that contain these curlers.
Particularly, the samples in category~1-3 contain many events with a large number of low transverse momentum tracks that occupy the inner part of the CDC.

\begin{figure*}[ht!]
     \centering
     \begin{subfigure}[b]{\thirdwidth\textwidth}
         \centering
         \includegraphics[width=\textwidth]{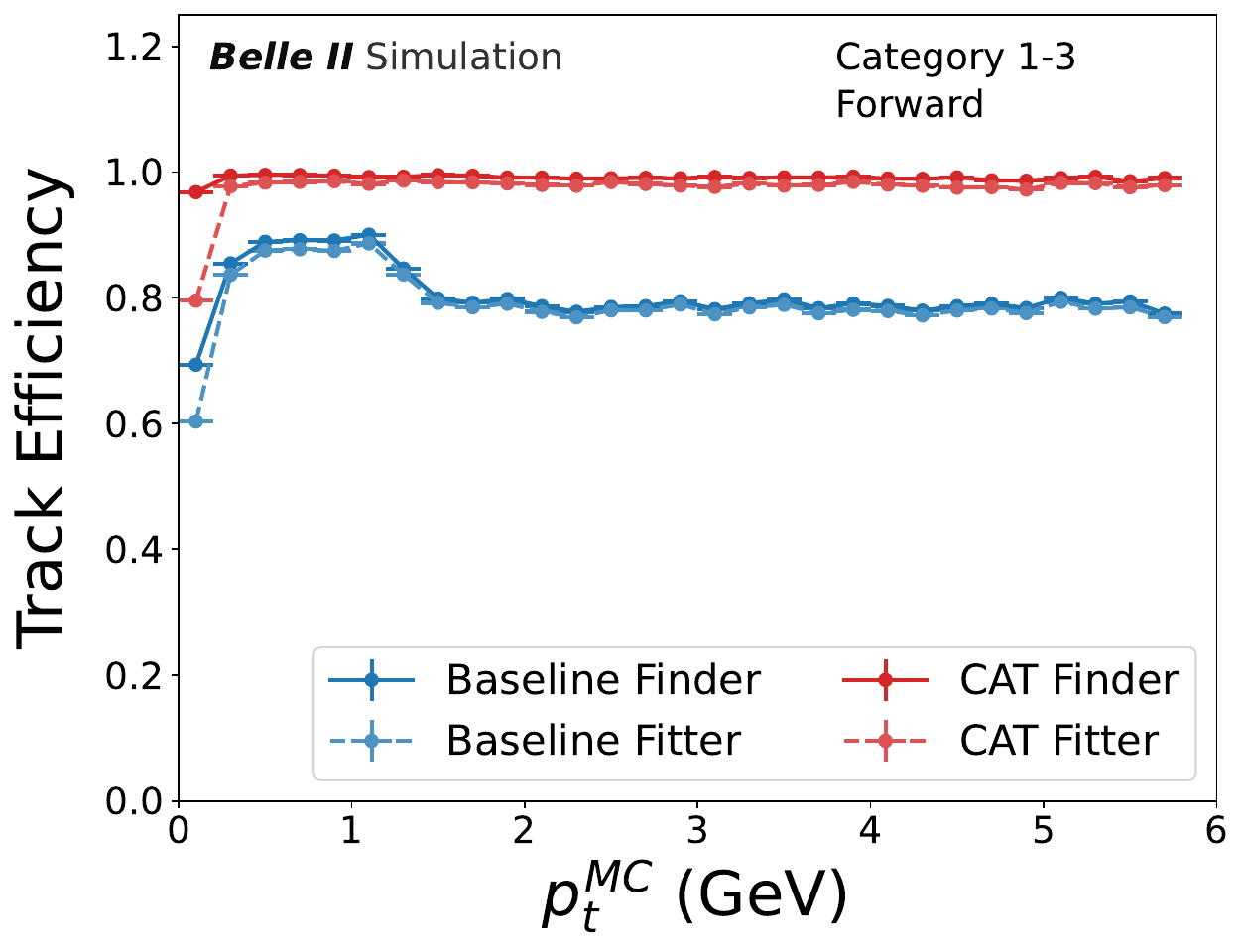}
         \caption{Forward endcap.}
         \label{fig:trainingsample_eff_pt:a}
     \end{subfigure}\hfill
        \begin{subfigure}[b]{\thirdwidth\textwidth}
         \centering
         \includegraphics[width=\textwidth]{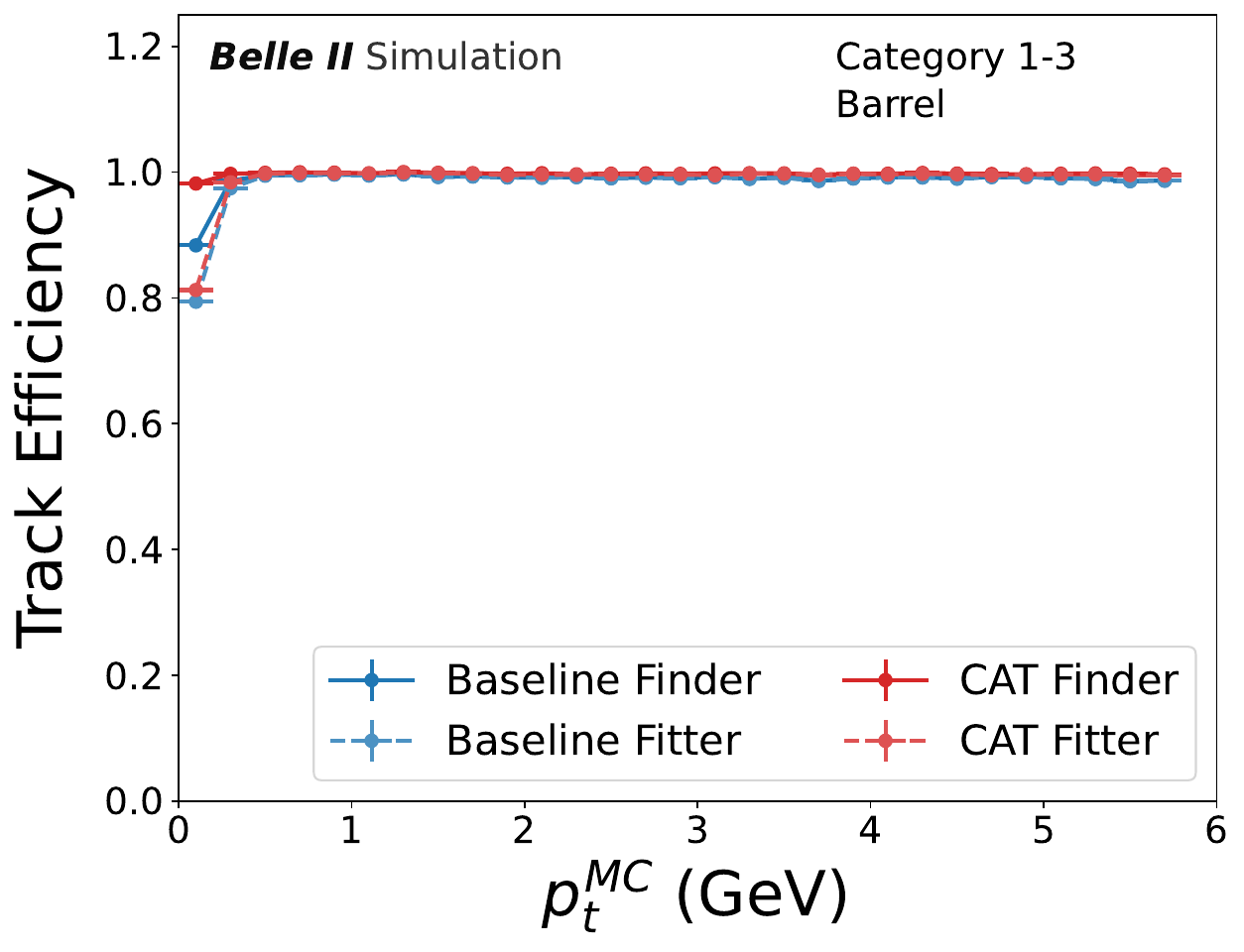}
         \caption{Barrel.}
         \label{fig:trainingsample_eff_pt:b}
     \end{subfigure}\hfill
        \begin{subfigure}[b]{\thirdwidth\textwidth}
         \centering
         \includegraphics[width=\textwidth]{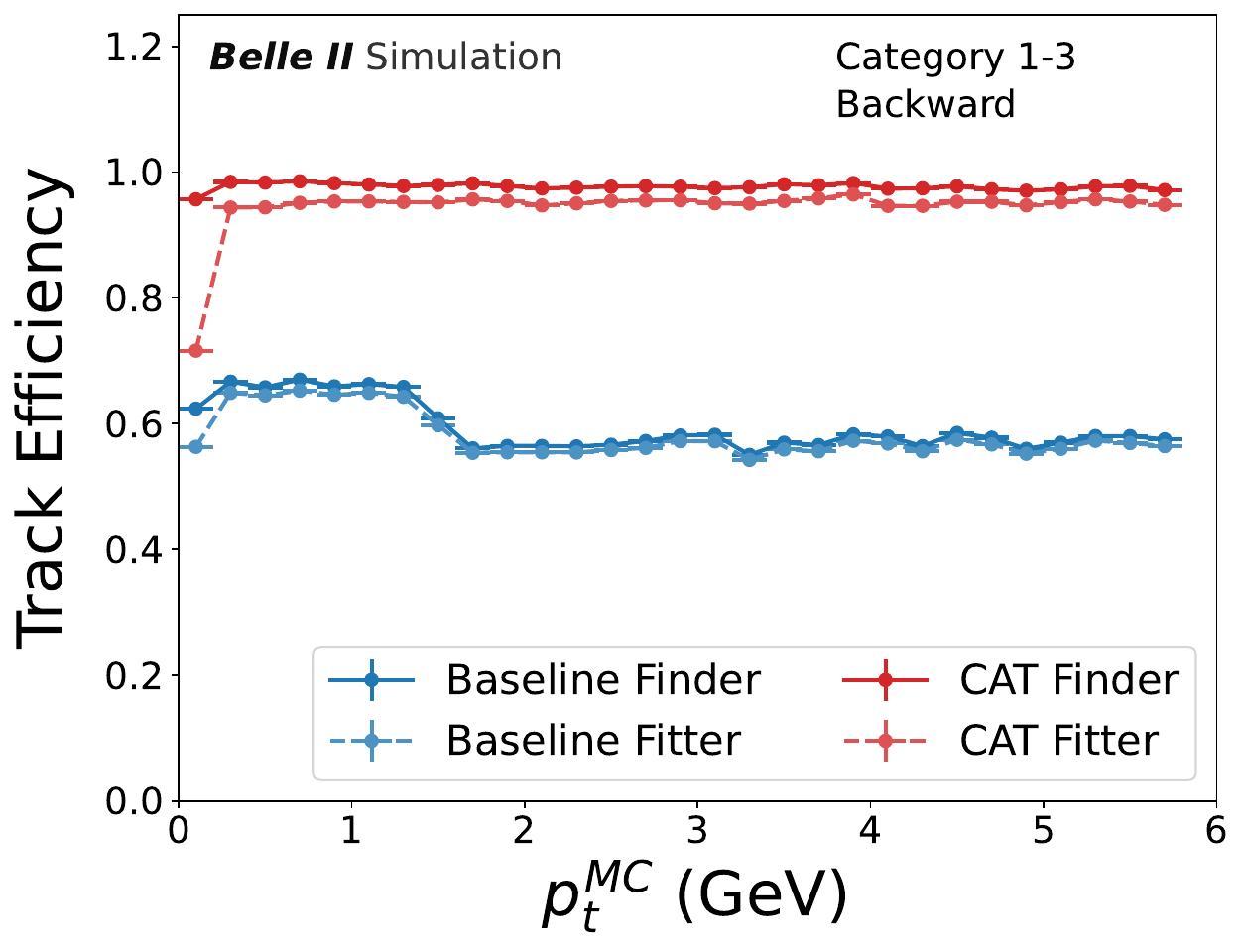}
         \caption{Backward endcap.}
         \label{fig:trainingsample_eff_pt:c}
     \end{subfigure}
\caption{Track finding~(markers connected by solid lines to guide the eye) and combined track finding and fitting efficiency~(markers connected by dashed lines to guide the eye) for the prompt evaluation samples (category 1-3, \databackground, see \cref{tab:samples}) with curler tracks removed, as function of simulated transverse momentum $p_t^{MC}$ for the \legendre~(blue) and the \cat~(red) in the (\subref{fig:trainingsample_eff_pt:a}) forward endcap, (\subref{fig:trainingsample_eff_pt:b}) barrel, and (\subref{fig:trainingsample_eff_pt:c}) backward endcap.
The vertical error bars that show the statistical uncertainty are smaller than the marker size.
The horizontal error bars indicate the bin width.
The uncertainties of the two track finding algorithms are correlated, since they use the same simulated events.}
\label{fig:trainingsample_eff_pt}
\end{figure*}

\paragraph{Non-curling tracks}
The track finding efficiencies, and the combined track finding and track fitting efficiency for the \legendre in comparison with the \cat are shown in \cref{fig:trainingsample_eff_pt}.
A similar comparison but for track charge efficiencies can be found in \cref{app:tceff}.

The performance metrics are summarized in \cref{tab:trainingsample_table_high_data_beambackground}.
The track finding efficiency of the \cat in the barrel is higher than the \legendre, but also features larger fake and clone rates.
In the endcaps, the \cat has a significantly higher efficiency and charge efficiency and again higher fake and clone rates.
Tracks that point towards the endcaps leave fewer hits, and those hits are in the inner CDC region that features a higher occupancy from background.
The \cat is able to efficiently reject beam background hits leading to much higher hit purity for tracks pointing towards the endcaps, and to a better track efficiency.
The combined track finding and fitting efficiency shows the same trend of significantly better performance in the endcaps and comparable performance in the barrel, but the fake rate is significantly smaller than for the \legendre, indicating that some fraction of the additionally found tracks by the \cat cannot be fitted.

\begin{table*}
    \fontsize{6pt}{6pt}\selectfont
    \centering
    \caption{The performance metrics for the prompt evaluation samples (category 1-3, \databackground, see \cref{tab:samples} and \cref{sec_dataset} for details) for non-curling tracks for \cat and \legendre in different detector regions. Uncertainties below $<$0.01\% are not shown in the table.}
    \begin{tabular}{r ccc cc}
         \toprule
        (in \%)& \trackeff & \fakerate & \clonerate &\trackchareff & \wrongchargerate\\
      \midrule
 \midrule
& \multicolumn{5}{c}{forward endcap} \\
\midrule
Baseline Finder & $80.1^{+0.1}_{-0.1}$ & $0.55^{+0.02}_{-0.02}$ & $0.01^{}_{}$ & $78.4^{+0.1}_{-0.1}$ & $2.06^{+0.04}_{-0.04}$ \\
CAT Finder & $98.94^{+0.03}_{-0.03}$ & $1.62^{+0.03}_{-0.03}$ & $0.21^{+0.01}_{-0.01}$ & $98.89^{+0.03}_{-0.03}$ & $0.06^{+0.01}_{-0.01}$ \\
\midrule
Baseline Fitter & $78.1^{+0.1}_{-0.1}$ & $0.49^{+0.02}_{-0.02}$ & $0.01^{}_{}$ & $77.1^{+0.1}_{-0.1}$ & $1.37^{+0.04}_{-0.04}$ \\
CAT Fitter & $95.93^{+0.06}_{-0.05}$ & $0.31^{+0.02}_{-0.02}$ & $0.06^{+0.01}_{-0.01}$ & $94.29^{+0.06}_{-0.06}$ & $1.71^{+0.04}_{-0.04}$ \\
\midrule
 \midrule
& \multicolumn{5}{c}{barrel} \\
\midrule
Baseline Finder & $97.97^{+0.04}_{-0.04}$ & $2.31^{+0.04}_{-0.04}$ & $0.05^{+0.01}_{-0.01}$ & $95.92^{+0.06}_{-0.06}$ & $2.09^{+0.04}_{-0.04}$ \\
CAT Finder & $99.61^{+0.02}_{-0.02}$ & $3.34^{+0.05}_{-0.05}$ & $0.59^{+0.02}_{-0.02}$ & $99.16^{+0.03}_{-0.03}$ & $0.46^{+0.02}_{-0.02}$ \\
\midrule
Baseline Fitter & $96.88^{+0.05}_{-0.05}$ & $1.83^{+0.04}_{-0.04}$ & $0.03^{}_{}$ & $95.5^{+0.06}_{-0.06}$ & $1.42^{+0.03}_{-0.03}$ \\
CAT Fitter & $97.6^{+0.04}_{-0.04}$ & $1.26^{+0.03}_{-0.03}$ & $0.16^{+0.01}_{-0.01}$ & $97.39^{+0.05}_{-0.04}$ & $0.22^{+0.01}_{-0.01}$ \\
\midrule
 \midrule
& \multicolumn{5}{c}{backward endcap} \\
\midrule
Baseline Finder & $60.5^{+0.1}_{-0.1}$ & $1.08^{+0.04}_{-0.04}$ & $0.03^{+0.01}_{-0.01}$ & $58.0^{+0.1}_{-0.1}$ & $4.08^{+0.07}_{-0.07}$ \\
CAT Finder & $97.64^{+0.04}_{-0.04}$ & $1.2^{+0.03}_{-0.03}$ & $0.14^{+0.01}_{-0.01}$ & $97.42^{+0.04}_{-0.04}$ & $0.22^{+0.01}_{-0.01}$ \\
\midrule
Baseline Fitter & $58.8^{+0.1}_{-0.1}$ & $0.92^{+0.03}_{-0.04}$ & $0.02^{}_{-0.01}$ & $56.8^{+0.1}_{-0.1}$ & $3.28^{+0.06}_{-0.06}$ \\
CAT Fitter & $92.43^{+0.07}_{-0.07}$ & $0.69^{+0.02}_{-0.02}$ & $0.03^{}_{-0.01}$ & $87.67^{+0.09}_{-0.09}$ & $5.16^{+0.06}_{-0.06}$ \\
\midrule

        \bottomrule
        \\
    \end{tabular}
    \label{tab:trainingsample_table_high_data_beambackground}
\end{table*}

\begin{figure*}[ht!]
     \centering
     \begin{subfigure}[b]{\thirdwidth\textwidth}
         \centering
         \includegraphics[width=\textwidth]{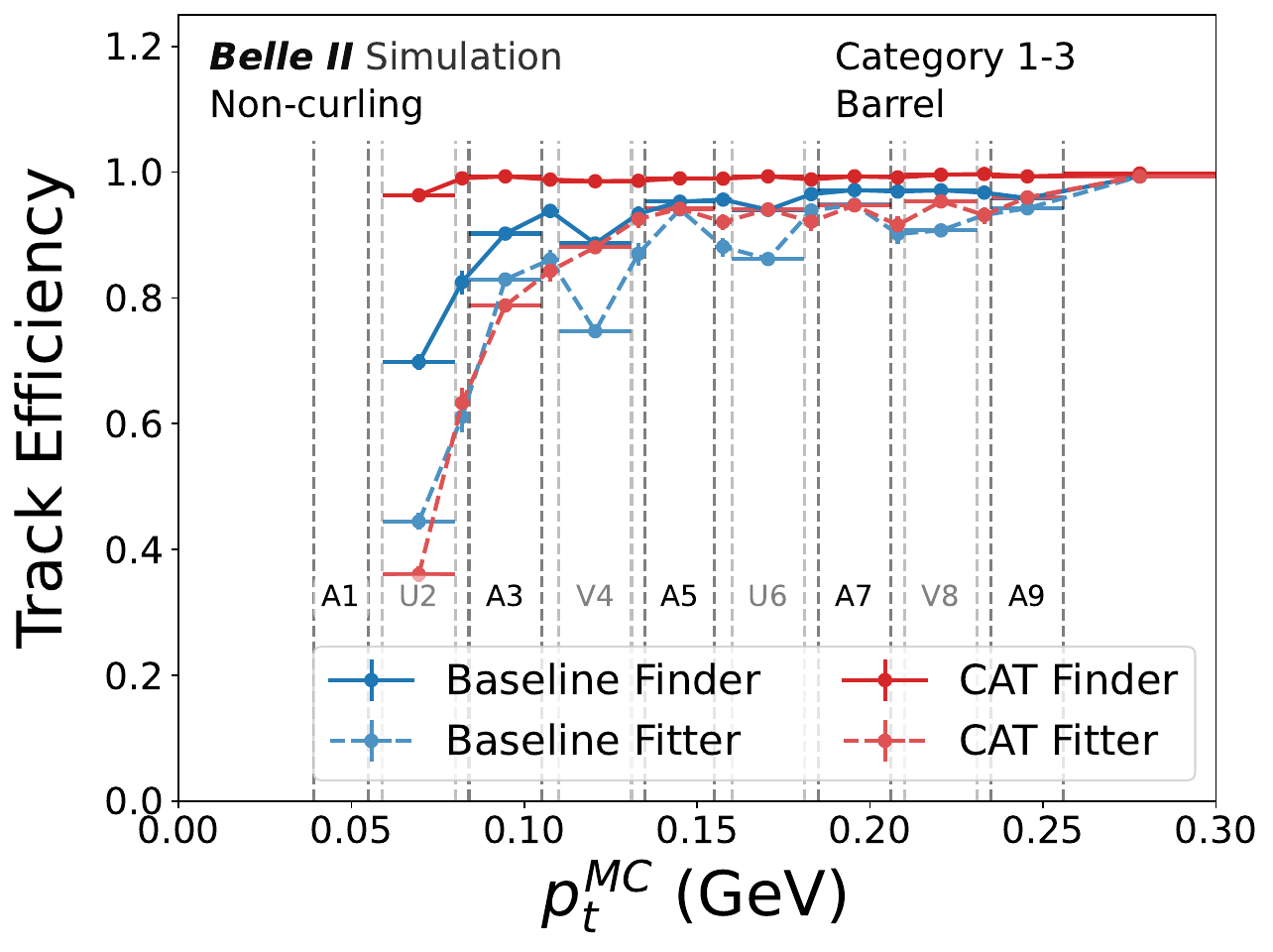}
         \caption{Track finding and fitting efficiency.}
         \label{fig:lowp_noculer_eff:a}
     \end{subfigure}\quad
     \begin{subfigure}[b]{\thirdwidth\textwidth}
         \centering
         \includegraphics[width=\textwidth]{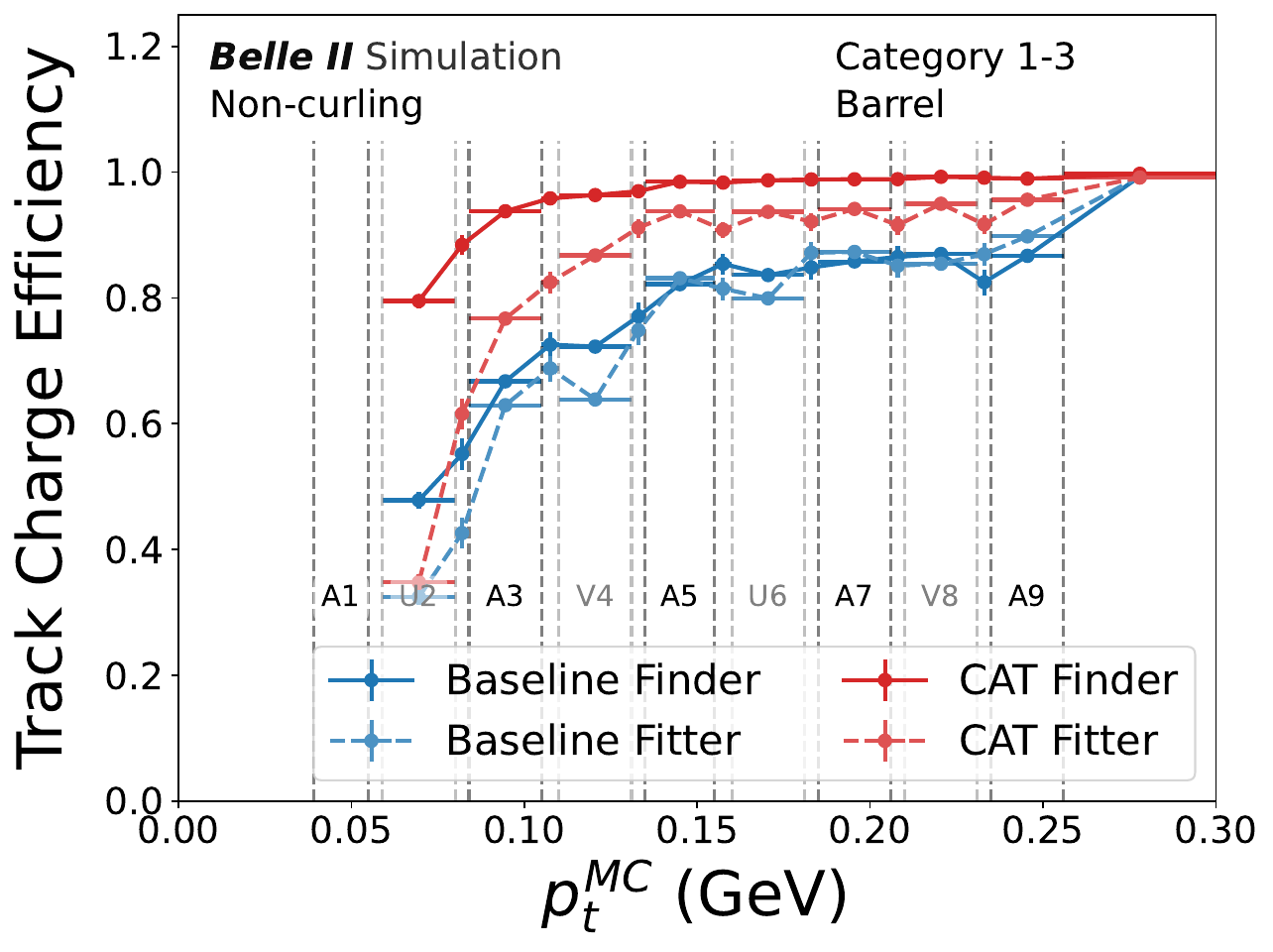}
         \caption{Track finding and fitting charge efficiency.}
         \label{fig:lowp_noculer_eff:b}
     \end{subfigure}\hfill
     \\
     \centering
     
     \begin{subfigure}[b]{\thirdwidth\textwidth}
         \centering
         \includegraphics[width=\textwidth]{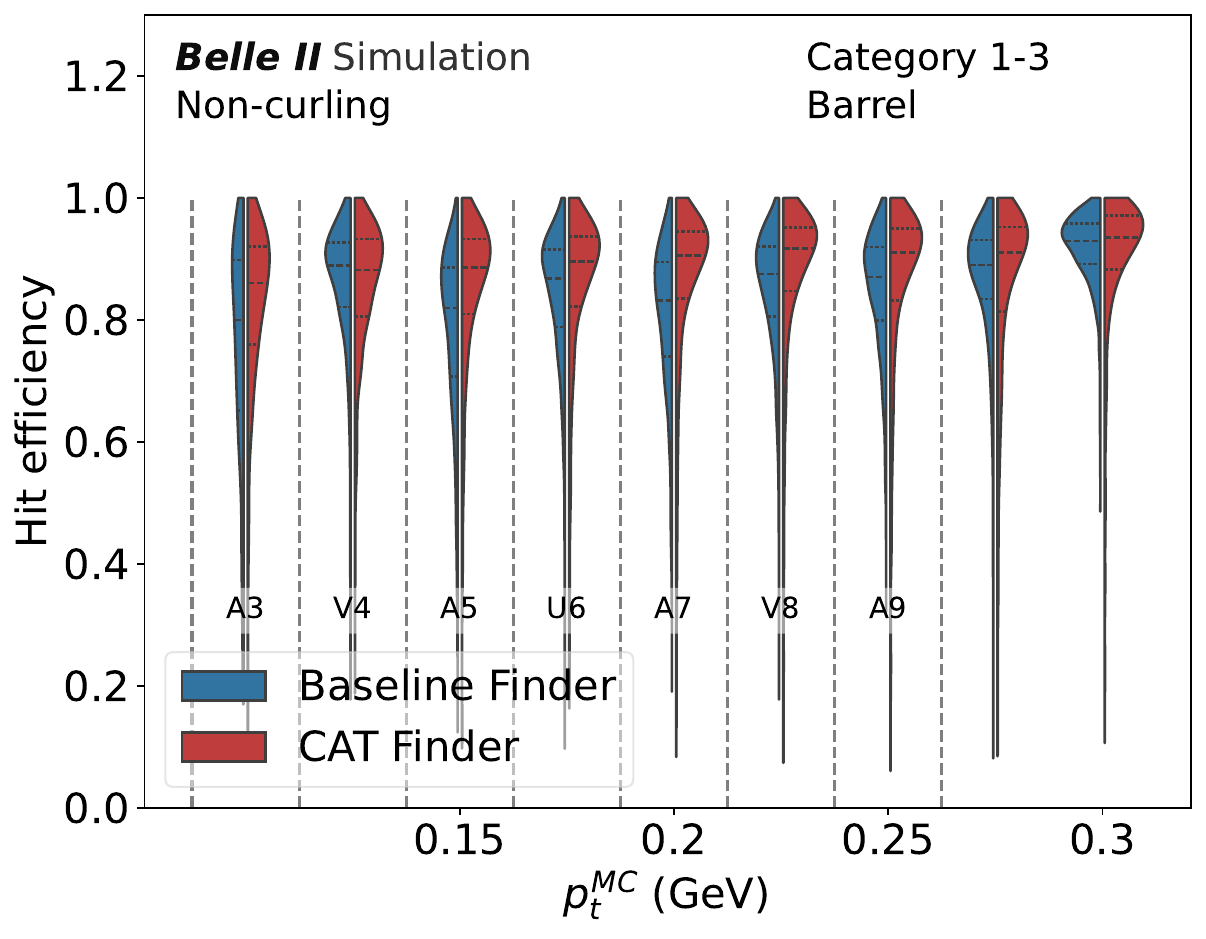}
         \caption{Hit efficiency for intersecting sample.}
         \label{fig:lowp_nocurler:e}
     \end{subfigure}\quad
        \begin{subfigure}[b]{\thirdwidth\textwidth}
         \centering
         \includegraphics[width=\textwidth]{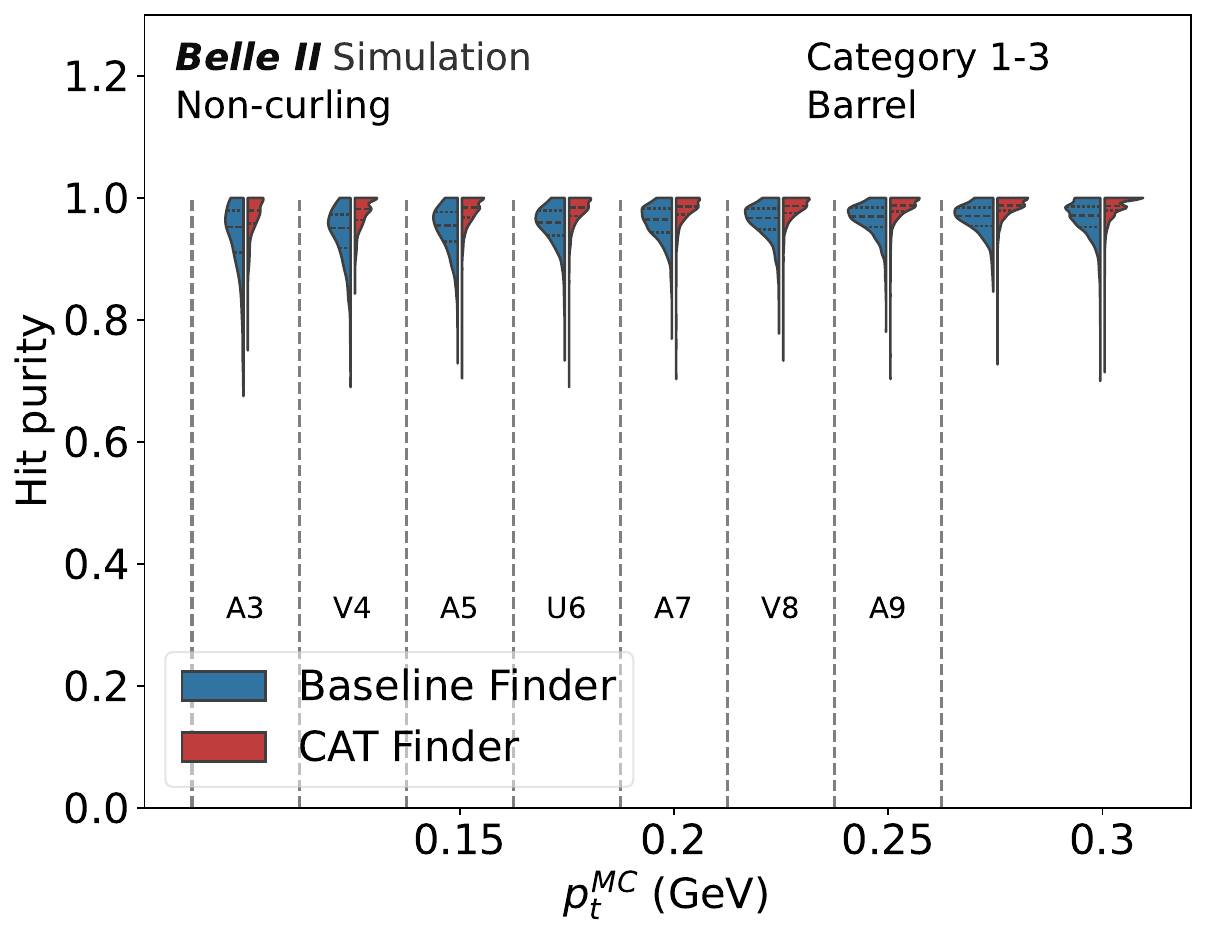}
         \caption{Hit purity for intersecting samples.}
         \label{fig:lowp_nocurler:f}
     \end{subfigure}\hfill
     \\
     \centering
       \begin{subfigure}[b]{\thirdwidth\textwidth}
         \centering
          \includegraphics[width=\textwidth]{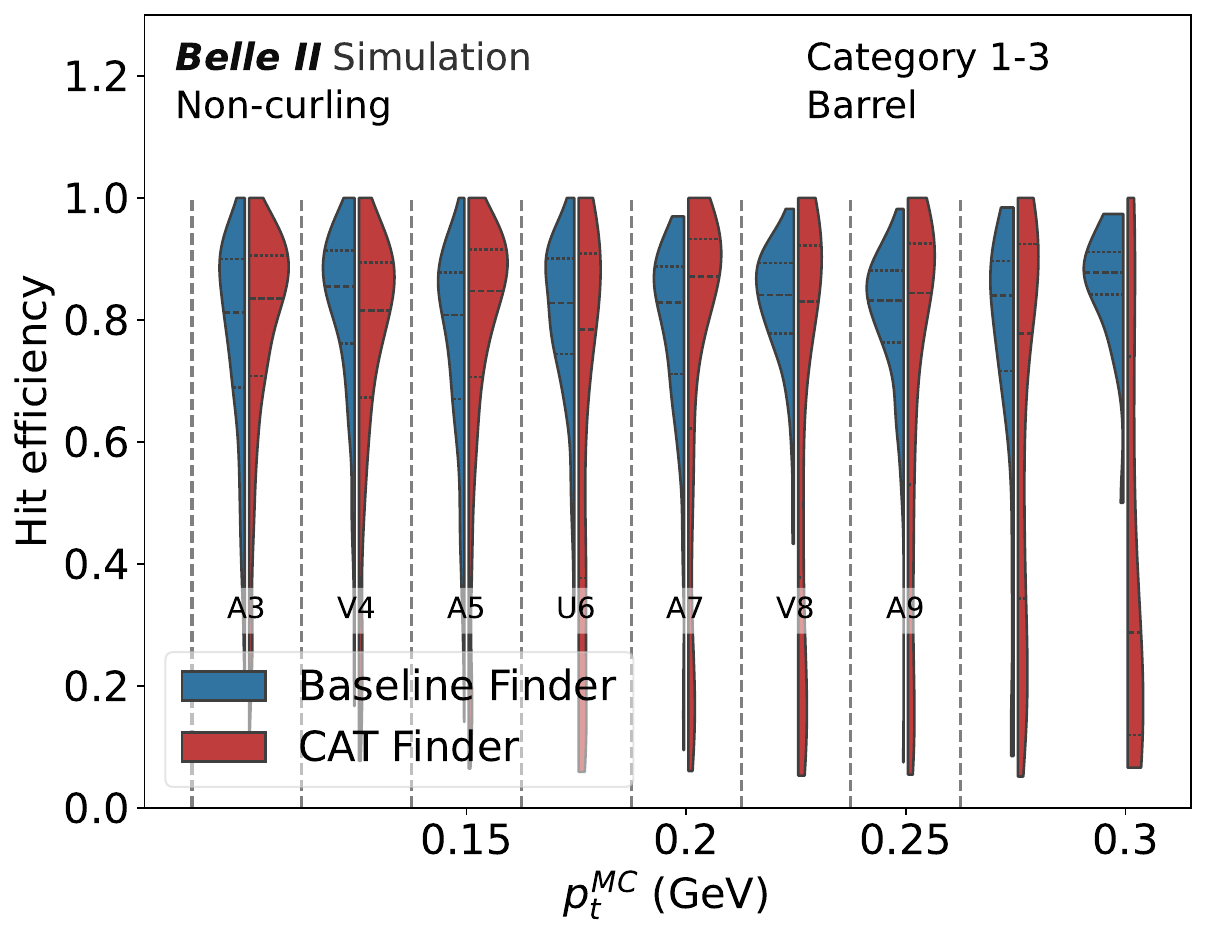}
          \caption{Hit efficiency for additional samples.}
         \label{fig:lowp_nocurler:g}
     \end{subfigure}\quad 
     \begin{subfigure}[b]{\thirdwidth\textwidth}
         \centering
         \includegraphics[width=\textwidth]{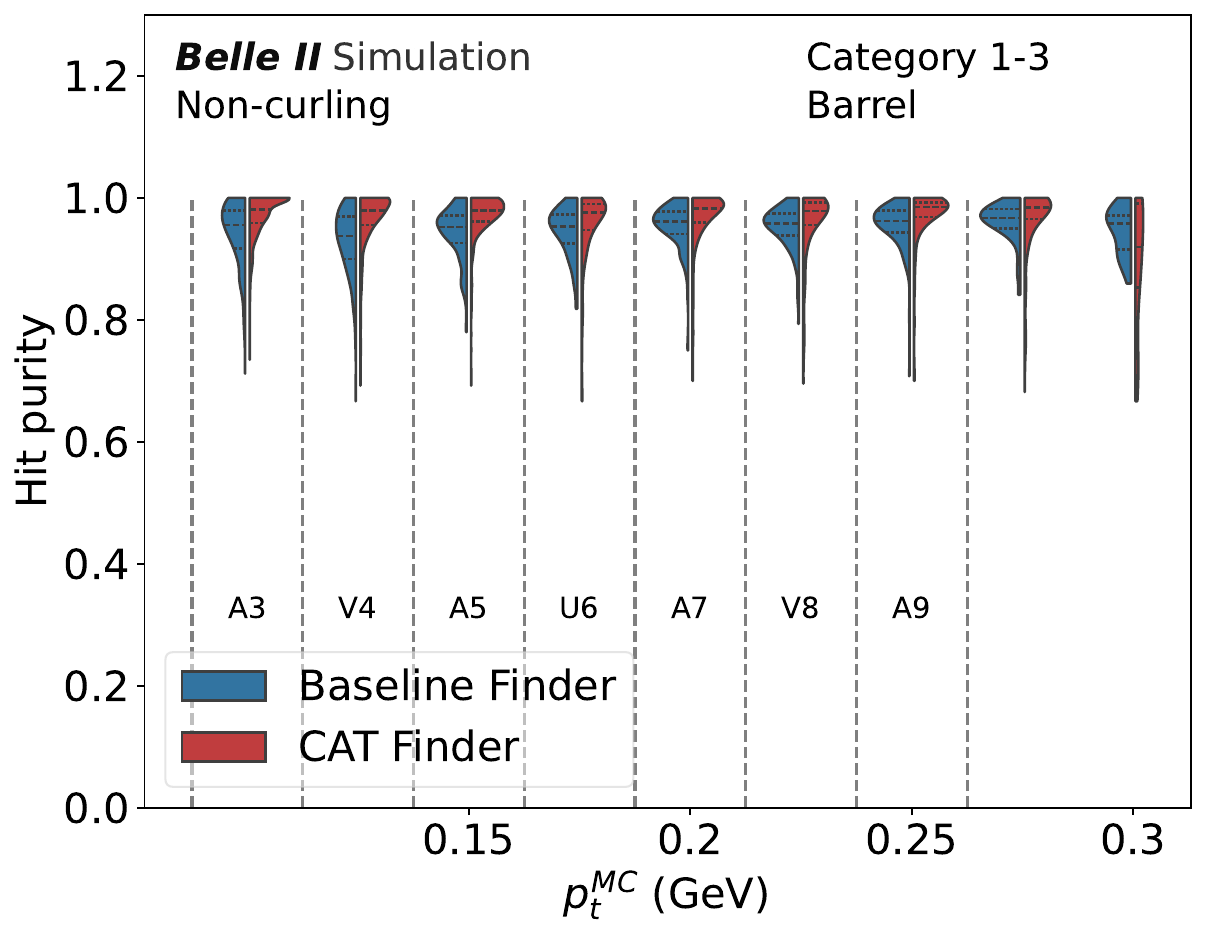}
         
          \caption{Hit purity for additional samples.\\}
         \label{fig:lowp_nocurler:h}
     \end{subfigure}\hfill
\caption{The top row shows the low momentum track finding~(empty markers, connected by lines to guide the eye) and combined track finding and fitting charge efficiency~(filled markers) (\subref{fig:lowp_noculer_eff:a}) and the track finding and combined track finding and fitting charge efficiency (\subref{fig:lowp_noculer_eff:b}) for the prompt evaluation samples (category 1-3, \databackground, see \cref{tab:samples}) for non-curling tracks. 
The middle row shows hit efficiency and hit purity for tracks found by both \cat and \legendre (intersecting sample) (\subref{fig:lowp_nocurler:e} and \subref{fig:lowp_nocurler:f}) and the bottom row for the additional found tracks (\subref{fig:lowp_nocurler:g} and \subref{fig:lowp_nocurler:h}).
The dashed horizontal dark (light) gray lines show the axial (stereo) superlayer boundaries how far the prompt track reaches with the given transverse momentum. 
}
\label{fig:lowp_nocurler}
\end{figure*}

Despite the removal of curling tracks in \cref{fig:trainingsample_eff_pt} and \cref{tab:trainingsample_table_low_beambackground}, one observes a significant drop in efficiency for both \cat and \legendre for $p_t\lesssim 0.3\,\gev$.
A detailed inspection of this region is shown in \cref{fig:lowp_nocurler}.
For the tracks with removed curlers, the \cat shows a very high track finding efficiency down to transverse momenta of about 50\,\mev that outperforms the \legendre\,(see \cref{fig:lowp_noculer_eff:a}).
The hit efficiency and hit purity for the tracks that are found both by \cat and \legendre (intersecting sample) are much higher for the \cat (see \cref{fig:lowp_nocurler:e}, \cref{fig:lowp_nocurler:f}).
For tracks that are only found by the \cat but not by the \legendre or vice versa (additional sample), the \cat has a higher hit efficiency and purity.
We note that the number of additional tracks found only by the \legendre is very small.
\cat has a higher track finding efficiency than the \legendre, but the combined track finding and fitting efficiency is comparable. 
This indicates that the \cat finds more complicated track topologies that cannot be fitted.
Improving the fitting efficiency of these tracks will require additional tuning of the track-fitting algorithms, which is beyond the scope of this work.
The track fitting charge efficiency for the \cat is again significantly better than for the \legendre since the charge and momentum direction prediction of the \cat is better than that of the \legendre, which in turn leads to a correct hit ordering and more successful fits for the \cat.

\begin{figure*}[ht!]
     \centering
     \begin{subfigure}[b]{\thirdwidth\textwidth}
         \centering
         \includegraphics[width=\textwidth]{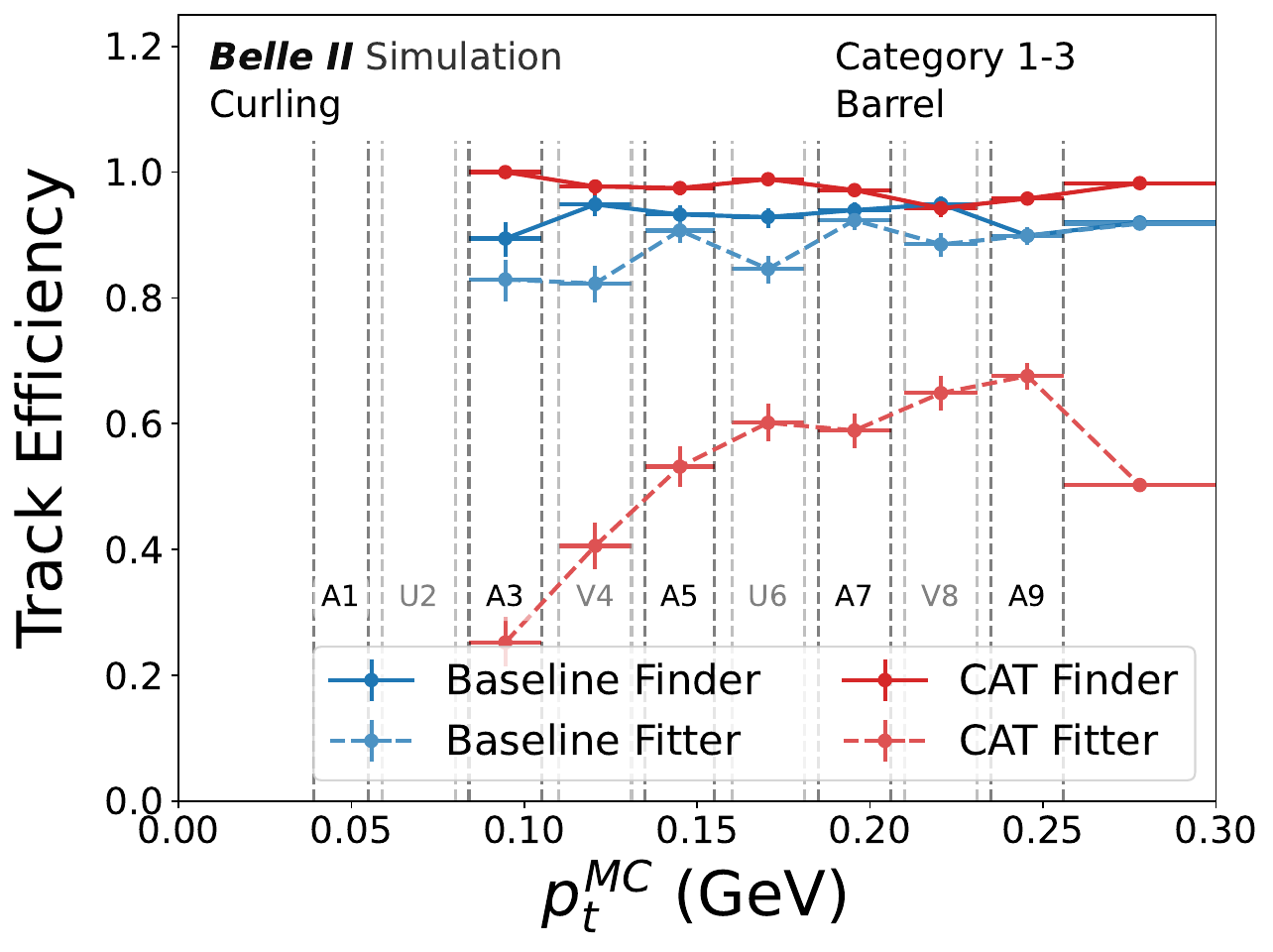}
         \caption{Track finding and fitting efficiency.}
         \label{fig:lowp_curler_eff:a}
     \end{subfigure}\quad
      \begin{subfigure}[b]{\thirdwidth\textwidth}
         \centering
         \includegraphics[width=\textwidth]{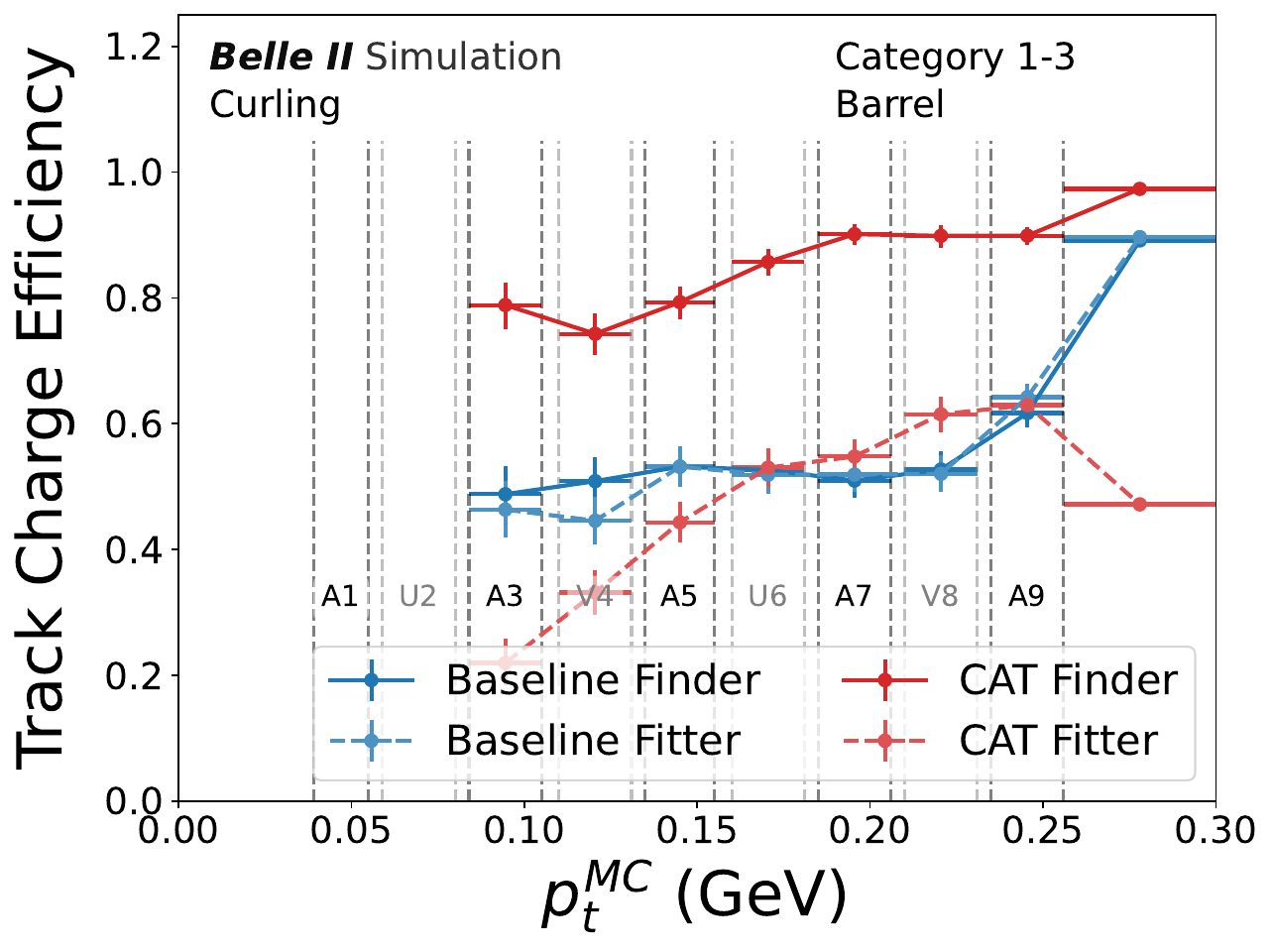}
         \caption{Track finding and fitting charge Efficiency.}
         \label{fig:lowp_curler_eff:b}
     \end{subfigure}\hfill
    \\ 
     \centering        
       
     \begin{subfigure}[b]{\thirdwidth\textwidth}
         \centering
         \includegraphics[width=\textwidth]{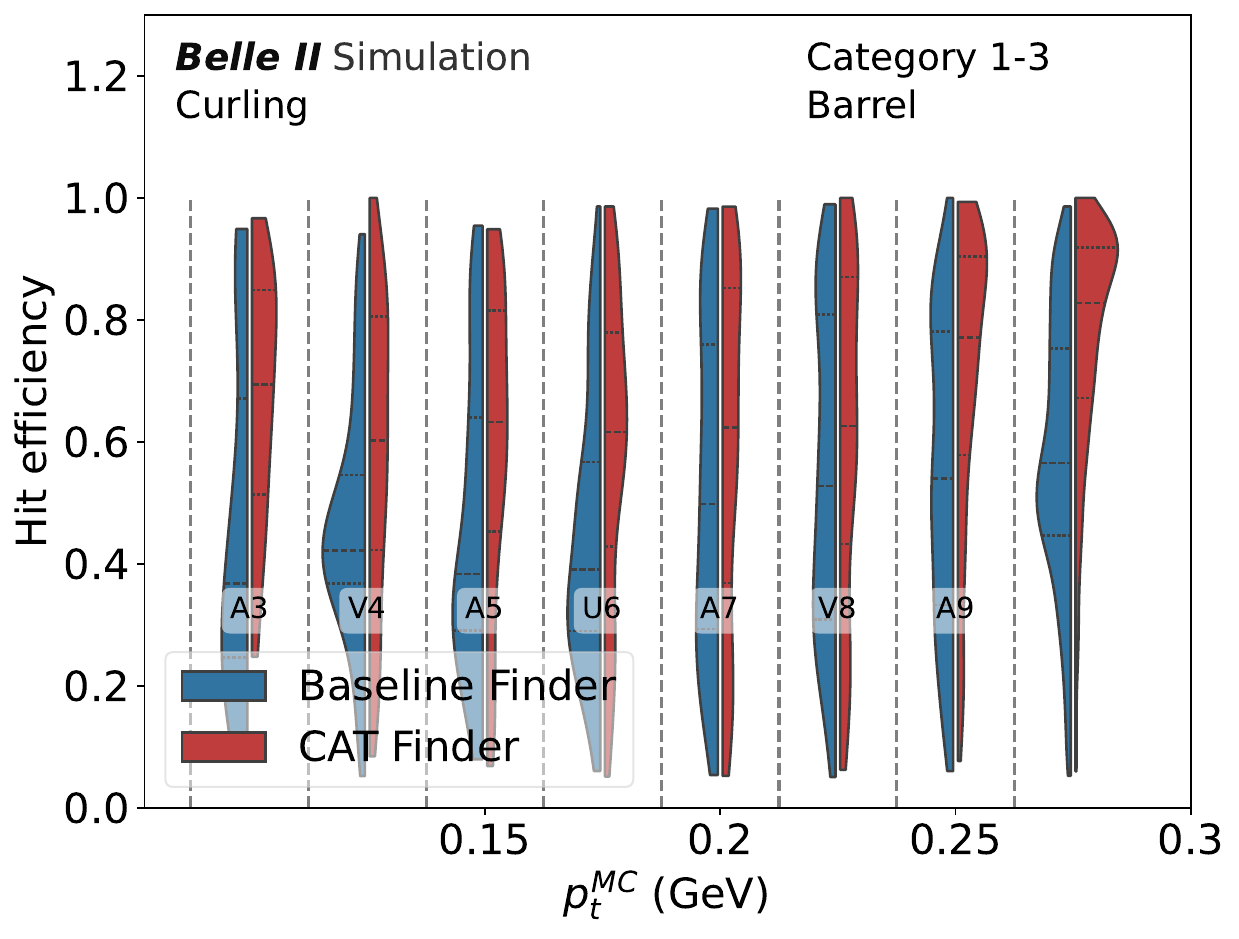}
         
         \caption{Hit efficiency for intersecting sample.}
         \label{fig:lowp_curler:a}
     \end{subfigure}\quad
        \begin{subfigure}[b]{\thirdwidth\textwidth}
         \centering
        \includegraphics[width=\textwidth]{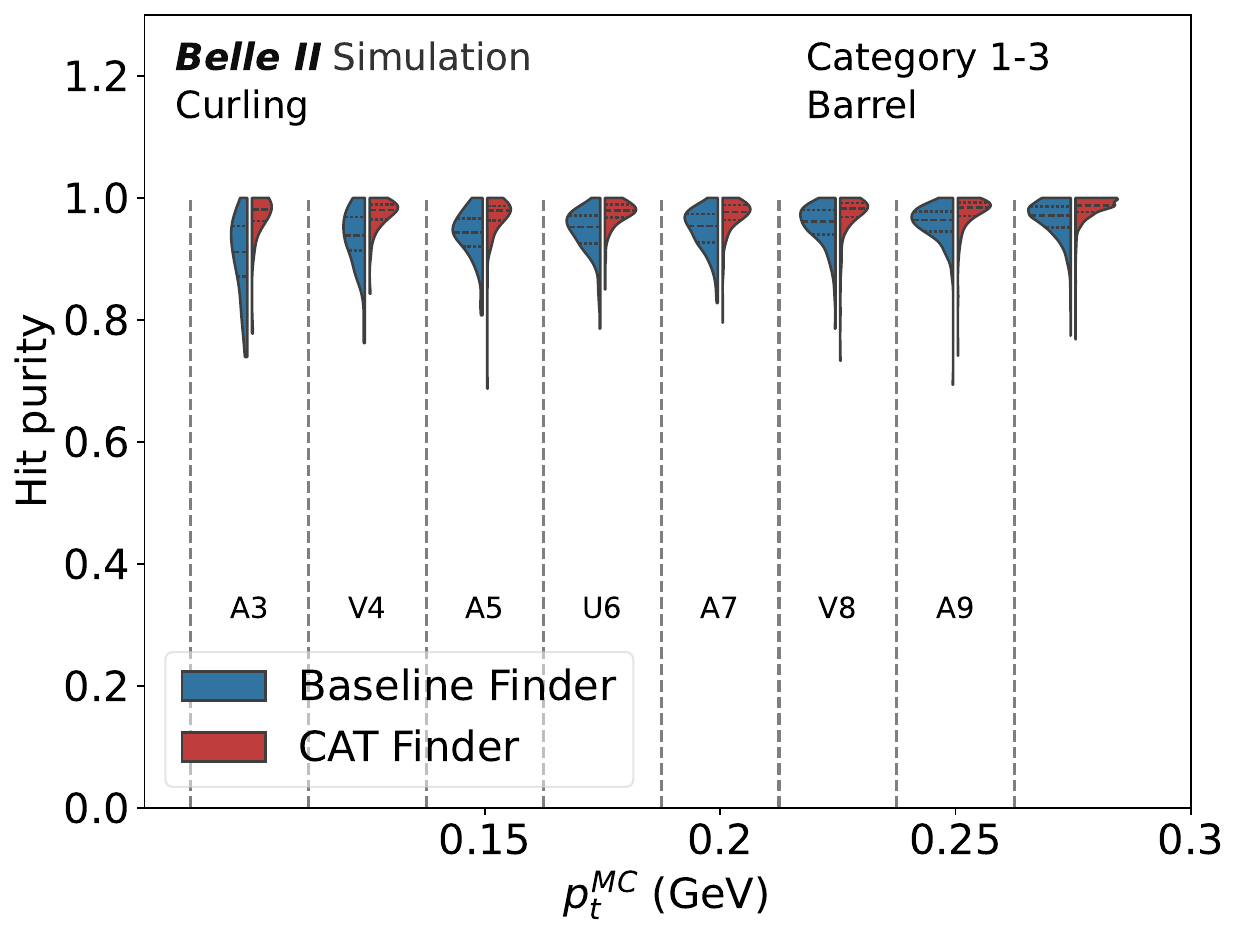}
         \caption{Hit purity for intersecting samples.}
         \label{fig:lowp_curler:b}
     \end{subfigure}\hfill \\
     \begin{subfigure}[b]{\thirdwidth\textwidth}
         \centering
          \includegraphics[width=\textwidth]{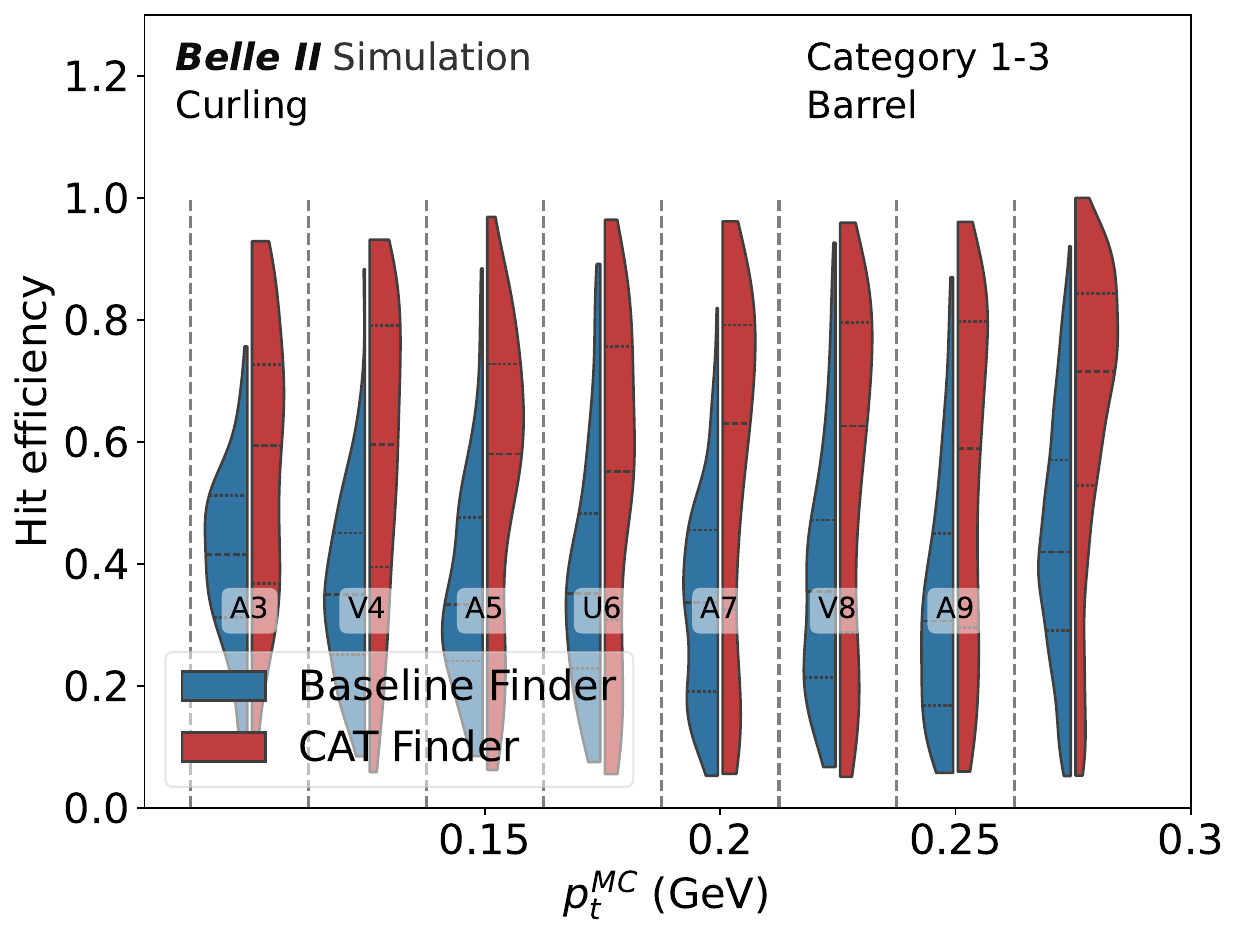}
         
         \caption{Hit efficiency for additional samples.}
         \label{fig:lowp_curler:c}
     \end{subfigure}\quad
     \begin{subfigure}[b]{\thirdwidth\textwidth}
         \centering
         \includegraphics[width=\textwidth]{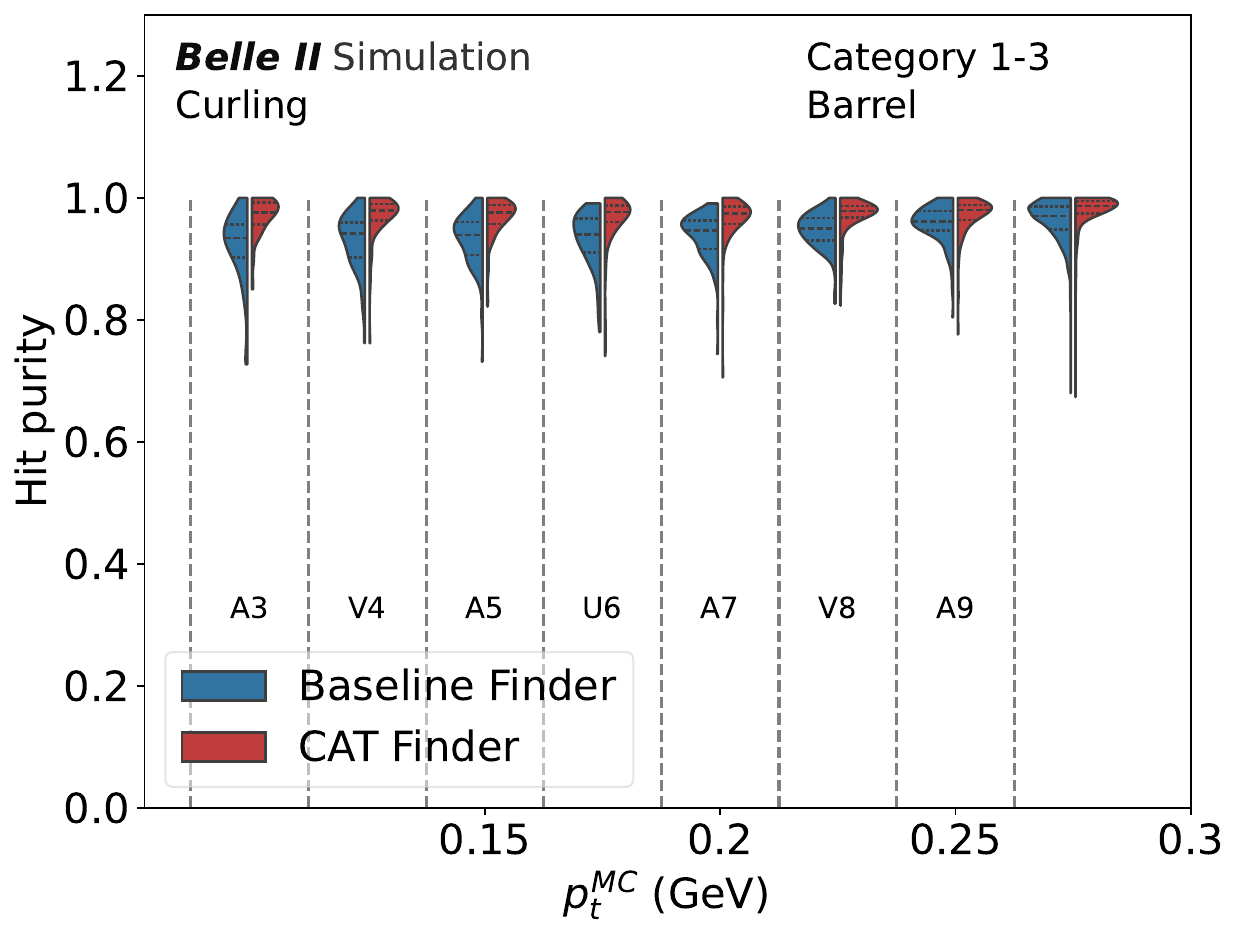}
         
         \caption{Hit purity for additional samples. \\}
         \label{fig:lowp_curler:d}
     \end{subfigure}\hfill
\caption{Low momentum 
 track finding~(empty markers, connected by lines to guide the eye) and combined track finding and fitting charge efficiency~(filled markers) (\subref{fig:lowp_curler_eff:a}) and the track finding and combined track finding and fitting charge efficiency (\subref{fig:lowp_curler_eff:b}) for curling tracks with \databackground. 
The middle row shows hit efficiency and hit purity for tracks found by both \cat and \legendre (intersecting sample) (\subref{fig:lowp_curler:a} and \subref{fig:lowp_curler:b}) and the bottom row for the additional found tracks (\subref{fig:lowp_curler:c} and \subref{fig:lowp_curler:d}).
The dashed horizontal dark (light) gray lines show the axial (stereo) superlayer boundaries how far the prompt track reaches with the given transverse momentum.}
\label{fig:lowp_curler}
\end{figure*}

\paragraph{Curling tracks}
The same comparisons as in \cref{fig:lowp_nocurler}, but for curling tracks, are shown in \cref{fig:lowp_curler}.
The track finding efficiency and especially the track finding charge efficiency of the \cat for curling tracks is very high (see \cref{fig:lowp_curler_eff:a} and \cref{fig:lowp_curler_eff:b}), 
showing that the \cat is able to assign curler hits even from multiple curls to the same track object (see \cref{fig:lowp_curler:a} and \cref{fig:lowp_curler:b}).

The track fitting algorithms on the other hand can not handle these signatures and fail for many of these \cat tracks (see \cref{fig:lowp_curler_eff:a}).
The \legendre on the other hand often only assigns the first curl to one object which produces tracks with very low hit efficiency, but that can be fitted successfully.
When comparing the track fitting charge efficiency (see \cref{fig:lowp_curler_eff:b}), both \cat and \legendre show similar efficiency again for all momenta but for $p_t\approx 0.3\,\gev$ which contains most of the \textit{outer curlers}, and the lowest momentum bin which contains very high number of \text{inner curlers}.
Overall, the very high \textit{track hit efficiency} and \textit{track hit purity} for \cat curling tracks that is even present for the additional samples, require significant adjustments of the track fitting algorithms to propagate the better track finding efficiency and the better track finding charge efficiency to the end of the tracking pipeline.
Additional steps are needed to choose the outermost first curl for the best momentum estimation and give this information to the fitter while keeping the additional hits on the track to minimize clones, which is beyond the scope of this work.

\subsubsection{Prompt tracks in \kkmc}
\label{sec_results_efficiency_kkmc}
In addition to the training and evaluation samples described above, we compare the track finding algorithms on simulated \kkmc events.
As one of the main calibration samples at \belletwo, the target track fitting charge efficiency in the barrel is 100\%.
Compared to the event samples from category 1-3 described above, these events almost always feature two isolated, prompt, high momentum tracks.
The \cat shows a significantly higher track finding efficiency in both endcaps but at the same time a slighter higher fake rate than the \legendre.
After track fitting, the combined track finding and fitting efficiency of the \cat is similar to the \legendre but with a significantly lower fake rate. 
In the barrel, both algorithms achieve a combined track finding and fitting charge efficiency of 99.8\% for the \cat and  99.4\% for the \legendre, while the \cat has the lower fake rate.
In the two endcaps, the combined track finding and fitting charge efficiency for forward and backward endcaps is 95.1\%  for the \cat compared to 73.2\% for the \legendre.
Additional plots and numerical results for \kkmc events are shown in \cref{app:kkmc}.

\FloatBarrier
\subsubsection{Displaced tracks}
\label{sec_results_efficiency_displaced}
\begin{figure*}[ht!]
     \centering

        \begin{subfigure}[b]{\thirdwidth\textwidth}
         \centering
         \includegraphics[width=\textwidth]{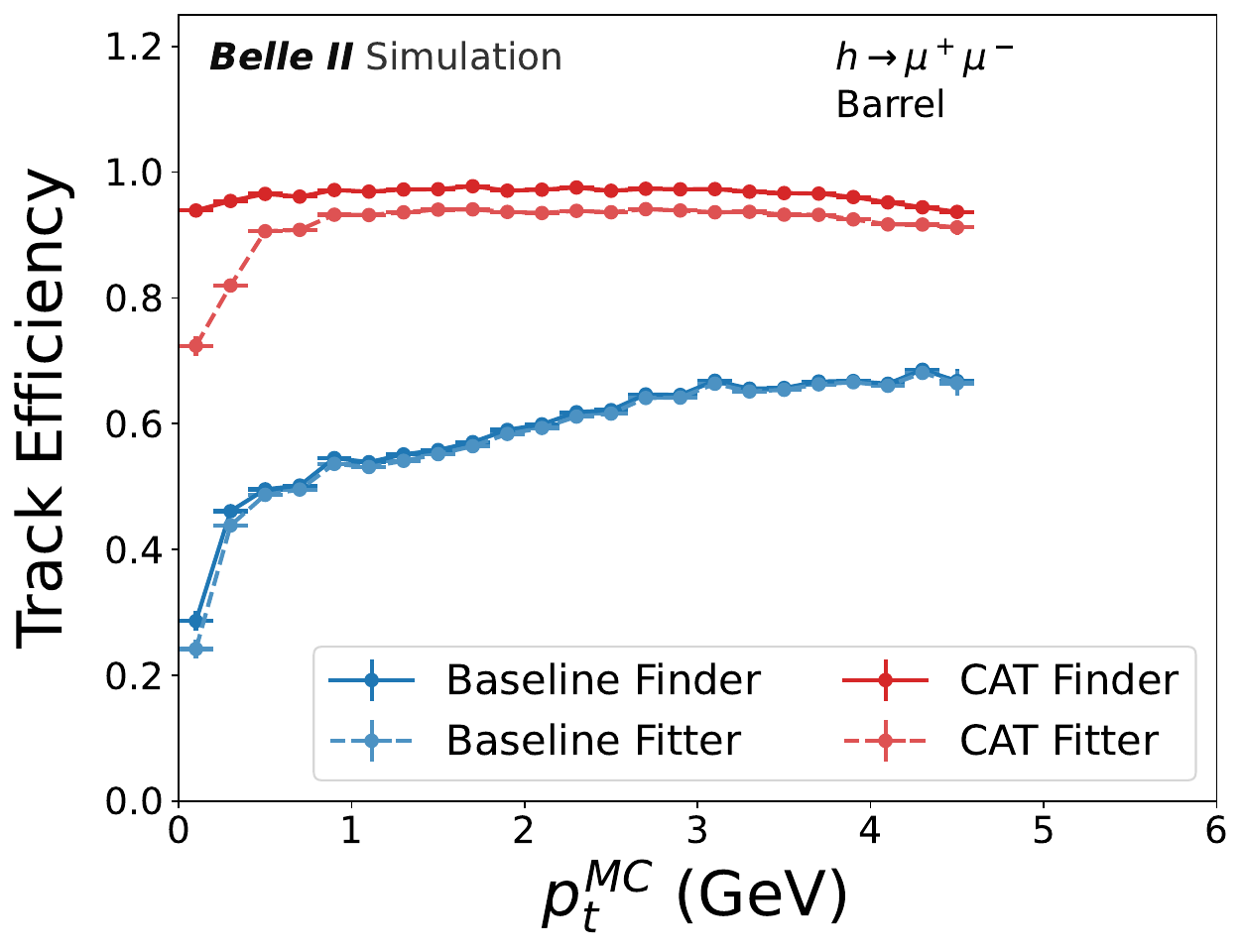}
         \caption{\dh, $p_t^{MC}$.}
         \label{fig:p_eff_darkhiggs:b}
     \end{subfigure}
     \quad
        \begin{subfigure}[b]{\thirdwidth\textwidth}
         \centering
         \includegraphics[width=\textwidth]{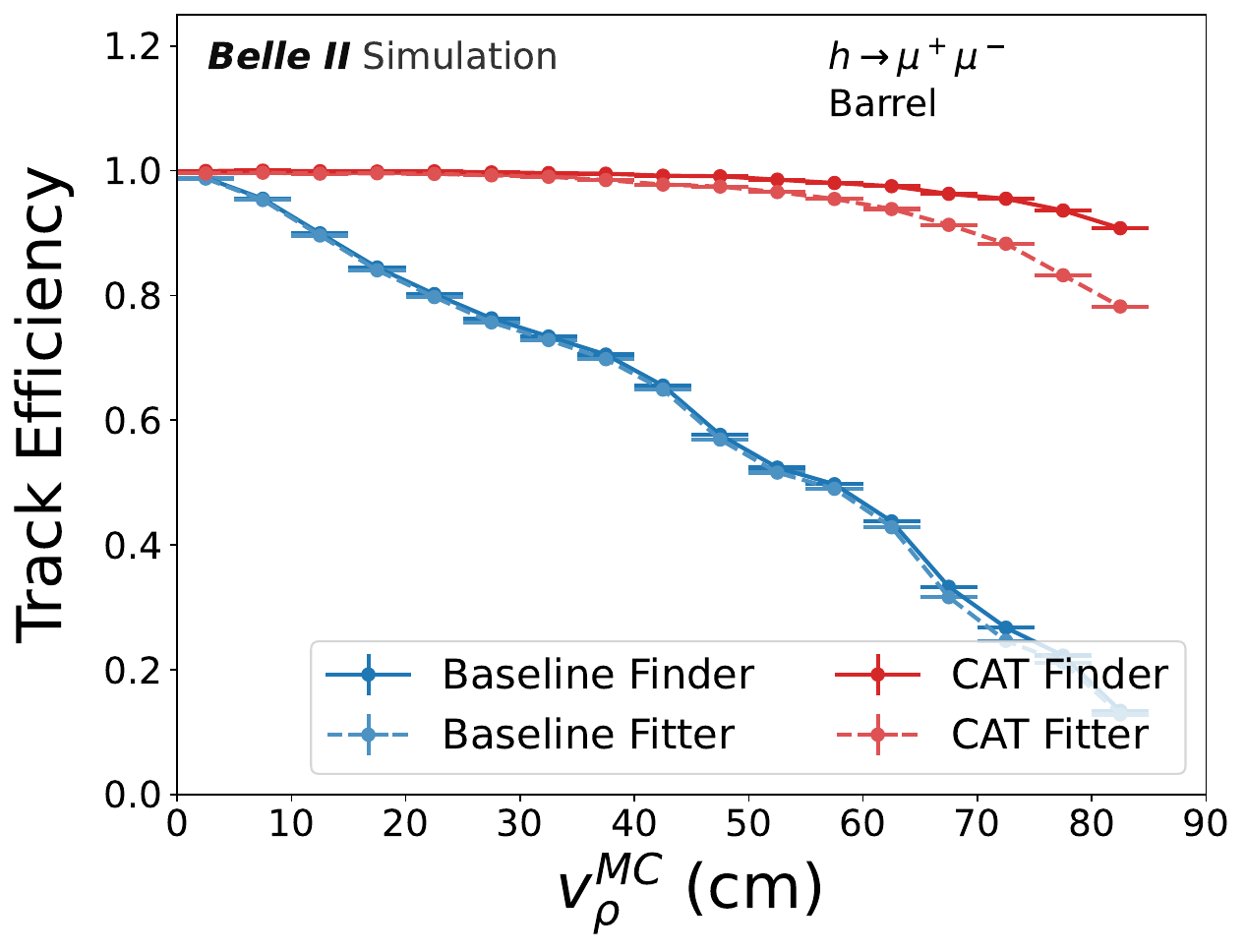}
         \caption{\dh, $v_{\rho}^{MC}$.}
         \label{fig:p_eff_darkhiggs:b2}
     \end{subfigure}\\
     
        \begin{subfigure}[b]{\thirdwidth\textwidth}
         \centering
         \includegraphics[width=\textwidth]{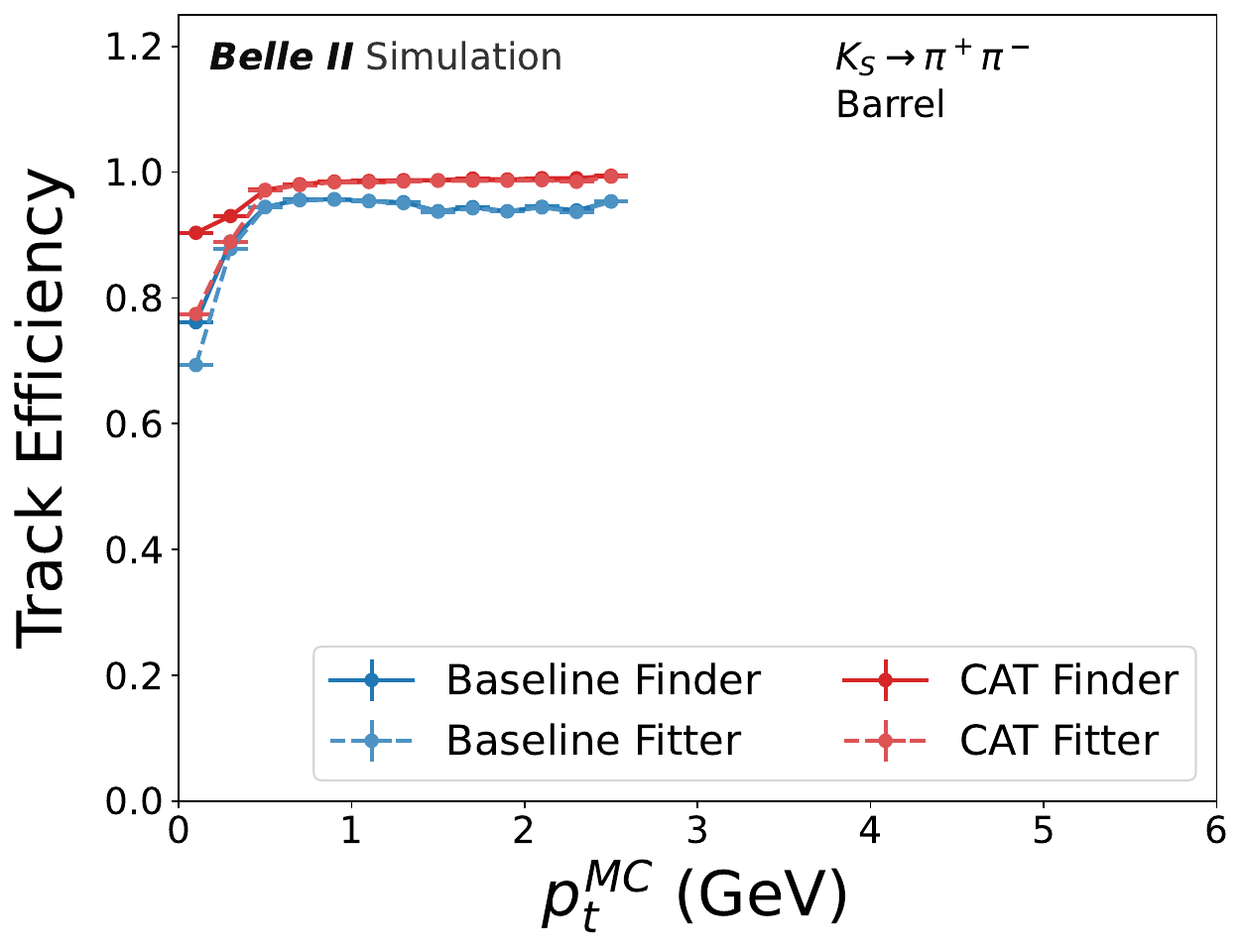}
         \caption{\kshort, $p_t^{MC}$.}
         \label{fig:p_eff_k0s:b}
     \end{subfigure}
     \quad
        \begin{subfigure}[b]{\thirdwidth\textwidth}
         \centering
         \includegraphics[width=\textwidth]{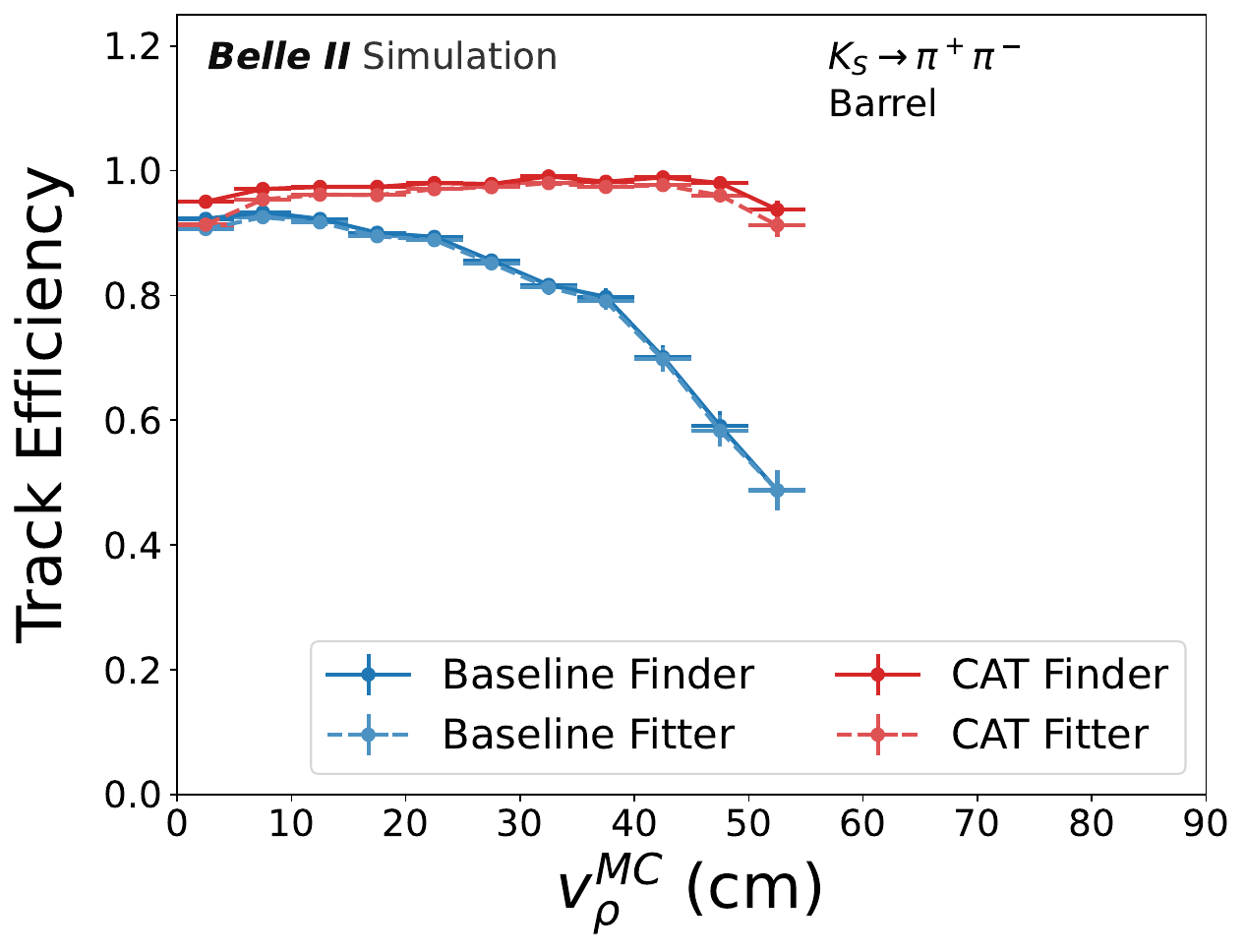}
         \caption{\kshort, $v_{\rho}^{MC}$.}
         \label{fig:p_eff_k0s:b2}
     \end{subfigure}\hfill

\caption{Track finding~(empty markers) and combined track finding and fitting efficiency~(filled markers) for (top) displaced tracks in \dh events and in (bottom) \kshort events with \databackground, as function of (left) the true simulated transverse momentum $p_t^{MC}$, and (right) the true simulated displacement $v_{\rho}^{MC}$ in the $x-y$ plane. }
\label{fig:eff_displaced}
\end{figure*}

We evaluate the track finding efficiency for displaced tracks using events with dark Higgs decays \darkhiggs with $m_h=[0.5,2.0,4.0]\,\gev$ where the dark Higgs decays uniformly along its flight direction into two charged particles. Additionally, we consider events with single \kshort decays, where the distribution of the decay distance of the displaced vertex  follows an exponential pattern.
We do not split the sample into curling and non-curling tracks.
The track finding efficiencies, and the combined track finding and track fitting efficiency for the \legendre in comparison with the \cat are shown in \cref{fig:eff_displaced}.
The same information but for track charge efficiencies can be found in \cref{app:displaced_ceff}.
The performance metrics integrated over the full $p_t$ range as shown in Fig.\,\ref{fig:eff_displaced}, are summarized in \cref{tab:dh_table_efficiencies} and \cref{tab:kshort_finding_efficiencies}.

The dark Higgs samples exhibit two types of decays: for large dark Higgs masses, decays into two muons occur with relatively large opening angles, resulting in distinct trajectories within the same superlayer. 
In contrast, for small dark Higgs masses, decays feature smaller opening angles, producing partially overlapping tracks.
Since the lifetime is uniformly distributed in the respective dark Higgs direction, the sample contains many very displaced tracks.
The \cat demonstrates significantly higher track-finding efficiencies both before and after track fitting in all detector regions, for all transverse momenta, and for all displacements. 
It also exhibits the lowest fake and clone rates, achieving a combined track-finding and fitting charge efficiency of 85.4\% per track, with a fake rate of 2.5\%, averaged over the full detector acceptance.
In comparison, the \legendre achieves 52.2\% efficiency and a fake rate of 4.1\%.

We finally evaluate the performance on a signal sample close to the expected \belletwo sensitivity for such a BSM scenario\,\cite{Duerr:2020muu} with a dark Higgs mass $m_h$ =1.5\,\gev and a mixing angle $\sin(\theta)=10^{-4}$, which corresponds to a lifetime of $c\tau=$21.5\,cm in the dark Higgs restframe.
The \cat efficiency to reconstruct both tracks in the event is 87.2\% compared to the baseline algorithm with only 44.9\%, with a smaller fake rate of 2.5\% for the \cat and 3.3\% for the \legendre; restricted to tracks that are in the barrel region, the efficiency is 90.0\% for the \cat and 52.2\% for the \legendre.
The fake rates are 2.1\% for the \cat and 3.0\% for the \legendre.

In contrast to the \darkhiggs decays, the \kshort decays occur on average at smaller displacement from the IP and they have a smaller total momentum.
Even though the \dh and the \ks samples probe rather different displacement kinematics, the general trend of the \cat for \kshort and dark Higgs is comparable: 
the \cat has a significantly higher track finding and fitting efficiency, and a comparable or even lower fake and clone rate than the \legendre.

\begin{table*}
    \fontsize{6pt}{6pt}\selectfont
    \centering
    \caption{The performance metrics per track for \dh ($m_h=$ [0.5,2.0,4.0]~GeV) samples with \databackground decaying uniformly along its flight direction into two charged particles (see \cref{sec_dataset} for details) for different track finding algorithms in different detector regions.}
     \begin{tabular}{r ccc cc}
         \toprule
        (in \%)& \trackeff & \fakerate & \clonerate &\trackchareff & \wrongchargerate\\
 \midrule
& \multicolumn{5}{c}{forward endcap} \\
\midrule
Baseline Finder & $36.2^{+0.4}_{-0.4}$ & $15.1^{+0.8}_{-0.8}$ & $0.3^{+0.1}_{-0.1}$ & $33.8^{+0.3}_{-0.4}$ & $6.5^{+0.3}_{-0.3}$ \\
CAT Finder & $88.1^{+0.2}_{-0.2}$ & $15.8^{+0.4}_{-0.4}$ & $0.64^{+0.09}_{-0.10}$ & $83.9^{+0.3}_{-0.3}$ & $4.8^{+0.2}_{-0.2}$ \\
\midrule
Baseline Fitter & $35.5^{+0.4}_{-0.4}$ & $17.5^{+0.7}_{-0.7}$ & $0.22^{+0.1}_{-0.13}$ & $34.3^{+0.4}_{-0.4}$ & $3.4^{+0.2}_{-0.2}$ \\
CAT Fitter & $79.7^{+0.3}_{-0.3}$ & $7.4^{+0.3}_{-0.3}$ & $0.1^{+0.03}_{-0.04}$ & $75.4^{+0.3}_{-0.3}$ & $5.4^{+0.2}_{-0.2}$ \\
\midrule
 \midrule
& \multicolumn{5}{c}{barrel} \\
\midrule
Baseline Finder & $59.5^{+0.1}_{-0.1}$ & $4.94^{+0.07}_{-0.07}$ & $0.53^{+0.02}_{-0.03}$ & $56.4^{+0.1}_{-0.1}$ & $5.13^{+0.08}_{-0.08}$ \\
CAT Finder & $96.89^{+0.05}_{-0.05}$ & $5.12^{+0.06}_{-0.06}$ & $1.56^{+0.03}_{-0.03}$ & $94.94^{+0.06}_{-0.06}$ & $2.01^{+0.04}_{-0.04}$ \\
\midrule
Baseline Fitter & $58.8^{+0.1}_{-0.1}$ & $3.57^{+0.06}_{-0.06}$ & $0.33^{+0.02}_{-0.02}$ & $57.4^{+0.1}_{-0.1}$ & $2.36^{+0.05}_{-0.05}$ \\
CAT Fitter & $92.75^{+0.07}_{-0.07}$ & $2.12^{+0.04}_{-0.04}$ & $0.54^{+0.02}_{-0.02}$ & $89.21^{+0.08}_{-0.08}$ & $3.81^{+0.05}_{-0.05}$ \\
\midrule
 \midrule
& \multicolumn{5}{c}{backward endcap} \\
\midrule
Baseline Finder & $17.2^{+0.4}_{-0.4}$ & $4.4^{+0.2}_{-0.2}$ & $0.32^{+0.06}_{-0.07}$ & $15.1^{+0.3}_{-0.3}$ & $12.1^{+0.7}_{-0.8}$ \\
CAT Finder & $71.0^{+0.4}_{-0.4}$ & $14.8^{+0.3}_{-0.3}$ & $0.74^{+0.07}_{-0.07}$ & $64.6^{+0.5}_{-0.5}$ & $9.1^{+0.3}_{-0.3}$ \\
\midrule
Baseline Fitter & $16.5^{+0.4}_{-0.4}$ & $4.8^{+0.3}_{-0.3}$ & $0.23^{+0.05}_{-0.07}$ & $15.6^{+0.3}_{-0.3}$ & $5.8^{+0.5}_{-0.6}$ \\
CAT Fitter & $58.0^{+0.5}_{-0.5}$ & $3.1^{+0.1}_{-0.1}$ & $0.08^{+0.02}_{-0.03}$ & $53.3^{+0.5}_{-0.5}$ & $8.2^{+0.3}_{-0.4}$ \\
\midrule
        \bottomrule
    \end{tabular}
    \label{tab:dh_table_efficiencies}
\end{table*}

\begin{table*}
    \fontsize{6pt}{6pt}\selectfont
    \centering
    \caption{The performance metrics per displaced pion track in \kshort samples with \databackground with a uniformly generated transverse momentum of $p_t(\kshortsingle)=[0.05-3]$~GeV. 
    The average transverse decay distance is $v_{\rho}=8.24$\,cm (see \cref{sec_dataset} for details) for different track finding algorithms in different detector regions.}
     \begin{tabular}{r ccc cc}
         \toprule
        (in \%)& \trackeff & \fakerate & \clonerate &\trackchareff & \wrongchargerate\\
  \midrule
& \multicolumn{5}{c}{forward endcap} \\
\midrule
Baseline Finder & $63.2^{+0.2}_{-0.2}$ & $3.5^{+0.1}_{-0.1}$ & $0.13^{+0.02}_{-0.03}$ & $62.5^{+0.2}_{-0.2}$ & $1.2^{+0.06}_{-0.06}$ \\
CAT Finder & $93.2^{+0.1}_{-0.1}$ & $6.8^{+0.1}_{-0.1}$ & $0.26^{+0.02}_{-0.02}$ & $92.7^{+0.1}_{-0.1}$ & $0.45^{+0.03}_{-0.03}$ \\
\midrule
Baseline Fitter & $61.9^{+0.2}_{-0.2}$ & $4.1^{+0.1}_{-0.1}$ & $0.1^{+0.02}_{-0.02}$ & $61.2^{+0.2}_{-0.2}$ & $1.23^{+0.06}_{-0.06}$ \\
CAT Fitter & $88.6^{+0.1}_{-0.1}$ & $3.01^{+0.08}_{-0.08}$ & $0.08^{+0.01}_{-0.01}$ & $86.6^{+0.2}_{-0.2}$ & $2.23^{+0.07}_{-0.07}$ \\
\midrule
 \midrule
& \multicolumn{5}{c}{barrel} \\
\midrule
Baseline Finder & $91.25^{+0.09}_{-0.09}$ & $6.89^{+0.08}_{-0.08}$ & $0.83^{+0.03}_{-0.03}$ & $88.5^{+0.1}_{-0.1}$ & $3.0^{+0.06}_{-0.06}$ \\
CAT Finder & $96.15^{+0.06}_{-0.06}$ & $11.52^{+0.09}_{-0.09}$ & $1.99^{+0.04}_{-0.04}$ & $95.56^{+0.07}_{-0.07}$ & $0.61^{+0.03}_{-0.03}$ \\
\midrule
Baseline Fitter & $90.05^{+0.1}_{-0.1}$ & $5.39^{+0.07}_{-0.07}$ & $0.59^{+0.02}_{-0.03}$ & $88.1^{+0.1}_{-0.1}$ & $2.16^{+0.05}_{-0.05}$ \\
CAT Fitter & $93.43^{+0.08}_{-0.08}$ & $5.13^{+0.07}_{-0.07}$ & $0.54^{+0.02}_{-0.02}$ & $92.99^{+0.08}_{-0.08}$ & $0.46^{+0.02}_{-0.02}$ \\
\midrule
 \midrule
& \multicolumn{5}{c}{backward endcap} \\
\midrule
Baseline Finder & $44.0^{+0.2}_{-0.2}$ & $2.51^{+0.08}_{-0.08}$ & $0.1^{+0.02}_{-0.02}$ & $42.7^{+0.2}_{-0.2}$ & $3.0^{+0.1}_{-0.1}$ \\
CAT Finder & $90.1^{+0.1}_{-0.1}$ & $9.6^{+0.1}_{-0.1}$ & $0.42^{+0.03}_{-0.03}$ & $89.4^{+0.1}_{-0.1}$ & $0.74^{+0.04}_{-0.04}$ \\
\midrule
Baseline Fitter & $42.6^{+0.2}_{-0.2}$ & $2.24^{+0.08}_{-0.08}$ & $0.07^{+0.01}_{-0.02}$ & $41.2^{+0.2}_{-0.2}$ & $3.2^{+0.1}_{-0.1}$ \\
CAT Fitter & $83.2^{+0.2}_{-0.2}$ & $2.35^{+0.07}_{-0.07}$ & $0.12^{+0.01}_{-0.02}$ & $79.3^{+0.2}_{-0.2}$ & $4.7^{+0.1}_{-0.1}$ \\
\midrule
        \bottomrule
    \end{tabular}
    \label{tab:kshort_finding_efficiencies}
\end{table*}

\FloatBarrier
\subsection{Track momentum resolution}
\label{subsec:resolution}
The \cat provides estimates of the track three-momentum for each condensation point.
We use these estimators as starting values for subsequent track fitting algorithms, but they can be used as end-to-end result of a complete single-step GNN-based track reconstruction algorithm.
The resolutions of the fitting step are based on the results of the full GENFIT2 algorithm (see \cref{sec_fitting}) for both the \cat and the \legendre.
A comparison of track helix parameter resolutions can be found in \cref{app:helix}.

\subsubsection{Prompt tracks}
\label{subsec:resolution_prompt}

\begin{figure*}[ht!]
     \centering
     \begin{subfigure}[b]{\thirdwidth\textwidth}
         \centering
         \includegraphics[width=\textwidth]{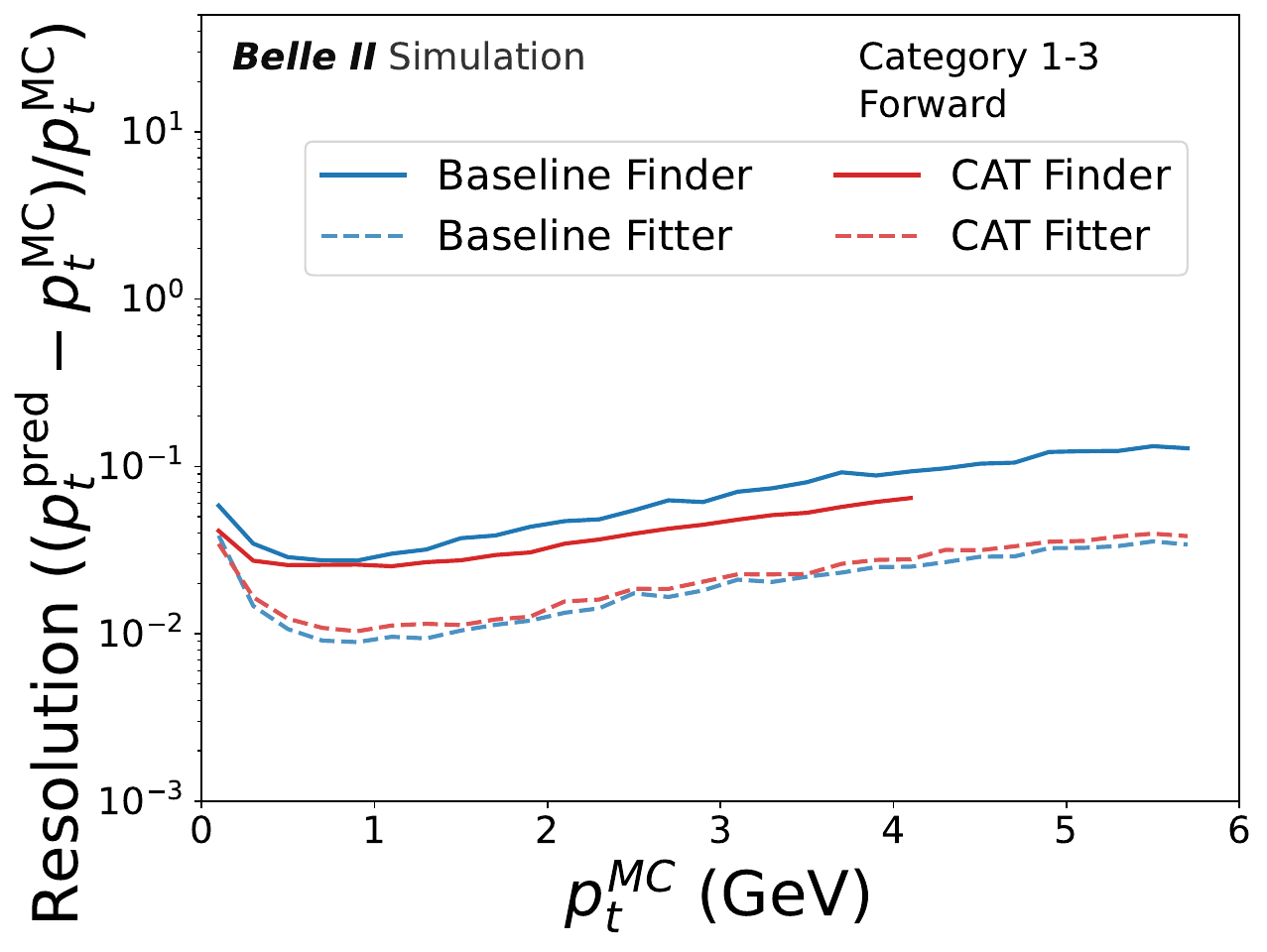}
         \caption{Forward endcap.}
         \label{fig:p_res:a}
     \end{subfigure}\hfill
        \begin{subfigure}[b]{\thirdwidth\textwidth}
         \centering
         \includegraphics[width=\textwidth]{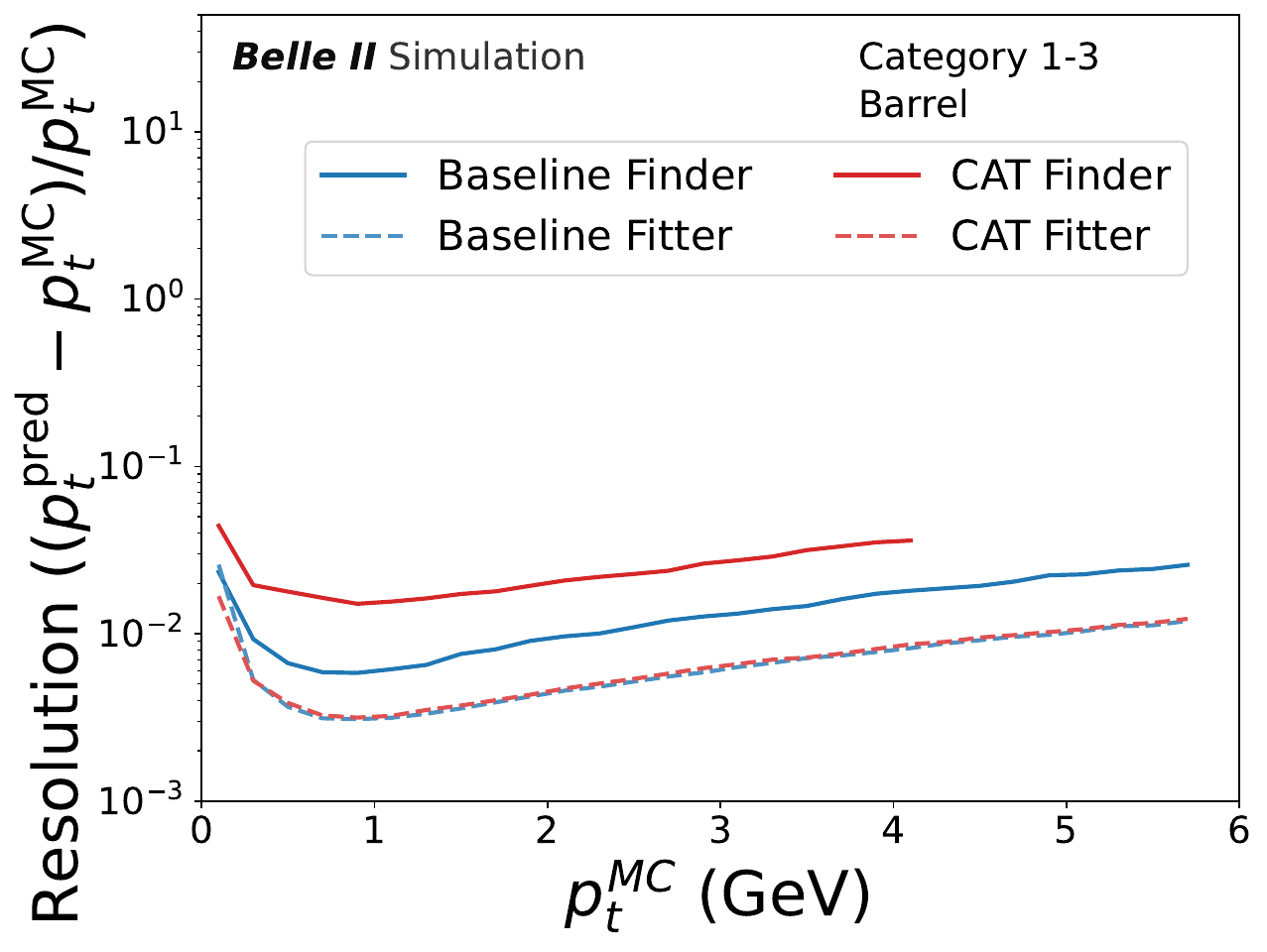}
         \caption{Barrel.}
         \label{fig:p_res:b}
     \end{subfigure}\hfill
        \begin{subfigure}[b]{\thirdwidth\textwidth}
         \centering
         \includegraphics[width=\textwidth]{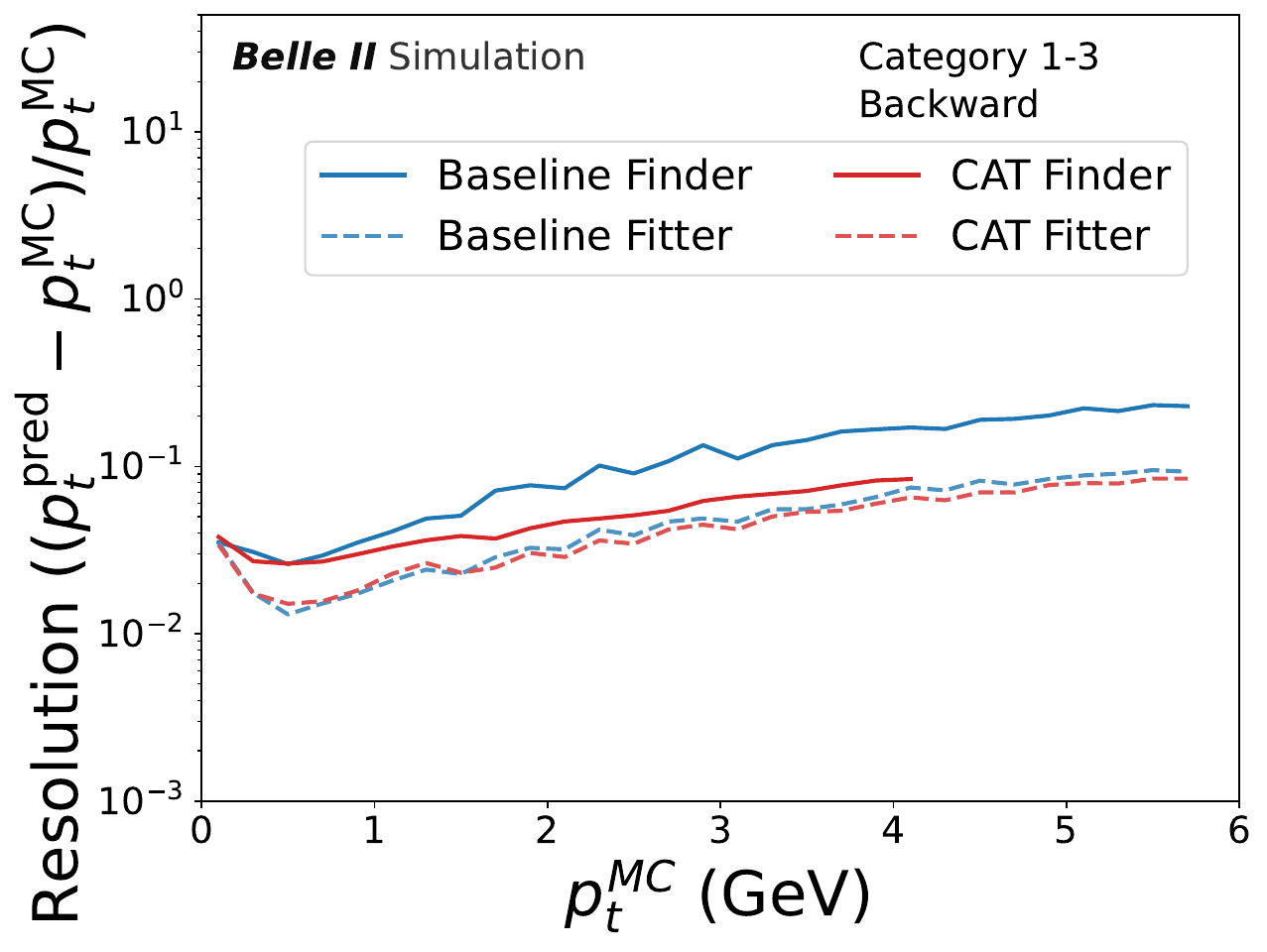}
         \caption{Backward endcap.}
         \label{fig:p_res:c}
     \end{subfigure}\hfill\\
      \centering
     \begin{subfigure}[b]{\thirdwidth\textwidth}
         \centering
         \includegraphics[width=\textwidth]{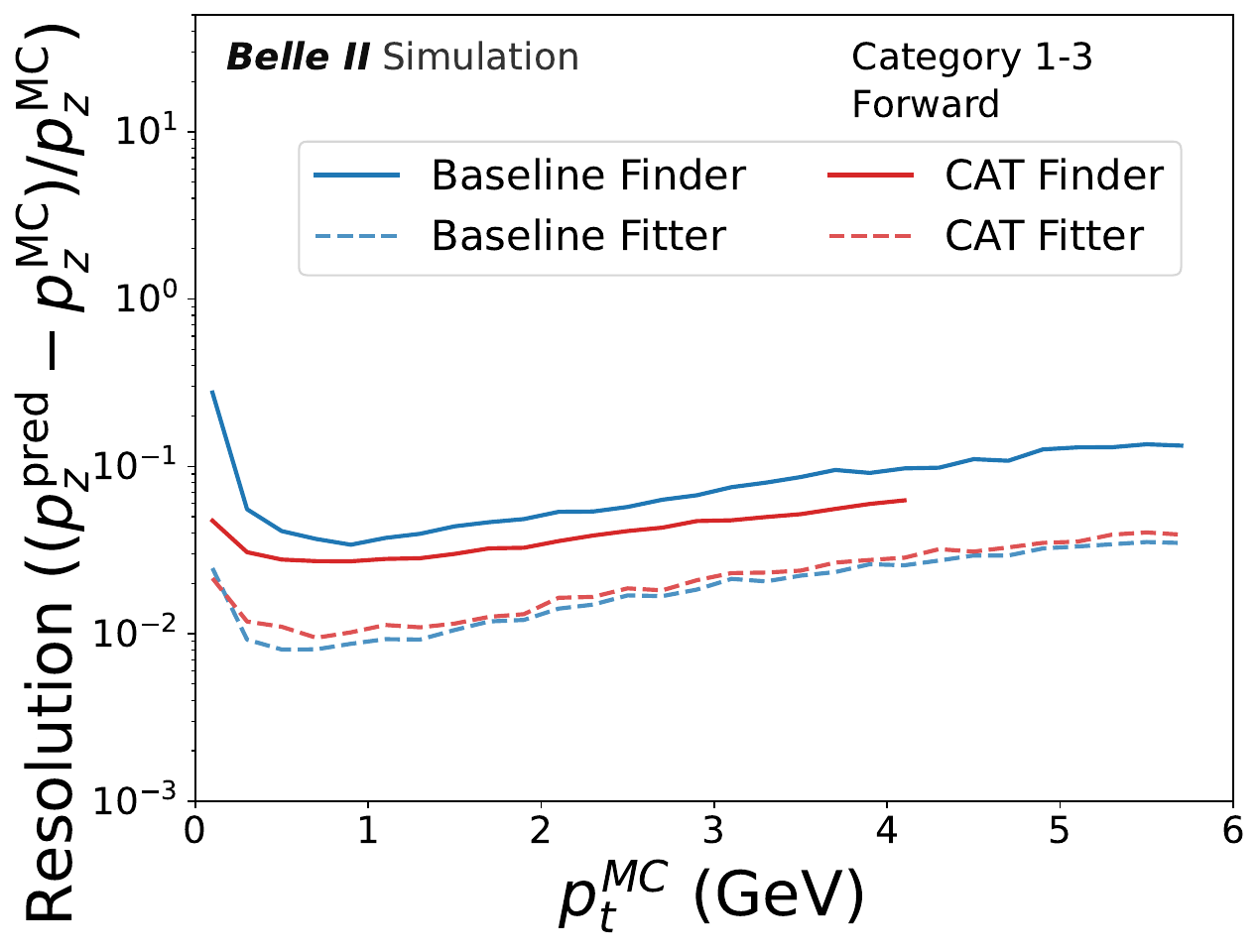}
         \caption{Forward endcap.}
         \label{fig:p_res:a2}
     \end{subfigure}\hfill
        \begin{subfigure}[b]{\thirdwidth\textwidth}
         \centering
         \includegraphics[width=\textwidth]{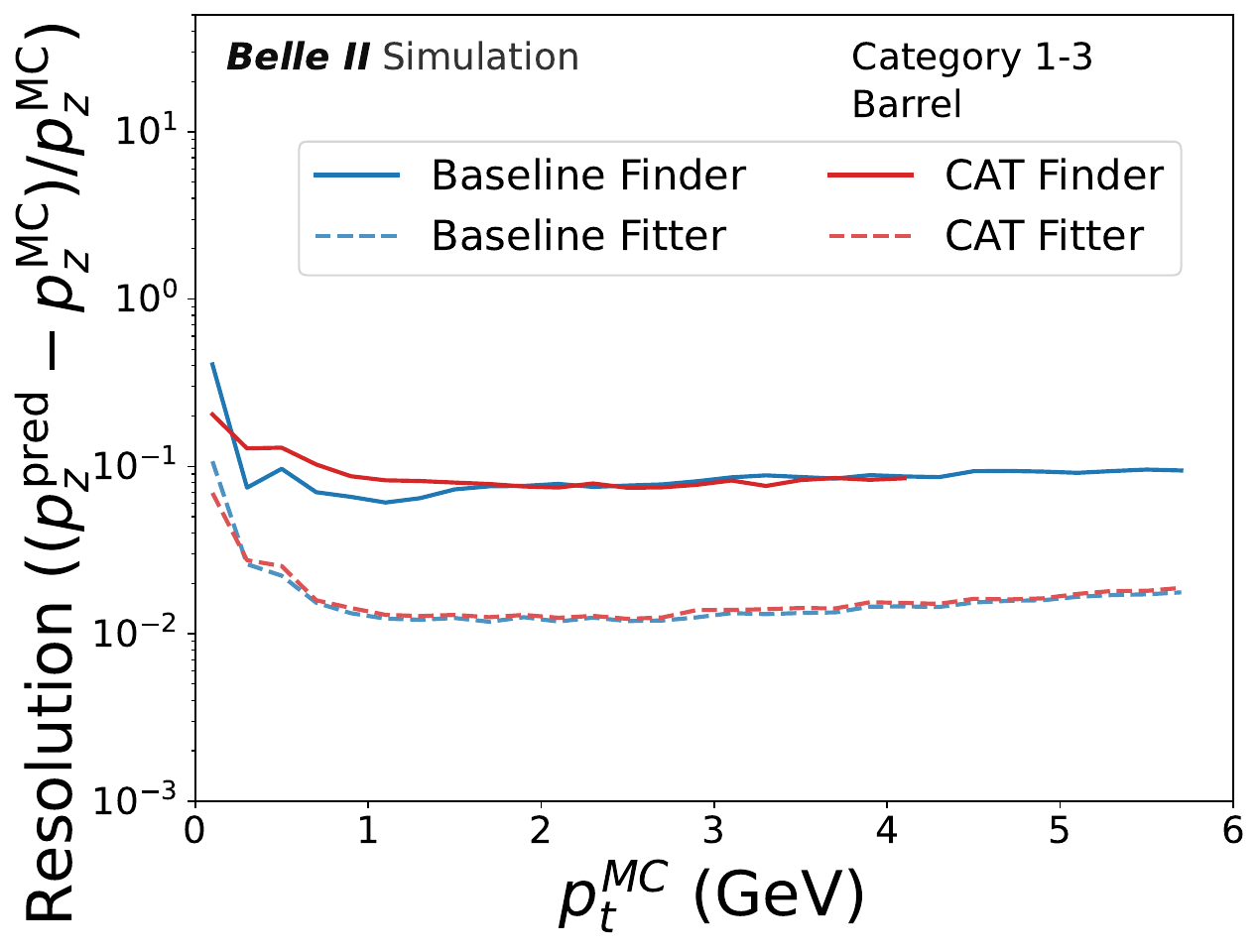}
         \caption{Barrel.}
         \label{fig:p_res:b2}
     \end{subfigure}\hfill
        \begin{subfigure}[b]{\thirdwidth\textwidth}
         \centering
         \includegraphics[width=\textwidth]{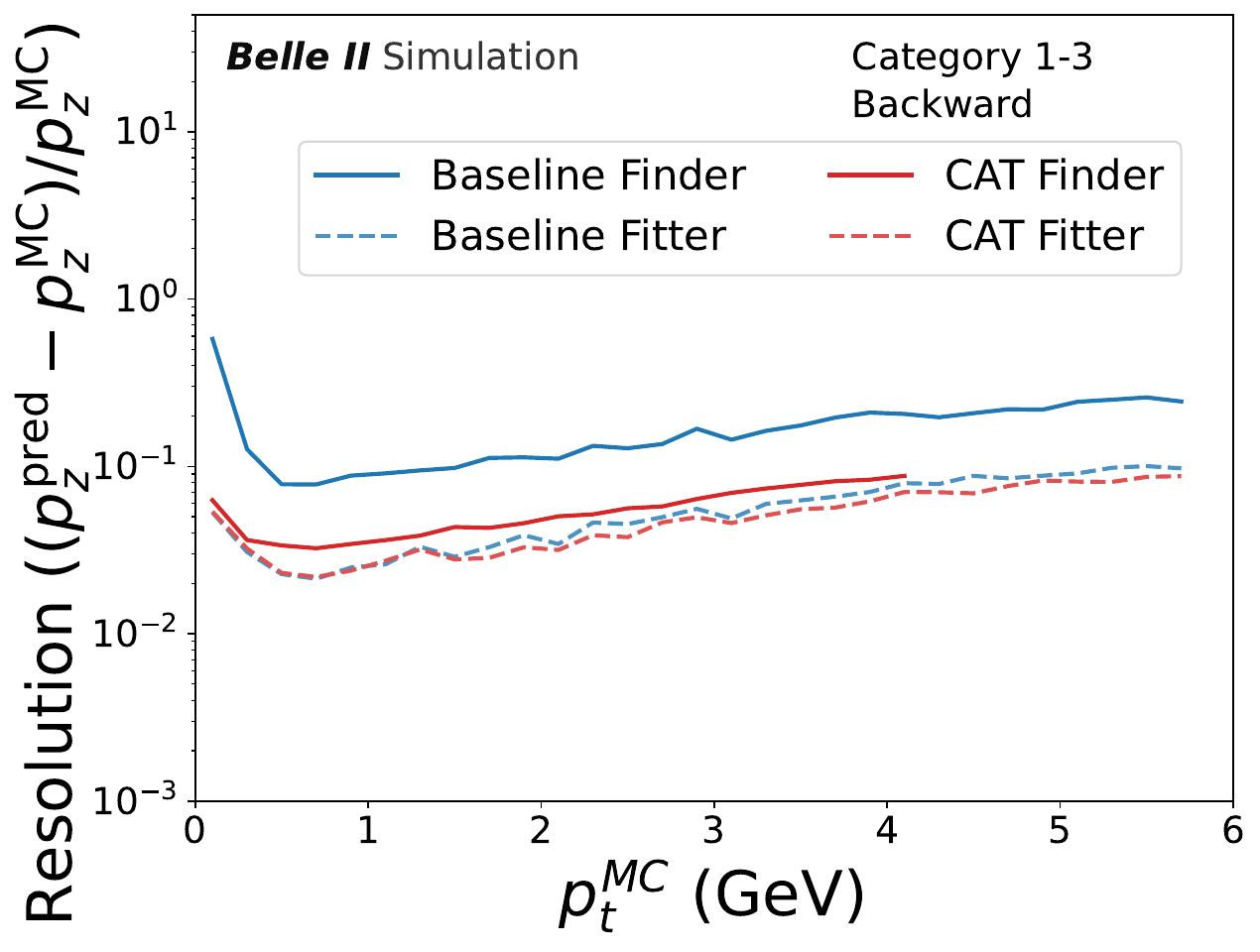}
         \caption{Backward endcap.}
         \label{fig:p_res:c2}
     \end{subfigure}

\caption{Relative (top) transverse and (bottom) longitudinal momentum resolution as function of simulated transverse momentum $p_t^{MC}$ for the intersecting prompt evaluation sample (category 1-3, \databackground, see \cref{tab:samples}) in the (left) forward endcap, (center) barrel, and (right) backward endcap for tracks found by both (red) \cat and (blue) \legendre. For the \cat the resolution is shown only for $p_t<$4\,\gev, see \cref{app:highptres} for details.}
\label{fig:intersect_testset_resolution_pt_ep3}
\end{figure*}

We evaluate the track momentum resolution for matched prompt tracks using the track categories 1-3 (see \cref{tab:samples}).
For prompt tracks we evaluate the resolution on the non-curling tracks (see \cref{sec_results_efficiency}) for tracks found by both the \cat and the \legendre. 
The transverse momentum resolution $\eta(p_t)$ and the longitudinal momentum resolution $\eta(p_z)$ for the \cat and the \legendre are shown in \cref{fig:intersect_testset_resolution_pt_ep3}.

Since our training samples do not include tracks with transverse momenta above 6\,\gev, the model does not predict transverse momentum values above around 6.5\,\gev which leads to a biased distribution of the momentum distribution (see \cref{app:highptres} for details).
For this reason we do not report the CAT Finder resolution for $p_t>4\,\gev$.

The $\eta(p_t)$ resolution before track fitting for the \cat is comparable for the different detector regions, and amounts to a few percent.
The \legendre performs better than the \cat in the barrel and reaches a relative resolution of better than 1\%.
It performs significantly worse than the \cat in both endcaps, due to the much lower hit efficiency and hit purity.
The $\eta(p_z)$ resolution before track fitting on the other hand is comparable for \cat and the \legendre in the barrel and plateaus around 1\%.
As for the transverse momentum resolution, the \cat performs significantly better in the endcaps.
After track fitting with GENFIT2, the transverse and the longitudinal momentum resolutions are very similar for the two algorithms in all detector regions.
We attribute this similarity to the comparable hit efficiency of the two algorithms for not too small transverse momenta.

\FloatBarrier
\subsubsection{Displaced tracks}
\label{subsec:resolution_displaced}
We evaluate the track momentum resolution for matched displaced tracks using events with dark Higgs decays and with single $\kshort$ decays as described in \cref{sec_dataset}.
We do not remove curling tracks, and we evaluate the resolution separately for the intersecting and the additional samples in the barrel only.

\begin{figure*}[ht!]
     \centering
     \begin{subfigure}[b]{\thirdwidth\textwidth}
         \centering
         \includegraphics[width=\textwidth]{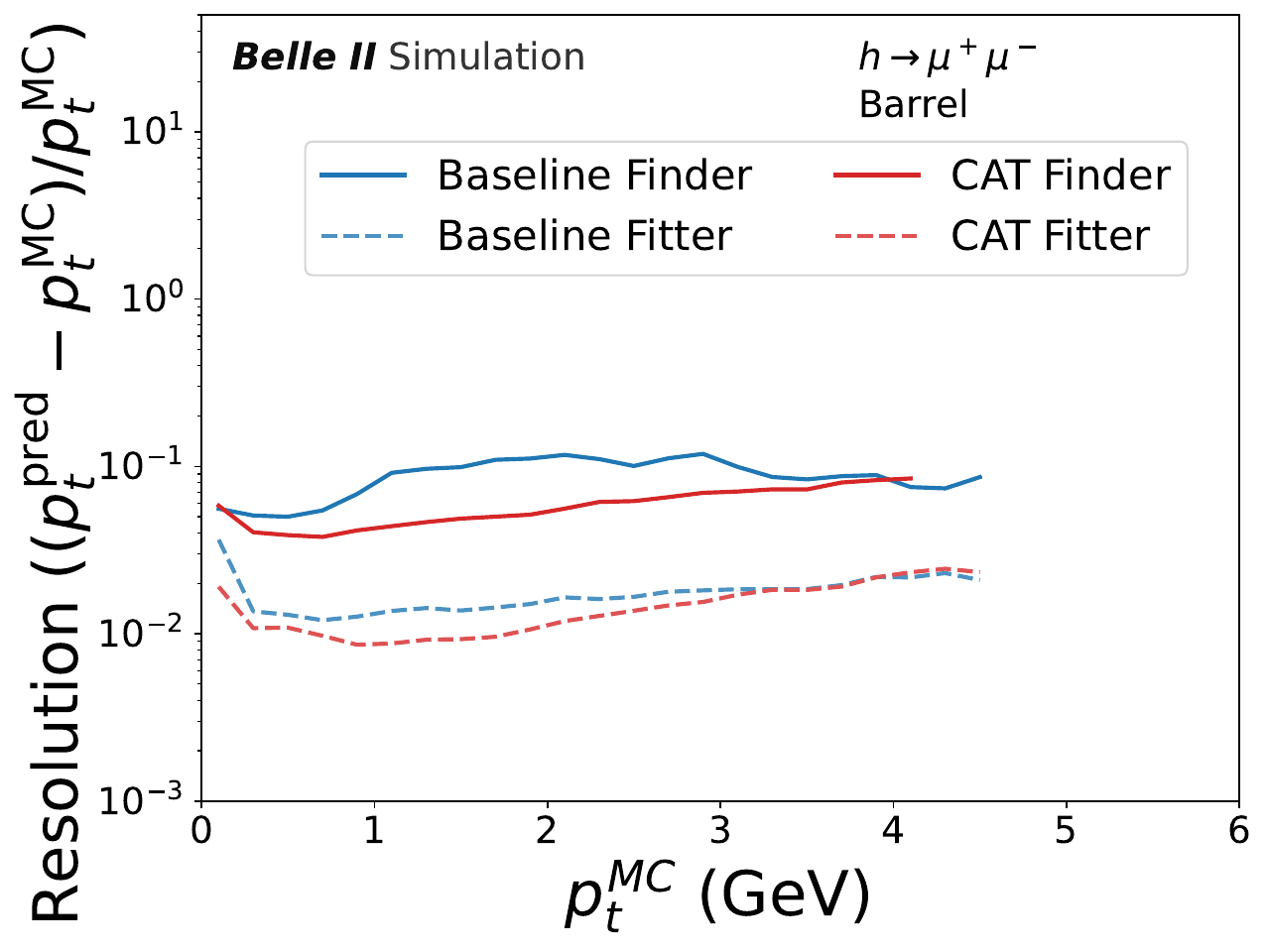}
         \caption{$\eta(p_t)$ as function of $p_t$ in the barrel.}

         \label{fig:res_dh_pt:a}
     \end{subfigure}\quad
        \begin{subfigure}[b]{\thirdwidth\textwidth}
         \centering
         \includegraphics[width=\textwidth]{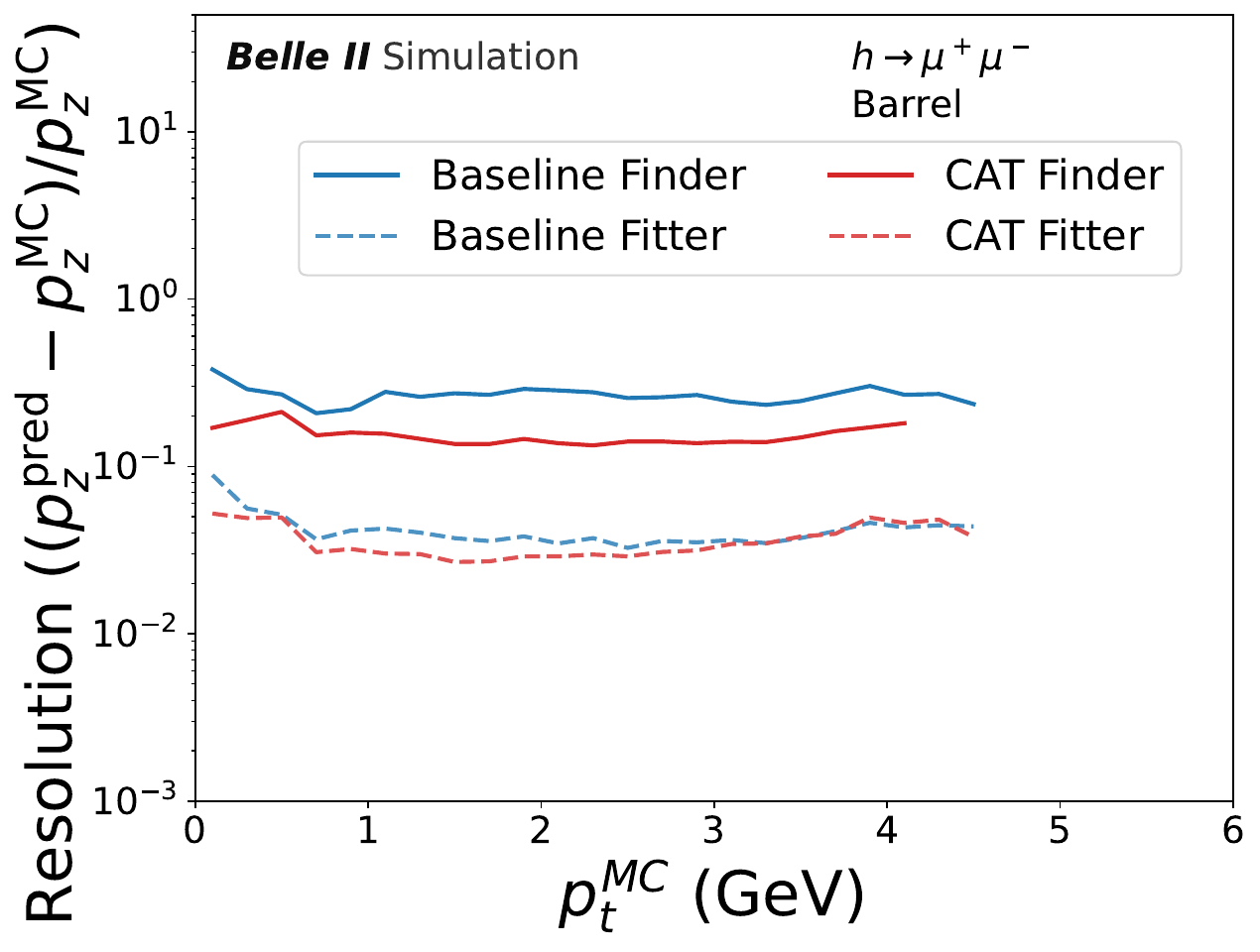}
         \caption{$\eta(p_z)$ as function of $p_t$ in the barrel.}
         \label{fig:res_dh_pt:b}
     \end{subfigure}\hfill \\ 
     \begin{subfigure}[b]{\thirdwidth\textwidth}
         \centering
         \includegraphics[width=\textwidth]{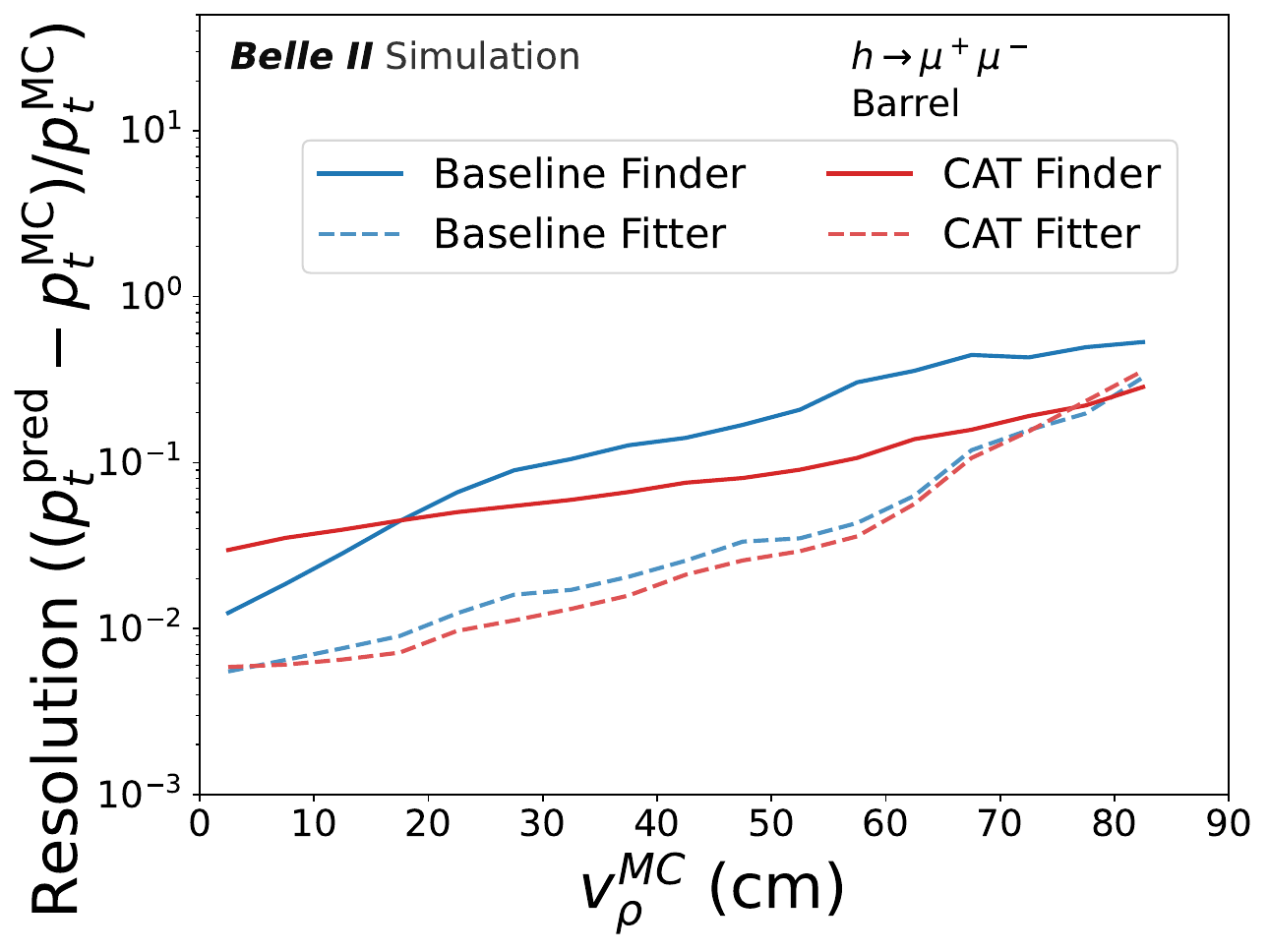}
         \caption{$\eta(p_t)$ as function of displacement $v_{\rho}$ in the barrel.}
         \label{fig:res_dh_pz:d}
     \end{subfigure}\quad
        \begin{subfigure}[b]{\thirdwidth\textwidth}
         \centering
         \includegraphics[width=\textwidth]{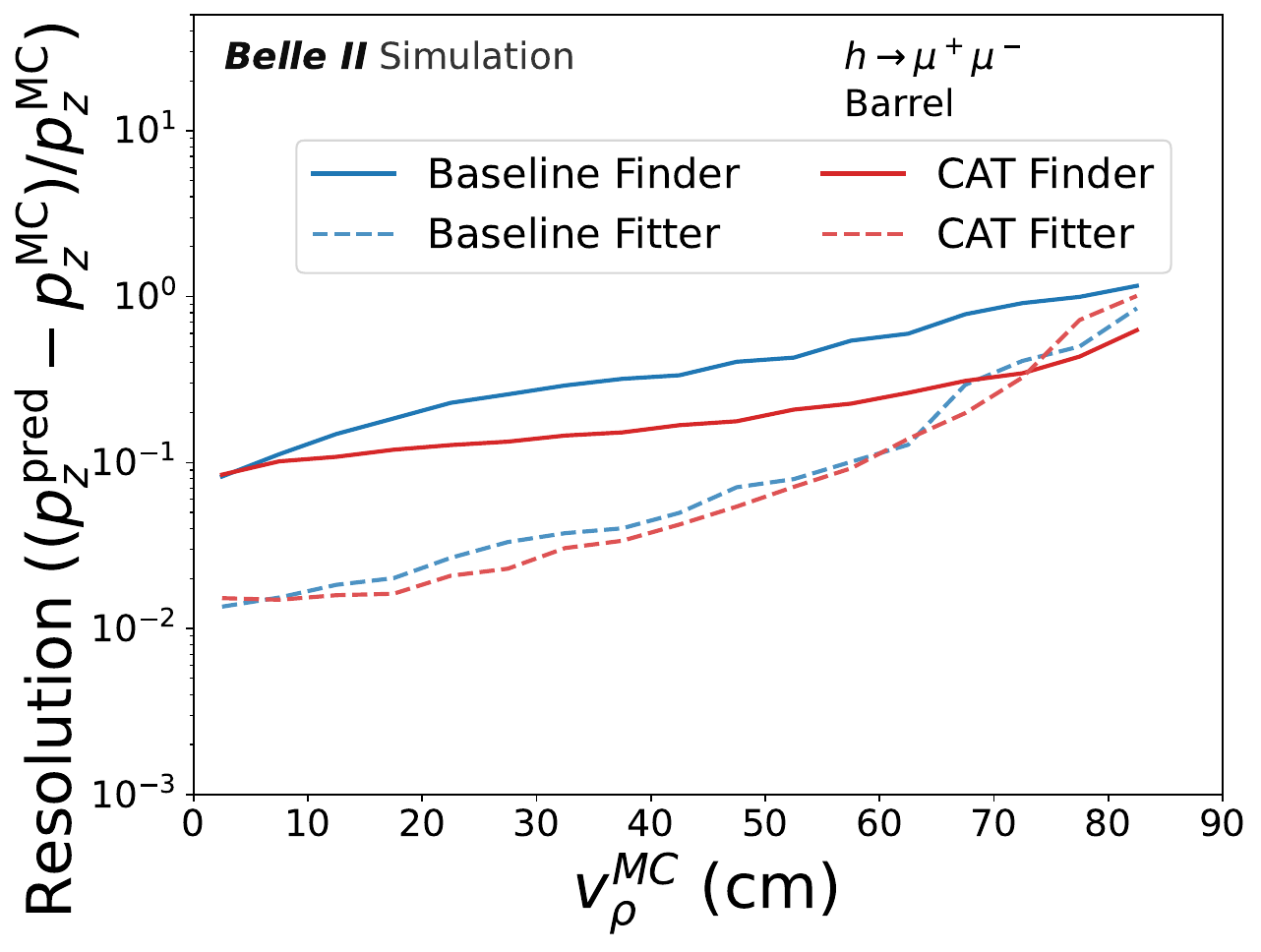}
         \caption{$\eta(p_z)$ as function of $v_{\rho}$ in the barrel.}
         \label{fig:res_dh_pz:e}
     \end{subfigure}\hfill
\caption{Relative resolution of (first column) transverse and (second column) longitudinal momentum as function of  simulated transverse momentum $p_t^{MC}$ (top row) and simulated displacement $v_{\rho}^{MC}$ (bottom row)  for displaced tracks from \dh decays with \databackground in the  barrel for tracks found by both (red) \cat and (blue) \legendre for the intersecting sample. For the \cat the resolution is shown only for $p_t<$4\,\gev, see \cref{app:highptres} for details.}
\label{fig:dh_resolution_pt_ep3}
\end{figure*}

The transverse momentum resolution $\eta(p_t)$ and the longitudinal momentum resolution $\eta(p_z)$ for the \cat and the \legendre are shown in \cref{fig:dh_resolution_pt_ep3}.
The corresponding information for the additional \cat sample for \dh decays and for the intersecting and the additional \cat sample for \kshort decays are shown in \cref{app:trackres_dh} and \cref{app:trackres_ks}.
For the displaced tracks, the additional \legendre sample is too small to provide meaningful information.

The $\eta(p_t)$ and $\eta(p_z)$ resolutions for dark Higgs tracks before track fitting is better for the \cat for all transverse momenta and displacement regions, but for very small displacements where the \legendre shows a better resolution. 

The \kshort sample features on average a much smaller displacement than the \dh sample which is visible in the better transverse and longitudinal momentum resolutions for both the \cat and the \legendre.

After track fitting with GENFIT2, the transverse and the longitudinal momentum resolutions are again similar for the two algorithms.

\FloatBarrier
\subsection{Position Reconstruction}
\label{subsec:position}

In this chapter, we evaluate only the position prediction of the \cat, which is directly derived from the GNN output.
Position information is not available from either the \legendre or post-fitting for any approach, as the tracks are described as a helix without a defined starting point.

\begin{figure*}[ht!]
    \centering
    \begin{subfigure}[b]{\thirdwidth\textwidth}
         \centering
         \includegraphics[width=\textwidth]{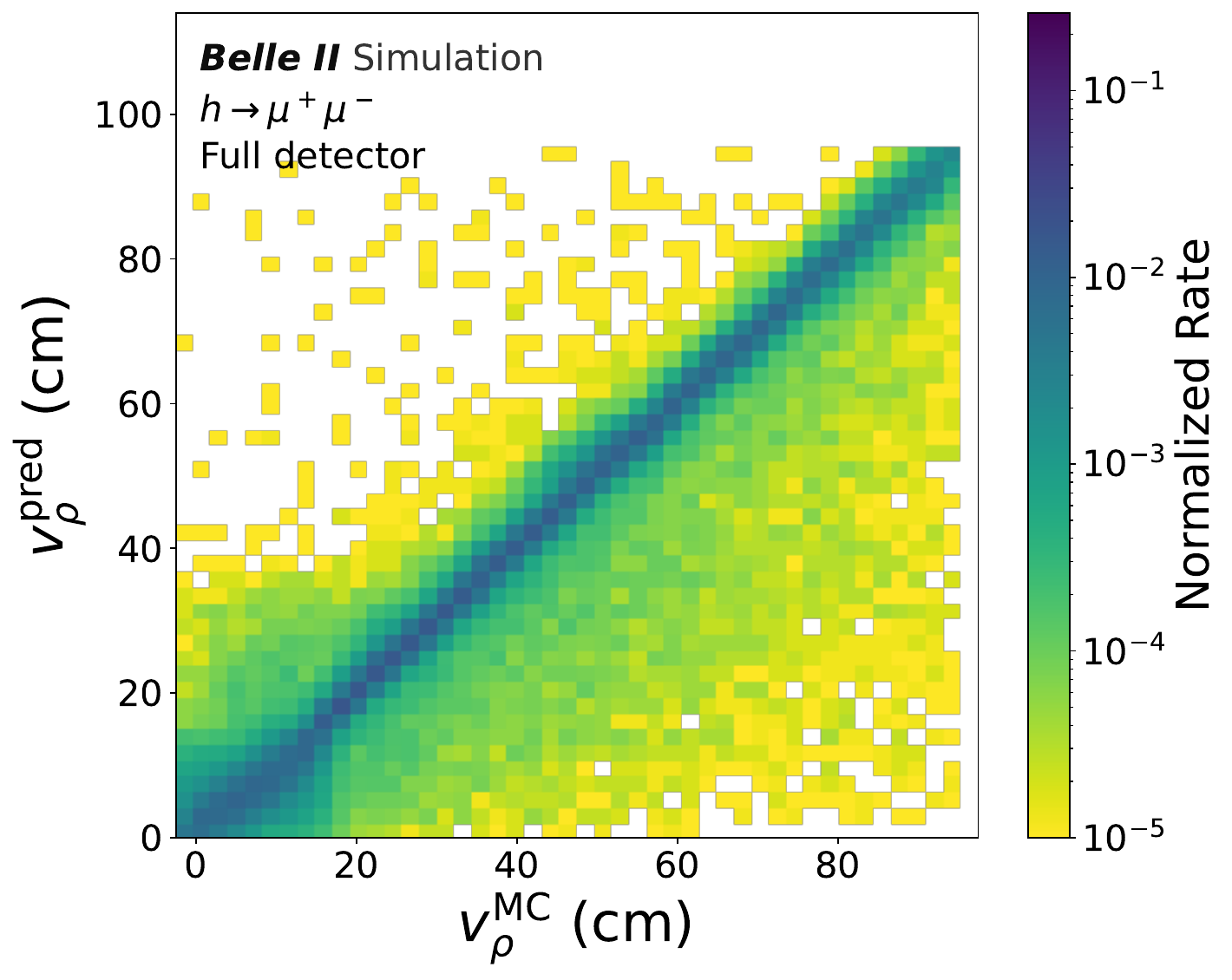}
         \caption{\dh.\newline{}}
         \label{fig:pos:a}
    \end{subfigure}\hfill
    \begin{subfigure}[b]{\thirdwidth\textwidth}
         \centering
         \includegraphics[width=\textwidth]{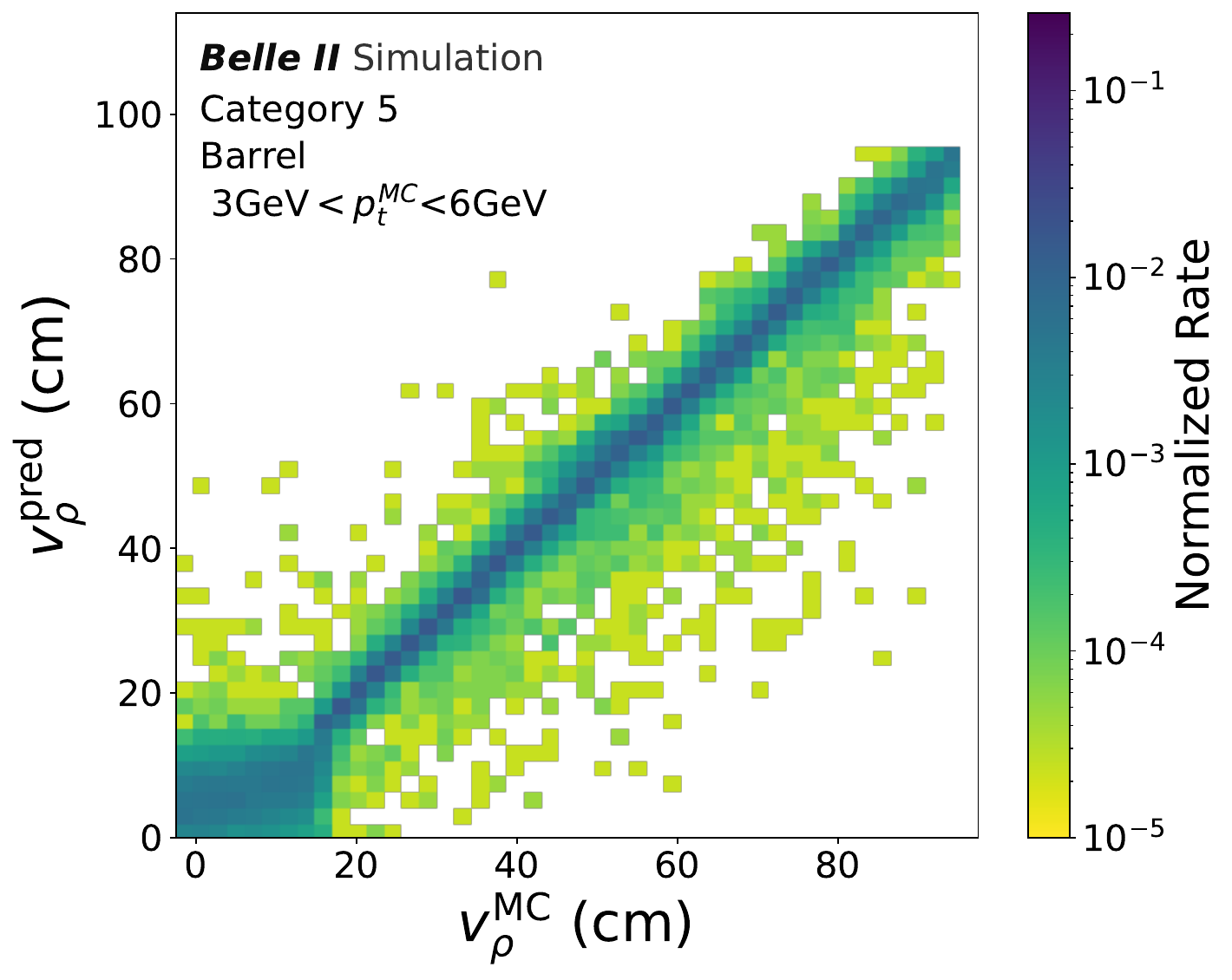}
         \caption{Displaced tracks (category 5), $3\,\gev < p_t < 6\,\gev$.}
         \label{fig:pos:b}
    \end{subfigure}\hfill
    \begin{subfigure}[b]{\thirdwidth\textwidth}
         \centering
         \includegraphics[width=\textwidth]{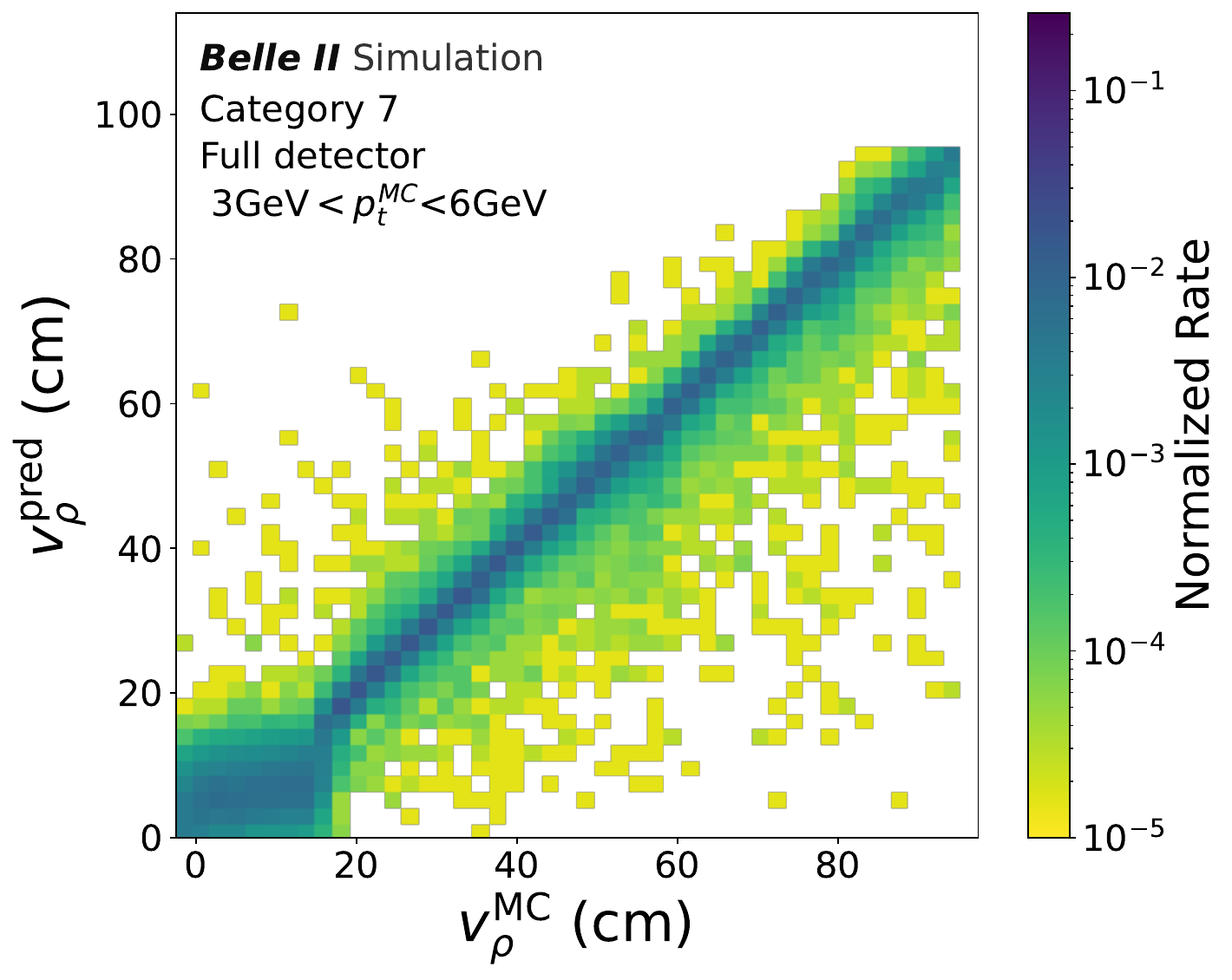}
         \caption{Displaced angled tracks (category 7), $3\,\gev < p_t < 6\,\gev$.}
         \label{fig:pos:c}
    \end{subfigure}\hfill\\
    \begin{subfigure}[b]{\thirdwidth\textwidth}
         \centering
         \includegraphics[width=\textwidth]{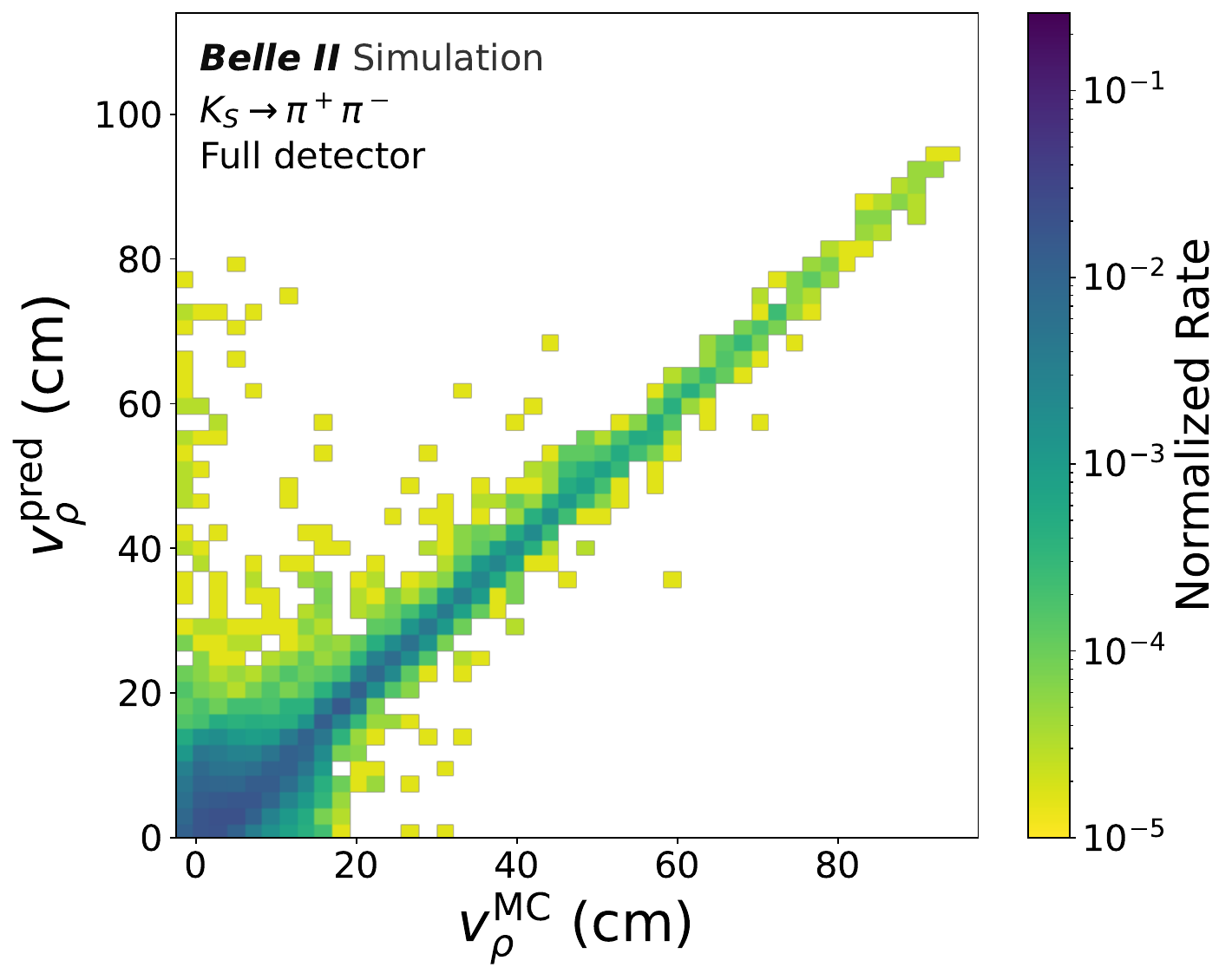}
         \caption{\kshort.\newline{}}
         \label{fig:pos:d}
    \end{subfigure}\hfill
     \begin{subfigure}[b]{\thirdwidth\textwidth}
         \centering
         \includegraphics[width=\textwidth]{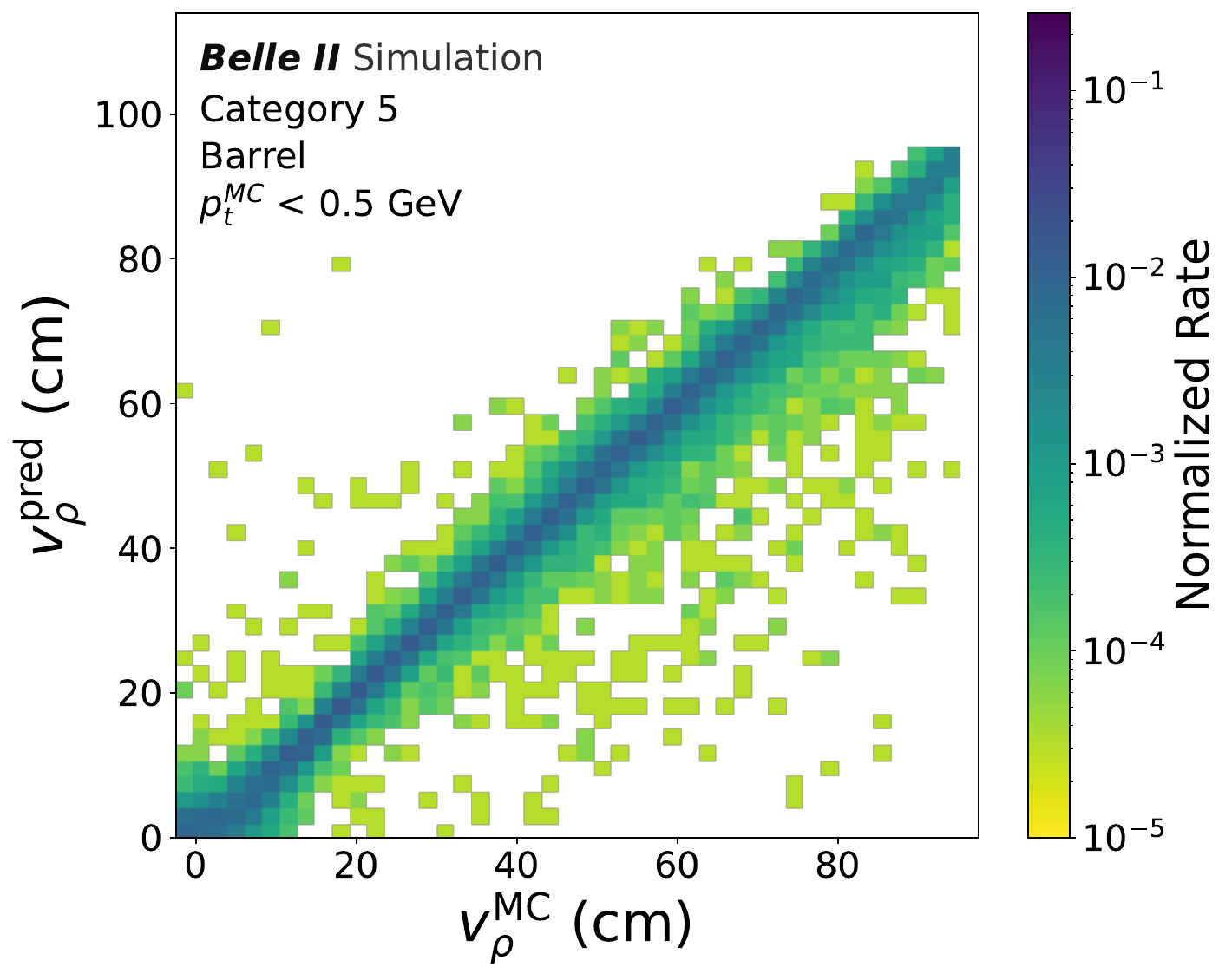}
         \caption{Displaced tracks (category 5), low $p_t < 0.5\,\gev$}
         \label{fig:pos:e}
     \end{subfigure}\hfill
        \begin{subfigure}[b]{\thirdwidth\textwidth}
         \centering
         \includegraphics[width=\textwidth]{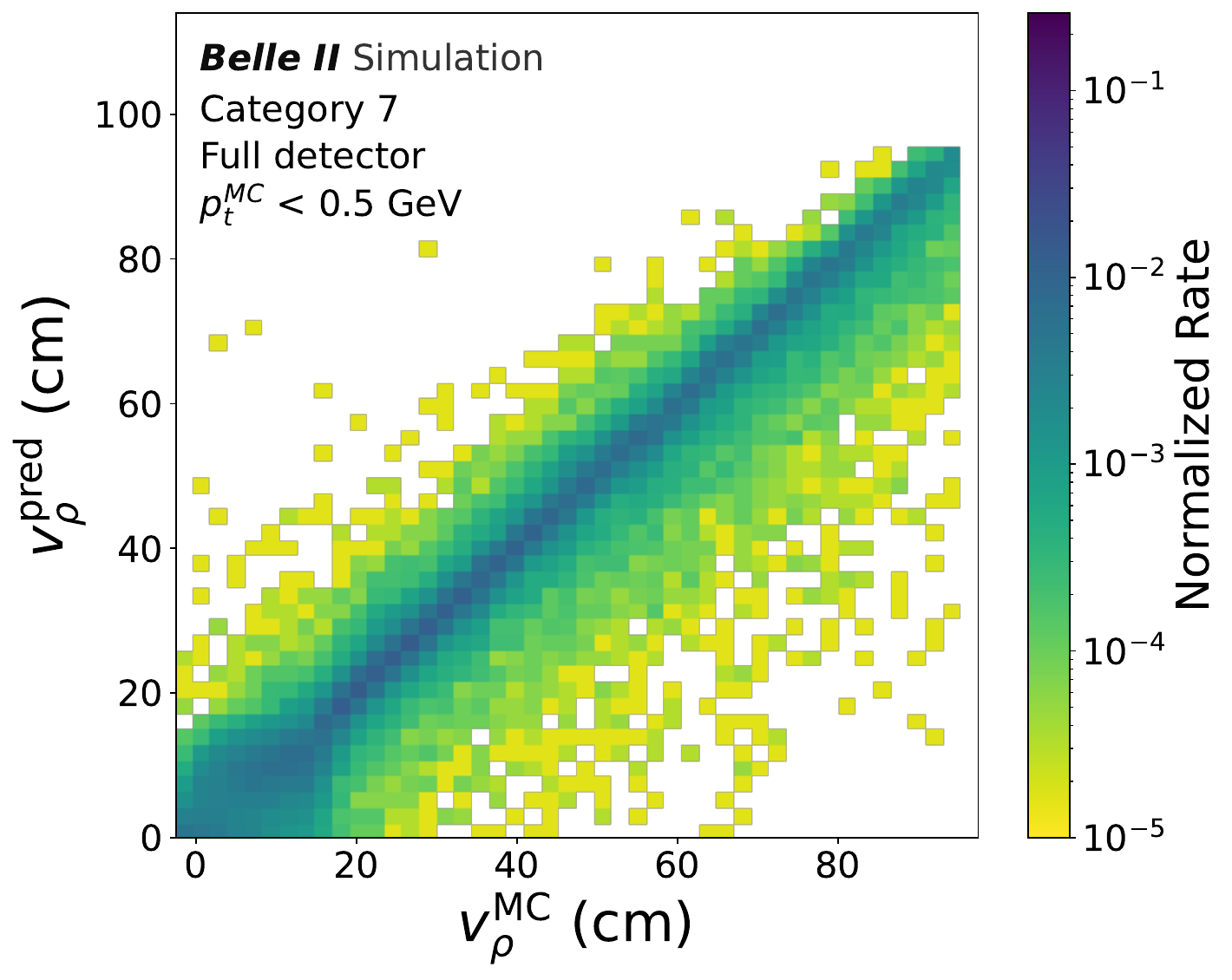}
         \caption{Displaced angled tracks (category 7), low $p_t < 0.5\,\gev$}
         \label{fig:pos:f}
     \end{subfigure}

\caption{Two-dimensional histograms showing the correlation between the reconstructed position of the \cat model output $v_{\rho}^{pred}$ in the $x-y$ plane and the simulated position $v_{\rho}^{MC}$ for (\subref{fig:pos:a})~\dh, (\subref{fig:pos:b})~displaced tracks with high transverse momentum, (\subref{fig:pos:c})~displaced angled tracks with high transverse momentum, (\subref{fig:pos:d})~\kshort, (\subref{fig:pos:e})~displaced tracks with low transverse momentum, and  (\subref{fig:pos:f})~displaced angled tracks with low transverse momentum, each with \databackground. }
\label{fig:position}
\end{figure*}

The position prediction of the \cat for truth-matched displaced tracks are shown in \cref{fig:position} for different displaced samples.\\

\begin{figure*}[ht!]
\centering
{\includegraphics[width=0.7\textwidth]{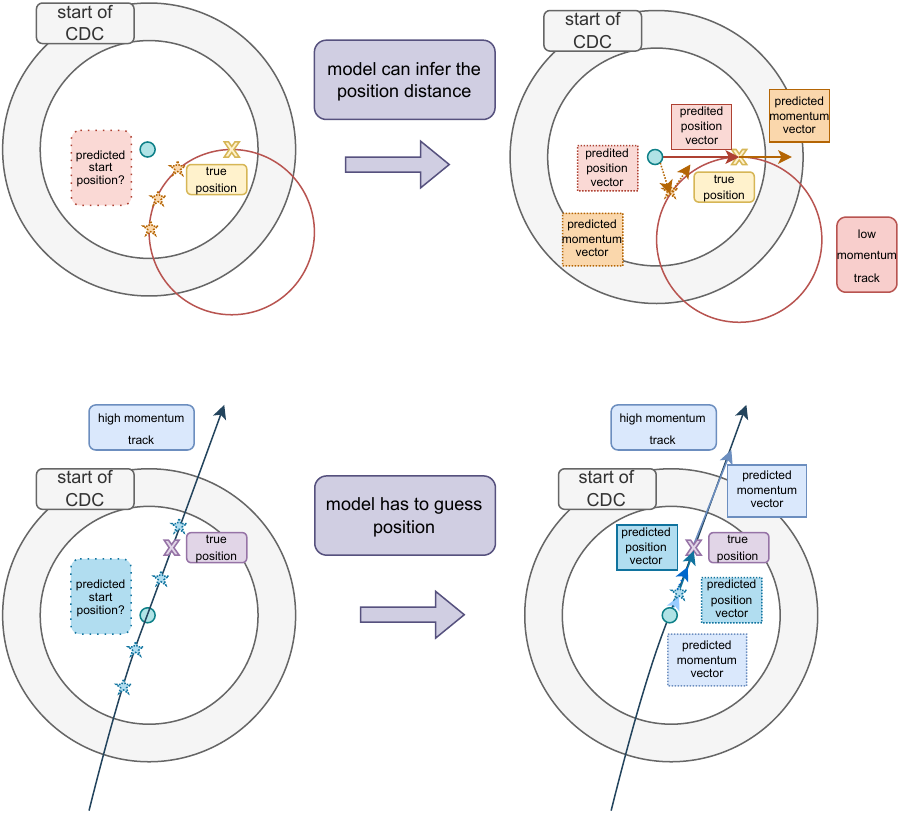}}
\caption{Illustration how the GNN learns to predict the starting position for particles with low transverse momentum even without the presence of another nearby track from a common vertex.}\label{fig:positionprediction}
\end{figure*}

For the sample with a displaced vertex (\dh and \kshort), the \cat is able to provide an unbiased prediction with reasonable resolution even in the inner part of the detector with no close-by CDC hits (see \cref{fig:pos:a} and \cref{fig:pos:d}).
This indicates that the \cat utilizes the information of nearby detector hits from the second track to actually infer a vertex position and not just the track starting position.
For individual tracks, we observe a rather complex and non-trivial behaviour of the GNN for particles with low transverse momentum:
For tracks originating within the CDC volume, the GNN learns a helical representation of the parameters, with the starting point located anywhere along the trajectory and the momentum vector tangential to the helix. 
Since the training samples are enhanced with prompt events and are biased towards tracks originating from the interaction point, with the negative momentum vector directed towards it, the network can infer the starting point by selecting it such that the negative momentum vector points back to the interaction point (see \cref{fig:positionprediction}). 
As a result, the \cat is able to infer the track starting point even for $v_{\rho}^{MC}\lesssim 20\,\text{cm}$ for tracks with low transverse momentum (see \cref{fig:pos:e}).
This is no longer the case for particles with higher $p_t$, because the trajectory increasingly approaches a straight line, with a constant momentum-direction vector along the particle trajectory almost everywhere from 0 to 16 cm where the CDC starts (see \cref{fig:pos:b}).

As a cross check, we tested the \cat performance on displaced tracks with non-pointing momentum vectors\footnote{The samples are similar to the \dh or \kshort samples but with only one track instead of a decay into two particles.} and observe no predictive power for $v_{\rho}^{MC}\lesssim 20\,\text{cm}$ as expected (see \cref{fig:pos:c} and \cref{fig:pos:f}).

\FloatBarrier
\subsection{Robustness to variable detector conditions}
\label{sec_results_beambackground}
The detector conditions are changing during \belletwo data taking due to changes in accelerator settings and during beam injections resulting in variations of the beam background conditions.
In addition, the hit position resolutions and wire efficiencies are not constant over long periods of data taking.
The reconstruction algorithms must be robust against moderate changes.
As described in \cref{sec_gnn}, we use a model pre-trained on lower background and finalize training for higher background.

While retraining models and re-optimizing hyperparameters may be required for optimal performance, we test the robustness of the \cat by comparing models trained and tested on the same beam background and wire efficiency maps, against models trained and tested on two different beam background environments and wire efficiency maps.

The performance metrics integrated over the full momenta range are summarized in \cref{tab:trainingsample_table_low_beambackground} for a model trained and evaluated on low simulated beam background. 
The same information for the model trained and evaluated on high data beam background is given in \cref{tab:trainingsample_table_high_data_beambackground}. 
For a model trained on low simulated beam background but evaluated on high data beam background, the information is given in \cref{tab:trainingsample_finding_rd_beambackground}.

The track performance when moving from low simulated beam background to high data beam background (i.e. comparing \cref{tab:trainingsample_table_low_beambackground} and \cref{tab:trainingsample_table_high_data_beambackground}) decreases significantly for the \legendre in the endcaps.
A performance reduction is also visible for the \cat, but to a much smaller extend.

When evaluating the model pretrained on low simulated beam background on high data beam background, the performance is slightly worse compared to the model trained on high data beam background (i.e. comparing \cref{tab:trainingsample_table_high_data_beambackground} and \cref{tab:trainingsample_finding_rd_beambackground}).
The track fitting efficiency in the barrel is reduced by about 1 percent point, in the endcaps by about 3-4 percent points.
The fake rate however is slightly lower for the pre-trained model, indicating that hyperparameter optimization may be enough to recover the efficiency loss.
We note that even for the non-optimal model, the \cat outperforms the \legendre in all detector regions.
This demonstrates a robust generalisation with respect to different levels of beam background without the need to retrain the GNNs frequently.

\begin{table*}
    \fontsize{6pt}{6pt}\selectfont
    \centering
    \caption{The performance metrics for the prompt evaluation samples (category 1-3, see \cref{tab:samples} and \cref{sec_dataset} for details) for non-curling tracks for \cat and \legendre in different detector regions for a model trained and evaluated on \textit{low simulated beam-background}. Uncertainties below $<$0.01\% are not shown in the table.}
    \begin{tabular}{r ccc cc}
         \toprule
         
        (in \%)& \trackeff & \fakerate & \clonerate &\trackchareff & \wrongchargerate\\
      \midrule
& \multicolumn{5}{c}{forward endcap} \\
\midrule
Baseline Finder & $87.05^{+0.09}_{-0.09}$ & $0.83^{+0.03}_{-0.03}$ & $0.01^{}_{}$ & $84.97^{+0.1}_{-0.1}$ & $2.39^{+0.05}_{-0.05}$ \\
CAT Finder & $99.26^{+0.02}_{-0.02}$ & $1.02^{+0.03}_{-0.03}$ & $0.15^{+0.01}_{-0.01}$ & $99.22^{+0.02}_{-0.02}$ & $0.03^{}_{-0.01}$ \\
\midrule
Baseline Fitter & $85.18^{+0.1}_{-0.1}$ & $0.78^{+0.03}_{-0.03}$ & $0.01^{}_{}$ & $84.3^{+0.1}_{-0.1}$ & $0.99^{+0.03}_{-0.03}$ \\
CAT Fitter & $97.12^{+0.05}_{-0.05}$ & $0.32^{+0.02}_{-0.02}$ & $0.04^{+0.01}_{-0.01}$ & $96.42^{+0.05}_{-0.05}$ & $0.72^{+0.02}_{-0.02}$ \\
\midrule
 \midrule
& \multicolumn{5}{c}{barrel} \\
\midrule
Baseline Finder & $98.71^{+0.03}_{-0.03}$ & $2.06^{+0.04}_{-0.04}$ & $0.03^{}_{}$ & $96.73^{+0.05}_{-0.05}$ & $2.0^{+0.04}_{-0.04}$ \\
CAT Finder & $99.72^{+0.01}_{-0.01}$ & $2.15^{+0.04}_{-0.04}$ & $0.42^{+0.02}_{-0.02}$ & $99.4^{+0.02}_{-0.02}$ & $0.33^{+0.02}_{-0.02}$ \\
\midrule
Baseline Fitter & $97.68^{+0.04}_{-0.04}$ & $1.75^{+0.04}_{-0.04}$ & $0.01^{}_{}$ & $96.27^{+0.05}_{-0.05}$ & $1.44^{+0.03}_{-0.03}$ \\
CAT Fitter & $98.13^{+0.04}_{-0.04}$ & $0.97^{+0.03}_{-0.03}$ & $0.13^{+0.01}_{-0.01}$ & $97.97^{+0.04}_{-0.04}$ & $0.17^{+0.01}_{-0.01}$ \\
\midrule
 \midrule
& \multicolumn{5}{c}{backward endcap} \\
\midrule
Baseline Finder & $69.5^{+0.1}_{-0.1}$ & $0.72^{+0.03}_{-0.03}$ & $0.02^{}_{}$ & $66.2^{+0.1}_{-0.1}$ & $4.66^{+0.07}_{-0.07}$ \\
CAT Finder & $98.54^{+0.03}_{-0.03}$ & $0.75^{+0.02}_{-0.02}$ & $0.11^{+0.01}_{-0.01}$ & $98.43^{+0.03}_{-0.03}$ & $0.11^{+0.01}_{-0.01}$ \\
\midrule
Baseline Fitter & $67.8^{+0.1}_{-0.1}$ & $0.63^{+0.03}_{-0.03}$ & $0.02^{}_{}$ & $65.8^{+0.1}_{-0.1}$ & $2.98^{+0.06}_{-0.06}$ \\
CAT Fitter & $95.12^{+0.06}_{-0.06}$ & $0.3^{+0.02}_{-0.02}$ & $0.03^{}_{-0.01}$ & $91.66^{+0.08}_{-0.08}$ & $3.63^{+0.05}_{-0.05}$ \\
\midrule

        \bottomrule
        \\
    \end{tabular}
    \label{tab:trainingsample_table_low_beambackground}
\end{table*}

\begin{table*}
    \fontsize{6pt}{6pt}\selectfont
    \centering
    \caption{The performance metrics for the evaluation samples for different track finding algorithms in different detector regions evaluated on \textit{high data beam background}, but trained on \textit{low simulated beam background} for non-curling tracks. 
    Uncertainties below $<$0.01\% are not shown in the table.}
    \begin{tabular}{r ccc cc}
         \toprule
        (in \%)& \trackeff & \fakerate & \clonerate &\trackchareff & \wrongchargerate\\
 \midrule
& \multicolumn{5}{c}{forward endcap} \\
\midrule
Baseline Finder & $80.1^{+0.1}_{-0.1}$ & $0.55^{+0.02}_{-0.02}$ & $0.01^{}_{}$ & $78.4^{+0.1}_{-0.1}$ & $2.06^{+0.04}_{-0.04}$ \\
CAT Finder & $97.73^{+0.04}_{-0.04}$ & $1.46^{+0.03}_{-0.03}$ & $0.59^{+0.02}_{-0.02}$ & $97.18^{+0.05}_{-0.05}$ & $0.56^{+0.02}_{-0.02}$ \\
\midrule
Baseline Fitter & $78.1^{+0.1}_{-0.1}$ & $0.49^{+0.02}_{-0.02}$ & $0.01^{}_{}$ & $77.1^{+0.1}_{-0.1}$ & $1.37^{+0.04}_{-0.04}$ \\
CAT Fitter & $93.89^{+0.07}_{-0.07}$ & $0.22^{+0.01}_{-0.01}$ & $0.19^{+0.01}_{-0.01}$ & $91.93^{+0.08}_{-0.08}$ & $2.09^{+0.04}_{-0.04}$ \\
\midrule
 \midrule
& \multicolumn{5}{c}{barrel} \\
\midrule
Baseline Finder & $97.97^{+0.04}_{-0.04}$ & $2.31^{+0.04}_{-0.04}$ & $0.05^{+0.01}_{-0.01}$ & $95.92^{+0.06}_{-0.06}$ & $2.09^{+0.04}_{-0.04}$ \\
CAT Finder & $99.5^{+0.02}_{-0.02}$ & $2.25^{+0.04}_{-0.04}$ & $1.94^{+0.04}_{-0.04}$ & $98.69^{+0.03}_{-0.03}$ & $0.82^{+0.03}_{-0.03}$ \\
\midrule
Baseline Fitter & $96.88^{+0.05}_{-0.05}$ & $1.83^{+0.04}_{-0.04}$ & $0.03^{}_{}$ & $95.5^{+0.06}_{-0.06}$ & $1.42^{+0.03}_{-0.03}$ \\
CAT Fitter & $96.89^{+0.05}_{-0.05}$ & $0.76^{+0.02}_{-0.02}$ & $0.45^{+0.02}_{-0.02}$ & $96.46^{+0.05}_{-0.05}$ & $0.44^{+0.02}_{-0.02}$ \\
\midrule
 \midrule
& \multicolumn{5}{c}{backward endcap} \\
\midrule
Baseline Finder & $60.5^{+0.1}_{-0.1}$ & $1.08^{+0.04}_{-0.04}$ & $0.03^{+0.01}_{-0.01}$ & $58.0^{+0.1}_{-0.1}$ & $4.08^{+0.07}_{-0.07}$ \\
CAT Finder & $95.02^{+0.06}_{-0.06}$ & $1.03^{+0.03}_{-0.03}$ & $0.4^{+0.02}_{-0.02}$ & $94.28^{+0.06}_{-0.06}$ & $0.78^{+0.02}_{-0.03}$ \\
\midrule
Baseline Fitter & $58.8^{+0.1}_{-0.1}$ & $0.92^{+0.03}_{-0.04}$ & $0.02^{}_{-0.01}$ & $56.8^{+0.1}_{-0.1}$ & $3.28^{+0.06}_{-0.06}$ \\
CAT Fitter & $87.55^{+0.09}_{-0.09}$ & $0.44^{+0.02}_{-0.02}$ & $0.16^{+0.01}_{-0.01}$ & $82.8^{+0.1}_{-0.1}$ & $5.4^{+0.07}_{-0.07}$ \\
\midrule
        \bottomrule
    \end{tabular}
    \label{tab:trainingsample_finding_rd_beambackground}
\end{table*}

\FloatBarrier

\subsection{Lessons learned}
\label{sec_results_lessonslearned}
During the training and evaluation of the GNN, we faced various challenges that impacted the model performance and interpretability of the results. 

\begin{itemize}
    \item The model exhibits overfitting to physical correlations when displaced vertex samples with physical constraints are included in the training dataset.
Early trainings included the \kshort and \dh evaluation samples (see Sec.\,\ref{sec_dataset}) also in the training data sets.
The model identified that only \dh events exhibited displaced vertices with significant displacements and captured the absolute momentum scale of the two decay particles, which together sum to half the collision energy at \belletwo.
This affected only the GNN's parameter inference but not the fit results of GENFIT2 which mostly rely on the correct momentum direction but not on the absolute value for the initialisation of the fit.
Removing physical samples from the training resolved that issue with negligible loss of performance.

\item When the model is trained solely on prompt tracks and displaced vertices, the training process is generally unstable and fails to achieve adequate performance.
The inclusion of displaced and displaced-angle tracks as intermediate samples, bridging the two event topologies, helps mitigate this issue.

\item The definition of the related and matched tracks~(see Sec.\ref{sec_Metrics}) if secondary particles are produced during the full simulation has a significant impact on the training stability and performance of the model.
A large number of secondary particles, primarily low-momentum electrons, produce highly localized, cluster-like energy depositions, which the model learned to interpret as track-like signatures.
This in turn significantly increased the fake and clone rates and complicated optimization metrics.
Removing secondary particles results in more stable training.
For both the \legendre and the \cat, including the secondary particles in the evaluation reduces the track finding efficiency.

\item Earlier versions of our trainings used separate training samples with events that either had only low transverse momentum particles or only high transverse momentum particles.
Evaluation on these samples showed good performance, but evaluation on samples that contained events with both low and high transverse momentum particles showed very low efficiency for low momentum particles.
We attribute this behaviour to the fact that the training sample with high transverse momentum signal tracks contained low transverse momentum beam background tracks, but not vice versa.
Enriching all samples with a rather large number of signal tracks with low transverse momentum mitigated this problem.

\item The network surprised us with the ability to predict the track starting point even if no detector hits or additional tracks forming a vertex were nearby.
We described the underlying mechanism in Sec.\,\ref{subsec:position} and conclude that this is an interesting feature with no real practical relevance, since physical samples will not contain such tracks that violate momentum conservation.

\item We observed up to a 10\% variation in validation loss due to differences in repeated training and model initialization, leading to a 0.5\% variation in track finding and fitting charge efficiency across specific samples and detector regions. 
Since the loss was not the primary target of our optimization, we accepted this variation. 
Potential mitigation strategies, such as improved model initialization, repeated trainings, or larger batch sizes, can be explored in future work.

\item The hyper-parameter optimization of the latent space parameters and in particular of the working points $t_D$ and $t_h$ (see Sec.\,\ref{sec_gnn}) is very challenging if one wants to balance optimal performance not only for prompt tracks, but also for different displaced signatures.
On the other hand, it is rather straight forward to optimize the GNN for specific signatures if physics requirements can be restricted to certain decay topologies or momentum ranges.

\end{itemize}

\section{Conclusion and Outlook}\label{sec_conclusion}
We have presented the implementation and a detailed study of the \cat~(\textit{CDC AI Tracking}), an end-to-end multi-track reconstruction algorithm utilizing graph neural networks (GNNs) for the \belletwo central drift chamber.
The \cat uses detector hits as inputs and simultaneously predicts the number of track candidates in an event, their associated hits, and their kinematic properties.
We have used a full detector simulation and included beam backgrounds from actual collision data, and compared the GNN-based algorithm to the baseline track finding algorithm currently used in \belletwo for a wide range of event signatures, including tracks from decays that are macroscopically displaced from the interaction point.
We find significant improvements in track finding efficiencies for displaced tracks in the barrel, and for both prompt and displaced tracks in the detector endcaps.
The combined track-finding and fitting efficiency, as well as the low fake track rates for prompt tracks, are both comparable to the existing \belletwo track-finding method in the barrel region.
To fully capitalize on the enhanced track-finding efficiency of the \cat, future adjustments to the track fitting procedure will be necessary.

This work represents a significant conceptual step towards enabling end-to-end real-time GNN-based tracking on FPGAs \cite{Neu:2023sfh}.
For real-time trigger applications, the clusterisation based on condensation points, the hit ordering in real space,  and the track fitting step can be omitted if the inferred kinematics from the GNN provide sufficient resolution for trigger decisions.
However, future work is needed on input quantization, pre-processing to remove beam background, as well as network size reduction.

To our knowledge, the \cat is the first end-to-end machine learning tracking algorithm that has been utilized in a realistic particle physics environment, and the first completely GNN-based track finding in a drift chamber detector.

\backmatter%
\bmhead{Data Availability Statement}%
The datasets generated and analysed during the current study are property of the \belletwo collaboration and not publicly available.
The Belle~II software is available at \cite{basf21, basf22}.
The instructions and code to replicate the studies in this paper are available at \cite{catfinder}.

\bmhead{Acknowledgments}
It is a great pleasure to thank (in alphabetical order) Isabel~Haide, Jan~Kieseler, and Yannis~Kluegl for discussions.\\

The training of the models was performed on the TOpAS GPU cluster at the Scientific Computing Center~(SCC) at KIT.
This work is funded by BMBF ErUM-Pro~05H24VK1.\\

\bmhead{Conflict of interest}
The  authors  declare  that  they  have  no  conflict  of  interest.

\clearpage
\bibliographystyle{unsrt} 
\bibliography{bibliography}

\FloatBarrier
\clearpage

\begin{appendices}
\crefalias{section}{appendix}

\section{Data set event displays}\label{app:displays}
\begin{figure*}[ht]
     \centering
     \begin{subfigure}[t]{\thirdwidth\textwidth}
         \centering
         \includegraphics[width=\textwidth]{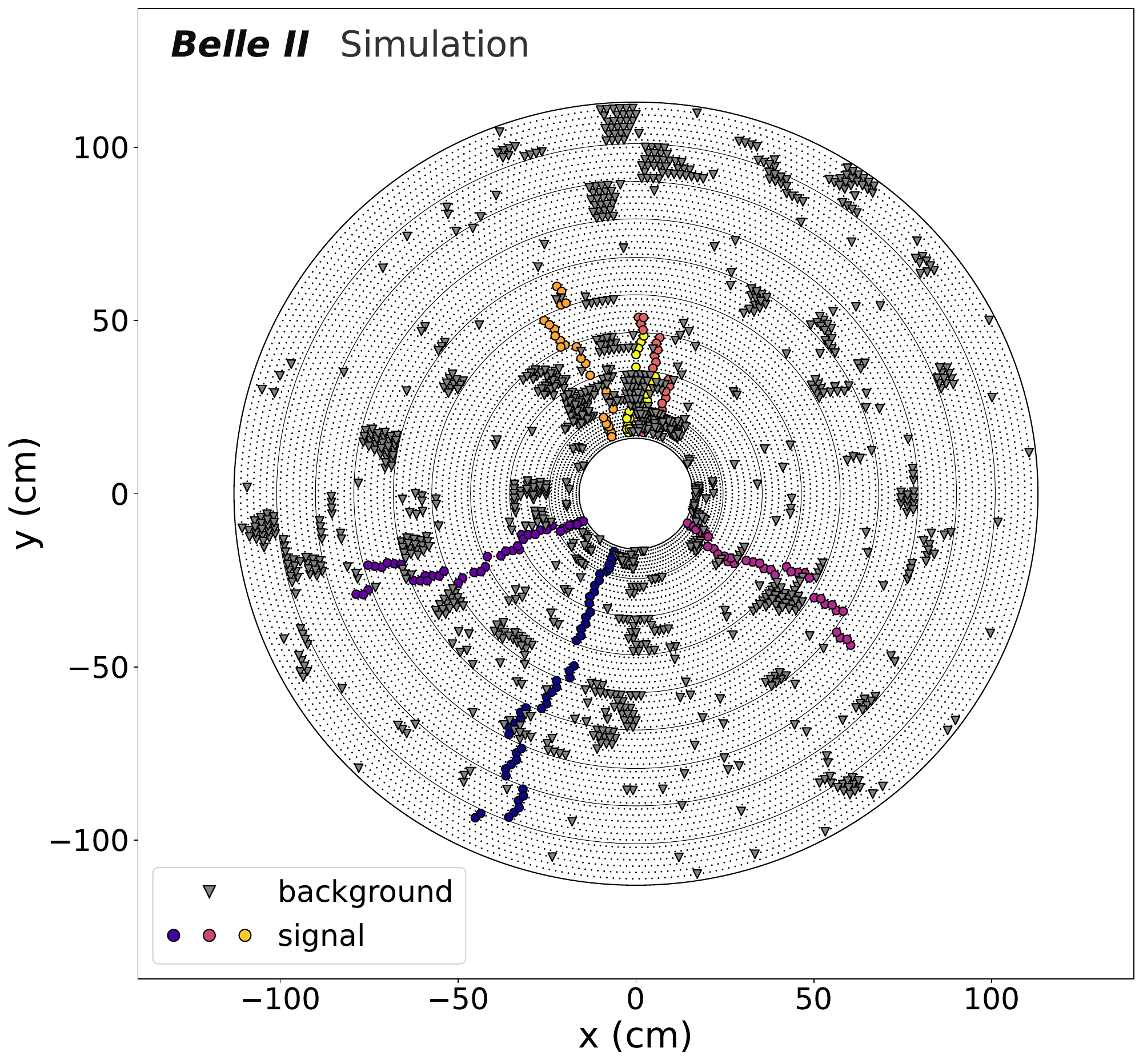}
         \caption{Prompt forward~(category~1)}
         \label{fig:prompt_barrel}
     \end{subfigure}
     \hfill
     \begin{subfigure}[t]{\thirdwidth\textwidth}
         \centering
         \includegraphics[width=\textwidth]{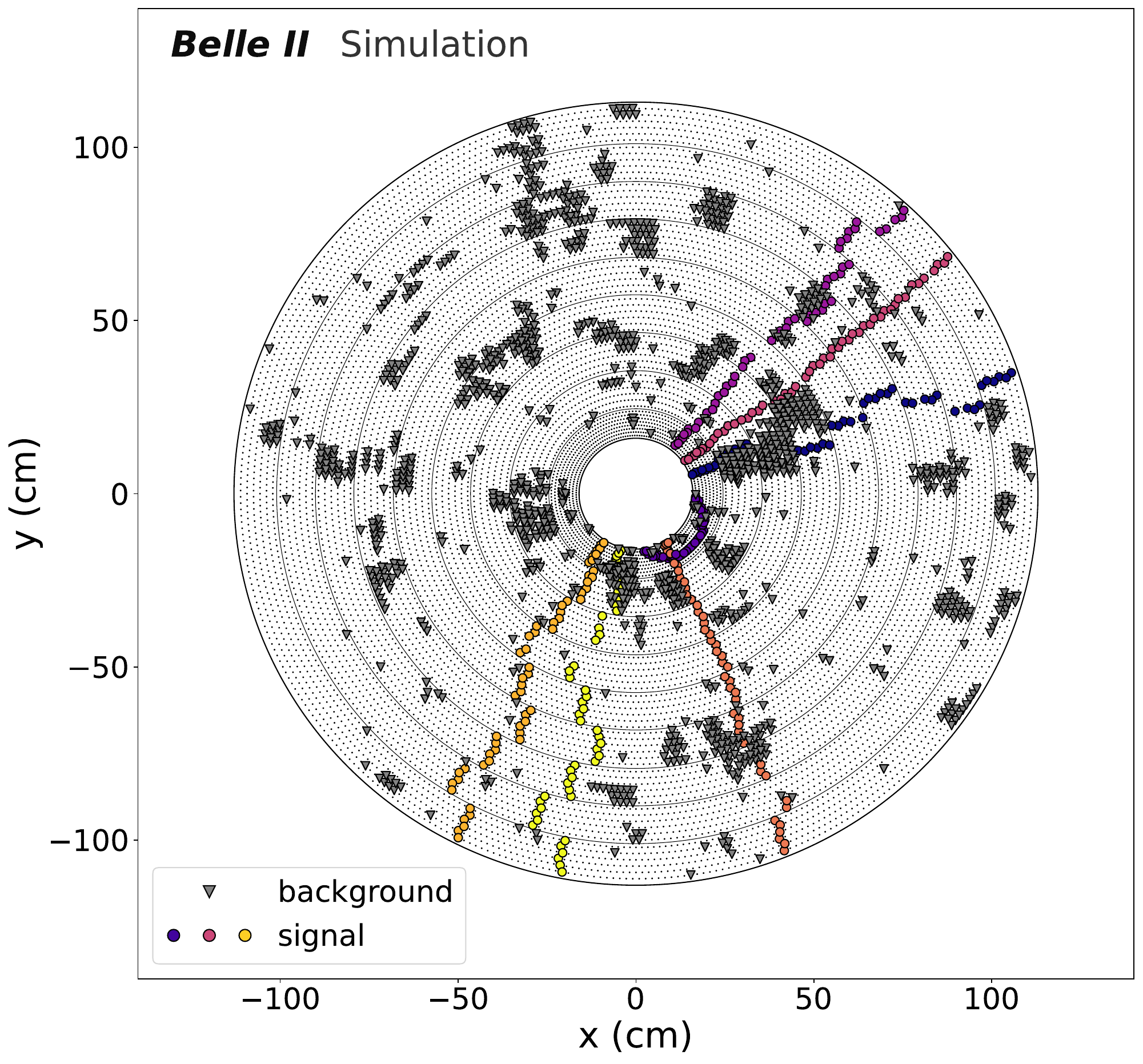}
         \caption{Prompt barrel~(category~2)}
         \label{fig:prompt_fwd}
     \end{subfigure}
     \hfill
     \begin{subfigure}[t]{\thirdwidth\textwidth}
         \centering
         \includegraphics[width=\textwidth]{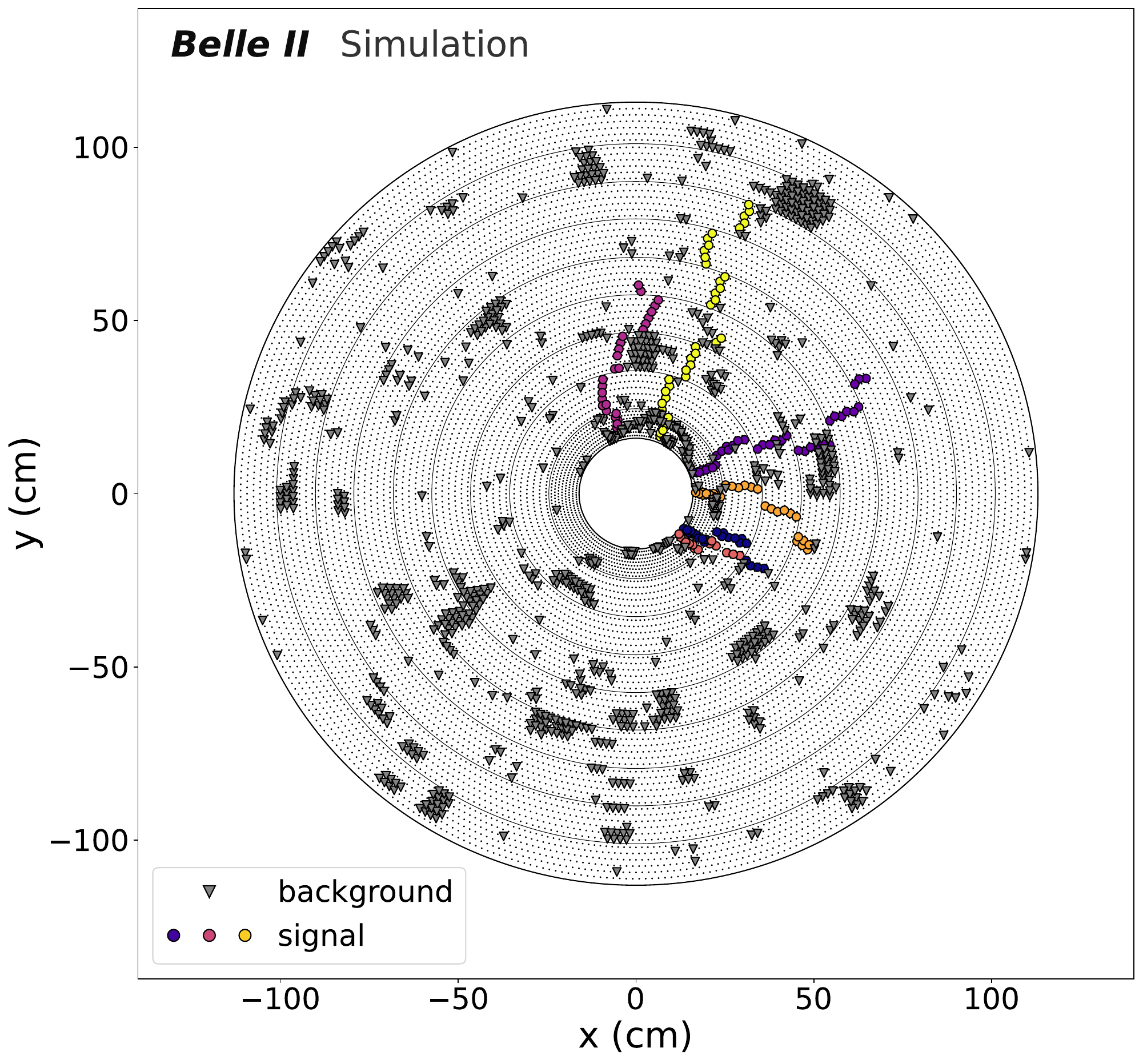}
         \caption{Prompt backward~(category~3)}
         \label{fig:prompt_bwd}
     \end{subfigure}\\

       \begin{subfigure}[t]{\thirdwidth\textwidth}
         \centering
         \includegraphics[width=\textwidth]{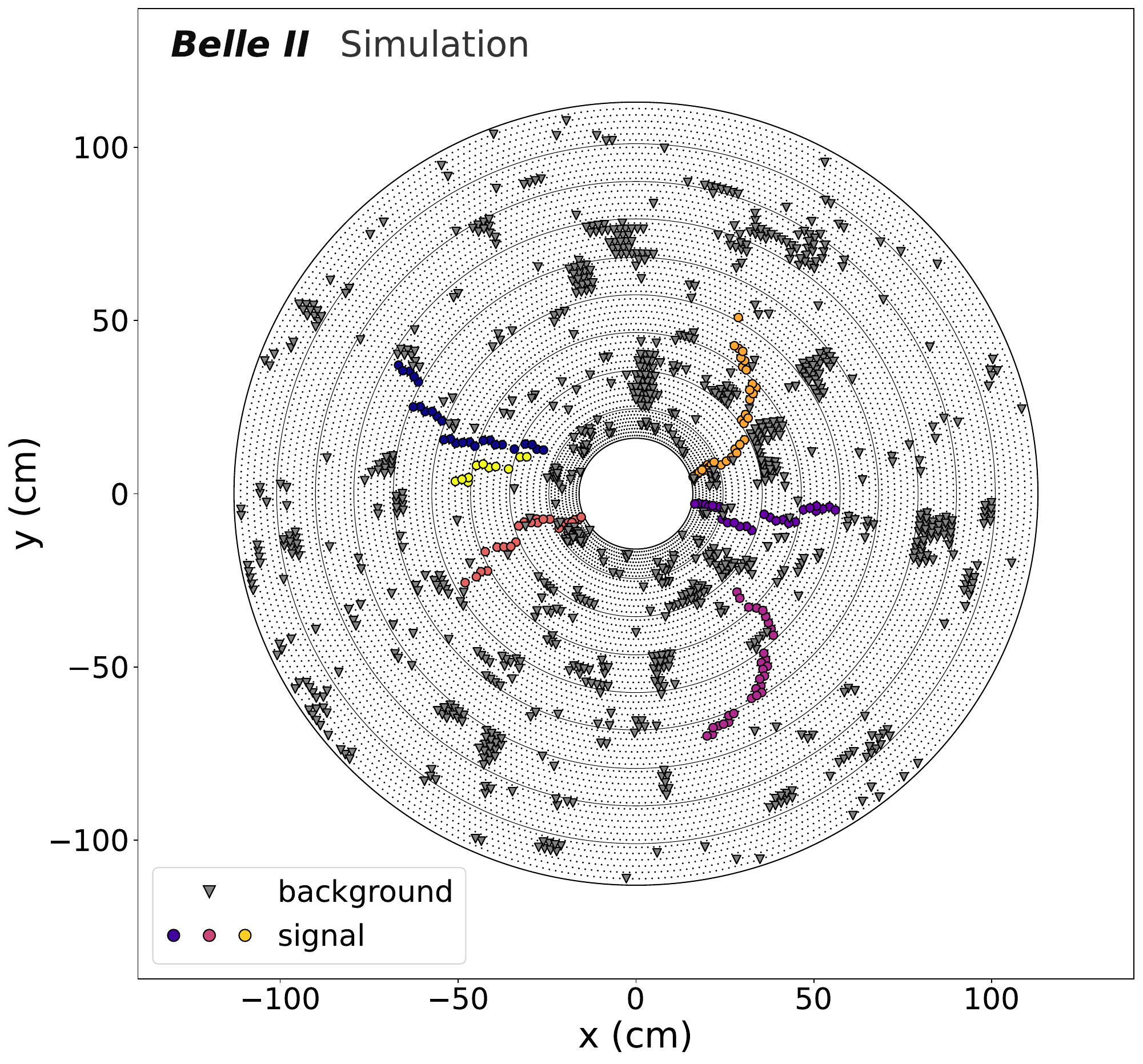}
         \caption{Displaced forward~(category~4)}
         \label{fig:displaced_barrel}
     \end{subfigure}
     \hfill
     \begin{subfigure}[t]{\thirdwidth\textwidth}
         \centering
         \includegraphics[width=\textwidth]{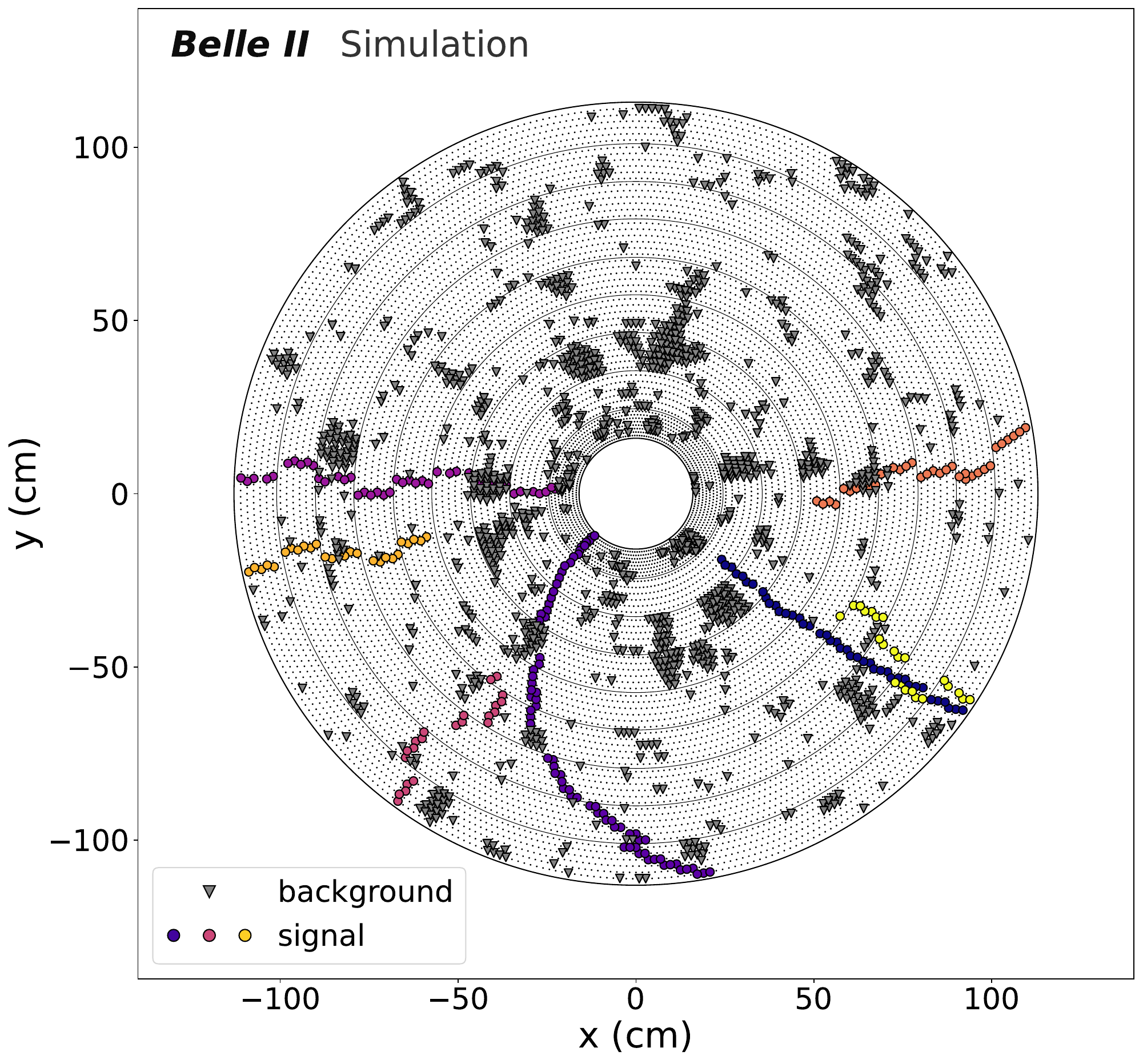}
         \caption{Displaced barrel~(category~5)}
         \label{fig:displaced_fwd}
     \end{subfigure}
     \hfill
     \begin{subfigure}[t]{\thirdwidth\textwidth}
         \centering
         \includegraphics[width=\textwidth]{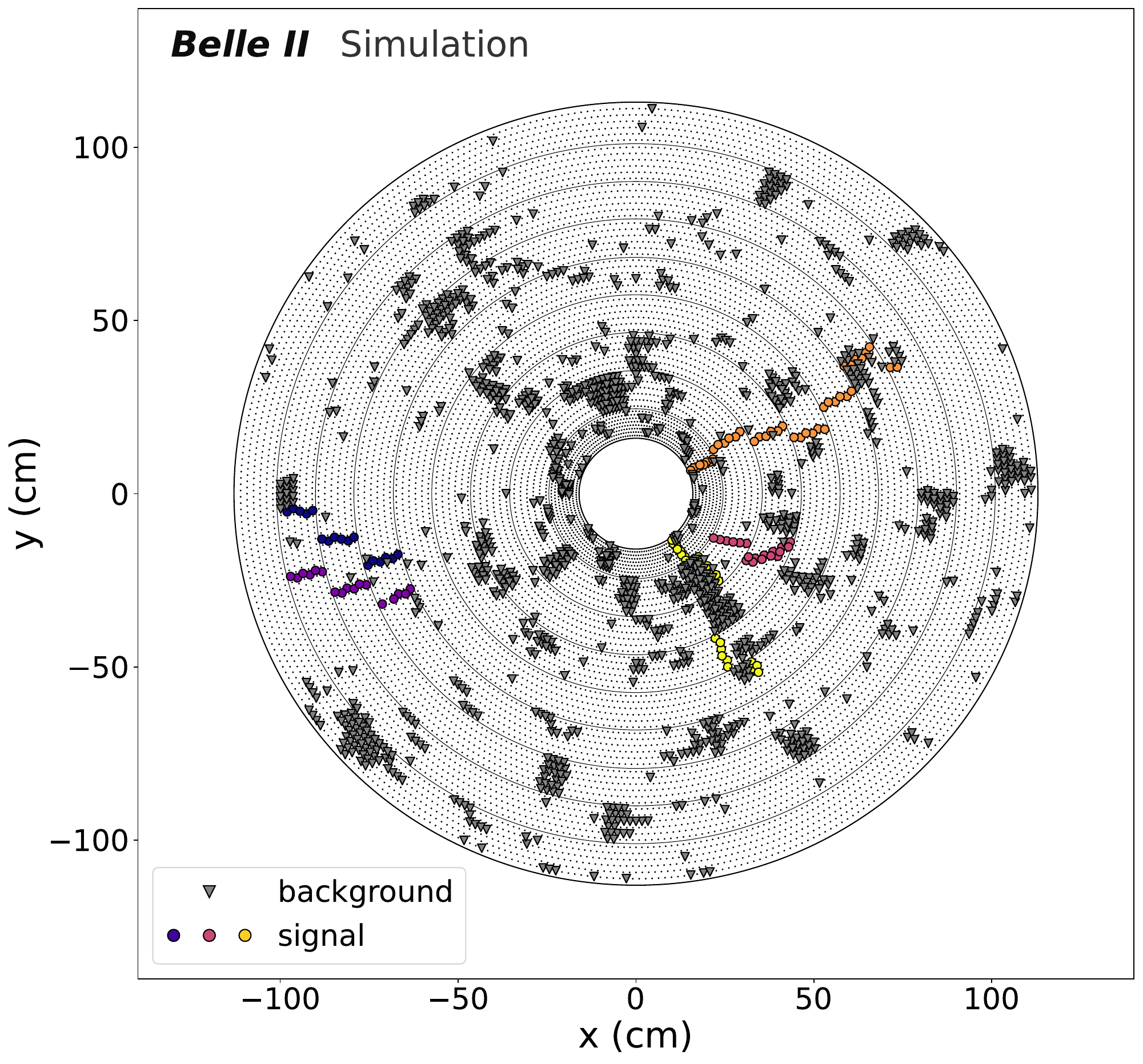}
         \caption{Displaced backward~(category~6)}
         \label{fig:displaced_bwd}
     \end{subfigure}\\

       \begin{subfigure}[t]{\thirdwidth\textwidth}
         \centering
         \includegraphics[width=\textwidth]{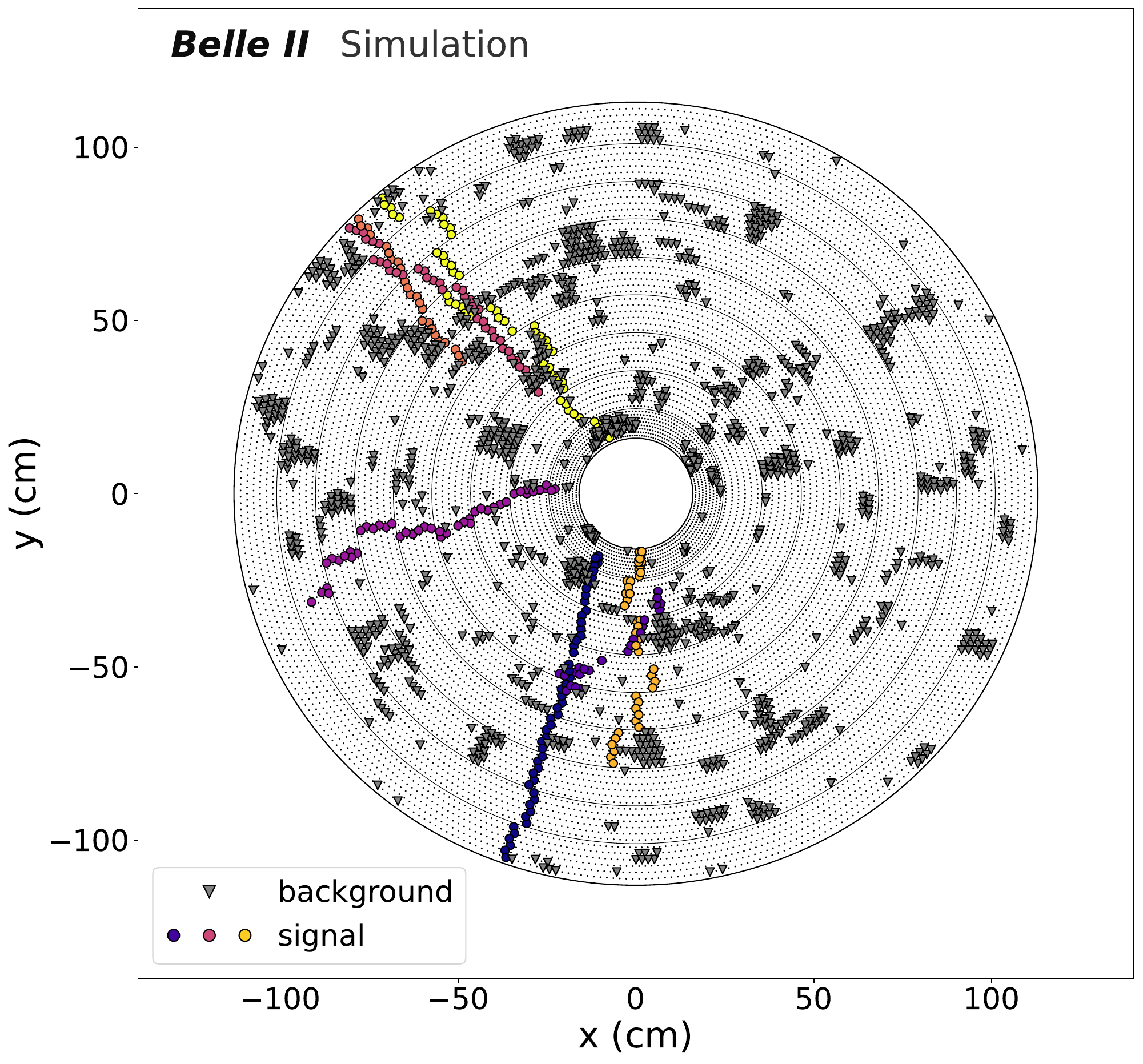}
         \caption{Displaced angled~(category~7)}
         \label{fig:prompt_angled}
     \end{subfigure}
     \hfill
     \begin{subfigure}[t]{\thirdwidth\textwidth}
         \centering
         \includegraphics[width=\textwidth]{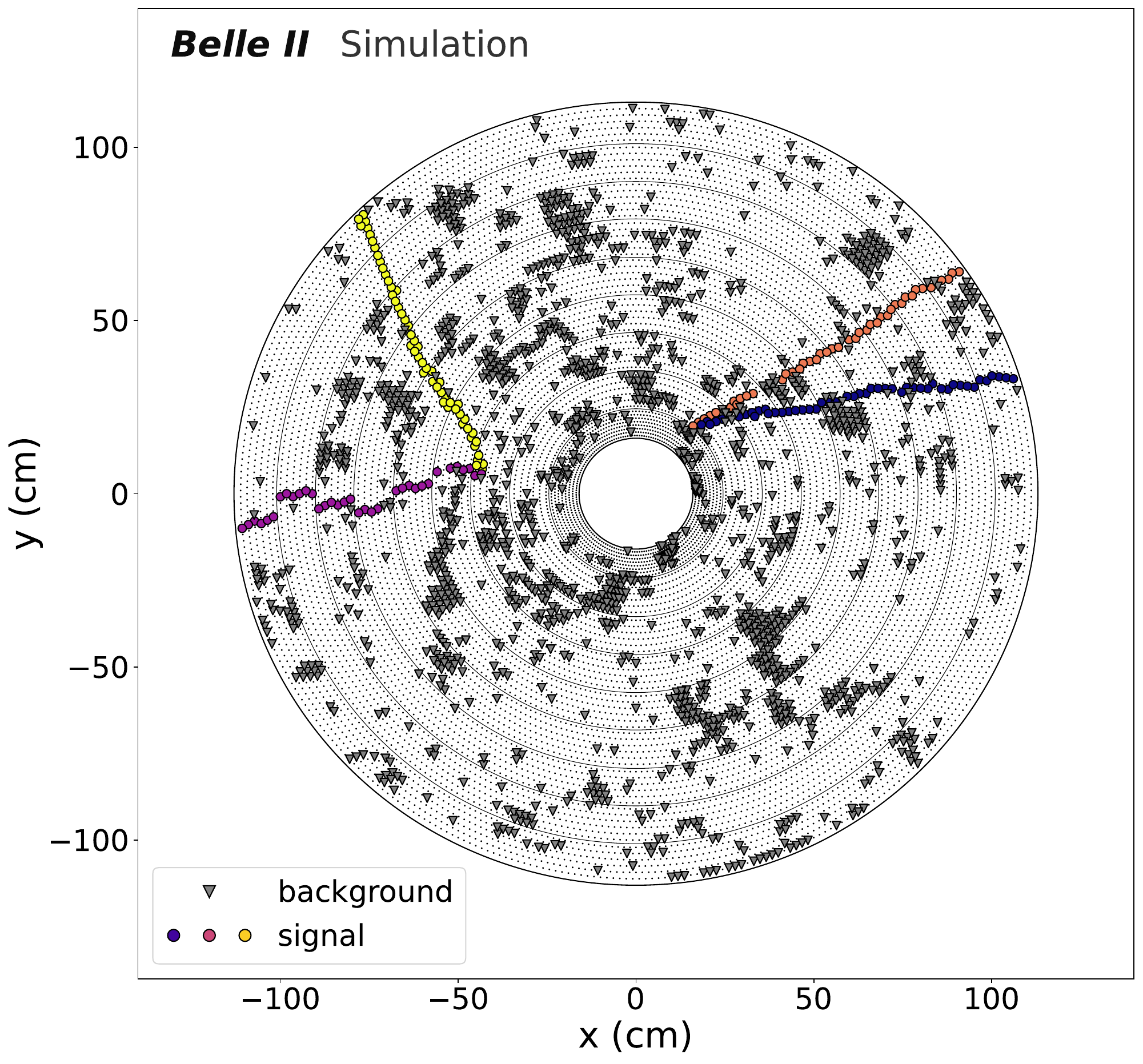}
         \caption{Vertex~(category~9+10)}
         \label{fig:prompt_zadj}
     \end{subfigure}
     \hfill
     \begin{subfigure}[t]{\thirdwidth\textwidth}
         \centering
         \includegraphics[width=\textwidth]{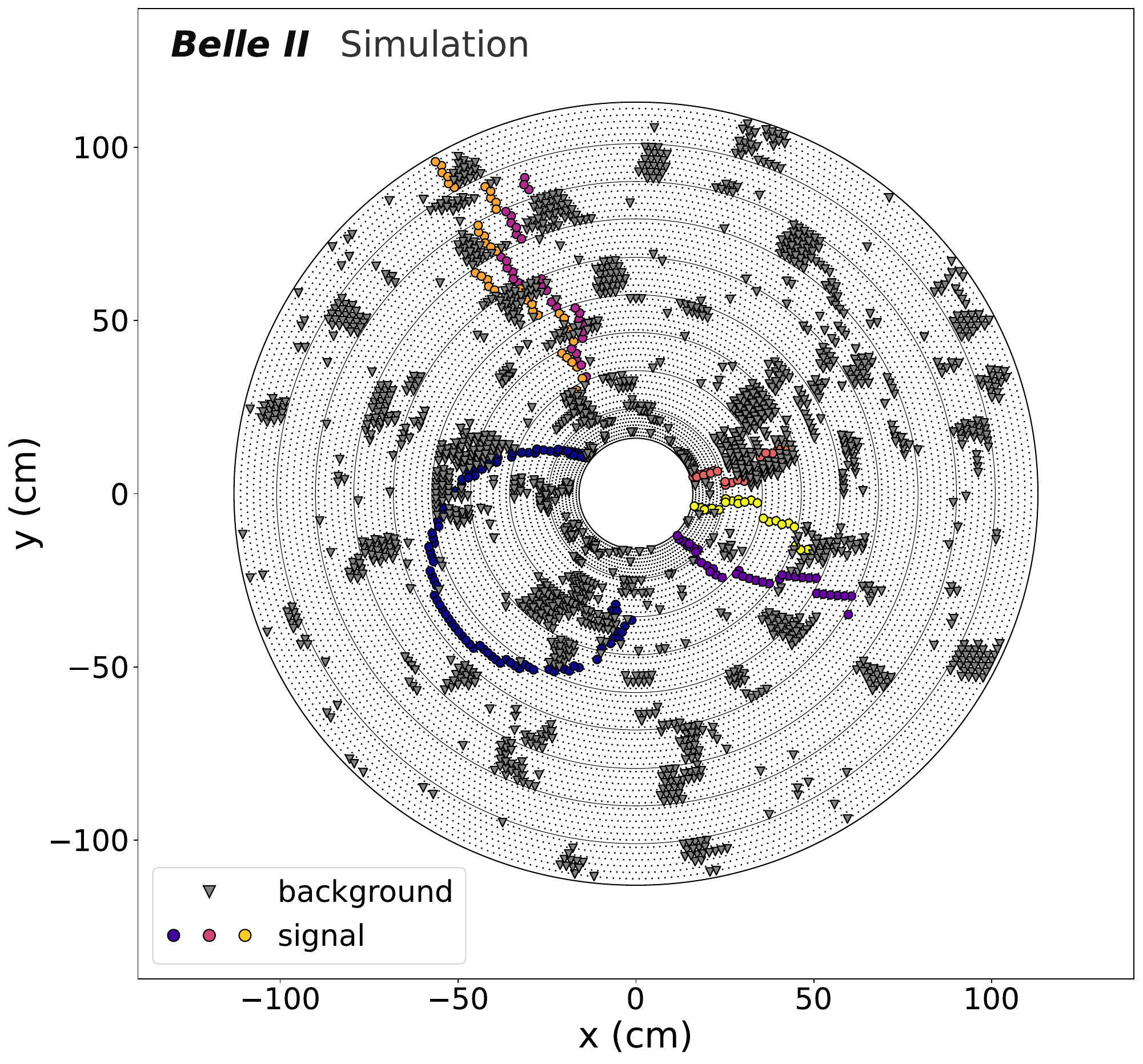}
         \caption{Mix~(category~11)}
         \label{fig:displaced_vertex}
     \end{subfigure}\\     
    \caption{Typical event displays showing examples of the different training samples for \textit{high data beam backgrounds}. 
Filled colored circular markers show signal hits, filled gray triangular markers show background hits. 
The markers correspond to the locations of the sense wires at the $z$ position of the center of the wire for the wires with recorded ADC signals.}
    \label{fig:trainingsamples}
\end{figure*}

Typical event displays showing examples of the different training samples as described in \cref{sec_dataset} and \cref{tab:samples} are shown in \cref{fig:trainingsamples}.

\clearpage
\FloatBarrier

\section{CDC wire inefficiencies}\label{app:wireeff}
\begin{figure}[ht!]
         \centering
         \includegraphics[width=\halfwidth\textwidth]{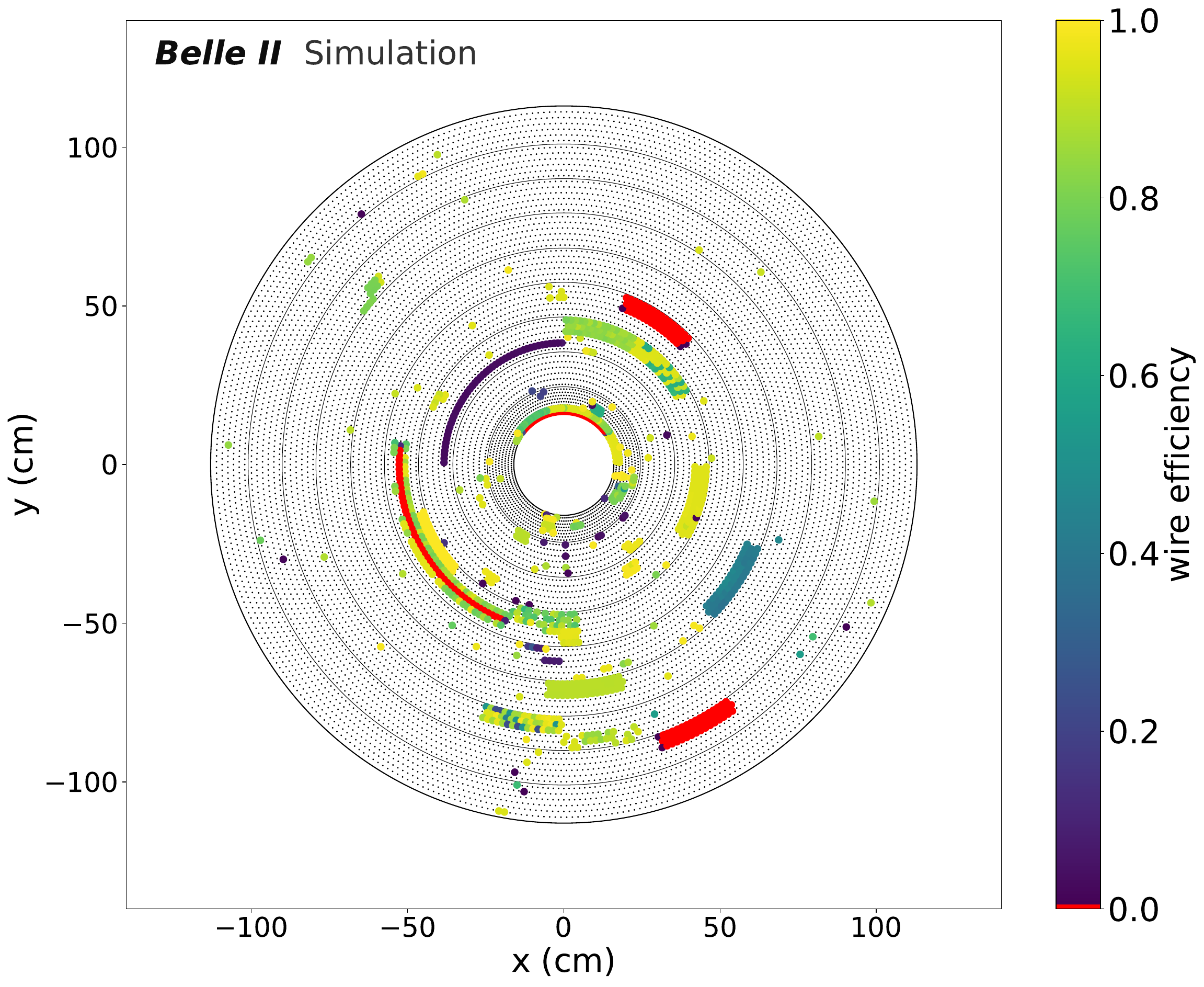}
\caption{Overview of the simulated wire efficiency for the default simulated samples. Coloured wires have an efficiency below 1. Red wires have an efficiency of 0.}
\label{fig:cdcwires}
\end{figure}

\begin{figure}[ht!]
         \centering
         \includegraphics[width=\halfwidth\textwidth]{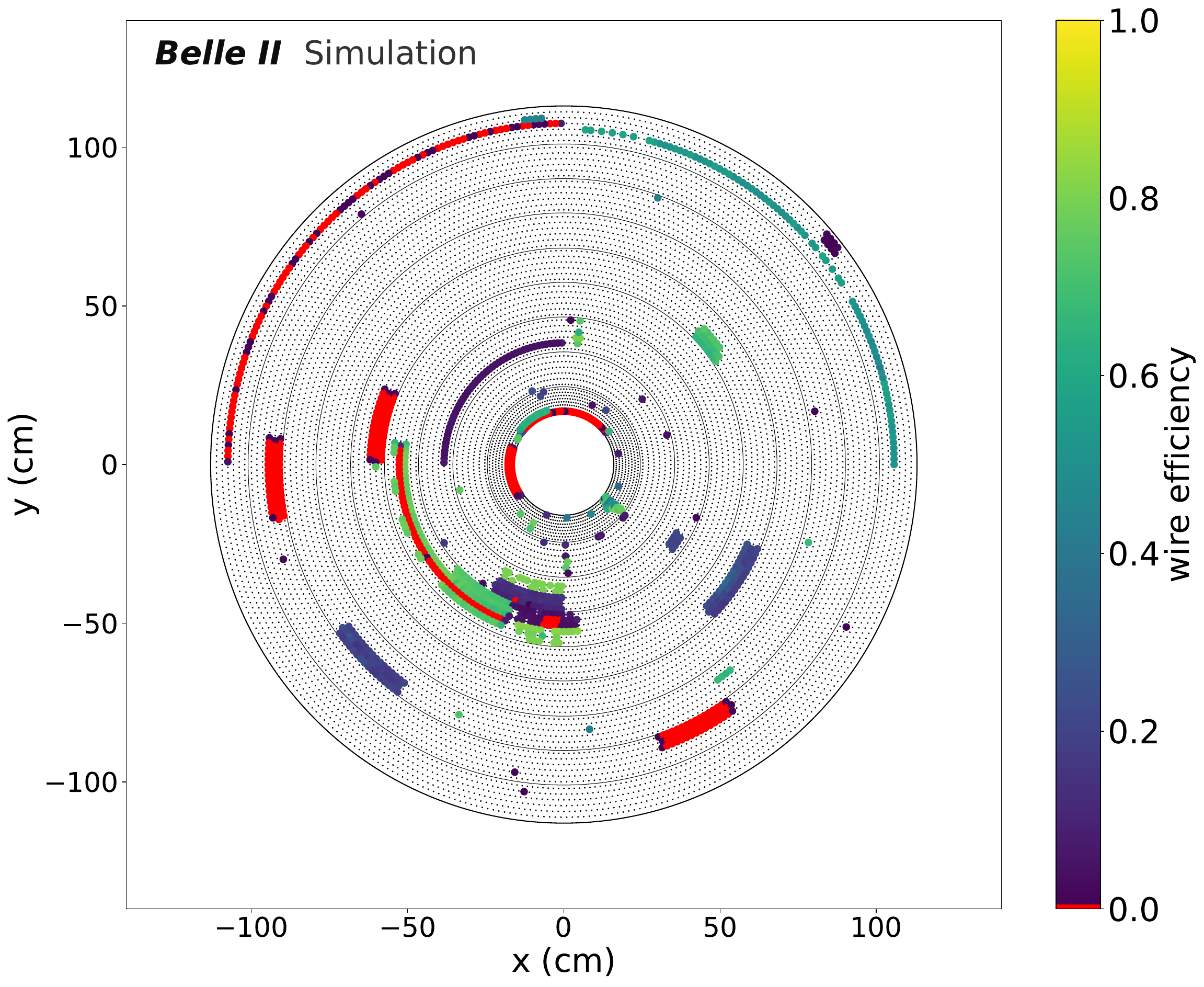}
\caption{Overview of the simulated wire efficiency for the simulated samples with beam background and detector conditions taken from data. Coloured wires have an efficiency below 1. Red wires have an efficiency of 0.}
\label{fig:cdcwires2}
\end{figure}

Displays of the wire efficiencies and dead wires used for the low simulated beam background conditions are shown in \cref{fig:cdcwires}.
Displays of the wire efficiencies and dead wires used for the high data beam background conditions are shown in \cref{fig:cdcwires2}.

\clearpage
\FloatBarrier

\section{Track charge efficiency for category~1-3}\label{app:tceff}
\begin{figure*}[ht!]
     \centering
     \begin{subfigure}[b]{\thirdwidth\textwidth}
         \centering
         \includegraphics[width=\textwidth]{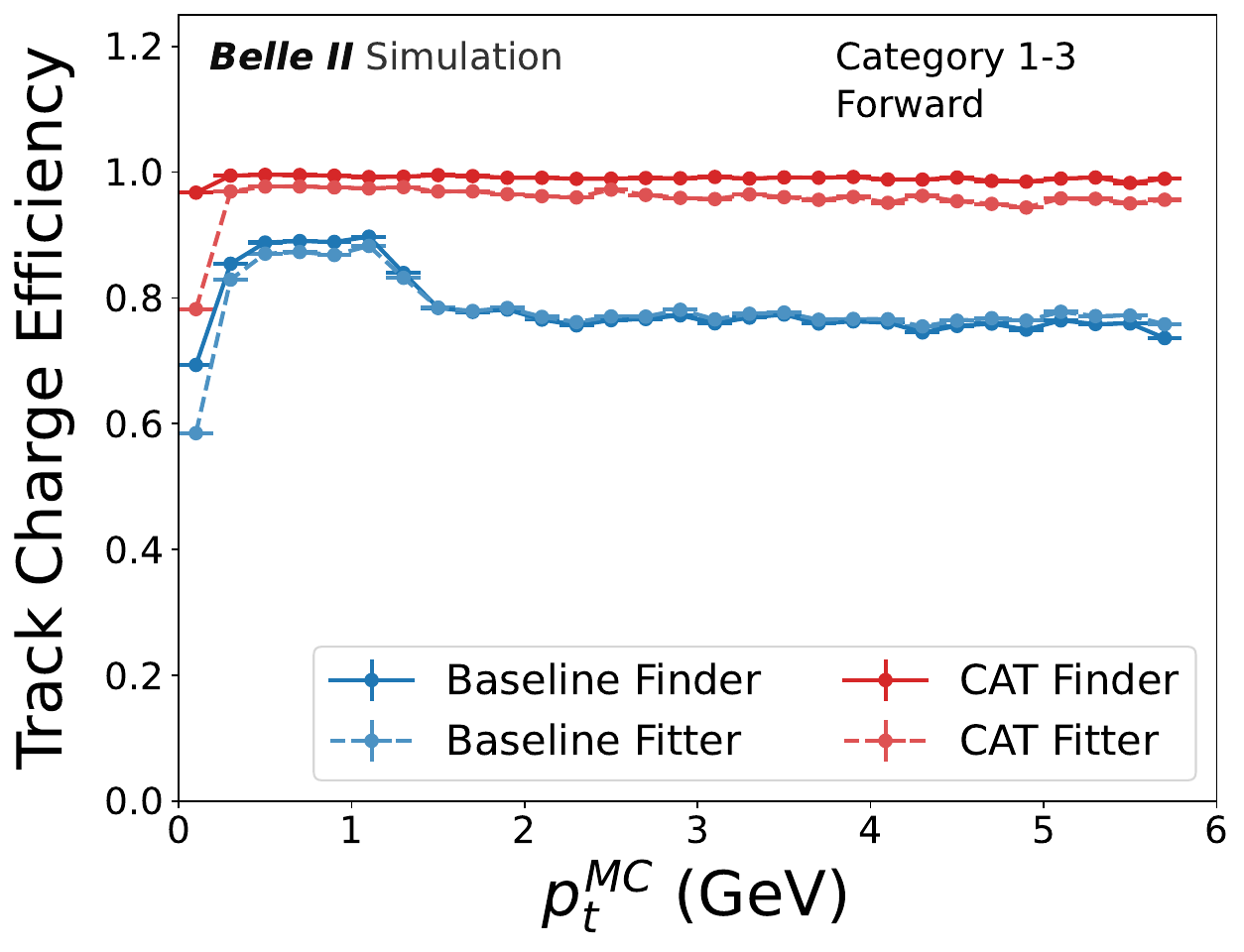}
         \caption{Forward endcap.}
         \label{fig:trainingsample_ceff_pt:a}
     \end{subfigure}\hfill
        \begin{subfigure}[b]{\thirdwidth\textwidth}
         \centering
         \includegraphics[width=\textwidth]{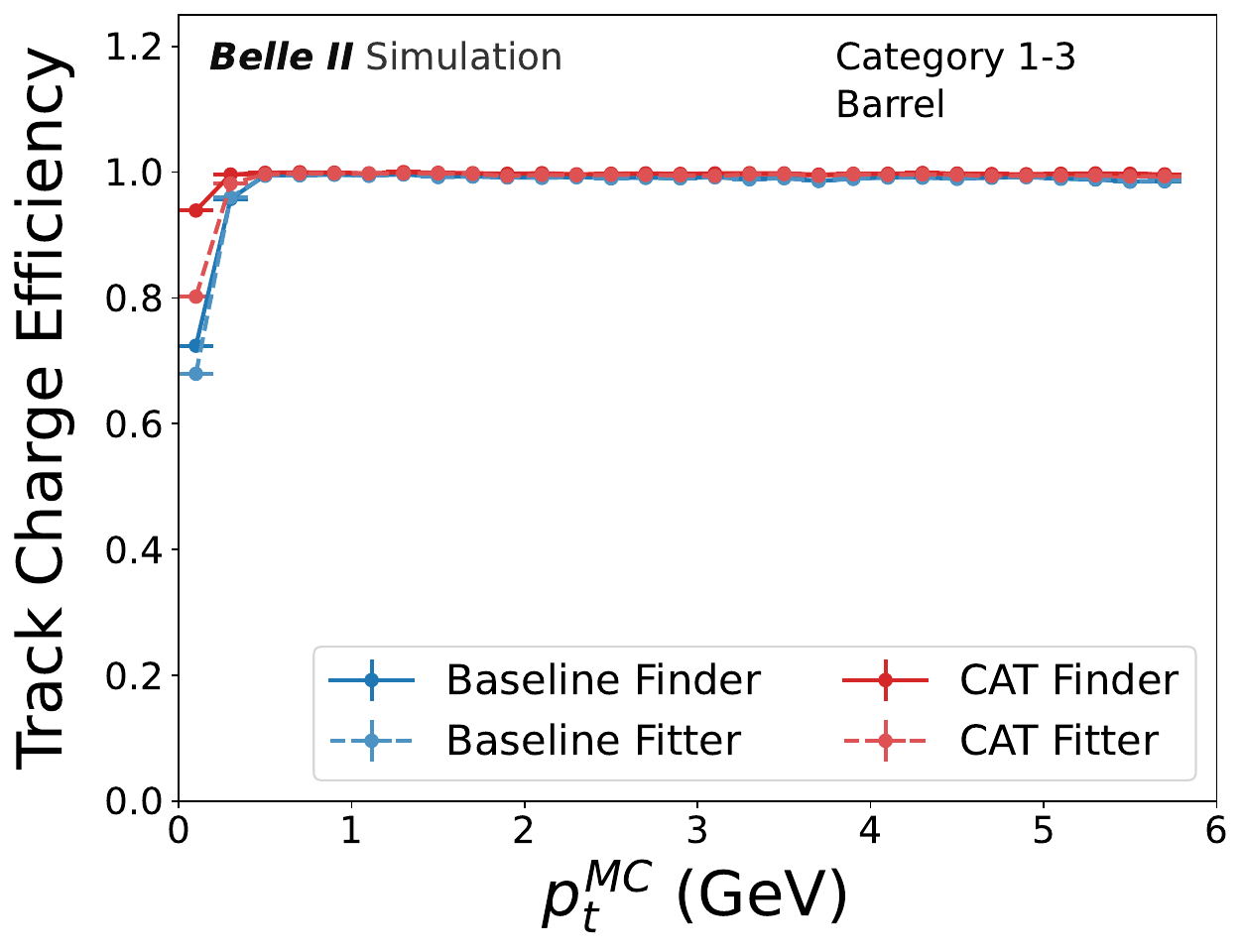}
         \caption{Barrel.}
         \label{fig:trainingsample_ceff_pt:b}
     \end{subfigure}\hfill
        \begin{subfigure}[b]{\thirdwidth\textwidth}
         \centering
         \includegraphics[width=\textwidth]{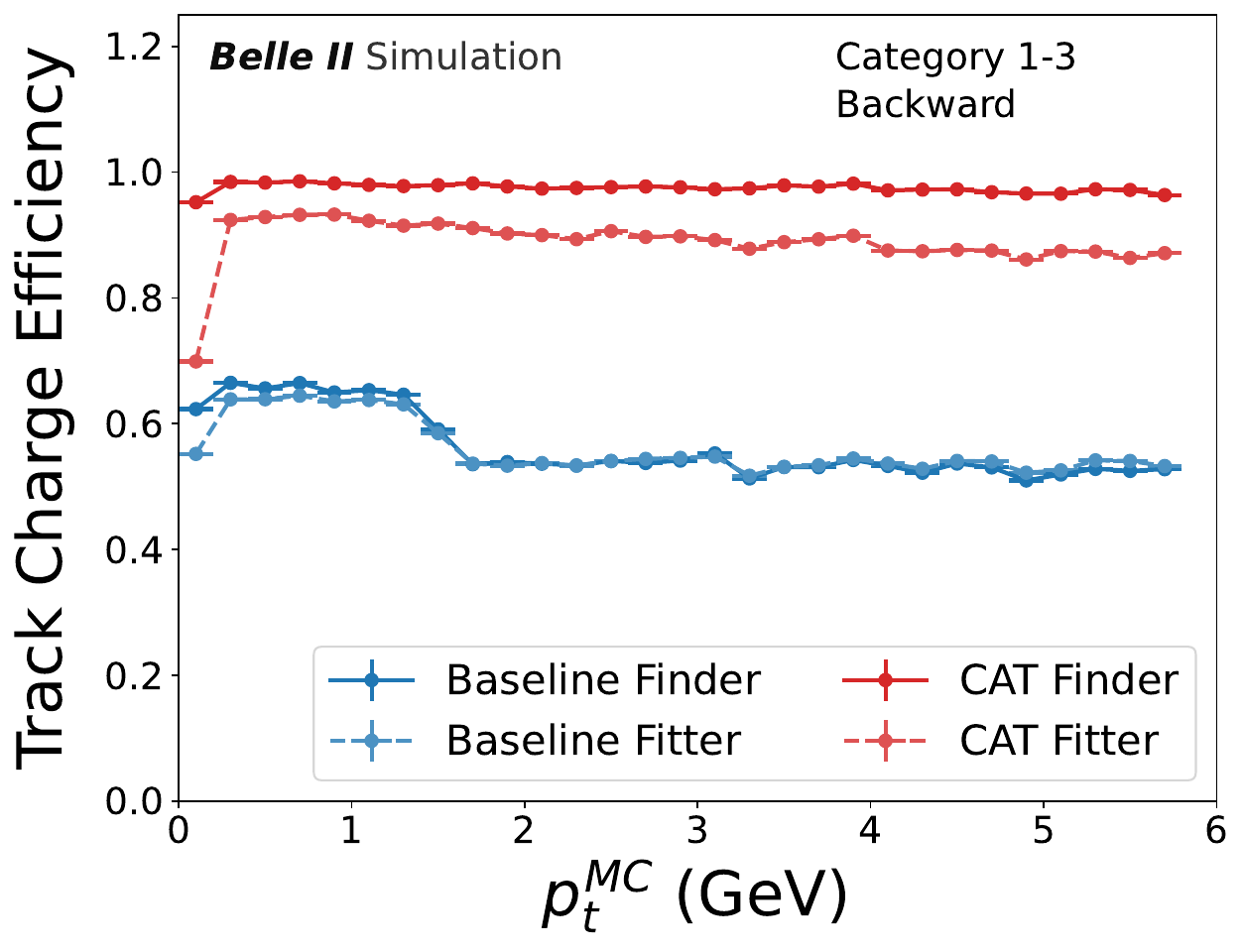}
         \caption{Backward endcap.}
         \label{fig:trainingsample_ceff_pt:c}
     \end{subfigure}
\caption{Track finding~(empty markers, connected by solid lines to guide the eye) and combined track finding and fitting charge efficiency~(filled markers, connected by dashed lines to guide the eye) for the prompt evaluation samples (category 1-3, see \cref{tab:samples}, \databackground) with curler tracks removed, as function of simulated transverse momentum $p_t^{MC}$ for the \legendre~(blue) and the \cat~(red) in the (\subref{fig:trainingsample_ceff_pt:a}) forward endcap, (\subref{fig:trainingsample_ceff_pt:b}) barrel, and (\subref{fig:trainingsample_ceff_pt:c}) backward endcap.
The vertical error bars that show the statistical uncertainty are smaller than the marker size.
The horizontal error bars indicate the bin width.
The uncertainties of the different track finding algorithms are correlated since they use the same simulated events.}
\label{fig:trainingsample_ceff_pt}
\end{figure*}

The track finding charge efficiencies, and the combined track finding and track fitting charge  efficiencies for the \legendre in comparison with the \cat are shown in \cref{fig:trainingsample_ceff_pt} for non-curling tracks from category~1-3.

\clearpage
\FloatBarrier

\section{Efficiencies, fake rates, and clone rates for \texorpdfstring{\kkmc}{muon pairs}}
\label{app:kkmc}
\begin{figure*}[ht!]
     \centering
     \begin{subfigure}[b]{\thirdwidth\textwidth}
         \includegraphics[width=\textwidth]{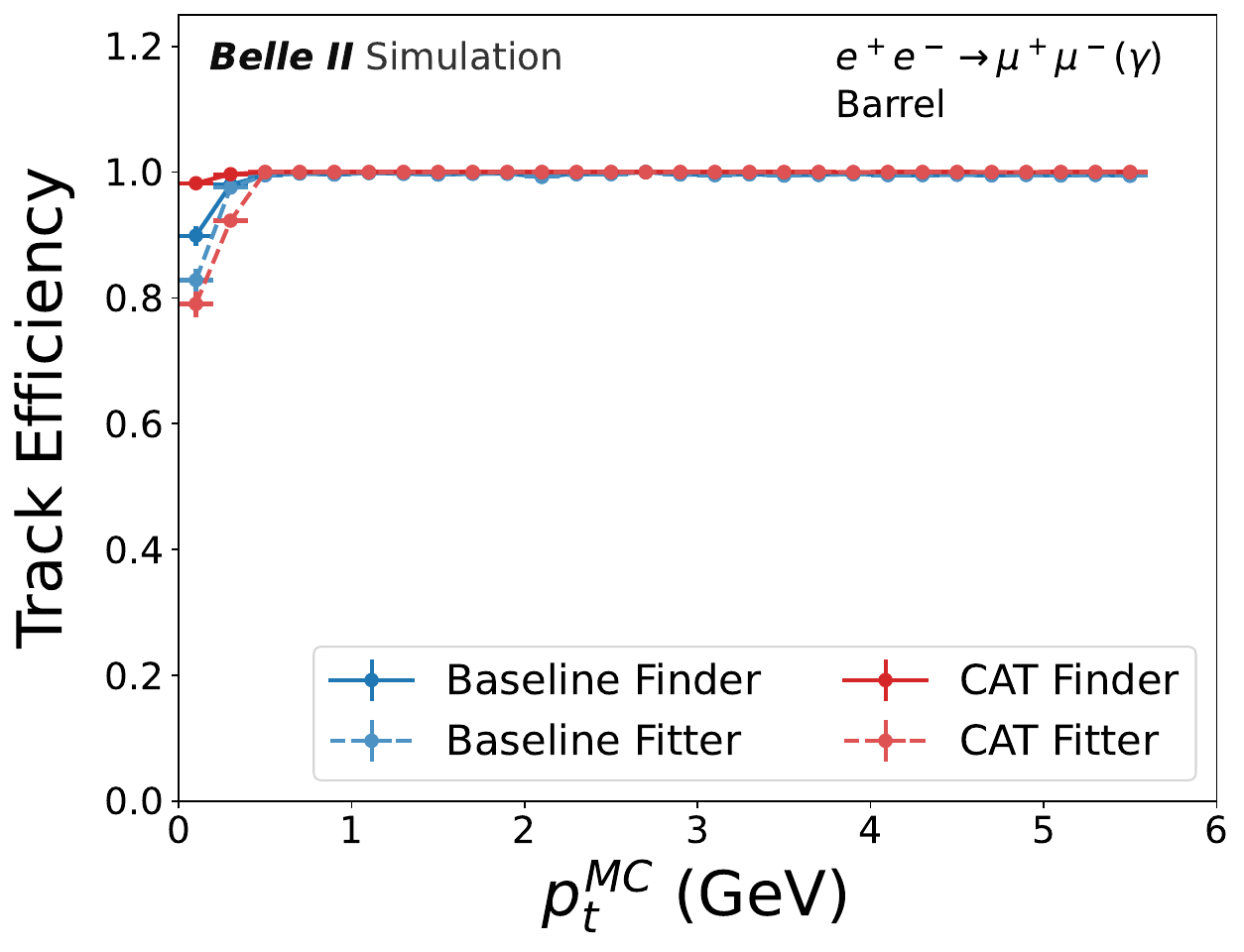}
         \caption{Track finding and fitting efficiency.}
         \label{fig:kkmc_eff:a}
     \end{subfigure}
     \quad
     \begin{subfigure}[b]{\thirdwidth\textwidth}
         \includegraphics[width=\textwidth]{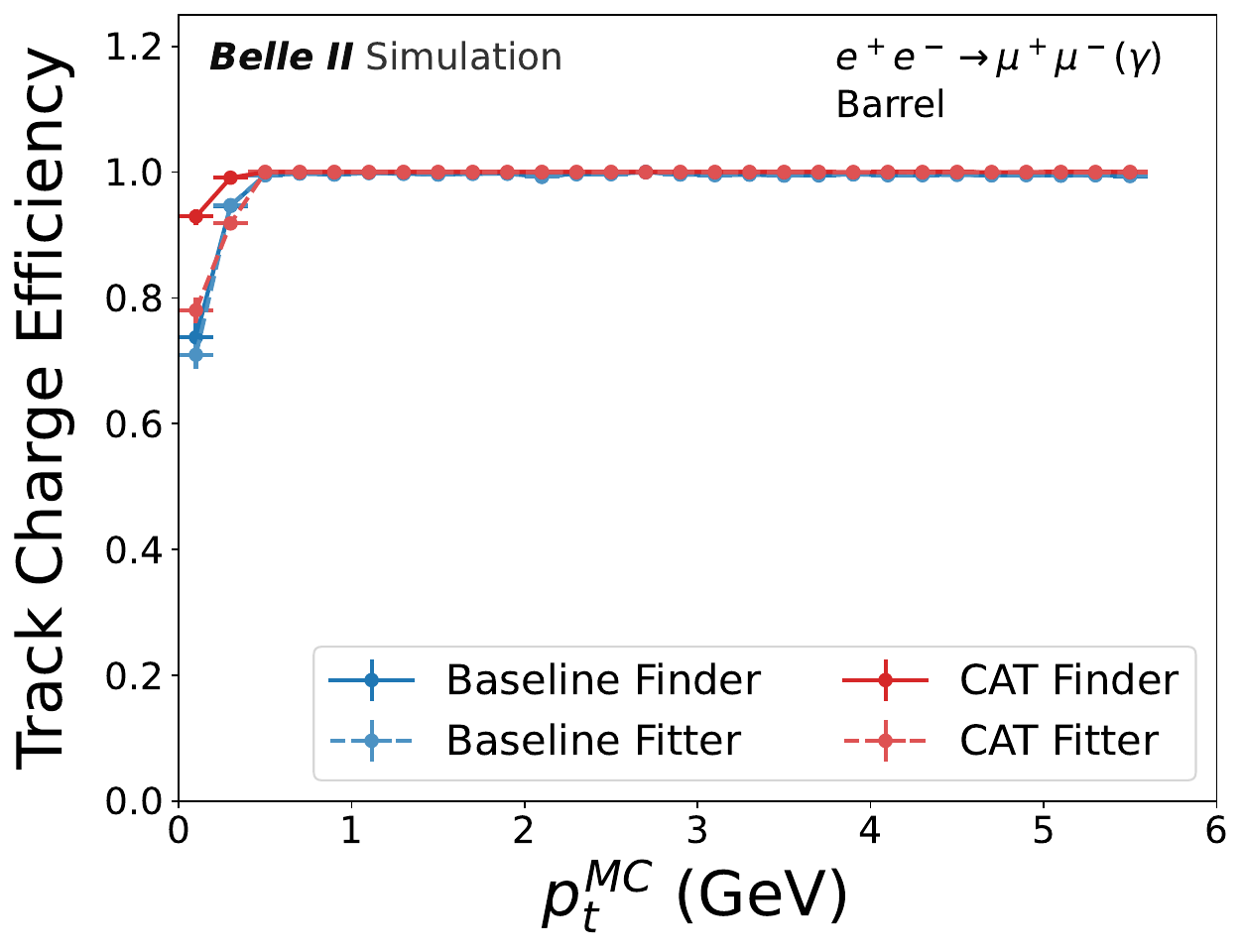}
         \caption{Track finding and fitting charge efficiency.}
         \label{fig:kkmc_ceff:b}
     \end{subfigure}\hfill
         
\caption{Track finding~(empty markers, connected by lines to guide the eye) and combined track finding and fitting charge efficiency~(filled markers) for \kkmc evaluation sample in the barrel with \databackground. 
See \cref{fig:trainingsample_eff_pt} caption for details. }
\label{fig:kkmc_eff_pt}
\end{figure*}

The track finding charge efficiencies, and the combined track finding and track fitting charge efficiency for the \legendre in comparison with the \cat are shown in \cref{fig:kkmc_eff_pt}.

Track finding and fitting efficiency \trackeff, fake rate \fakerate, clone rate \clonerate, track charge efficiency \trackchareff and wrong charge rate \wrongchargerate integrated over the full $p_t$ for \kkmc events are shown in \cref{tab:kkmc_finder_low_beambackground}.

\begin{table*}
    \fontsize{6pt}{6pt}\selectfont
    \centering
    \caption{The performance metrics for the \kkmc evaluation samples for different track finding algorithms in different detector regions for low beam background. Uncertainties below $<$0.01\% are not shown in the table. }

    \begin{tabular}{r ccc cc}
         \toprule
        (in \%)& \trackeff & \fakerate & \clonerate &\trackchareff & \wrongchargerate\\
 \midrule
& \multicolumn{5}{c}{forward endcap} \\
\midrule
Baseline Finder & $84.0^{+0.2}_{-0.2}$ & $1.31^{+0.07}_{-0.08}$ & $0.01^{+0.01}_{-0.01}$ & $82.1^{+0.2}_{-0.2}$ & $2.25^{+0.1}_{-0.1}$ \\
CAT Finder & $99.91^{+0.02}_{-0.02}$ & $8.1^{+0.2}_{-0.2}$ & $0.01^{+0.01}_{-0.01}$ & $99.89^{+0.02}_{-0.02}$ & $0.02^{+0.01}_{-0.01}$ \\
\midrule
Baseline Fitter & $83.5^{+0.2}_{-0.2}$ & $0.98^{+0.06}_{-0.07}$ & $0.01^{+0.01}_{-0.01}$ & $82.3^{+0.2}_{-0.2}$ & $1.4^{+0.08}_{-0.08}$ \\
CAT Fitter & $99.34^{+0.05}_{-0.05}$ & $1.21^{+0.07}_{-0.07}$ & $0.0^{}_{-0.01}$ & $97.21^{+0.1}_{-0.1}$ & $2.14^{+0.09}_{-0.09}$ \\
\midrule
 \midrule
& \multicolumn{5}{c}{barrel} \\
\midrule
Baseline Finder & $99.51^{+0.02}_{-0.02}$ & $2.91^{+0.05}_{-0.05}$ & $0.05^{+0.01}_{-0.01}$ & $99.4^{+0.02}_{-0.02}$ & $0.11^{+0.01}_{-0.01}$ \\
CAT Finder & $99.99^{}_{}$ & $5.11^{+0.07}_{-0.07}$ & $0.04^{+0.01}_{-0.01}$ & $99.96^{+0.01}_{-0.01}$ & $0.02^{}_{-0.01}$ \\
\midrule
Baseline Fitter & $99.48^{+0.02}_{-0.02}$ & $2.11^{+0.04}_{-0.04}$ & $0.05^{+0.01}_{-0.01}$ & $99.4^{+0.02}_{-0.02}$ & $0.08^{+0.01}_{-0.01}$ \\
CAT Fitter & $99.85^{+0.01}_{-0.01}$ & $1.78^{+0.04}_{-0.04}$ & $0.01^{}_{}$ & $99.84^{+0.01}_{-0.01}$ & $0.02^{}_{}$ \\
\midrule
 \midrule
& \multicolumn{5}{c}{backward endcap} \\
\midrule
Baseline Finder & $66.4^{+0.3}_{-0.3}$ & $4.1^{+0.2}_{-0.2}$ & $0.02^{+0.01}_{-0.01}$ & $62.8^{+0.3}_{-0.3}$ & $5.4^{+0.2}_{-0.2}$ \\
CAT Finder & $99.73^{+0.03}_{-0.03}$ & $7.4^{+0.2}_{-0.2}$ & $0.0^{}_{}$ & $99.68^{+0.04}_{-0.03}$ & $0.05^{+0.01}_{-0.02}$ \\
\midrule
Baseline Fitter & $65.7^{+0.3}_{-0.3}$ & $3.6^{+0.1}_{-0.1}$ & $0.02^{+0.01}_{-0.01}$ & $63.3^{+0.3}_{-0.3}$ & $3.6^{+0.1}_{-0.1}$ \\
CAT Fitter & $97.88^{+0.09}_{-0.09}$ & $4.0^{+0.1}_{-0.1}$ & $0.0^{}_{}$ & $92.9^{+0.2}_{-0.2}$ & $5.1^{+0.1}_{-0.1}$ \\
\midrule

\bottomrule
        \\
    \end{tabular}
 
    \label{tab:kkmc_finder_low_beambackground}
\end{table*}

\clearpage
\FloatBarrier

\section{Track charge efficiency for \dh and \kshort}\label{app:displaced_ceff}
\begin{figure*}[ht!]
     \centering

        \begin{subfigure}[b]{\thirdwidth\textwidth}
         \centering
         \includegraphics[width=\textwidth]{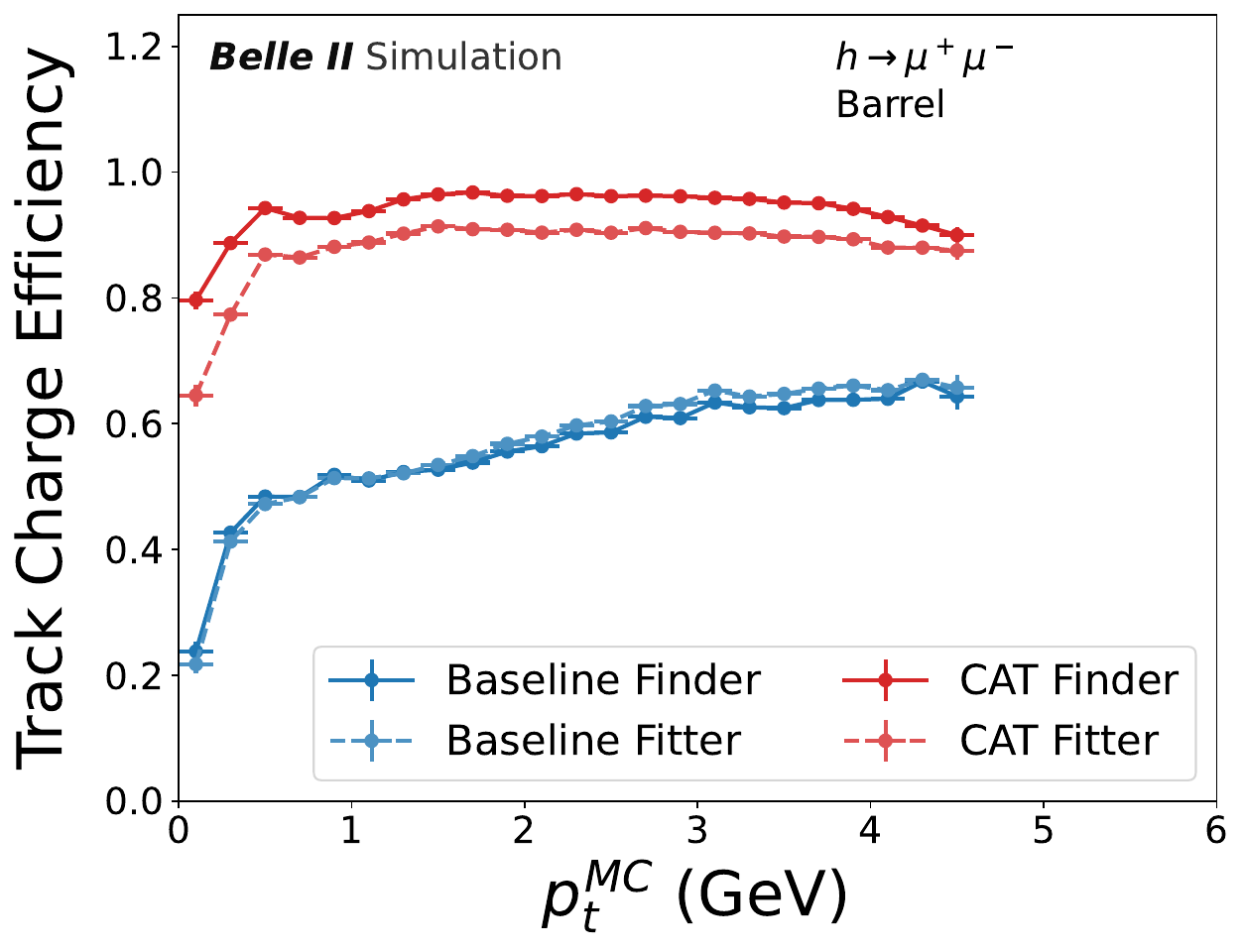}
         \caption{\dh, $p_t^{MC}$.}
         \label{fig:p_ceff_darkhiggs:b}
     \end{subfigure}
     \quad
        \begin{subfigure}[b]{\thirdwidth\textwidth}
         \centering
         \includegraphics[width=\textwidth]{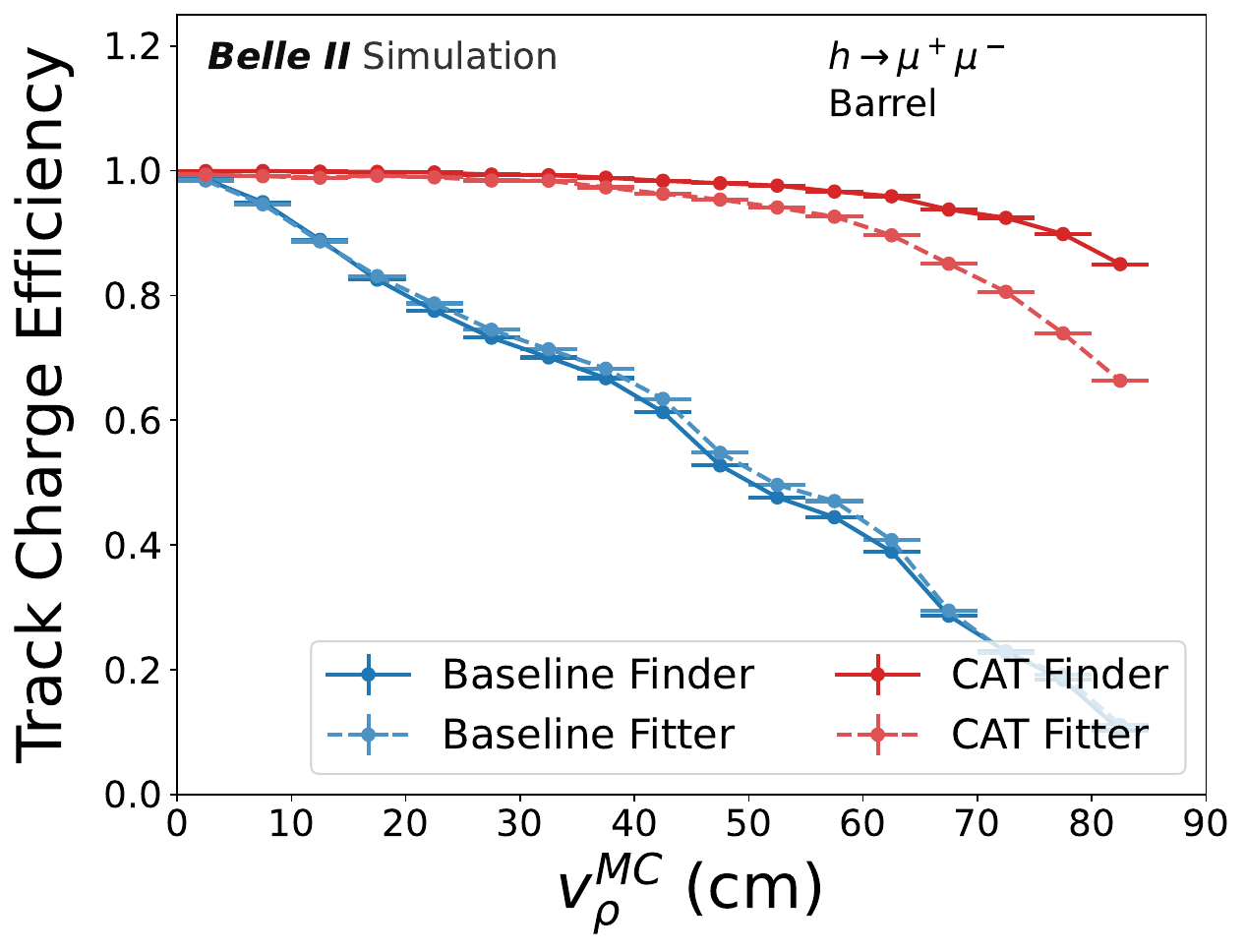}
         \caption{\dh, $v_{\rho}^{MC}$.}
         \label{fig:p_ceff_darkhiggs:b2}
     \end{subfigure}\\
     
        \begin{subfigure}[b]{\thirdwidth\textwidth}
         \centering
         \includegraphics[width=\textwidth]{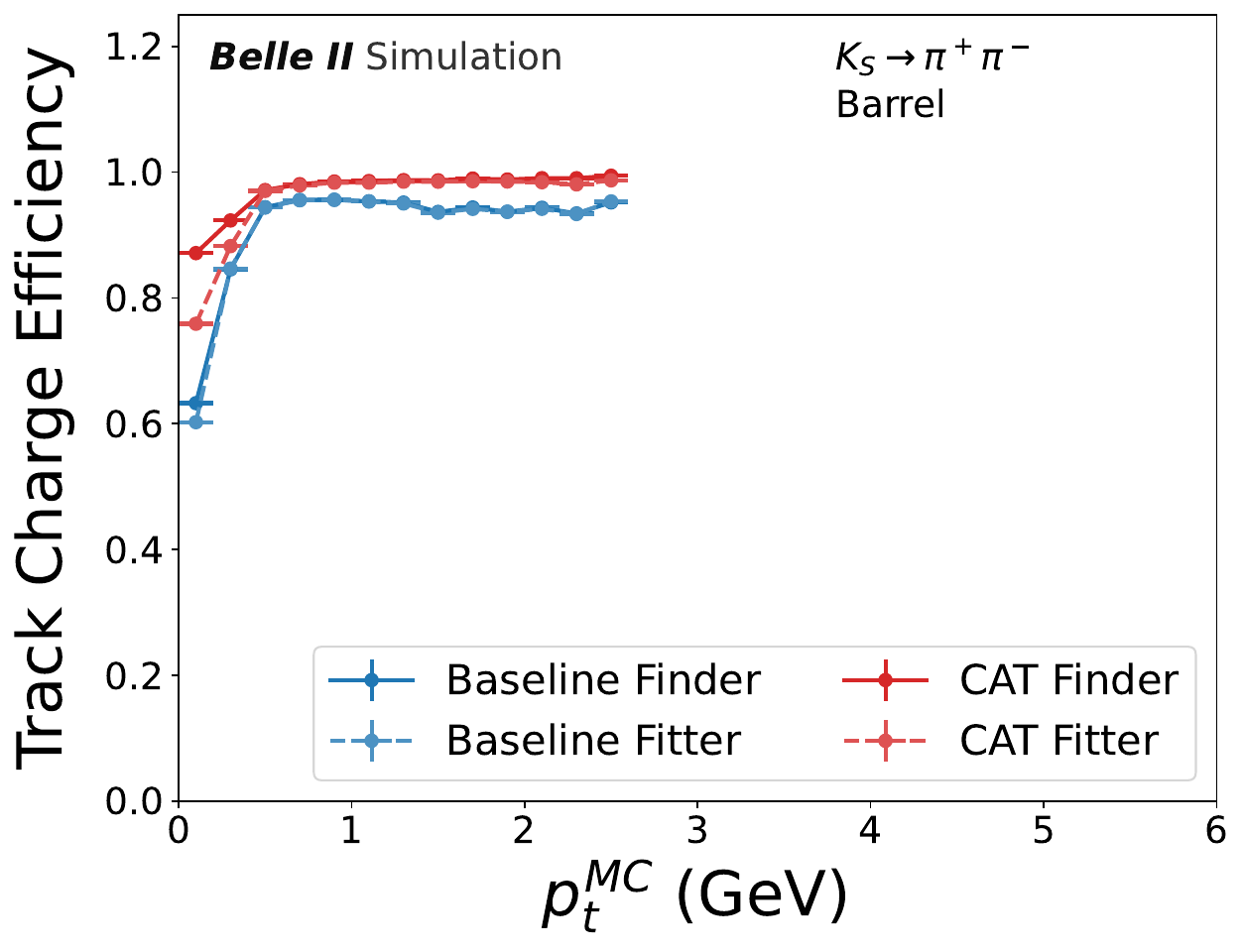}
         \caption{\kshort, $p_t^{MC}$.}
         \label{fig:p_ceff_k0s:b}
     \end{subfigure}
     \quad
        \begin{subfigure}[b]{\thirdwidth\textwidth}
         \centering
         \includegraphics[width=\textwidth]{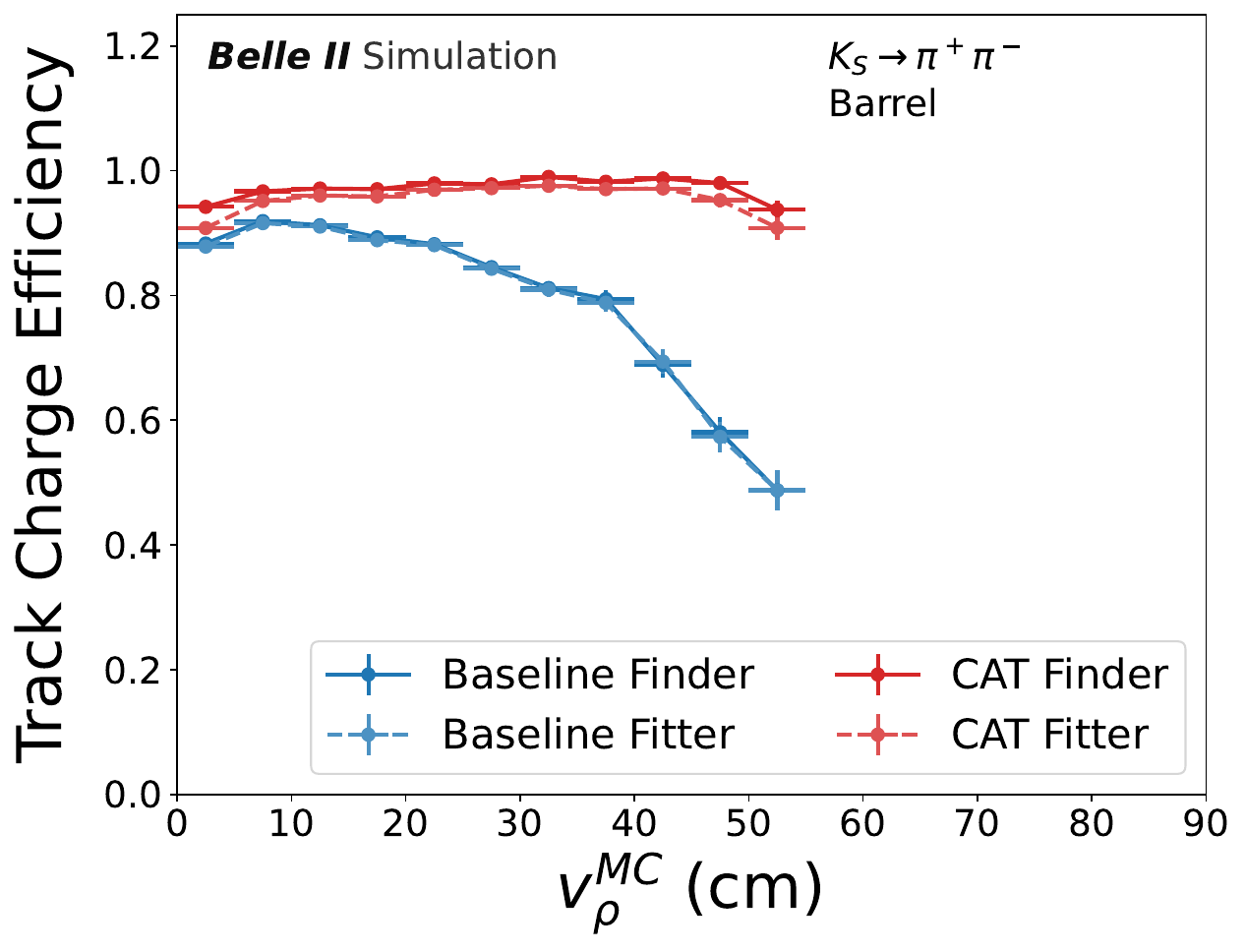}
         \caption{\kshort, $v_{\rho}^{MC}$.}
         \label{fig:p_ceff_k0s:b2}
     \end{subfigure}\hfill

\caption{Track finding~(empty markers) and combined track finding and fitting charge efficiency~(filled markers) for (top) displaced tracks in \dh events and in (bottom) \kshort events with \databackground, as function of (left) the true simulated transverse momentum $p_t^{MC}$, and (right) the true simulated displacement $v_{\rho}^{MC}$. }
\label{fig:ceff_displaced}
\end{figure*}

The track finding efficiencies, and the combined track finding and track fitting efficiency for the \legendre in comparison with the \cat for \dh and \kshort events are shown in \cref{fig:ceff_displaced}.

\clearpage
\FloatBarrier

\section{High transverse momentum track resolution}\label{app:highptres}
\begin{figure*}[ht!]
     \centering
     \begin{subfigure}[b]{\thirdwidth\textwidth}
         \centering
         \includegraphics[width=\textwidth]{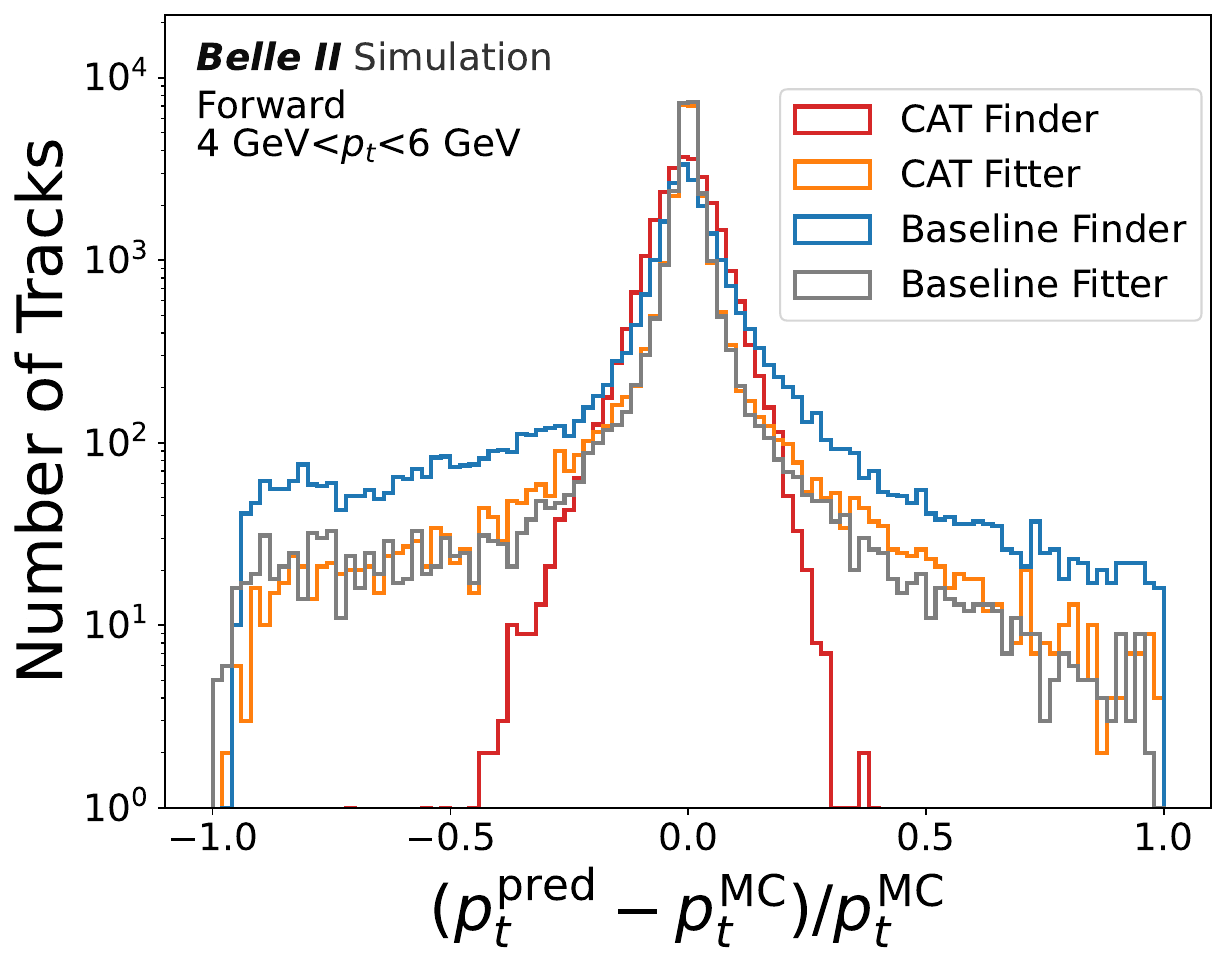}
         \caption{Forward endcap.}
         \label{fig:pt_res:a}
     \end{subfigure}\hfill
        \begin{subfigure}[b]{\thirdwidth\textwidth}
         \centering
         \includegraphics[width=\textwidth]{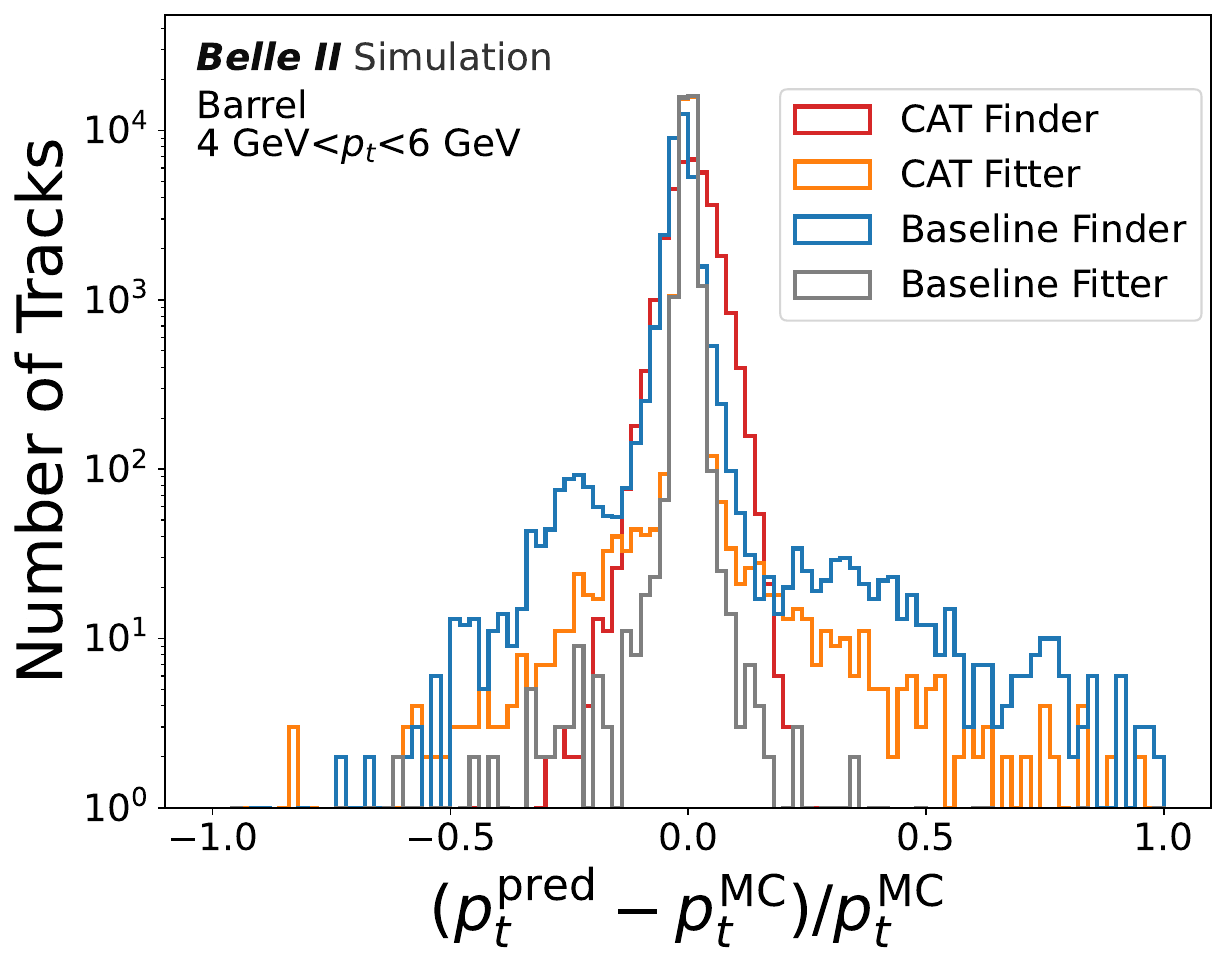}
         \caption{Barrel.}
         \label{fig:pt_res:b}
     \end{subfigure}\hfill
        \begin{subfigure}[b]{\thirdwidth\textwidth}
         \centering
         \includegraphics[width=\textwidth]{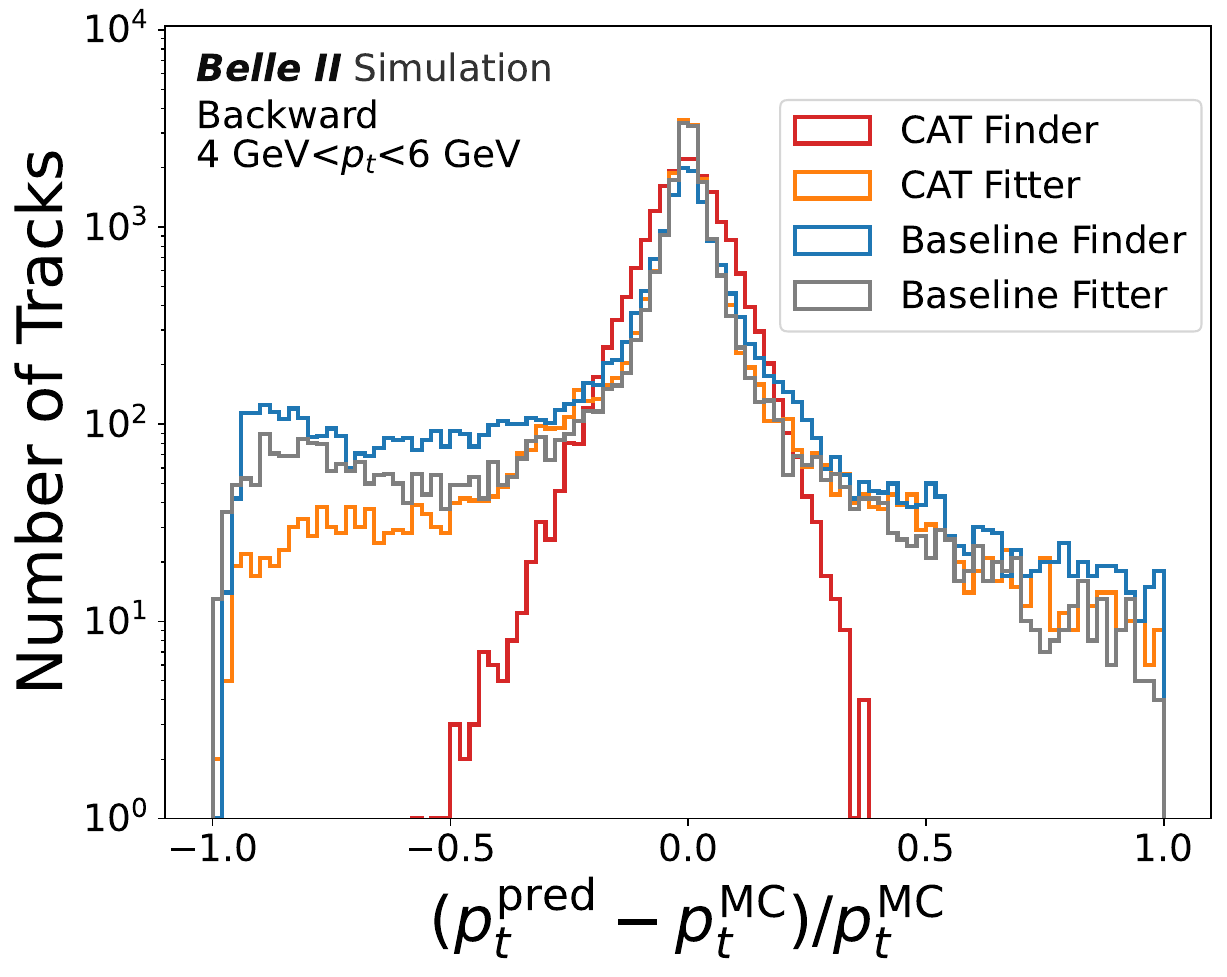}
         \caption{Backward endcap.}
         \label{fig:pt_res:c}
     \end{subfigure}\hfill\\
      \centering
     \begin{subfigure}[b]{\thirdwidth\textwidth}
         \centering
         \includegraphics[width=\textwidth]{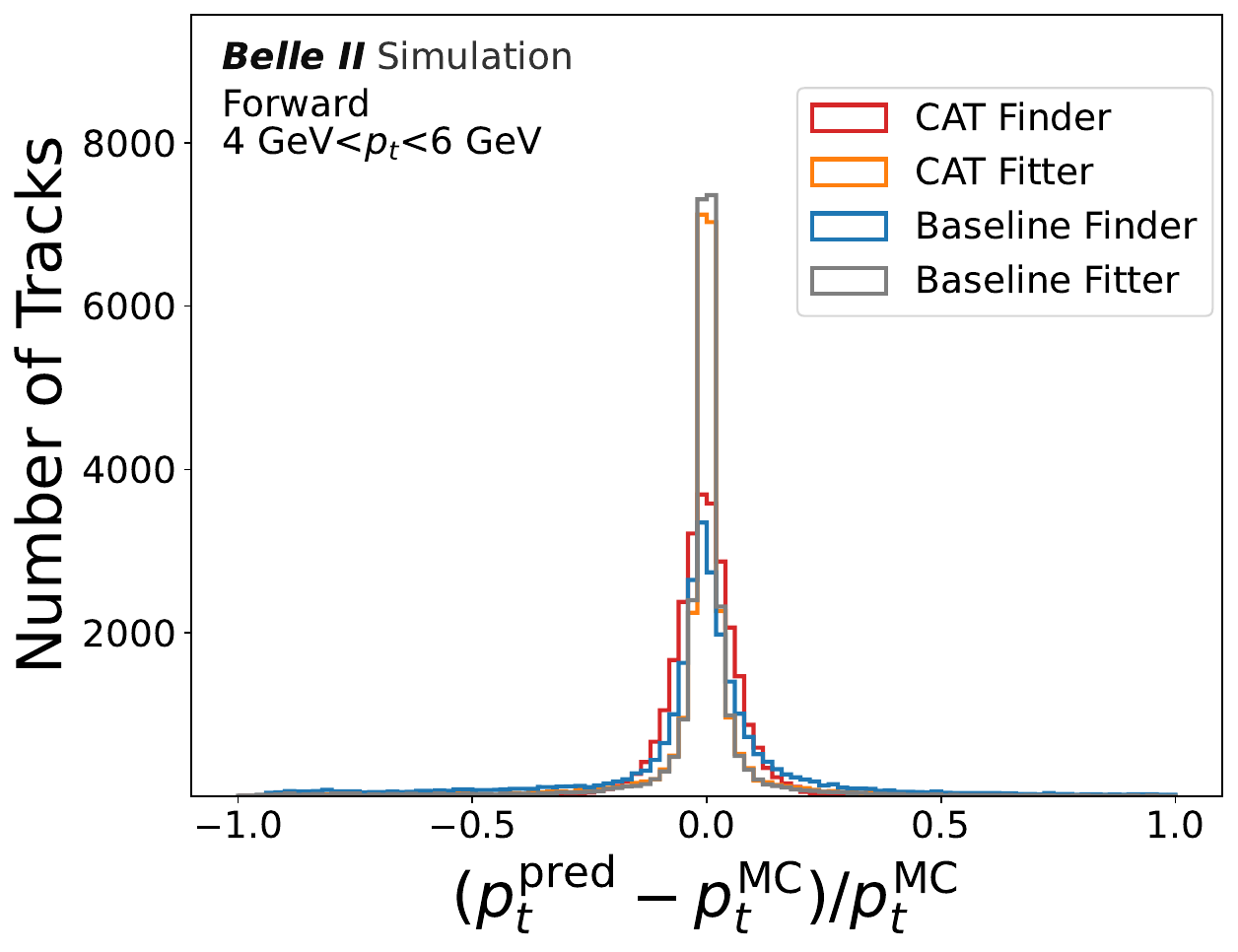}
         \caption{Forward endcap.}
         \label{fig:pt_res:a2}
     \end{subfigure}\hfill
        \begin{subfigure}[b]{\thirdwidth\textwidth}
         \centering
         \includegraphics[width=\textwidth]{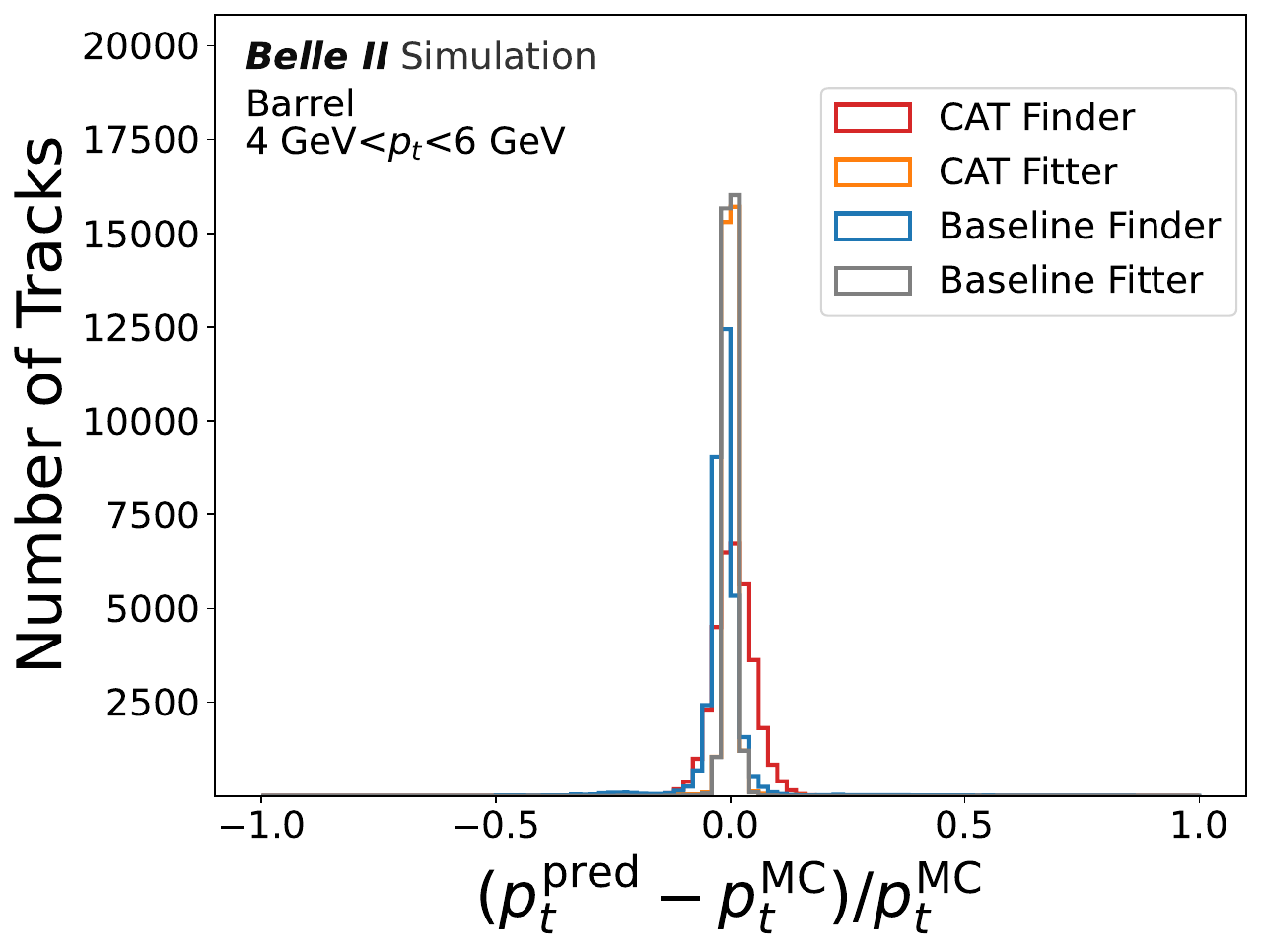}
         \caption{Barrel.}
         \label{fig:pt_res:b2}
     \end{subfigure}\hfill
        \begin{subfigure}[b]{\thirdwidth\textwidth}
         \centering
         \includegraphics[width=\textwidth]{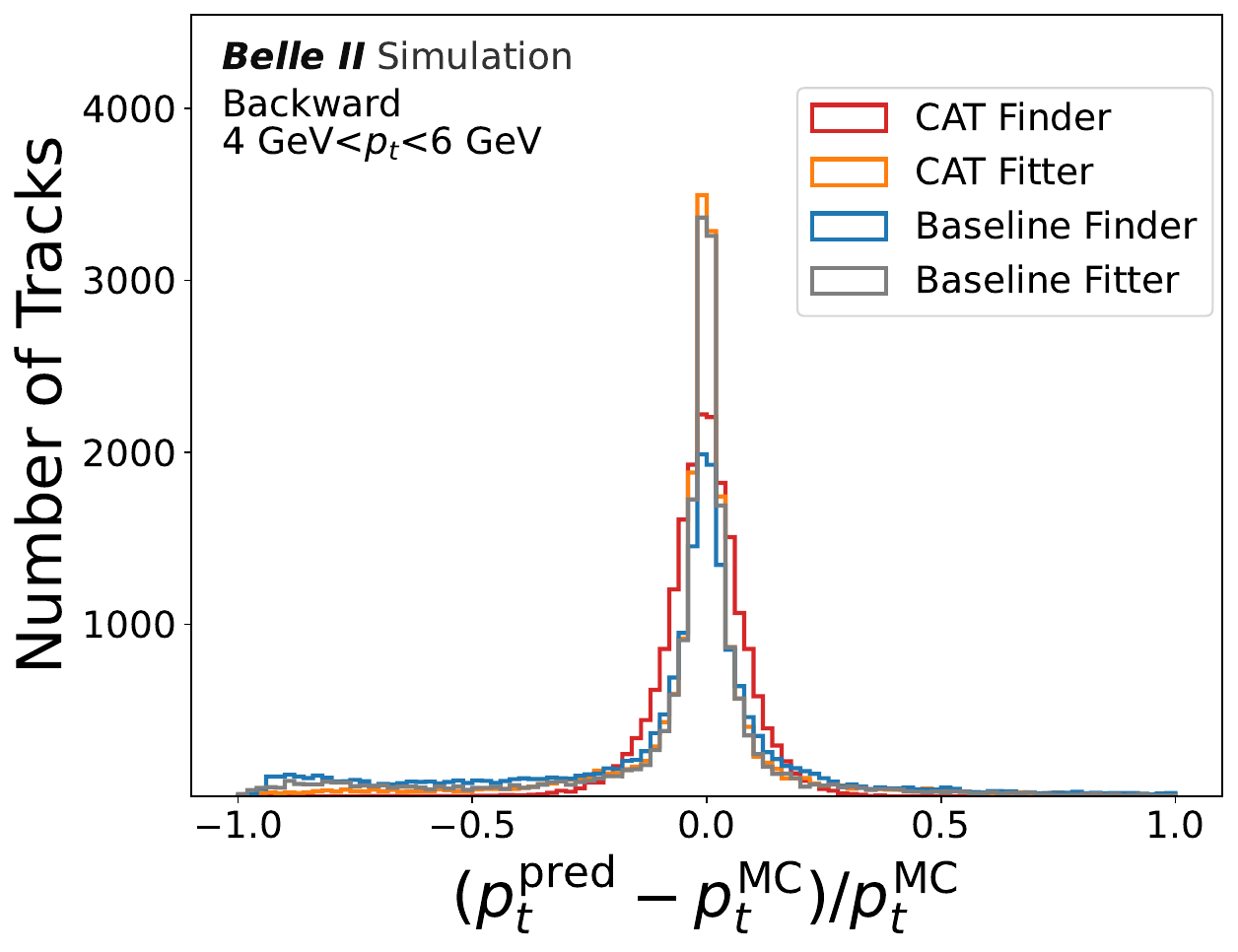}
         \caption{Backward endcap.}
         \label{fig:pt_res:c2}
     \end{subfigure}\\

          \centering
     \begin{subfigure}[b]{\thirdwidth\textwidth}
         \centering
         \includegraphics[width=\textwidth]{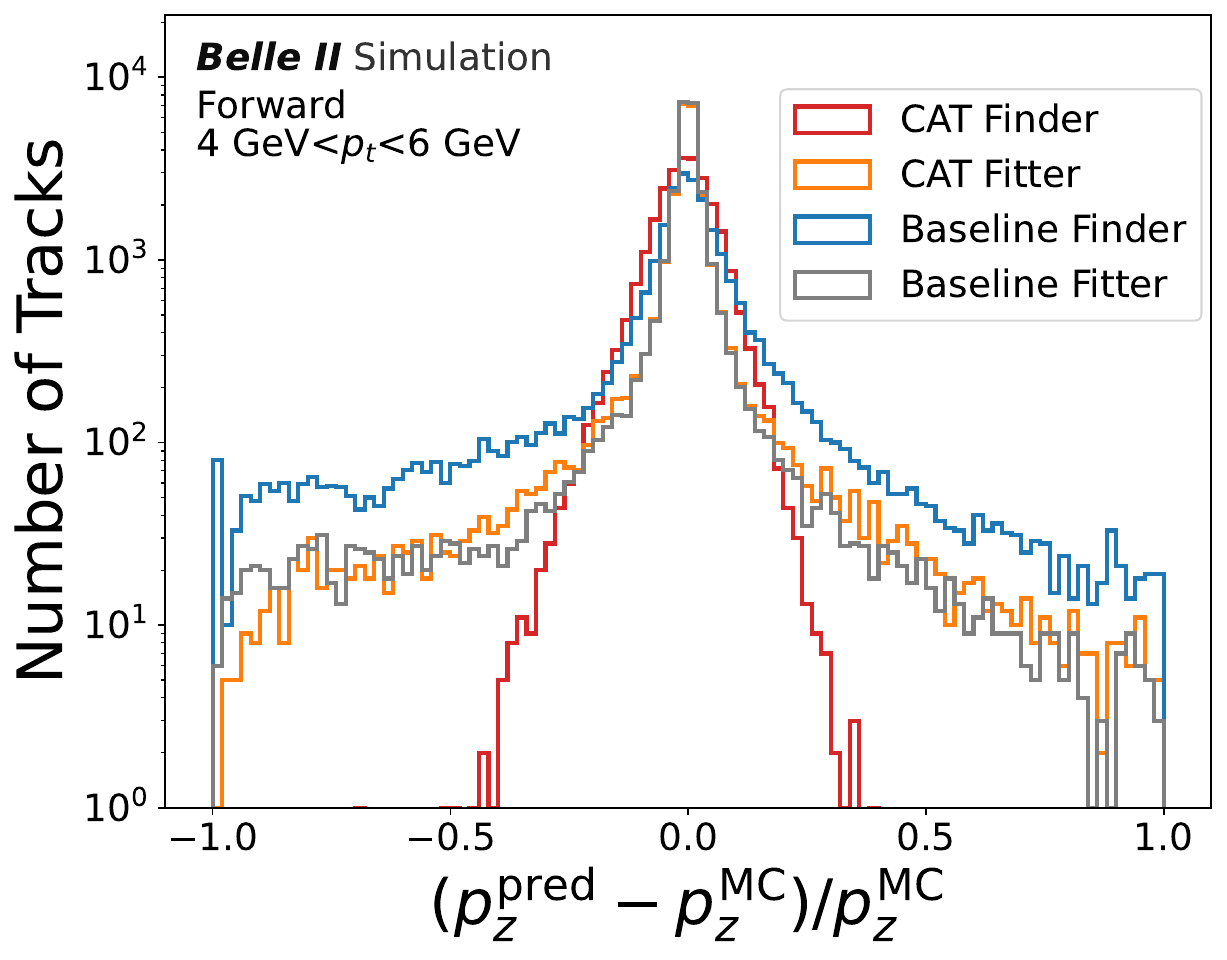}
         \caption{Forward endcap.}
         \label{fig:pz_res:a}
     \end{subfigure}\hfill
        \begin{subfigure}[b]{\thirdwidth\textwidth}
         \centering
         \includegraphics[width=\textwidth]{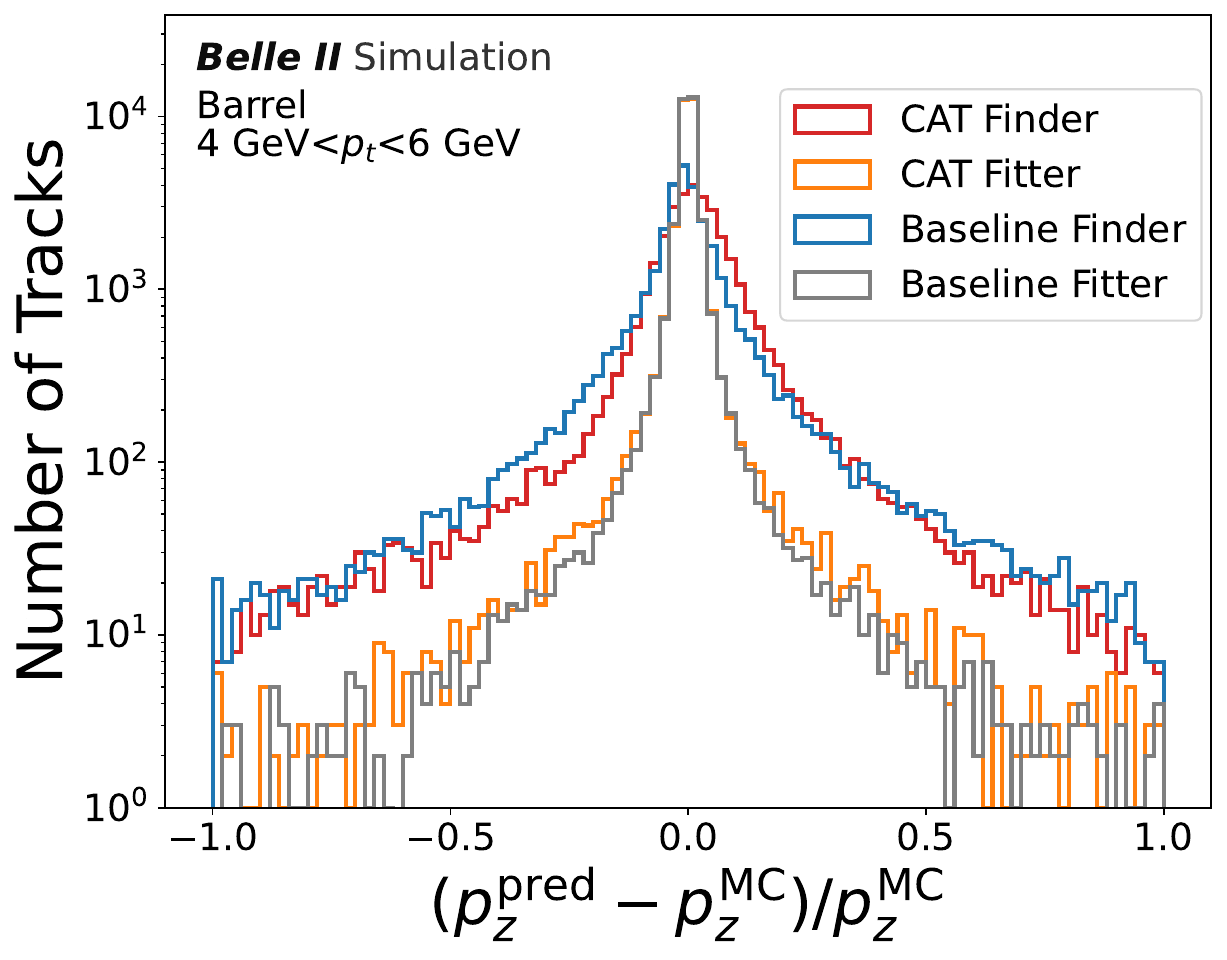}
         \caption{Barrel.}
         \label{fig:pz_res:b}
     \end{subfigure}\hfill
        \begin{subfigure}[b]{\thirdwidth\textwidth}
         \centering
         \includegraphics[width=\textwidth]{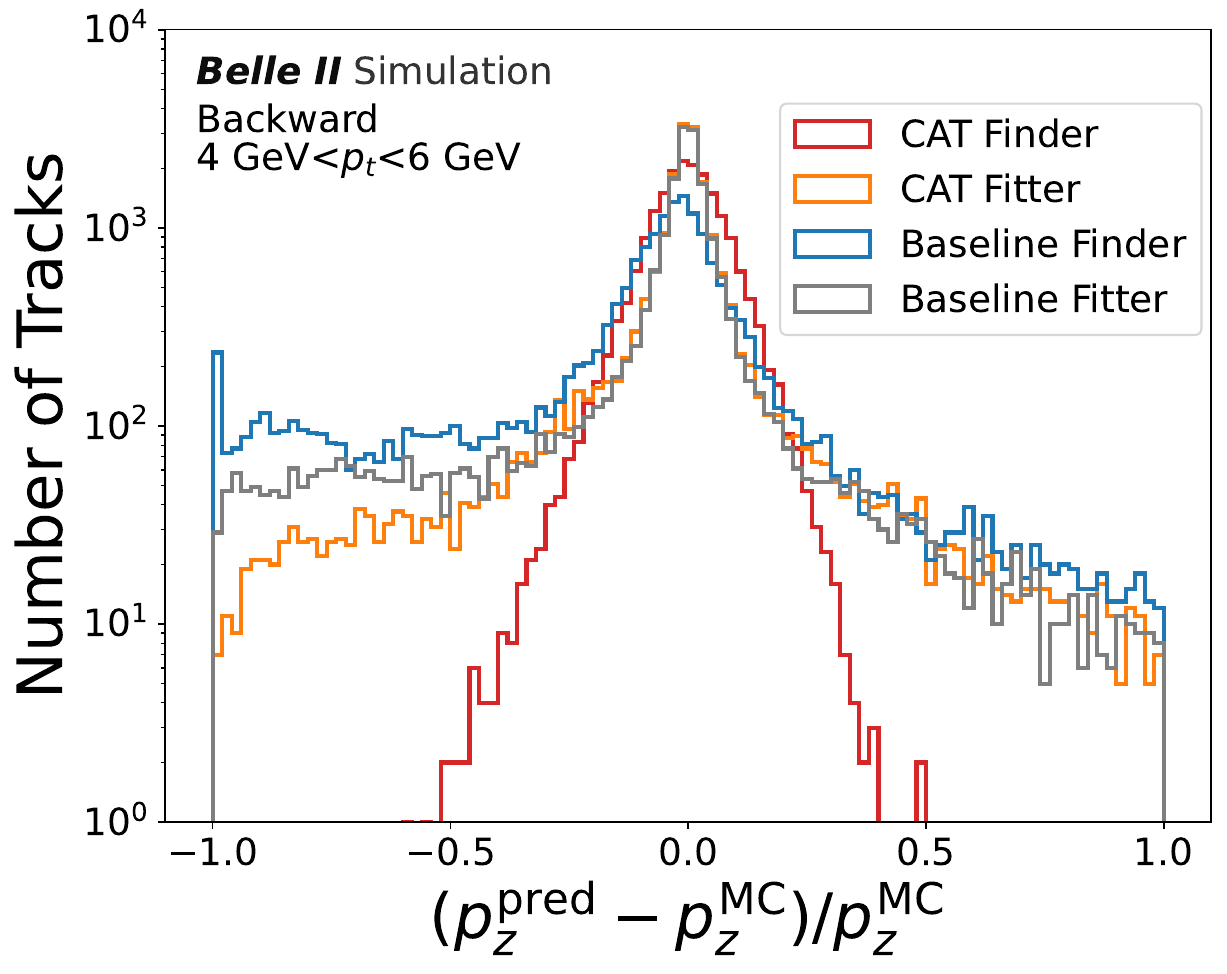}
         \caption{Backward endcap.}
         \label{fig:pz_res:c}
     \end{subfigure}\hfill\\
      \centering
     \begin{subfigure}[b]{\thirdwidth\textwidth}
         \centering
         \includegraphics[width=\textwidth]{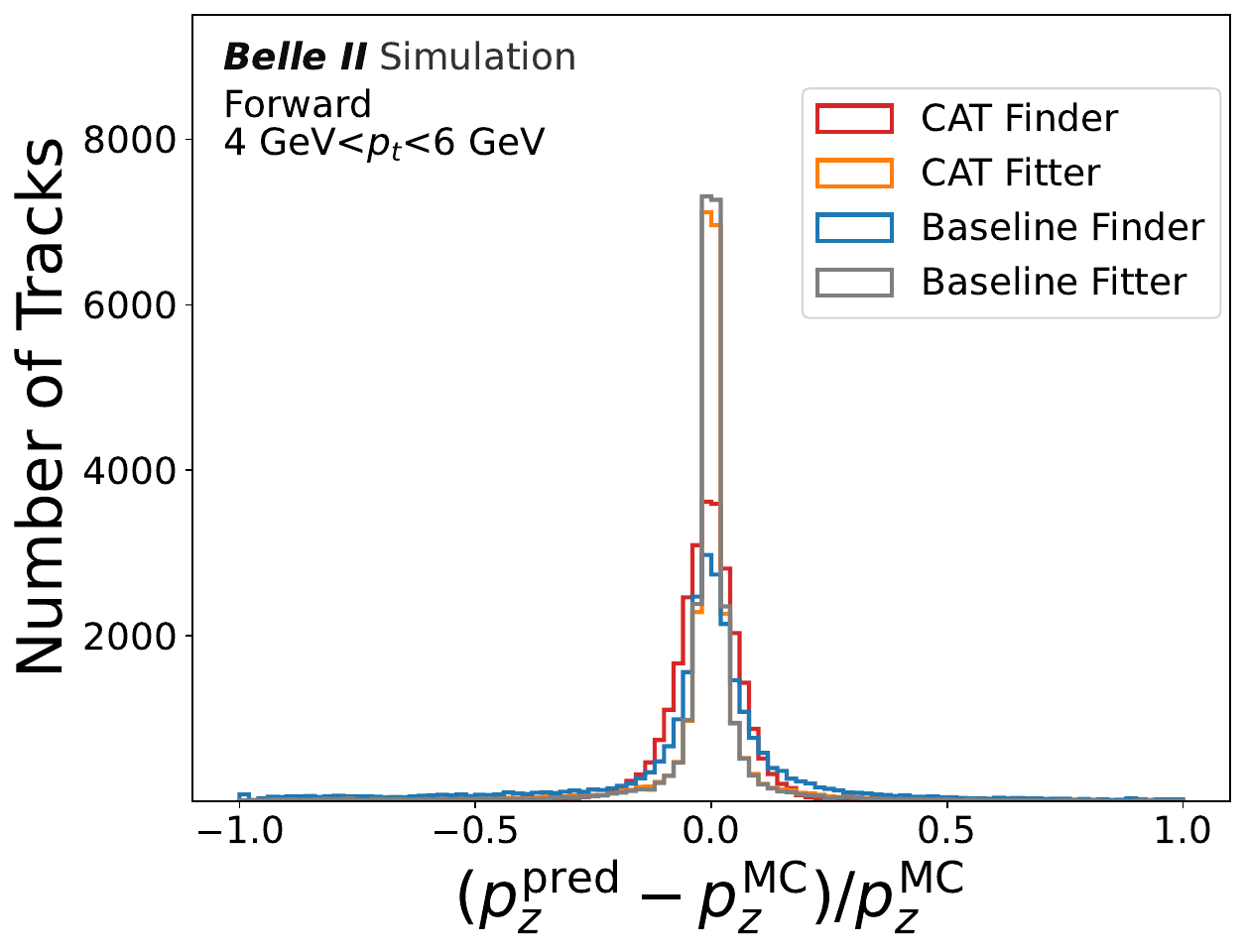}
         \caption{Forward endcap.}
         \label{fig:pz_res:a2}
     \end{subfigure}\hfill
        \begin{subfigure}[b]{\thirdwidth\textwidth}
         \centering
         \includegraphics[width=\textwidth]{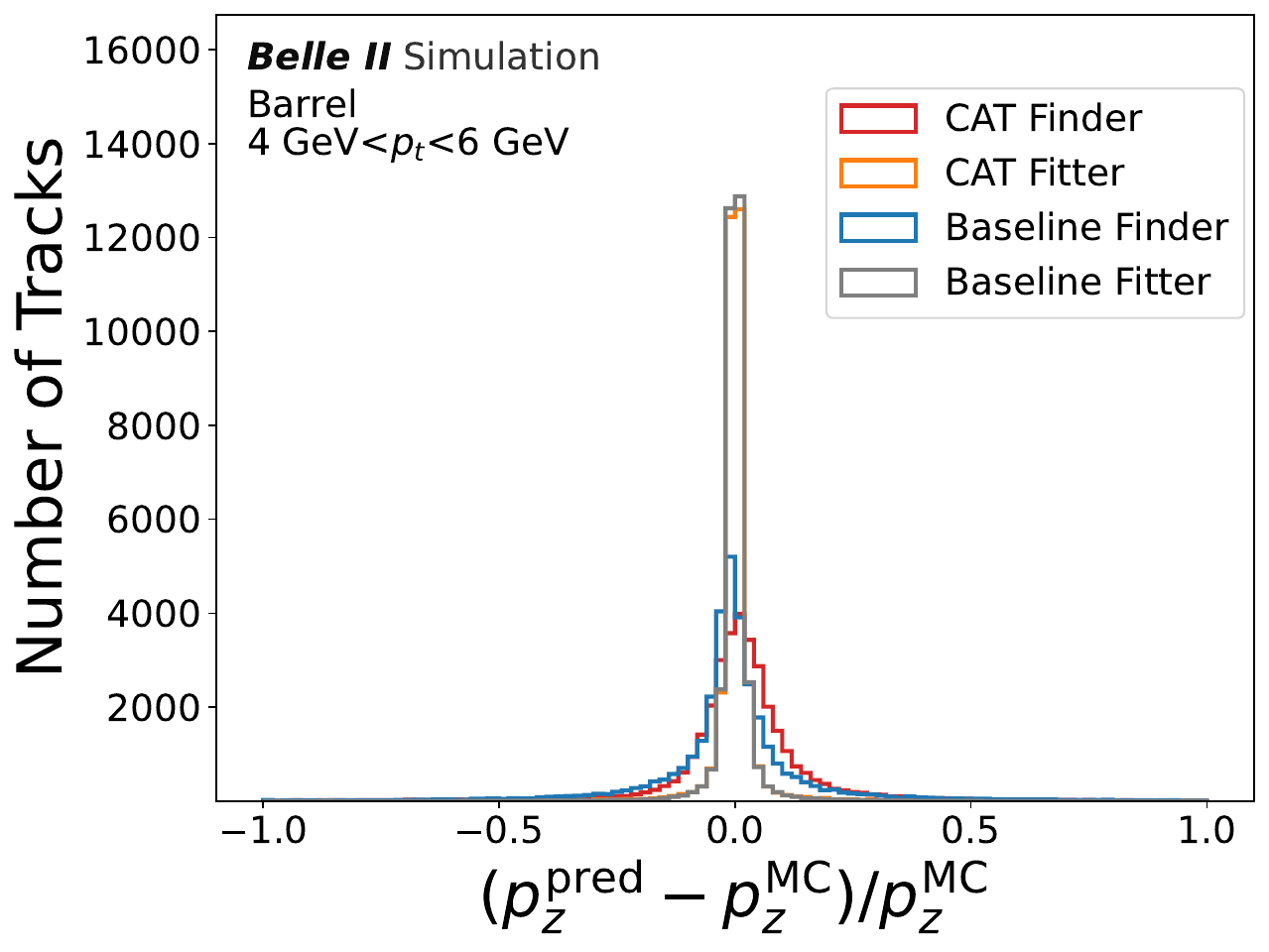}
         \caption{Barrel.}
         \label{fig:pz_res:b2}
     \end{subfigure}\hfill
        \begin{subfigure}[b]{\thirdwidth\textwidth}
         \centering
         \includegraphics[width=\textwidth]{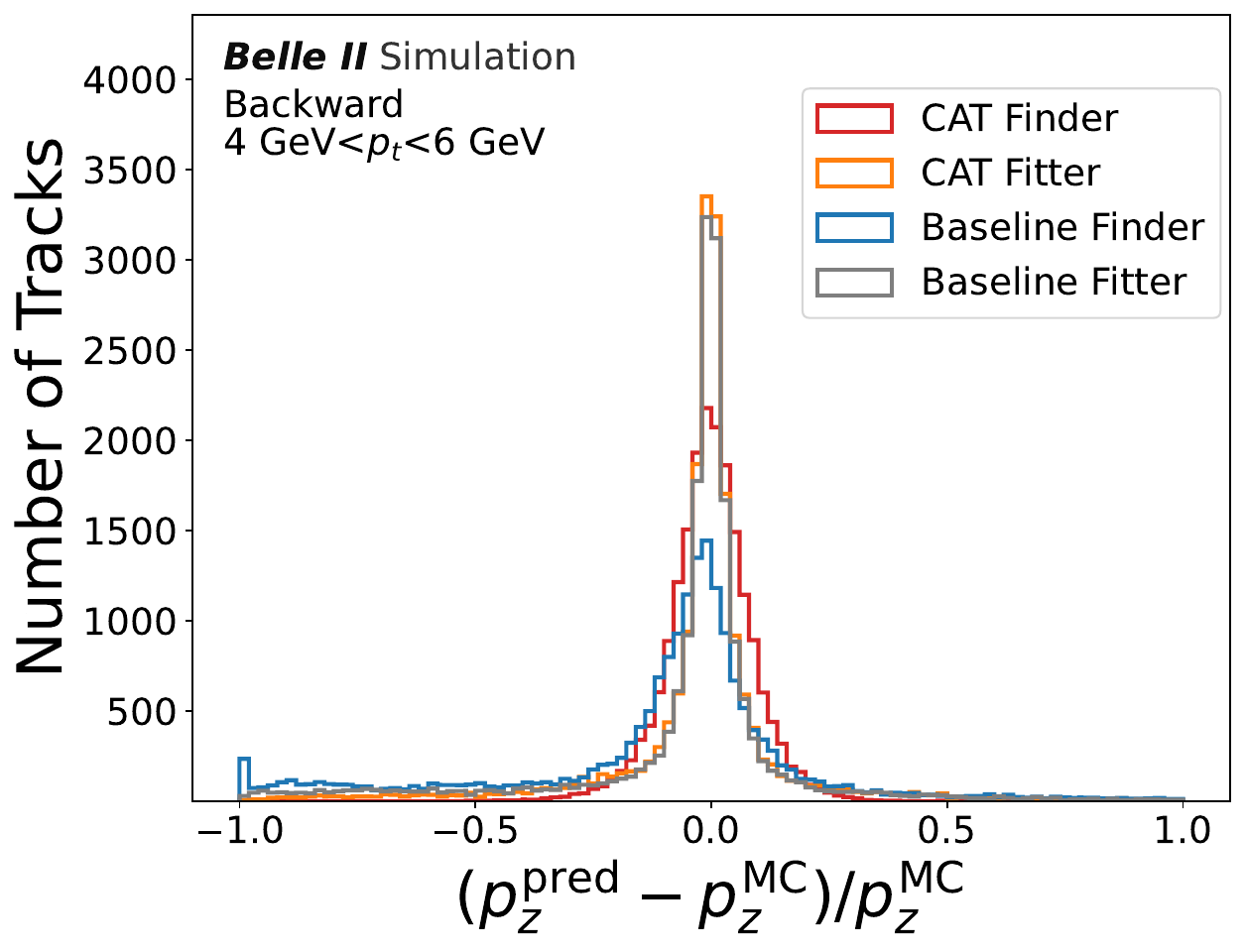}
         \caption{Backward endcap.}
         \label{fig:pz_res:c2}
     \end{subfigure}\\

\caption{Relative (\cref{fig:pt_res:a}-\cref{fig:pt_res:c2}) transverse and (\cref{fig:pz_res:a}-\cref{fig:pz_res:c2}) longitudinal momentum resolution as function of simulated transverse momentum $p_t^{MC}$ for the intersecting prompt evaluation sample (category 1-3, see \cref{tab:samples}) in the (left) forward endcap, (center) barrel, and (right) backward endcap for tracks found (red) and fitted (orange) by both the \cat and (blue and grey) the \legendre for the high transverse momentum bin of 4\,GeV$<p_t<$\,6 GeV.}
\label{fig:highptres}
\end{figure*}

\cref{fig:highptres} shows the relative transverse momentum resolution for tracks found and fitted by both the \cat and the \legendre in \cref{fig:pt_res:a}-\cref{fig:pt_res:c2}) and the relative  longitudinal momentum resolution in the transverse momentum bin of 4\gev$<p_t<$6\,\gev. 
This can be observed in \cref{fig:pt_res:c}, where there is no significant tail on the right side.
We note that even if the actual momentum was significantly higher than 6\,\gev, the initial prediction from the \cat would be sufficient as starting value for the subsequent track fitting algorithm. 
While the central part of the resolution distribution for the CAT Finder is broader compared to the other cases, the distribution has significantly smaller tails.

\clearpage
\FloatBarrier

\section{Track helix parametrization resolution}\label{app:helix}

The track parametrization follows a helix model, computed at the point of closest approach (POCA) to the collision point. 
We define the distance between the POCA and the collision point on the transverse (longitudinal) plane as $d_0$ ($d_z$), and the angle defined by the transverse (longitudinal) momentum at the POCA as $\phi$ ($\theta$).
We evaluate the absolute residuals for these track features $\phi$, $\theta$, $d_0$, and $d_z$ 
\begin{align}\label{eq:etap_trackparam}
     \eta(\phi,\theta,d_0,d_z) & = (\phi,\theta,d_0,d_z)_{\text{rec}}-(\phi,\theta,d_0,d_z)_{\text{simulated}}
\end{align}
for matched tracks.

We then define the resolution $r(\phi,\theta,d_0,d_z)$ for  the absolute residuals $\eta(\phi,\theta,d_0,d_z)$  as the 68\% coverage 
\begin{align}\label{eq:resolution_trackparam}
     r(p_{\phi,\theta,d_0,d_z}) = P_{68\%} \left(\left|\eta - P_{50\%}(\eta)\right|\right),
\end{align}
where $P_q$ is the $q$--th quantile of the distribution of $p_{t,z}$, and $P_{50\%}$ is the median of $\eta(p_{t,z})$~\cite{BelleIITrackingGroup:2020hpx}. 

\begin{figure*}[ht!]
     \centering
     \begin{subfigure}[b]{\thirdwidth\textwidth}
         \centering
         \includegraphics[width=\textwidth]{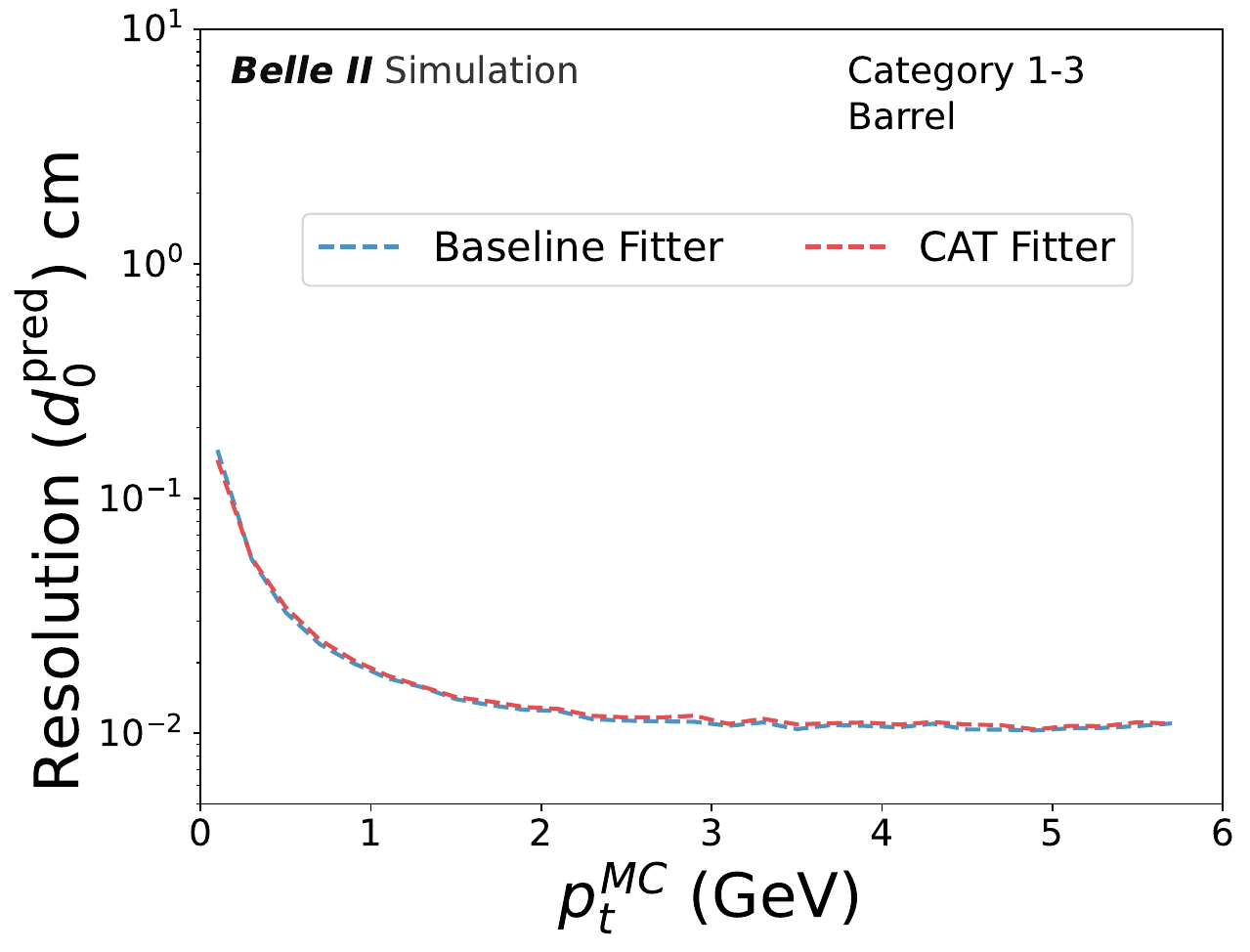}
         \caption{$d_0$ Resolution.}
         \label{fig:track_res:a}
     \end{subfigure}
     \quad
        \begin{subfigure}[b]{\thirdwidth\textwidth}
         \centering
         \includegraphics[width=\textwidth]{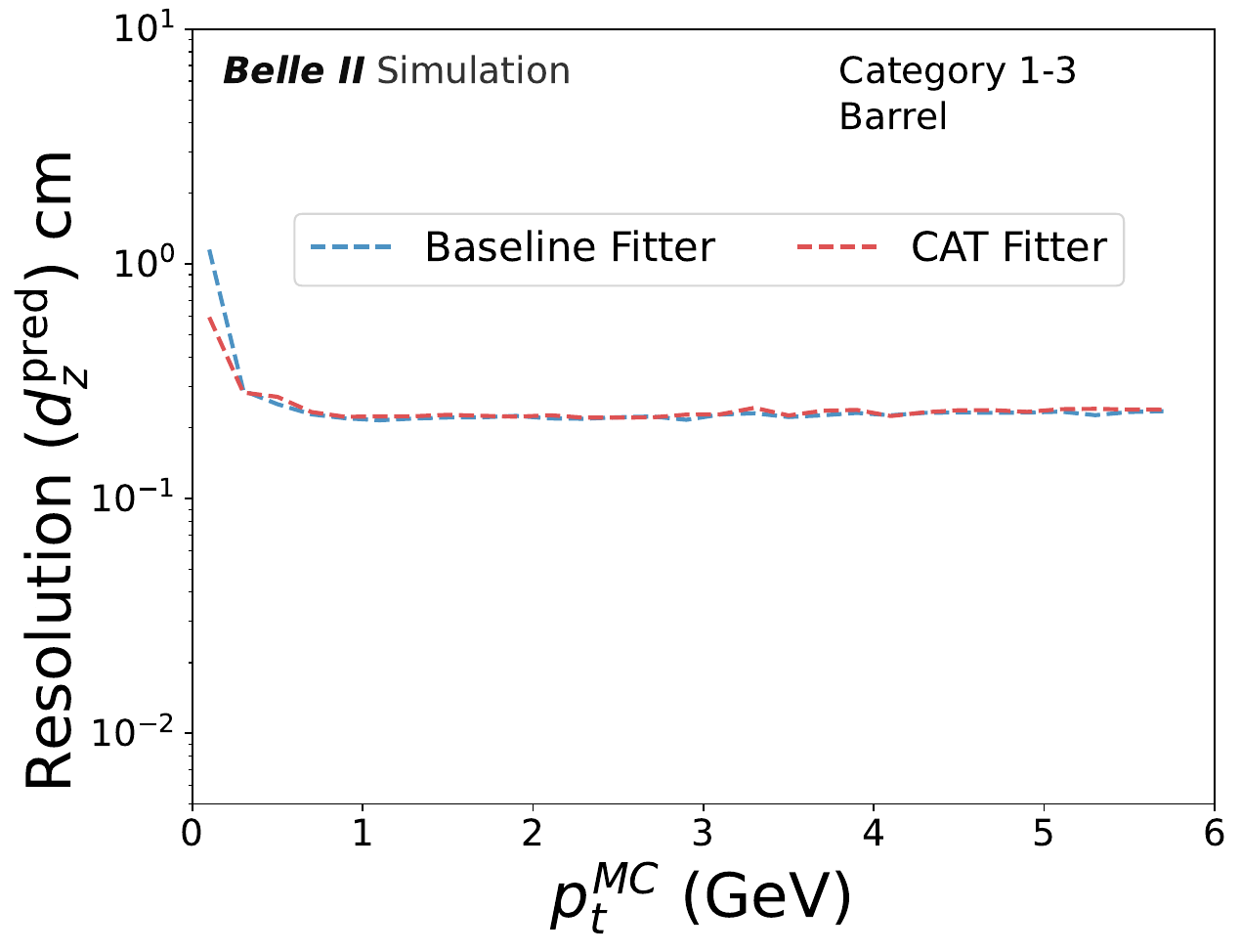}
         \caption{$d_z$ resolution.}
         \label{fig:track_res:b}
     \end{subfigure}
     \hfill\\
        \begin{subfigure}[b]{\thirdwidth\textwidth}
         \centering
         \includegraphics[width=\textwidth]{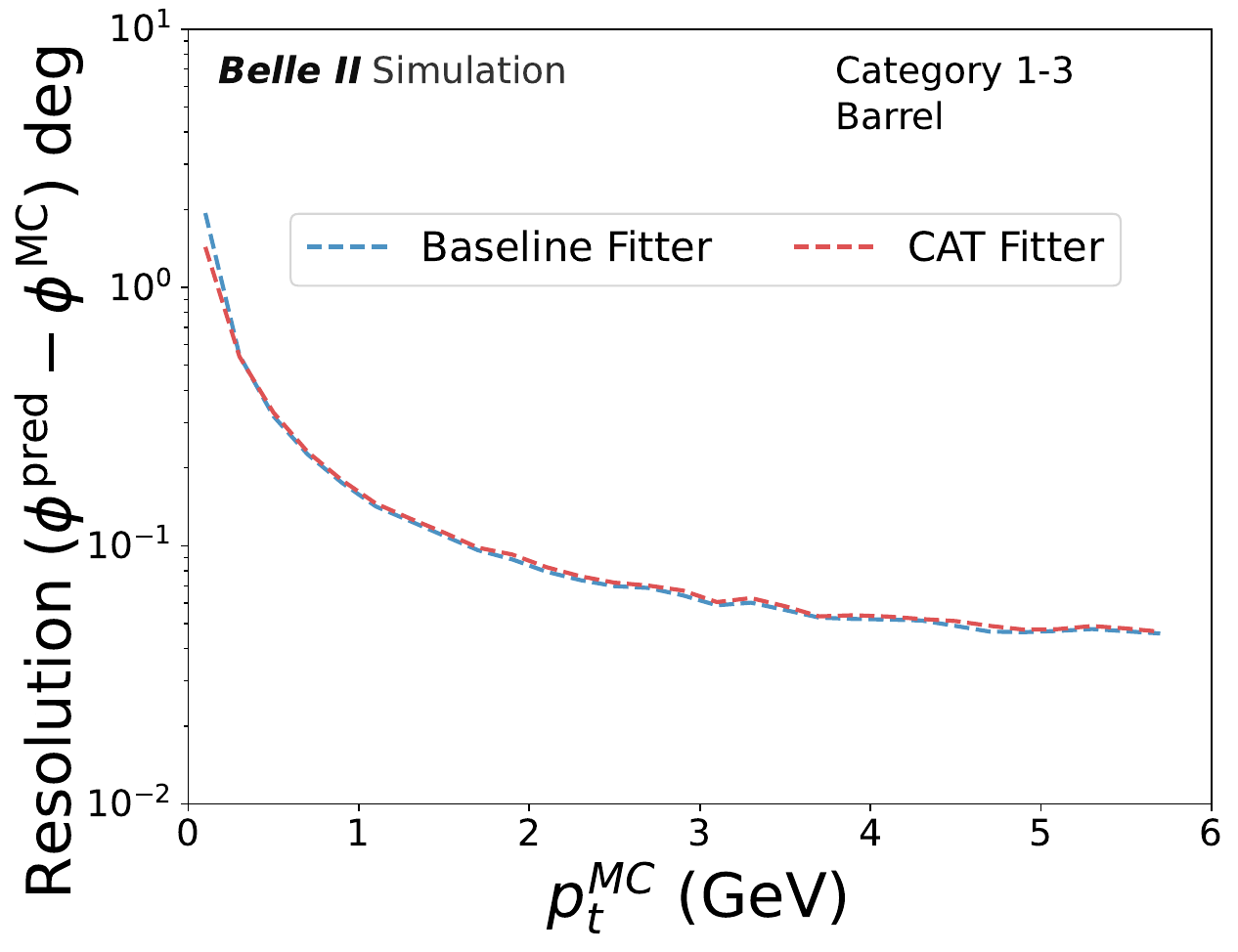}
         \caption{$\phi$ resolution.}
         \label{fig:track_res:c}
     \end{subfigure}
     \quad
      \centering
     \begin{subfigure}[b]{\thirdwidth\textwidth}
         \centering
         \includegraphics[width=\textwidth]{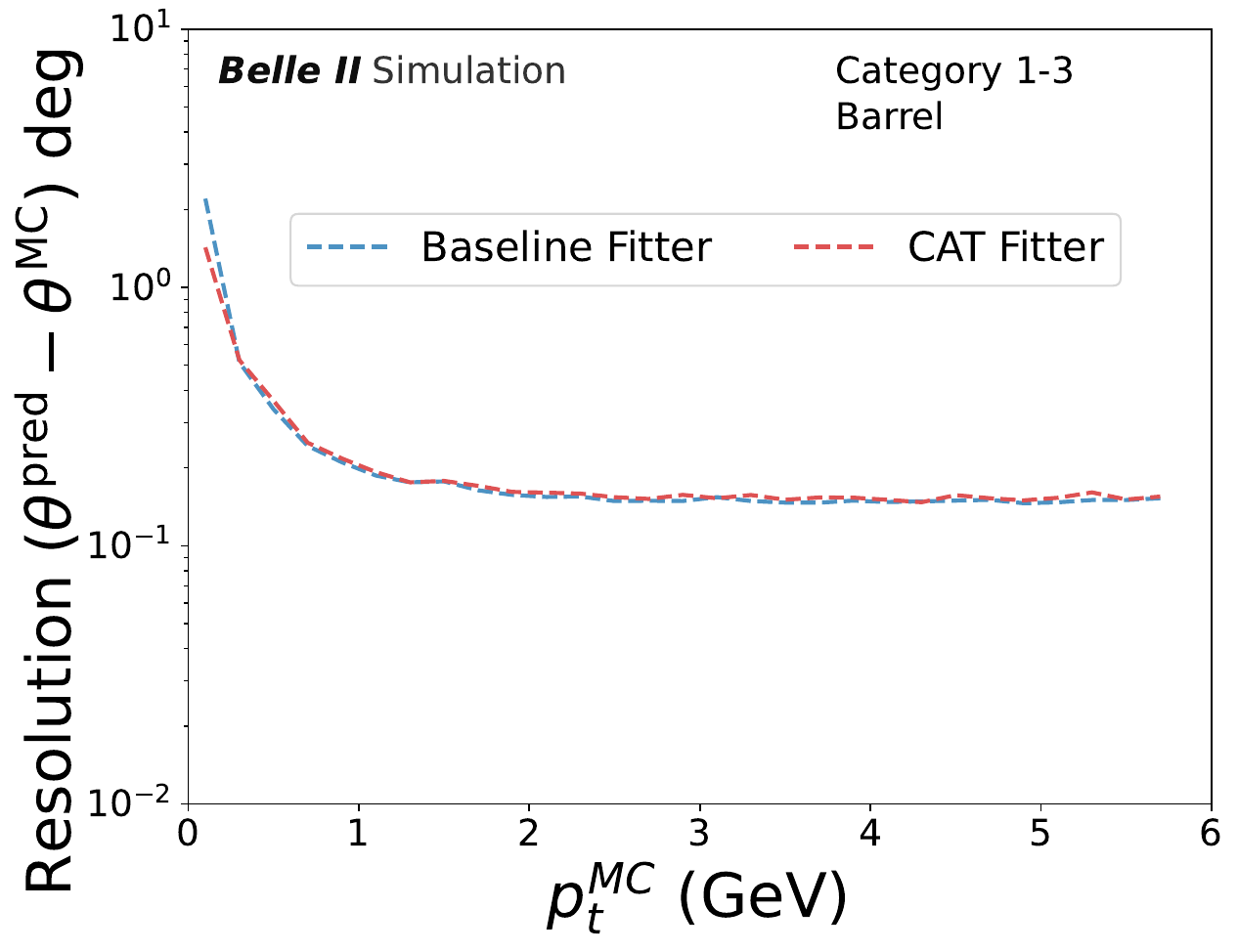}
         \caption{$\theta$ resolution.}
         \label{fig:track_res:d}
     \end{subfigure}\hfill

\caption{Fitting parameter resolution as function of simulated transverse momentum $p_t^{MC}$ for the intersecting prompt evaluation sample (category~1-3, see \cref{tab:samples}) barrel.}
\label{fig:testset_track_res}
\end{figure*}

The fitting parameter resolutions as function of simulated transverse momentum $p_t^{MC}$ are shown in \cref{fig:testset_track_res}.

\clearpage
\FloatBarrier

\section{Track momentum resolution for additional \cat samples in \dh events}\label{app:trackres_dh}
\begin{figure*}[ht!]
     
     \centering
     \begin{subfigure}[b]{\thirdwidth\textwidth}
         \centering
         \includegraphics[width=\textwidth]{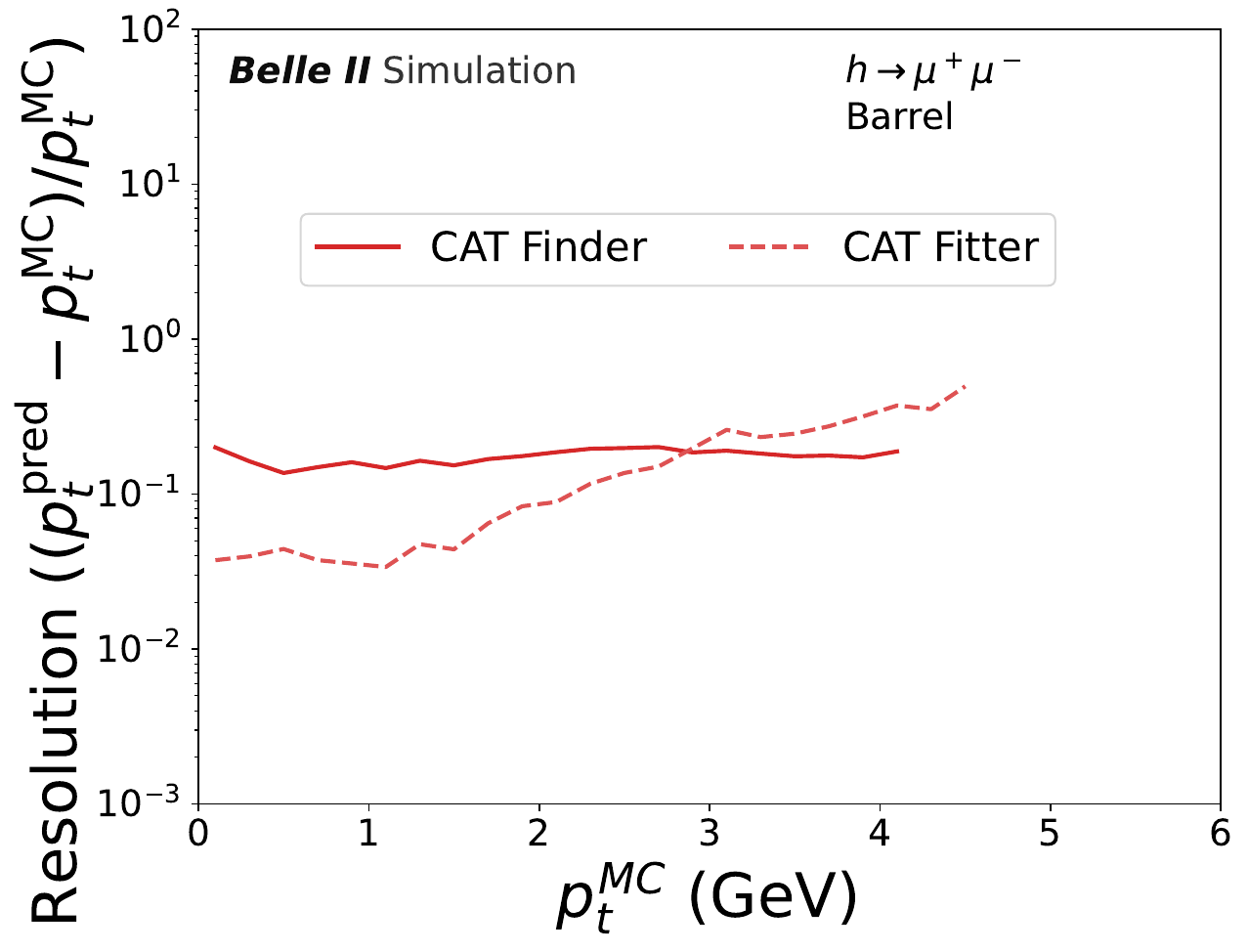}
         \caption{$\eta(p_t)$~Barrel.}
         \label{fig:res_dh_pt_add:a}
     \end{subfigure}\quad
        \begin{subfigure}[b]{\thirdwidth\textwidth}
         \centering
         \includegraphics[width=\textwidth]{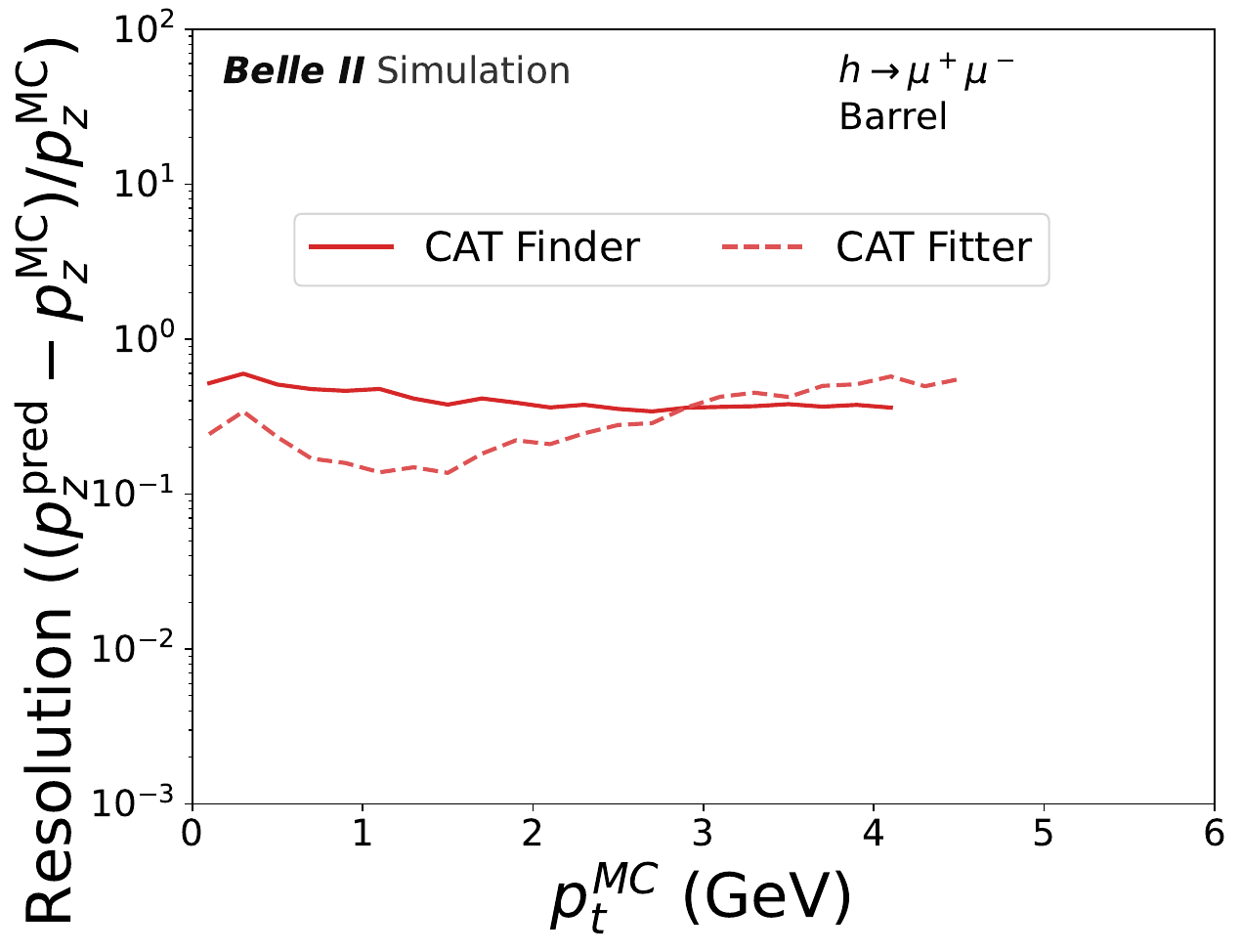}
         \caption{$\eta(p_z)$~Barrel.}
         \label{fig:res_dh_pz_add:b}
     \end{subfigure}\hfill\\
     \begin{subfigure}[b]{\thirdwidth\textwidth}
         \centering
         \includegraphics[width=\textwidth]{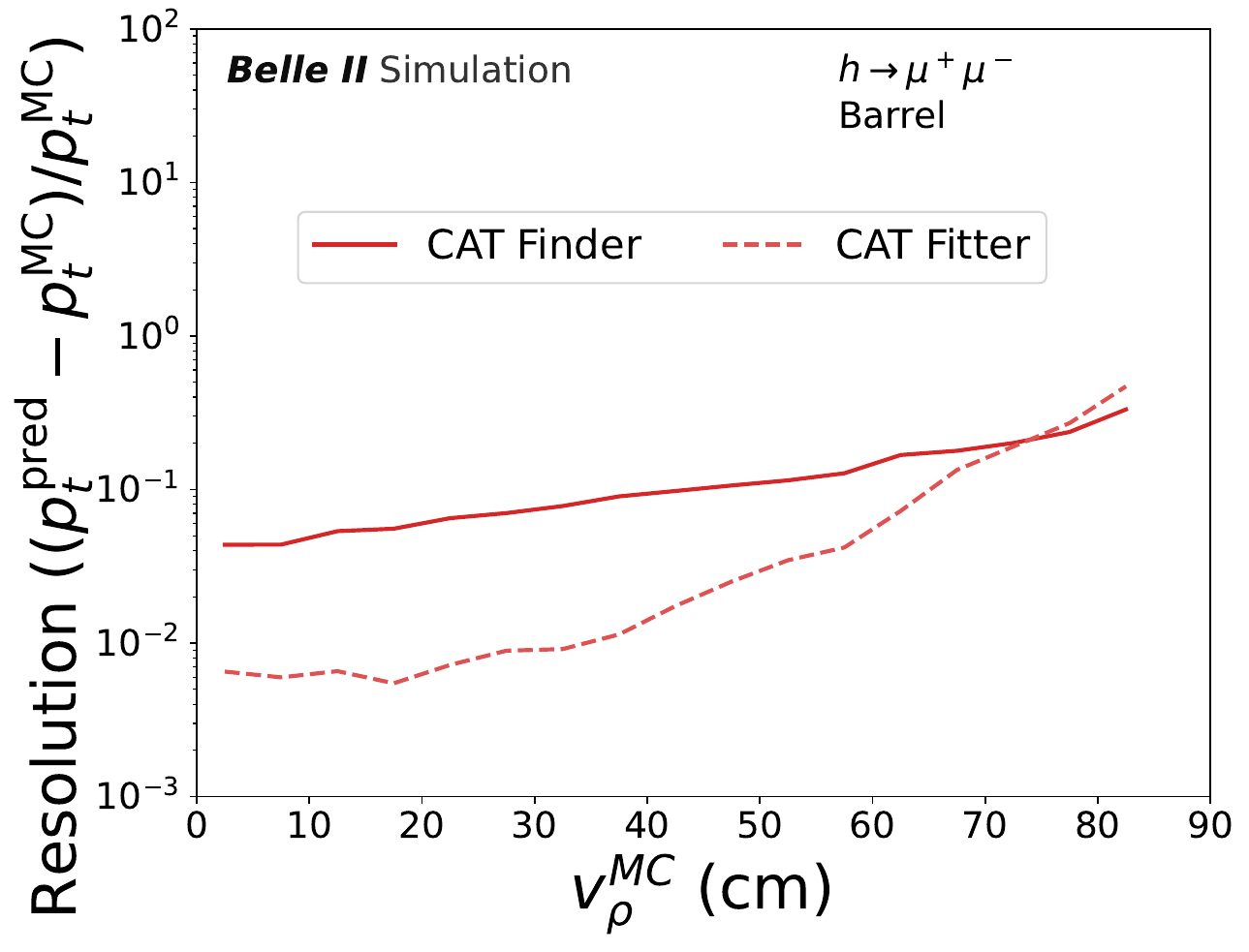}
         \caption{$\eta(p_t)$~Barrel.}
         \label{fig:res_dh_pt_add:c}
     \end{subfigure}\quad
        \begin{subfigure}[b]{\thirdwidth\textwidth}
         \centering
         \includegraphics[width=\textwidth]{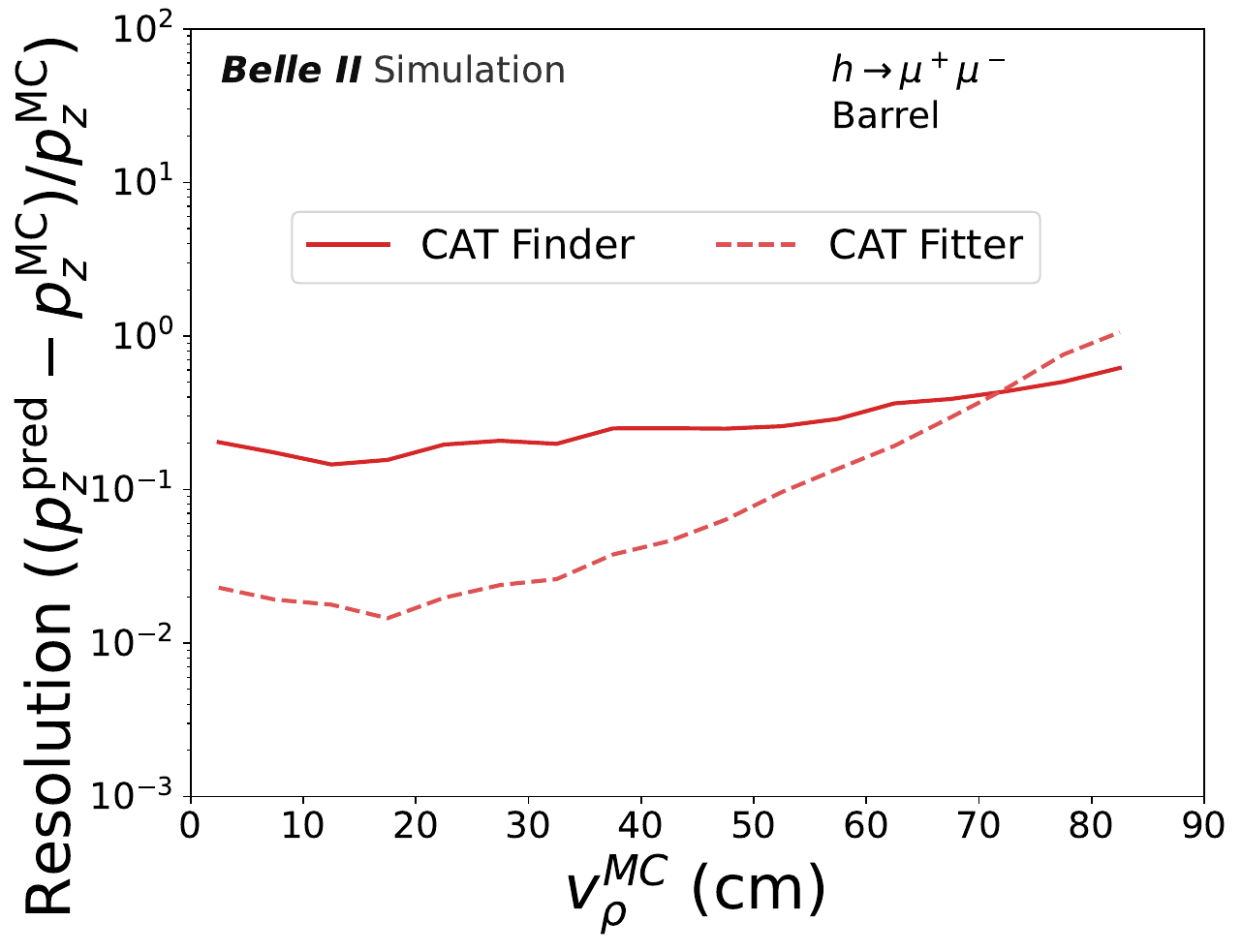}
         \caption{$\eta(p_z)$~Barrel.}
         \label{fig:res_dh_pz_add:d}
     \end{subfigure}\hfill
\caption{Relative resolution of (first column) transverse and (second column) longitudinal momentum as function of (top row) simulated transverse momentum $p_t^{MC}$ and (bottom row) simulated displacement $v_{\rho}^{MC}$ for displaced tracks from \dh decays in the barrel for tracks only found by \cat.}
\label{fig:dh_resolution_pt_add}
\end{figure*}

The relative momentum resolutions for displaced tracks from \dh decays in the barrel for tracks only found by \cat are shown in \cref{fig:dh_resolution_pt_add}.

\clearpage
\FloatBarrier

\section{Track momentum resolution in \kshort events}\label{app:trackres_ks}
\begin{figure*}[ht!]
\centering
     \begin{subfigure}[b]{\thirdwidth\textwidth}
         \centering
         \includegraphics[width=\textwidth]{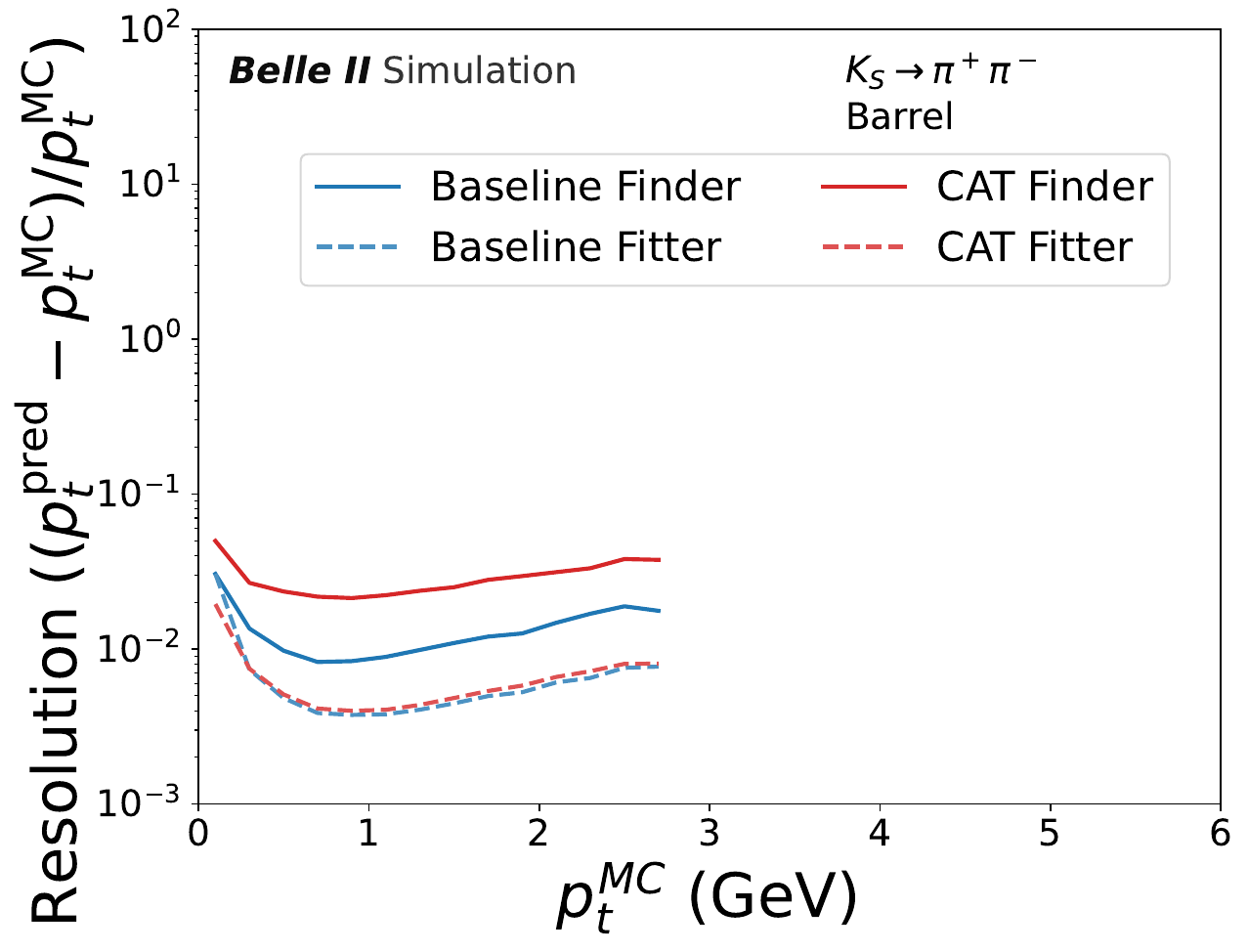}
         \caption{$\eta(p_t)$~Barrel.}
         \label{fig:res_ks_pt:a}
     \end{subfigure}\quad
        \begin{subfigure}[b]{\thirdwidth\textwidth}
         \centering
         \includegraphics[width=\textwidth]{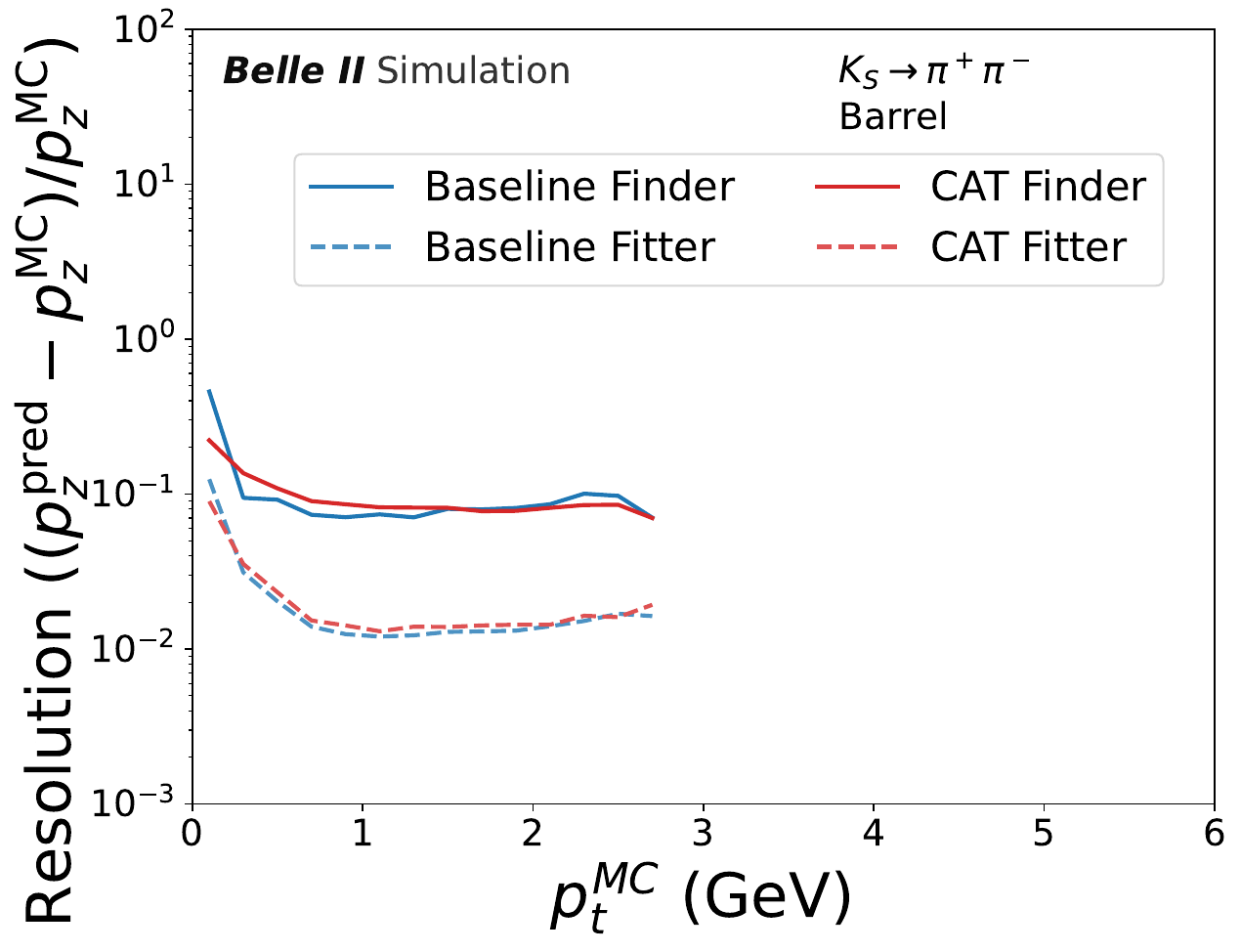}
         \caption{$\eta(p_z)$~Barrel.}
         \label{fig:res_ks_pt:b}
     \end{subfigure}\hfill\\
     \begin{subfigure}[b]{\thirdwidth\textwidth}
         \centering
         \includegraphics[width=\textwidth]{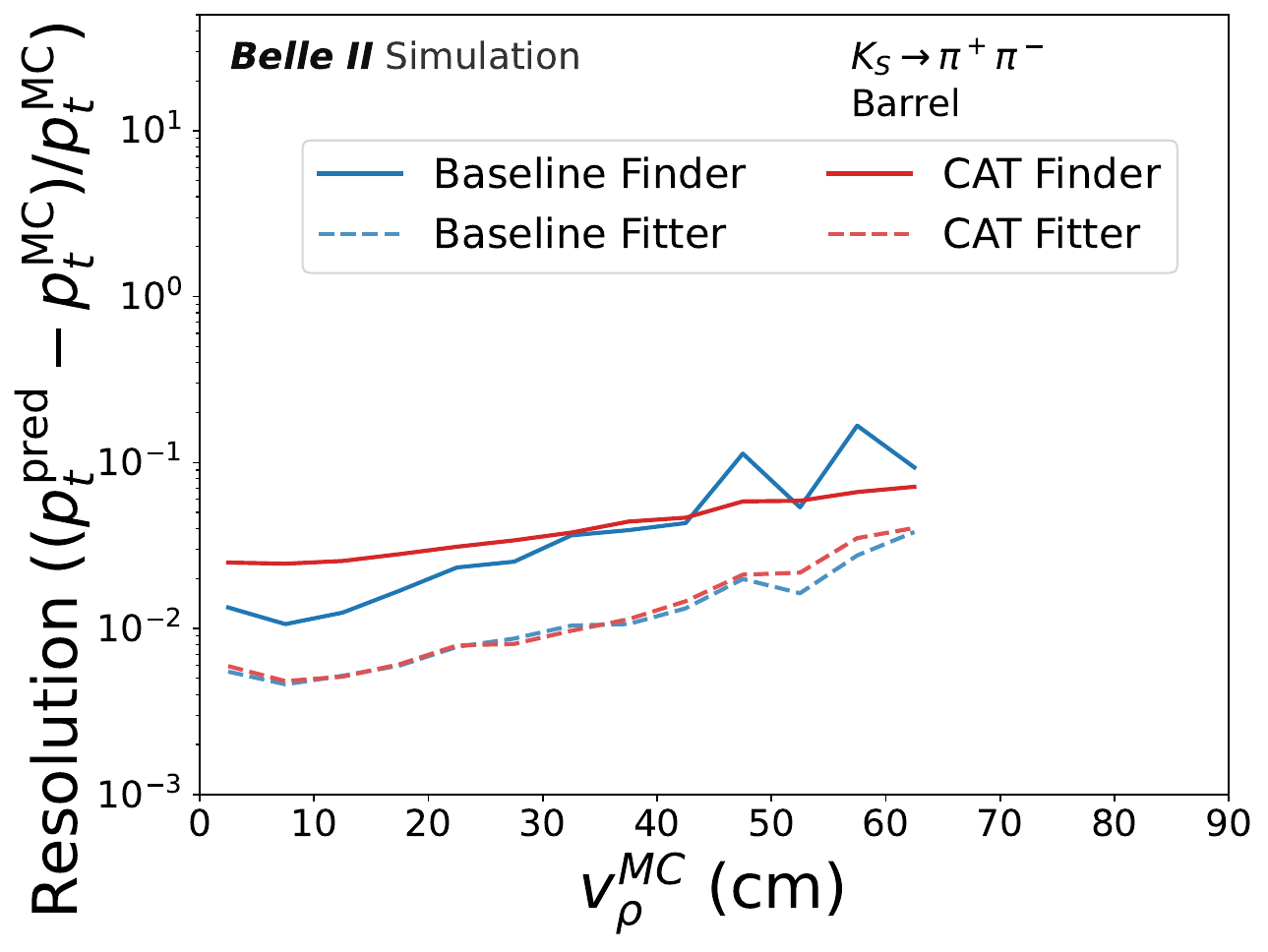}
         \caption{$\eta(p_t)$~Barrel.}
         \label{fig:res_ks_vrho:c}
     \end{subfigure}\quad
        \begin{subfigure}[b]{\thirdwidth\textwidth}
         \centering
         \includegraphics[width=\textwidth]{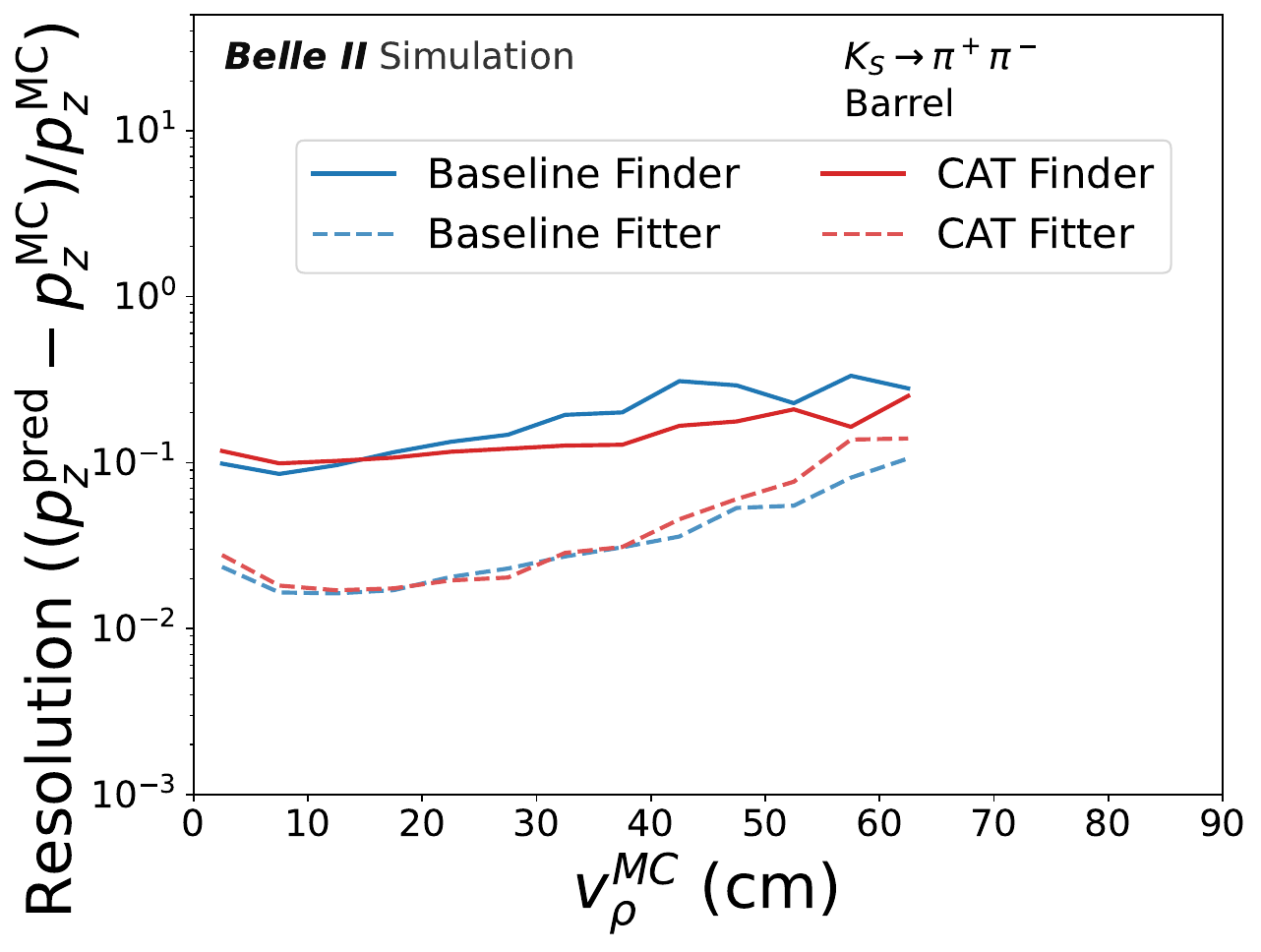}
         \caption{$\eta(p_z)$~Barrel.}
         \label{fig:res_ks_vrho:d}
     \end{subfigure}\hfill\\
      \centering
     \begin{subfigure}[b]{\thirdwidth\textwidth}
         \centering
         \includegraphics[width=\textwidth]{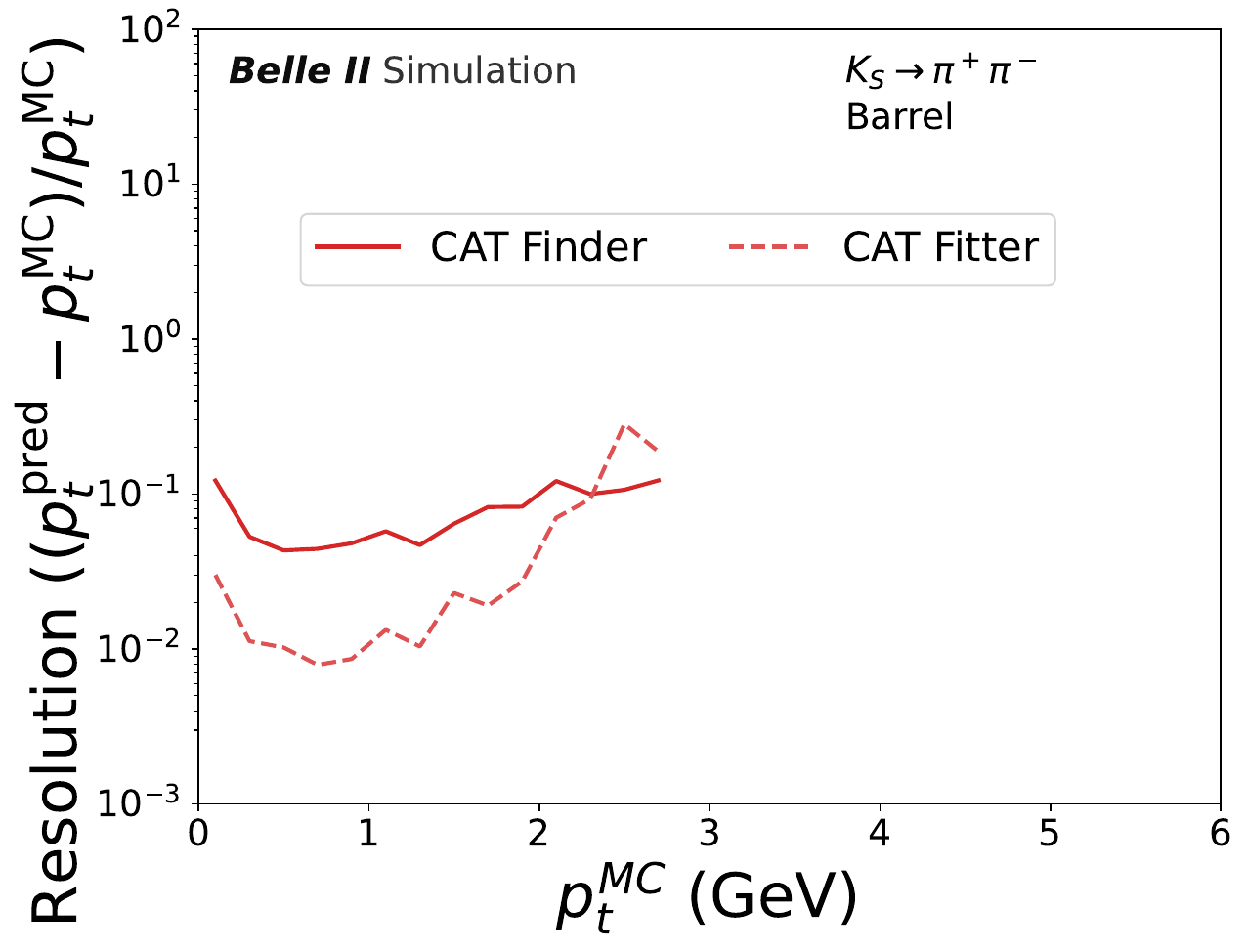}
         \caption{$\eta(p_t)$~Barrel.}
         \label{fig:res_ks_pt:a2}
     \end{subfigure}\quad
        \begin{subfigure}[b]{\thirdwidth\textwidth}
         \centering
         \includegraphics[width=\textwidth]{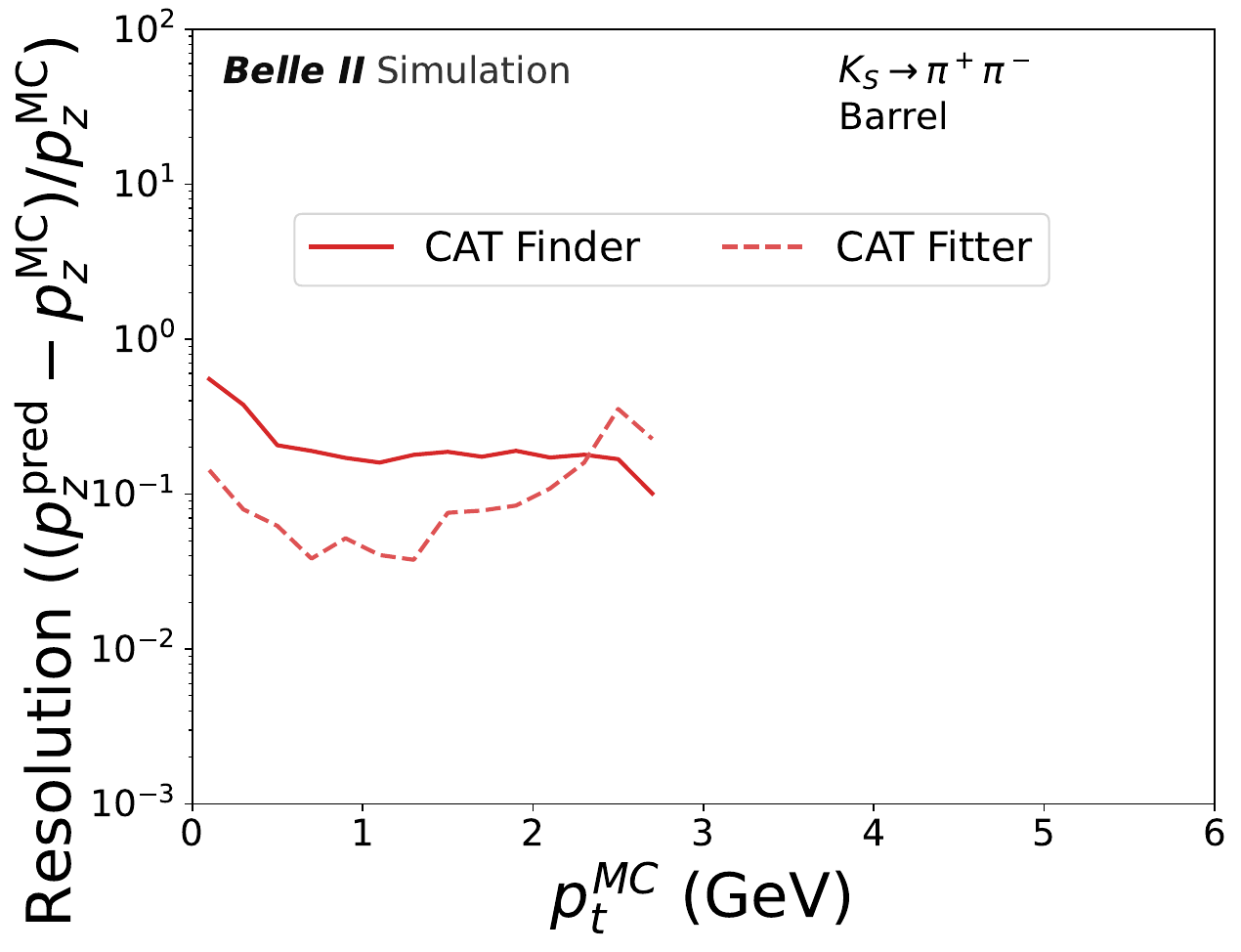}
         \caption{$\eta(p_z)$~Barrel.}
         \label{fig:res_ks_pt:b2}
     \end{subfigure}\hfill\\
     \begin{subfigure}[b]{\thirdwidth\textwidth}
         \centering
         \includegraphics[width=\textwidth]{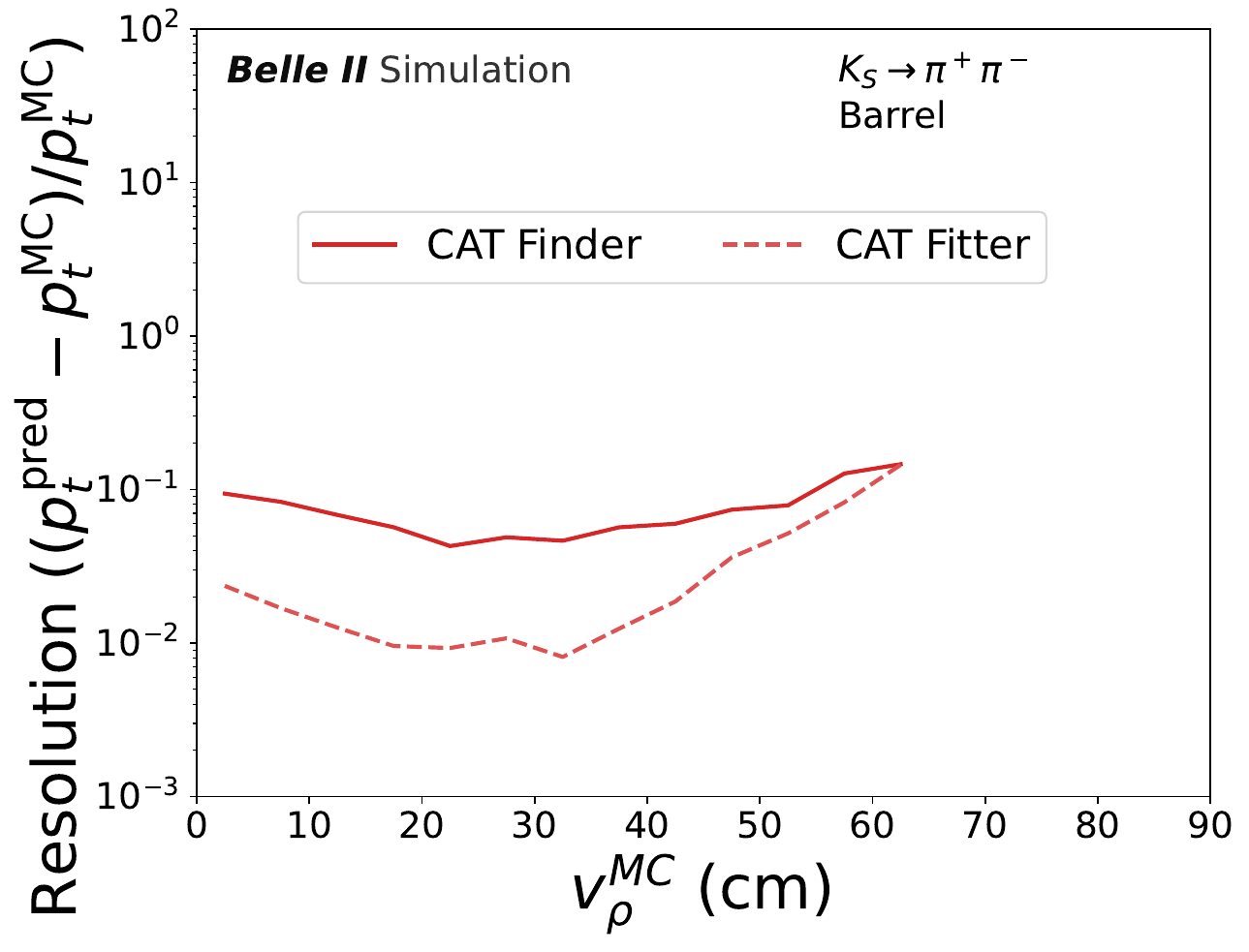}
         \caption{$\eta(p_t)$~Barrel.}
         \label{fig:res_ks_vrho:c2}
     \end{subfigure}\quad
        \begin{subfigure}[b]{\thirdwidth\textwidth}
         \centering
         \includegraphics[width=\textwidth]{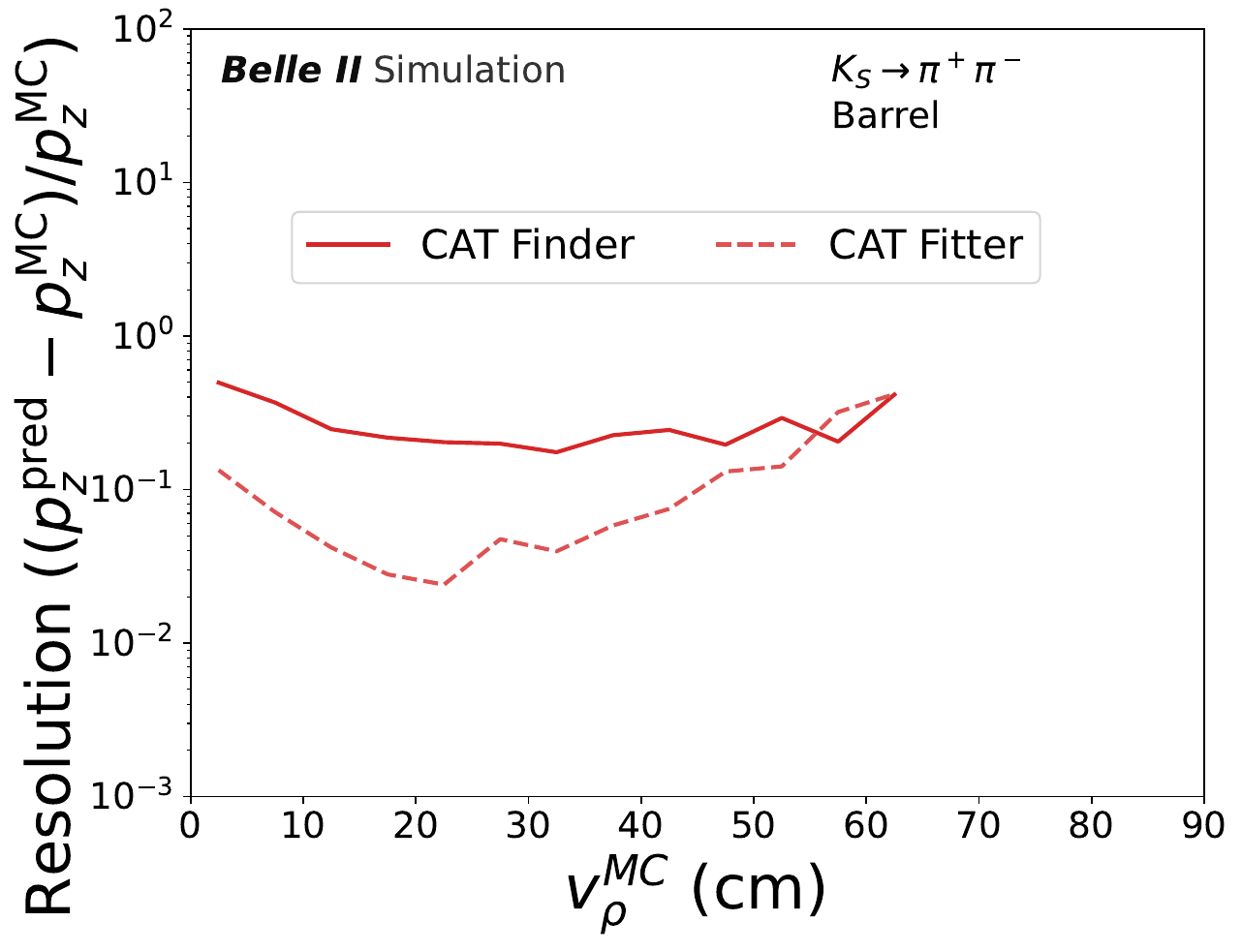}
         \caption{$\eta(p_z)$~Barrel.}
         \label{fig:res_ks_vrho:d2}
     \end{subfigure}\hfill
        
\caption{Relative resolution of (first column) transverse and (second column) longitudinal momentum as function of (top row) simulated transverse momentum $p_t^{MC}$ and (bottom row) simulated displacement $v_{\rho}^{MC}$ for displaced tracks from \kshort decays. Top (\subref{fig:res_ks_pt:a}-\subref{fig:res_ks_vrho:d}) row shows the resolution for tracks found by both \cat (red) and \legendre (blue), and bottom (\subref{fig:res_ks_pt:a2}-\subref{fig:res_ks_vrho:d2}) row for tracks only found by \cat. }
\label{fig:ks_resolution_pt_ep3}
\end{figure*}

The relative momentum resolutions for displaced tracks from \kshort decays in the barrel are shown in \cref{fig:ks_resolution_pt_ep3}.

\end{appendices}

\end{document}